\documentclass[%
aip,
amsmath,amssymb,
preprint,
groupedaddress,
]{revtex4-1}

\usepackage{graphicx}
\usepackage{dcolumn}
\usepackage{bm}

\usepackage[utf8]{inputenc}
\usepackage[T1]{fontenc}
\usepackage{mathptmx}
\usepackage{etoolbox}
\usepackage{xcolor}
\usepackage{subfigure}

\makeatletter
\def\@email#1#2{%
	\endgroup
	\patchcmd{\titleblock@produce}
	{\frontmatter@RRAPformat}
	{\frontmatter@RRAPformat{\produce@RRAP{*#1\href{mailto:#2}{#2}}}\frontmatter@RRAPformat}
	{}{}
}%
\makeatother

\begin{document}
	
	\preprint{AIP/123-QED}
	
	\title[Physics of Fluids]{Physics-informed Fourier Basis Neural Network for Fluid Mechanics}
	\author{Chao Wang}
	
	\author{Shilong Li}%
	
	\author{Zelong Yuan\textsuperscript{*}}
	\email{yuanzelong@hrbeu.edu.cn}
	
	\author{Chunyu Guo}
	
	\affiliation{%
		\textsuperscript{1}College of Shipbuilding Engineering, Harbin Engineering University, Harbin, 150001, P.R. China
	}%
	\affiliation{\textsuperscript{2}Qingdao Innovation and Development Base, Harbin Engineering University, Qingdao, 266000, P.R. China
	}%
	\affiliation{\textsuperscript{3}Nanhai Institute of Harbin Engineering University, Harbin Engineering University, Sanya, 572024, P.R. China
	}%
	\affiliation{\textsuperscript{4}National Key Laboratory of Hydrodynamics, Harbin Engineering University, Harbin, 150001, P.R. China
	}%

	\date{\today}

	\begin{abstract}
		Solving partial differential equations (PDEs) is an important yet challenging task in fluid mechanics. In this study, we embed an improved Fourier series into neural networks and propose a physics-informed Fourier basis neural network (FBNN) by incorporating physical information to solve canonical PDEs in fluid mechanics. The results demonstrated that the proposed framework exhibits a strong nonlinear fitting capability and exceptional periodic modeling performance. In particular, our model shows significant advantages for the Burgers equation with discontinuous solutions and Helmholtz equation with strong periodicity. By directly introducing sparse distributed data to reconstruct the entire flow field, we further intuitively validated the direct superiority of FBNN over conventional artificial neural networks (ANN) as well as the benefits of incorporating physical information into the network. By adjusting the activation functions of networks and comparing with an ANN and conventional physics-informed neural network, we proved that performance of the proposed FBNN architecture is not highly sensitive to the choice of activation functions. The nonlinear fitting capability of FBNN avoids excessive reliance on activation functions, thereby mitigating the risk of suboptimal outcomes or training failures stemming from unsuitable activation function choices.hese results highlightthe potential of PIFBNN as a powerful tool in computational fluid dynamics.
	\end{abstract}
	
	

	\keywords{} 
	
	\maketitle
	
\section{\label{sec:level1}Introduction}

Computational fluid dynamics (CFD) has been extensively used in aerospace engineering and naval and ocean engineering\cite{zhang2006application,phillips2010use,zhang2017aeroacoustic,kou2021data,mani2023perspective,xia2025skin} owing to its powerful predictive capabilities and high accuracy.CFD primarily solves the governing partial differential equations of fluid flow either directly or through modeling. The key numerical methods employed include the finite difference method (FDM), finite volume method (FVM), and finite element method (FEM)\cite{jeong2014comparison}. However, owing to chaotic features and complex inter-scale interactions of vortices, direct numerical simulations require prohibitively large computational resources, rendering some turbulence problems unsolvable with current capabailities\cite{moin1998direct,scardovelli1999direct,zhu2025influences}. 

In recent years, the surge in computing power has enabled the application of machine learning methods to fluid dynamics\cite{duraisamy2019turbulence,brunton2020machine,lei2025generating}, including the establishment of turbulence models and solving governing PDEs\cite{huang2025partial}. C Wu et al.\cite{wu2023enhancing} utilized symbolic regression to derive analytical mapping between the corrected Reynolds stress discrepancy and local flow variables, and enhanced the predictive capabilities of the shear-stress-transport model for separated flows across various test cases while addressing the interpretability and generalization limitations of conventional machine learning models. Ling et al.\cite{ling2016reynolds} developed a Galilean invariant tensor basis neural network that predicts the anisotropic component of Reynolds stress.
Gamahara et al.\cite{gamahara2017searching} applied artificial neural networks (ANN) to establish a functional mapping between grid-scale flow field and subgrid-scale (SGS) stress as a tool for searching a new subgrid model for SGS stress in LES.
Xu et al.\cite{xu2021deep} proposed a hybrid machine-learning framework that integrated neural networks, genetic algorithms, and stepwise methods for discovering PDEs from sparse and noisy measurement data.
Yang et al.\cite{yang2024implicit}introduced a modified implicit factorized transformer model (IFactFormer), employing parallel factorized attention for effective surrogate modeling of nonlinear dynamic systems governed by PDEs, achieving superior long-term predictions for turbulent channel flows.
Fan et al. \cite{fan2025neural} proposed an innovative neural differentiable modeling framework that significantly improved the predictability and computational efficiency of spatiotemporal turbulence simulations by combining deep neural networks with numerical PDE solvers.
W Suo et al.\cite{suo2025novel} proposed a novel multiscale neural computing paradigm for solving PDEs by combining neural networks to efficiently capture global features with finite difference methods for localized details. This helped achieve significant improvements in accuracy and efficiency while ensuring stable convergence and rigorous boundary condition satisfaction.
Most neural network models are expert at learning mappings between finite-dimensional Euclidean spaces, and their generalization is limited by the different discretizations, initial conditions, or boundary conditions of differential equations\cite{wu2020data}. Li et al.\cite{li2020fourier} proposed a Fourier neural operator (FNO) framework that can directly learn mapping functions between infinite dimensional spaces from a series of input--output pairs, parameterized the integral kernel directly in Fourier space, enabling efficient learning of mappings between function spaces for PDEs. FNO has demonstrated strong capability to model turbulent flows with zero-shot super-resolution, achieving improved performance and accuracy compared to conventional learning-based solvers. Subsequently, more neural operators based on FNO for fluid mechanics have been successively developed.\cite{peng2022attention,peng2023linear,wen2022u,you2022learning,li2023long,li2024transformer} M Lei et al.\cite{lei2025generating}established the first end-to-end, bidirectional mapping between natural language descriptions and parametric airfoil geometries using a CLIP-inspired architecture and semantically conditioned decoder. This approach significantly enhanced the accessibility of airfoil design through natural language interfaces, highlighting the potential for human--AI collaboration in aerospace engineering.

However, conventional machine learning models heavily rely on high-fidelity dataset for training, limiting their applicability to PDEs with no analytical solutions, such as the Navier--Stokes equations. Conventional neural network learning frameworks often lack the integration of physical information\cite{agatonovic2000basic,lu2021learning},which consequently limits their ability to effectively leverage prior physical knowledge when addressing specialized physical problems.\cite{de2018greedy}. 

To address this issue, Raissi et al.\cite{raissi2019physics} introduced a Physics-Informed Neural Network (PINN) by embedding PDEs into the network loss function. PINN can reduce reliance on labeled data, and incorporate constraints from physical conservation laws. This sparked a widespread adoption of PINN to embed physical information into neural networks for more accurate physical predictions\cite{blechschmidt2021three}.
W. Chen et al.\cite{chen2021physics} developed a reduced basis method using a physics-informed machine learning framework to efficiently model parameterized PDEs,achieving significantly faster online evaluations.
L Yang et al.\cite{yang2021b} proposed a Bayesian PINN framework to effectively solve both forward and inverse nonlinear problems governed by PDEs and noisy data, demonstrating their prediction accuracy to be better than conventional PINNs, particularly in high-noise scenarios.
Zhang et al.\cite{zhang2025mrf} proposed a multiple receptive field convolutional physics-informed neural network (MRF-PINN) that improved the accuracy and convergence of PDEs by incorporating high-order finite-difference methods and a dimensional balance approach.  
Zhao et al.\cite{zhao2025lesnets} developed physics-informed neural operators, specifically LESnets, by integrating LES equations into data-driven models to efficiently simulate three-dimensional incompressible turbulent flows, demonstrating comparable accuracy and faster performance compared to conventional approaches without the need for labeled datasets.
J Song et al.\cite{song2025fenn} proposed a feature-enhanced neural network for PINNs that incorporated geometric and physical features into the input, leading to improved accuracy and efficiency in solving strongly nonlinear PDEs in fluid dynamics. Moreover, J Song et al.\cite{song2025vw} addressed the convergence challenges faced by PINNs with nonuniform collocation points by introducing volume weighting PINNs, and obtained faster and more efficient convergence for both forward and inverse problems. 
At present, innovative improvements to neural networks can be broadly categorized into three aspects. The first involves advancements in training network methods, exploring network frameworks, and selecting hyperparameters.\cite{schwenk1997training,srinivasulu2006comparative,dai2019adversarial,li2019improved,donoghue2015framework,smithson2016neural} For instance, Zhang W et al.\cite{zhang2024physics} analyzed the limitations of PINNs in solving PDEs. They identified three main issues: (1) poor multiscale approximation ability and ill-conditioning; (2) insufficient convergence and error analysis; and (3) inadequate integration of physical information. 
W Cao et al.\cite{cao2025analysis}established a strong connection between the ill-conditioning of PINNs and the Jacobian matrix of the PDE system, which led to faster convergence and higher accuracy in PINN.
The second aspect is the practical application of neural network architectures, such as the use of physical information networks to solve specific PDEs by encoding the appropriate physical models.
Jin X et al.\cite{jin2021nsfnets} developed Navier--Stokes flow networks (NSFnets) using PINNs to effectively simulate incompressible laminar and turbulent flows by directly integrating governing equations into neural networks, thereby overcoming challenges related to data requirements and computational efficiency.
Wang et al.\cite{wang2023long} incorporated physical information into a self-supervised deep-learning framework (DeepONet) and applied it to learn evolutionary operators for PDE, achieving accurate long-term predictions through iterative time-domain decomposition without requiring paired training data. The method demonstrated robustness across various systems, including wave propagation, reaction--diffusion, and stiff chemical kinetics.
The third aspect involves innovation in the network layers or in a specific computing module within the neural network. 
Rao et al.\cite{rao2023encoding} proposed a deep learning framework that encoded a given physical structure in a recurrent convolutional neural network to effectively model complex spatiotemporal dynamical systems, demonstrating its high accuracy and robustness through extensive numerical experiments. They proposed the use of convolution to solve differentiation, which innovatively replaced automatic differentiation in PINN and could specify the accuracy of the differentiation solution.
Li et al.\cite{li2024transformer} introduced a transformer-based neural operator that successfully predicted the large-scale dynamics of three-dimensional turbulence, demonstrating an accuracy comparable to that of conventional LES models while outperforming previous neural operators and achieving faster predictions with fewer parameters.
Wang et al.\cite{wang2025gradient} proposed theoretical and practical approaches to address gradient conflicts in multitask learning. They demonstrated the effectiveness of their methods in PINNs through second-order optimization, achieving state-of-the-art results on challenging PDE problems. Additionally, they introduced a novel gradient alignment metric to analyze optimization dynamics.
Xiao Y et al.\cite{xiao2024least} proposed a least-squares finite-difference-based PINN to simulate steady incompressible flows. Their approach efficiently computes derivatives using the least-squares finite difference method, without requiring matrix operations or additional virtual collocation points. 
Liu et al.\cite{liu2024kan} introduced Kolmogorov--Arnold networks (KANs) as innovative alternatives to conventional MLPs by replacing fixed nodal activations with learnable spline-based edge-activation functions. The proposed architecture demonstrated superior accuracy with smaller network sizes for both data fitting and PDE solving, while offering enhanced interpretability through intuitive visualization capabilities. C Guo et al.\cite{guo2024physics} replaced the basis functions of KAN with orthogonal Chebyshev polynomials, achieving better fitting ability, and proposed ChebPIKAN by incorporating physical information, achieving excellent results in solving multiple PDE equations.

Conventional ANN, face difficulties in fitting periodic and strongly nonlinear functions, although periodicity is a fundamental characteristic of many problems in fluid dynamics. Therefore, the existence of mathematical structures within network layers that can accurately capture periodicity and fit nonlinearity is essential for the accurate prediction of periodic PDEs.
In this study, the proposed physics-informed Fourier basis neural network (PIFBNN) utilizes Fourier analysis methods and innovatively applies an improved Fourier series to the network layers by introducing Fourier nodes alongside standard fully connected nodes. This not only ensures universal approximation capability of the neural network, but also enhances its ability to learn periodic features, overcoming limitations commonly observed in conventional neural networks. Additionally, the predictive performance of the model can be improved by incorporating physical information.

The remainder of this paper is organized as follows. Section \ref{sec:level1/2} introduces the basic concepts of the Fourier series, followed by a discussion on how to embed the improved Fourier series into the meural network to establish the Fourier basis neural network (FBNN) framework. The section also outlines the basic concepts of PINN and presents the final PIFBNN model. Section \ref{sec:level1/3} compares the predictive performance of PIFBNN with PINN on five fundamental PDEs from fluid mechanics and other physical domains without using label data to directly predict flow fields. Section \ref{sec:level1/4} evaluates the generalization performance of the proposed FBNN framework by analyzing the influence of different activation functions on flow field reconstruction. Section \ref{sec:level1/5} highlights the distinct differences between the FBNN and conventional ANN frameworks through sparse data reconstruction and emphasizes the role of  physical information in enhancing the model accuracy. Finally, Section \ref{sec:level1/6} gives the conclusion.
\bibliographystyle{unsrt}

\section{\label{sec:level1/2}Methodology}
This section describesthe preliminary knowledge of Fourier series, the development of Fourier basis neural network and physics-informed Fourier basis neural network, respectively.
\subsection{\label{sec:level2}Preliminary knowledge}
The Fourier series, a fundamental concept in Fourier analysis, states that periodic functions satisfying Dirichlet's conditions can be represented as an infinite weighted sum of sine and cosine functions. It can be expanded using trigonometric functions as 
\begin{equation}
	f ( x ) = A _ { 0 } + \sum _ { n = 1 } ^ { \infty } \left( A _ { n } \cos \left( \frac { 2 \pi n x } { L } \right) + B _ { n } \sin \left( \frac { 2 \pi n x } { L } \right) \right),
	\label{eq:fourier_series}
\end{equation}
where $L$ is the period of the function. $A_0$ , $ A_n$ and $B_n$ are Fourier coefficients, which can be obtained by integrating as 
\begin{equation}
	A_n = \frac{1}{L} \int_0^L f(x) \cos\left( \frac{2\pi n x}{L} \right) dx, \quad
	B_n = \frac{1}{L} \int_0^L f(x) \sin\left( \frac{2\pi n x}{L} \right) dx.
	\label{eq:fourier_final}
\end{equation}
Through periodic extension, the Fourier series can also be used to represent nonperiodic functions\cite{gallant1984fourier}. By expanding a function through Fourier series, the domain is effectively mapped to frequency domain, enabling a more direct grasp of the frequency characteristics of the function. Furthermore most neural network-based methods only focus on reconstructing nonlinear mappings of flow fields in the physical domain, the FNO framework can learn mappings of high-dimensional data in the frequency domain. In this architecture, nonlinear operators can approximate by learning the relationship between Fourier coefficients. Inspired by this idea, we integrated the finite-term function approximation property of Fourier series expansion into the network architecture to improve its capacity of capturing frequency characteristics and modeling periodic patterns. Consequently, we propose a Fourier-Based Neural Network (FBNN).
\subsection{\label{sec:level2}Fourier basis neural network(FBNN)}
Given a task involving input--output pairs: $\{ x_i, y_i \}$, the objective of neural networks is to learn and optimize ltrainable parameters for them to establish an optimal mapping that effectively captures the relationship between these input-output pairs, $f_\theta(x):\mathbb{R}^{dx} \rightarrow \mathbb{R}^{dy}$,represented as a function expression $y_i = f(x_i) \approx f_\theta(x_i)$.  $f_\theta(x_i)$ is a fitted neural network using the optimized parameter $\theta$, and $f(x_i):\mathbb{R}^{dx} \rightarrow \mathbb{R}^{dy}$ is the standard mapping of the input--output pairs for the task.
In this architecture, neural networks are used to approximate the Fourier series expansion of the original function as
\begin{equation}
	f(x_i) \approx  A _ { 0 } + \sum _ { n = 1 } ^ {N} \left( A _ { n } \cos \left( \frac { 2 \pi n x_i } { L } \right) + B _ { n } \sin \left( \frac { 2 \pi n x_i } { L } \right) \right) \approx f_\theta(x_i),
	\label{eq:fit}
\end{equation}
where $N$ is the order of Fourier series.
To represent Eq.(\ref{eq:fit}) in the form of a neural network layer, the coefficients $a_n^l$ and $b_n^l$  represent the Fourier coefficients of the $l-th$ ($l \in [0, L]$)layer of Fourier basis neural network and a weight matrix $\bm{W}_p^l=[w_1^l,w_2^l,w_3^l,\dots,w_n^l]^T$ represent the angular frequencies $\frac { 2 \pi n } { L } $in the Fourier series
\begin{align}
	f^{l}_\theta(x_i) &\approx A_0 + \sum_{n=1}^{N} \left( a_n^l \cos \left( w_n^l x \right) + b_n^l \sin \left( w_n^l x \right) \right) \notag \\
	&= A_0 + [a_1^l, a_2^l, \dots, a_N^l] \cos([w_1^l, w_2^l, \dots, w_N^l]^T 	x_i) \notag \\
	&\quad + [b_1^l, b_2^l, \dots, b_N^l] \sin([w_1^l, w_2^l, \dots, w_N^l]^T x_i) \notag \\
	&= A_0 + \bm{A}^l \cos(\bm{W}_P^l x_i) + \bm{B}^l \sin(\bm{W}_P^l x_i), \label{eq4}
\end{align}
 where $[...]^T$ and $[...]$ denotes the concatenation along the first and
 second dimension, respectively. To enrich the fitting possibilities of the network, the neural network embedded with Fourier series is extended and uses different angular frequency weight matrices:$\bm{W_{a}}^l=[w_{1a}^l,w_{2a}^l,w_{3a}^l,\dots,w_{na}^l]^T$, and $\bm{W_{b}}^l=[w_{1b}^l,w_{2b}^l,w_{3b}^l,\dots,w_{nb}^l]^T$ , namely
\begin{equation}
	f^{l}_\theta(x_i) \approx  A_0 + A^l \cos(W_{a}^l x_i) + B^l \sin(W_{b}^l x_i)  
	=A_0+\bm{\Theta}^l[cos(\bm{W_a}^l x_i),sin(\bm{W_b}^l x_i)]^T=l^l\circ(x_i),
	\label{eq5} 
\end{equation}
where $\bm{\Theta}^l=[\bm{A}^l,\bm{B}^l]$. Inspired by Fourier analysis networks (FANs)\cite{dong2024fan}, if we stack the multiple Fourier layers and the architecture of FBNN structure is expressed as 
\begin{equation}
	\left. \begin{array} { l }{ f_\theta ( \bm{x} ) = l _ { L } ( l _ { L - 1 } \circ l _ { L - 2 } \circ \cdots \circ l _ { 1 } \circ \bm{x} ) } \ { = \bm{A}^L +  \bm{\Theta} ^ { L } [ \cos ( \bm{W _ { \text { a } }} ^ { L } ( l _ { 1 : L - 1 } \circ \bm{x} ) \| \sin ( \bm{W _ { \text { b } }} ^ { L } ( l _ { 1 : L - 1 } \circ \bm{x} ) ) ] }, \end{array} \right.
	\label{eq6}
\end{equation}
where $\bm{x}=[x_1,x_2...x_i...x_n]$, $dx = n$. This results in very poor learning of the coefficient matrix $\bm{\Theta} ^ { L }$, and it will not undergo more refined learning as the depth of the network increases. So in each layer, we follow the strategy of FAN and parallelize the coefficient matrix $\bm{\Theta} ^ { l }$ and angular frequency matrix $\bm{W_a}^l\in \mathbb{R} ^ { d _  { p }\times d_x  }$  and  $ \bm{W_b}^l\in \mathbb{R} ^ { d _  { p }\times d_x  }$ in the same layer of neural network for learning, namely
\begin{equation}
	\phi ^{l}( \bm{x} ) \triangleq [ \cos ( \bm{W _ { a }}^l\bm{ x} ) , \sin ( \bm{W _ { b }} ^l\bm{x} ) , \sigma (\bm{A}_{c}^l+\bm{\Theta}_{c}^l\bm{x} ) ]^T,
	\label{eq7}
\end{equation}
where $\sigma$ is the nolinear activation function,$\bm{A}_{c}^l \in \mathbb{R} ^ { d _  {\overline{p}} }$ and $\bm{\Theta}_{c}^l\in \mathbb{R} ^ { d _  {\overline{p}}\times d_x}$ are learnable parameters(with the hyper-parameters $\overline{p}$ indicating the first dimension of $\bm{\Theta}_c$, $p$ is half the number of Fourier neural nodes(divide $cosine$ and $sine$ calculation equally) ). We refer trigonometric nodes that perform $cosine$ and $sine$ operations as Fourier nodes. The fitting capability of a neural network is closely related to the diversity of basis functions at each node, therefore, two learnable bias matrices are added to the trigonometric operations of the Fourier nodes, namely, $\bm{B _ {a}} \in \mathbb{R} ^ { d _  { p }  },\bm{B _ {b}} \in \mathbb{R} ^ { d _  { p }  }$. According to the Angle Addition Formulas, it can be concluded that the trigonometric function with added bias provides fitting possibilities for Fourier nodes such as $cosine ^ 2, sine ^ 2 $, and $cosine \ times sine $. Additional learnable coefficients $C_a$  and  $C_b \in \mathbb{R}$ are added at the Fourier nodes to increase the construction complexity of the neural network layers. Therefore, our final FBNN network layer is
\begin{equation}
	L^{l}(\bm{x}) = \left[ C_a^l \cos(\bm{W_a}^l \bm{x} + \bm{B_a}^l) , C_b^l \sin(\bm{W_b}^l \bm{x} + \bm{B_b}^l) , \sigma(\bm{A_c}^l + \bm{\Theta}_c^l \bm{x}) \right]^T, \quad \text{if } l < L.
	\label{eq8}
\end{equation}
The complete FBNN layer is shown above in Figure \ref{architecture}.
The input and output layers of FBNN are mapped through simple linear layers.

\subsection{\label{sec:level2}Physics-informed Fourier basis neural network (PIFBNN)}
\begin{figure}[!htbp] 
	\centering
	\subfigure[]{
		\includegraphics[height=18cm]{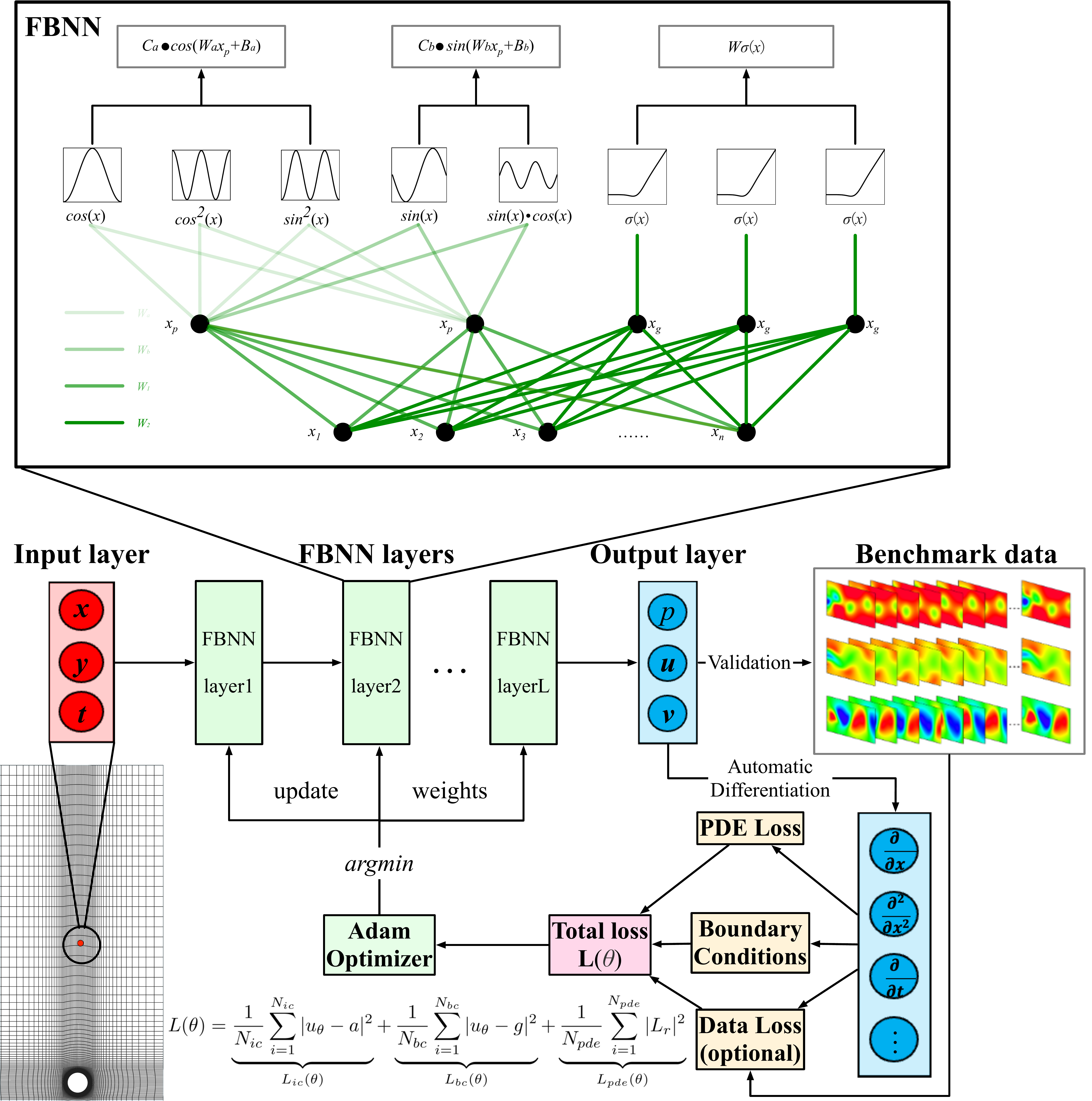}
	}
	
	\caption{The schematic diagram of Physics-informed Fourier Basis Neural Network.}
	\label{architecture}
\end{figure}

We proposed PIFBNN by adding physical information to FBNN based on PINN.
Consider the general form of a PDE subject to any boundary conditions defined as
\begin{align}
	\frac{\partial u}{\partial t} & = R(u; t,x),  &\text{in } \Omega \times (T_0, T), \label{eq11} \\ 
	\mathcal{B}(u) & = g(t,x),  &\text{on } \partial \Omega \times (T_0, T), \label{eq12} \\ 
	\mathcal{I}(u) & = a(t,x),  &\text{in } \Omega \times \{ T_0 \}, \label{eq13} 
\end{align}
where $\Omega$ is a bounded spatial domain,$(T_0,T)$ represents the time domain,$T_0$ and $T$ represent the initial and end time of time domain, respectively. $	\mathcal{B}(u)$ and $	\mathcal{I}(u)$ represents the boundary and initial condition operator, respectively. Eq.(\ref{eq11}) is the general expression of PDEs. Eq.(\ref{eq12}) describes the boundary condition, where $  \partial \Omega$ represents the boundary of the spatial domain. Eq.(\ref{eq13}) describes the initial condition.
As shown in Figure\ref{architecture}, the optimization of PIFBNN parameters is achieved by minimizing loss function. In a general neural network, the loss term is typically defined based on labeled data as 
\begin{equation}
\begin{split}
	\mathcal{L} ( \theta ) =& \underbrace { \frac { 1 } { N_{ic} } \sum _ { i = 1 } ^ { N_{ic} } | \mathcal{I}[u_{ \theta } ( 0 , x ^ { i } )] - a (0, x ^ { i } ) | ^ { 2 } } _ { L_{ic} ( \theta ) } +
	\underbrace { \frac { 1 } { N_{bc} } \sum _ { i = 1 } ^ { N_{bc} } | 	\mathcal{B}[u_{ \theta } ( t , x_{bc} ^ { i } )] - g (t, x_{bc} ^ { i } ) | ^ { 2 } } _ { L_{bc} ( \theta ) } \\+ &
	\underbrace { \frac { 1 } { N_{data} } \sum _ { i = 1 } ^ { N_{data} } | u_{ \theta } ( t , x ^ { i } ) - u_g (t, x ^ { i } ) | ^ { 2 } } _ { L_{data} ( \theta ) }  ,
	\label{14}
\end{split}
\end{equation}
where $x \in \Omega$, $x_{bc} \in \partial \Omega$, $t \in (0,T)$, $u_{\theta}$ represents the outputs of neural networks, $u_g$ represents the ground truth. $N_{ic},N_{bc}$ and $N_{data}$ represent the number of points corresponding to the initial conditions, boundary conditions, and label data, respectively. Note that boundary and initial conditions can also be considered as labeled data.

The architecture of PIFBNN  is shown in the Figure \ref{architecture}. By substituting the input point $(t_r,x_r)$ and model prediction $u_\theta(t_r,x_r)$ into Eq. (\ref{eq11}), the residual of the equation can be obtained through automatic differentiation defined as
\begin{equation}
	L_r ( u _ { \theta }  ,t_r , x_r ) = \frac { \partial u _ { \theta } } { \partial t } ( t _ { r } , x _ { r } ) - R ( u _ { \theta } ,t _ { r } , x _ { r } )]),
	\label{eq15}
\end{equation}
where $(t_r,x_r) \in D \times (0,T)$. By using this residual to replace the original label data loss term, the total loss of PINN without labeled data can be obtained as
\begin{equation}
\begin{split}
	\mathcal{L} ( \theta ) =& \underbrace { \frac { 1 } { N_{ic} } \sum _ { i = 1 } ^ { N_{ic} } | \mathcal{I}[u_{ \theta } ( 0 , x ^ { i } )] - a (0, x ^ { i } ) | ^ { 2 } } _ { L_{ic} ( \theta ) } +
	\underbrace { \frac { 1 } { N_{bc} } \sum _ { i = 1 } ^ { N_{bc} } | 	\mathcal{B}[u_{ \theta } ( t , x_{bc} ^ { i } )] - g (t, x_{bc} ^ { i } ) | ^ { 2 } } _ { L_{bc} ( \theta ) } \\+& 
	\underbrace { \frac { 1 } { N_{pde} } \sum _ { i = 1 } ^ { N_{pde} } |L_r ( u _ { \theta }  ,t_r , x_r ) | ^ { 2 } } _ { L_{pde} ( \theta ) }.
	\label{16}
\end{split}
\end{equation}
When labeled data are available, even in small quantities, a corresponding loss term can be added to the total loss to assist the network learning. This study adopts this approach for sparse data field reconstruction, as detailed in Section \ref{sec:level1/5}.

\section{\label{sec:level1/3} Prediction of PDEs}

This section compares the proposed PIFBNN and standard PINN to predict the solutions of five representative partial differential equations. We only use the given initial and boundary conditions of PDEs as constraints, without employing any labeled data in any of these cases. The neural networks are designed to predict solutions by minimizing the sum of squared errors, as specified in Eq.(\ref{16}), where the loss term $L_{pde}$ is calculated by solving the governing PDEs for each case. Unless otherwise specified, the ratio of Fourier nodes to hidden layer neurons is set to 0.3.  The networks employ the GELU activation function and are trained for $2\times(10^4)$ epochs using standard Adam optimizer.All experiments were conducted on an RTX 4090 GPU, with the learning rate is fixed to 0.001.
\subsection{Allen--Cahn equation}

\begin{figure}[h!] 
	\centering
	{
		\includegraphics[height=5.5cm]{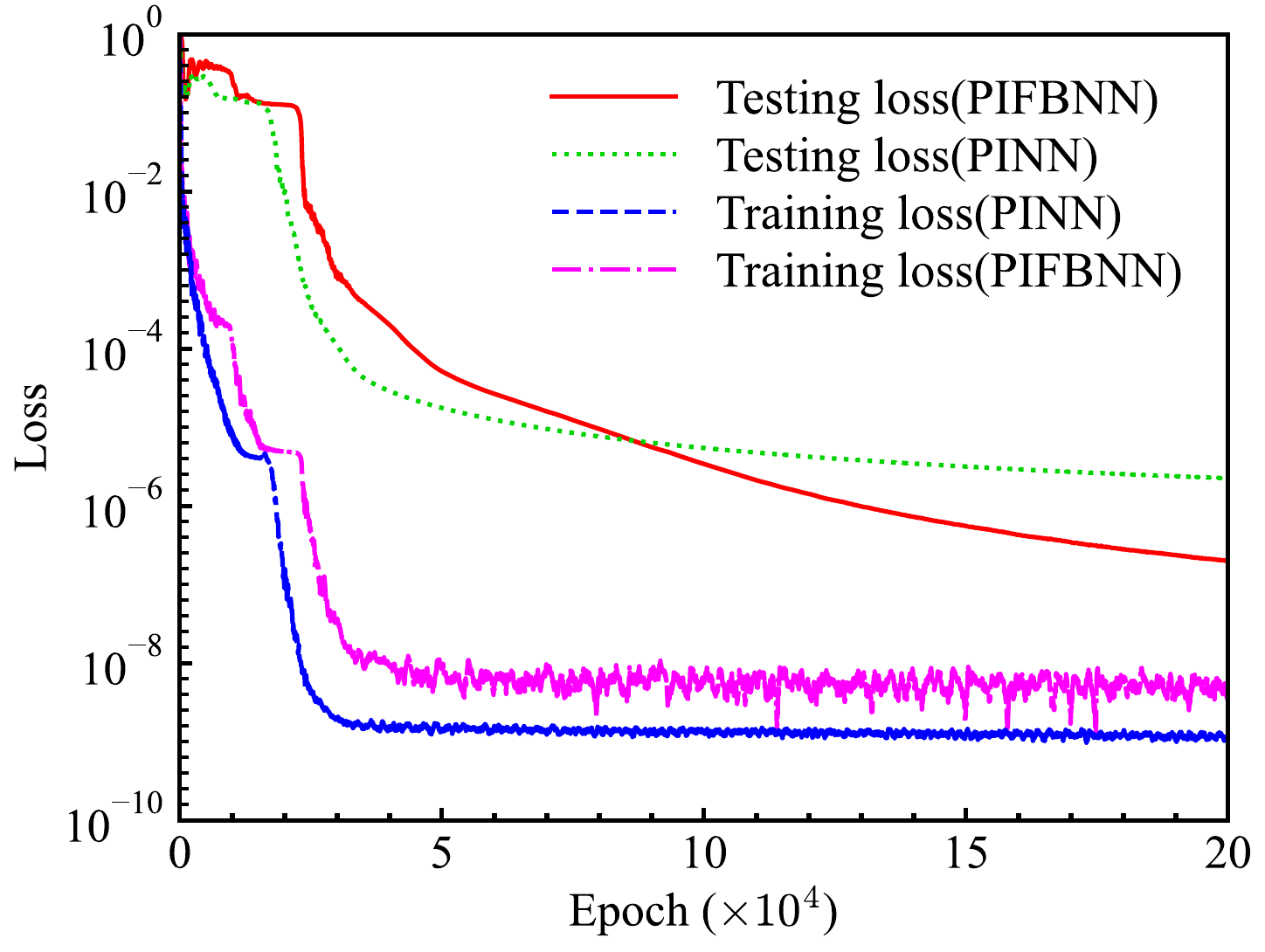}
	}
	
	\caption{Learning curves of PINN and PIFBNN models for Allen-Cahn equation. }
	\label{acpredloss}
\end{figure}

The general form of the Allen--Cahn (AC) equation is defined by
\begin{equation}
	\frac{\partial u}{\partial t}=d \frac{\partial^{2} u}{\partial x^{2}}+f(u),
\end{equation}
where $u$ is the phase field, $t$ is time, $x$ is the spatial coordinate, and $f(u)$ is the reaction driven term of $u$, $d=1$ is the diffusion coefficient, representing the diffusion rate of the medium. The specific operating conditions of the AC equation are investigated with the specific functional formula and parameter range given by
\begin{equation}
	\frac{\partial u}{\partial t}=d \frac{\partial^{2} u}{\partial x^{2}}+5\left(u-u^{3}\right), \quad x \in[-1,1], \quad t \in[0,1].
\end{equation}
The initial and boundary conditions are respectively defined by
\begin{equation}
	u(x, 0)=x^{2} \cos (\pi x), \quad u(-1, t)=u(1, t)=-1.
\end{equation}
The AC equation is a physical reaction--diffusion equation that describes the process of phase separation in multicomponent alloy systems, including the transition from order to disorder.
The training and testing datasets used in this example are simulated using conventional spectral methods \cite{raissi2019physics}. A fourth-order Runge--Kutta time integrator is employed up to a final time t = 1.0. At t = 0, 200 data points are extracted as initial condition and 100 points at boundary condition are randomly selected using Latin hypercube sampling strategy. The objective is to predict  the whole field using only the given initial and boundary conditions as constraints on the governing PDE. Both the proposed PIFBNN and baseline PINN are established with 10 hidden layers, each containing 50 neurons. The input layer receives two coordinate variables ($x$ and $t$), and the output layer generates a single predicted variable ($u$). 1000 points at the initial and boundary conditions are randomly sampled in the dataset to test and evaluate the performance of the PIFBNN.
The learning curves of PIFBNN and PINN are shown in Figure \ref{acpredloss}. The loss function is the mean square error (MSE) of the true and predicted values of the neural networks, namely
\begin{equation}
	Er_{\text{mse}} = \frac{1}{n} \sum_{i=1}^{n} (u_{\text{true},i} - u_{\text{pred},i})^2.
\end{equation}

\begin{figure}[h!] 
	\centering
	{
		\includegraphics[height=5.5cm]{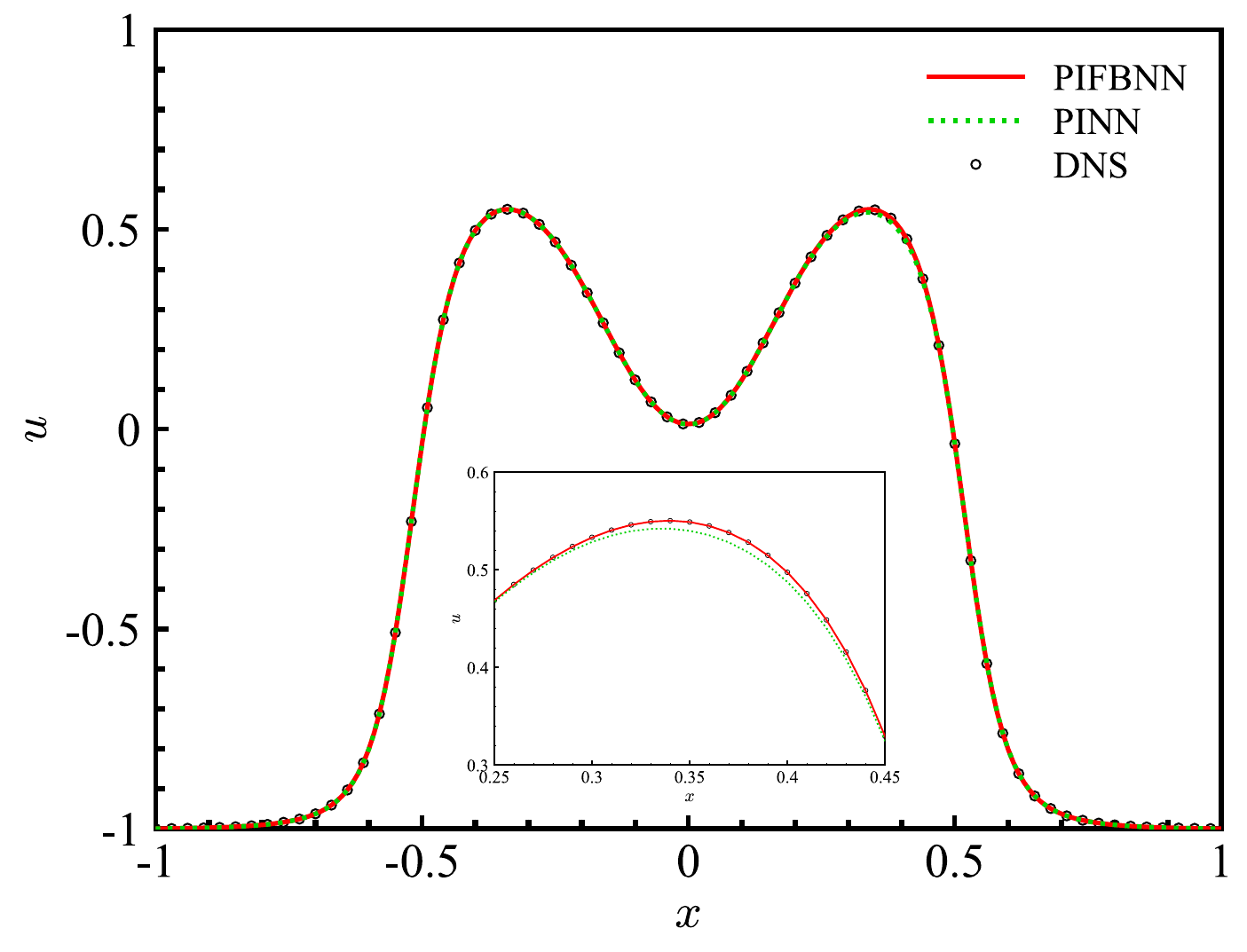}
	}
	
	\caption{Comparisons of velocity profiles using PINN and PIFBNN models for Allen-Cahn equation at t=0.5.}
	\label{acpred0.5t}
\end{figure}

\begin{figure}[h!] 
	\centering
	{
		\includegraphics[height=5.5cm]{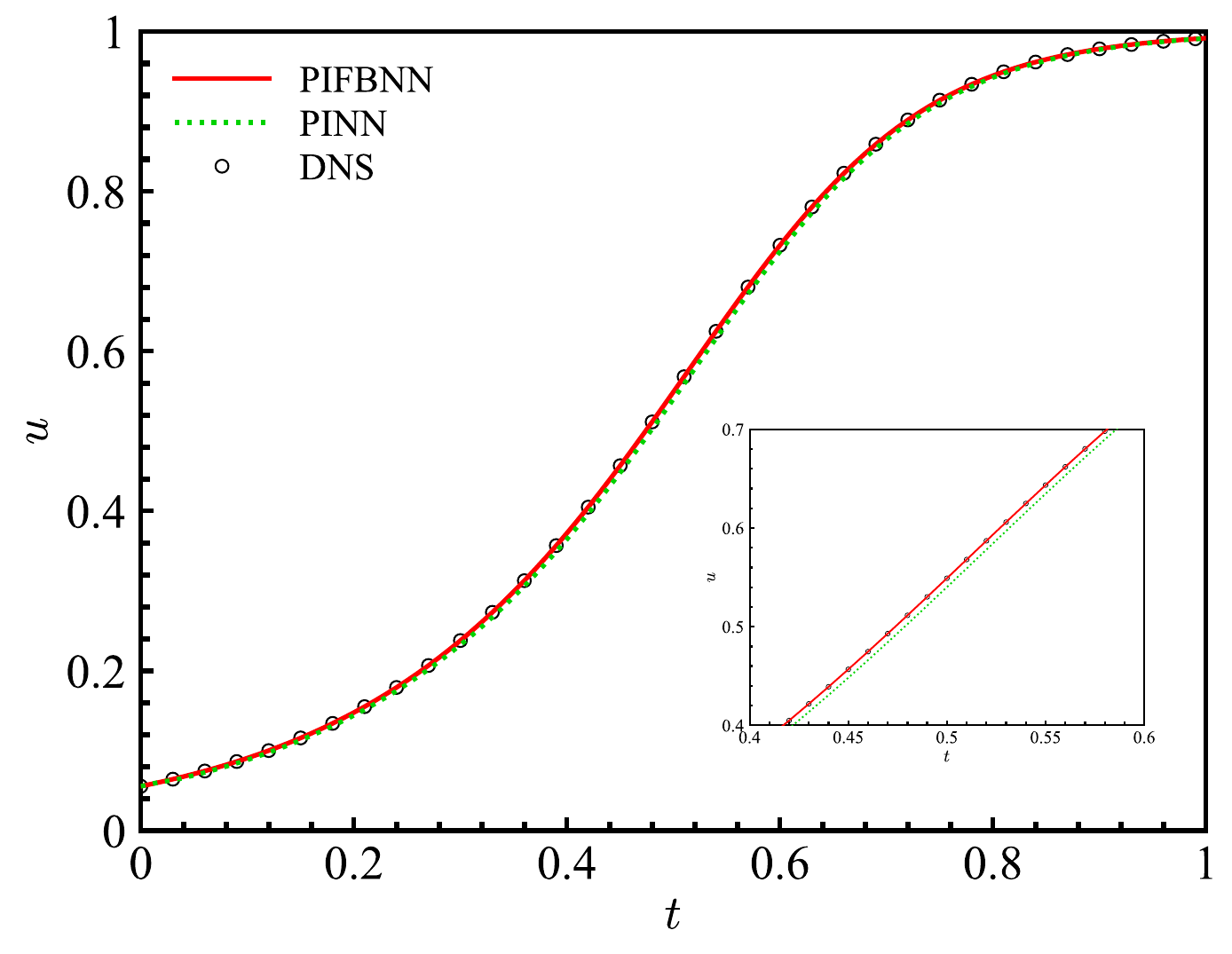}
	}
	
	\caption{Comparisons of velocity profiles using PINN and PIFBNN models for Allen-Cahn equation at x=0.35. }
	\label{acpred0.35x}
\end{figure}

The training loss reflects error associated only with the training data, namely, the sample points at boundary and initial conditions. By contrast, testing loss measures the prediction error on unseen data points over the entire spatiotemporal domain and represents the actual prediction performance of the model.
Notably, PIFBNN exhibits a significantly lower training loss compared to PINN, demonstrating stronger learning constraints on the initial and boundary conditions, which consequently results in a reduced testing loss. Prior to reaching $5\times 10^4$ training epochs, the training loss of PIFBNN is already significantly lower than that of PINN, but the testing loss during this period remains higher than PINN. As the number of iterations increases, the training loss remains in a convergent state, while the testing loss of PIFBNN continues to decrease and eventually becomes significantly lower than PINN. This indicates that physical information loss plays an important role in driving the network to continue learning, which indirectly confirms that PIFBNN using FBNN as the network architecture has a stronger ability to capture and learn physical information.
The ground truth and predictions of PINN and PIFBNN for the Allen--Cahn equation at t = 0.5 are shown in Figure \ref{acpred0.5t}.
Both PINN and PIFBNN can predict the AC equation well at t = 0.5. However, when the fitting curve is enlarged, it is evident that the prediction of PINN in regions with large curvatures is deviated, whereas PIFBNN perfectly fits the ground truth, indicating that PIFBNN has better nonlinear learning ability.
The velocity profile of PINN and PIFBNN for the Allen--Cahn equation at x = 0.35 are shown in Figure \ref{acpred0.35x}. Similar to the previous instance, both PINN and PIFBNN yield good prediction results. However, in the enlarged image, the predictions of the PINN are always slightly smaller than the ground truth.

\begin{figure}[h!] 
	\centering
	{
		\includegraphics[height=5.5cm]{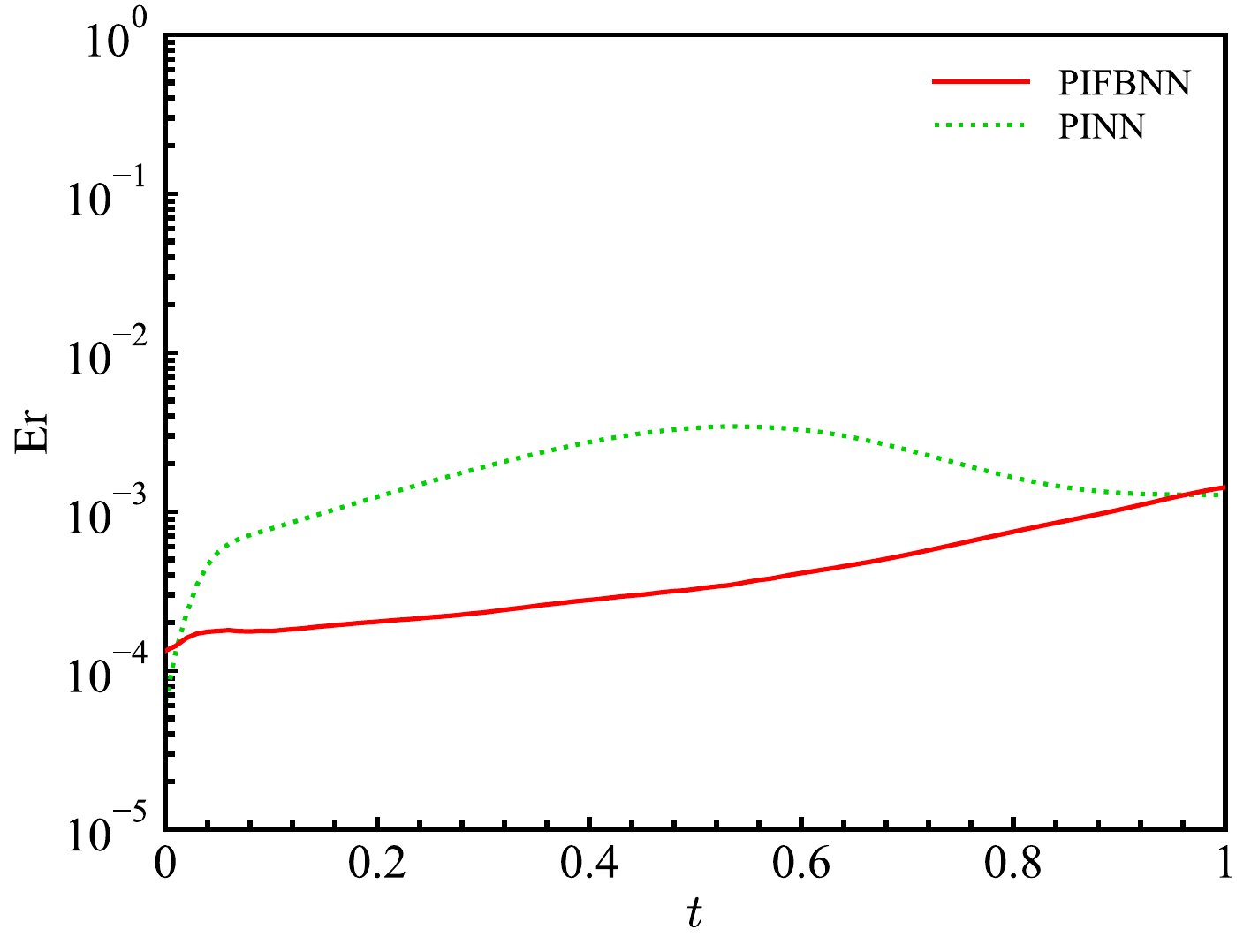}
	}
	
	\caption{Relative errors of PINN and PIFBNN models for Allen-Cahn equation.}
	\label{acpreder}
\end{figure}

The evolutions of predicted RMS relative errors of PINN and PIFBNN are shown in Figure \ref{acpreder}. The RMS relative error is defined by
\begin{equation}
	Er_{\text{rms}} = \frac{\sqrt{\frac{1}{N} \sum_{i=1}^{N} (u_{\text{true}, i} - u_{\text{pred}, i})^2}}{\sqrt{\frac{1}{N} \sum_{i=1}^{N} (u_{\text{true}, i})^2}} ,
\end{equation}
where $N$ is the number of sample points; $u_{\text{true},i}$ and $u_{\text{pred},i}$ are the ground truth and the predicted values modeled by the neural network, respectively. Owing to the constraints imposed by the initial conditions, both PINN and PIFBNN exhibit small prediction errors at t = 0. However, as time advances,  PIFBNN demonstrates significantly better predictive performance than PINN.

\begin{figure}[!htbp]
	\centering
	\subfigure[Ground truth]{
		\includegraphics[height=4cm]{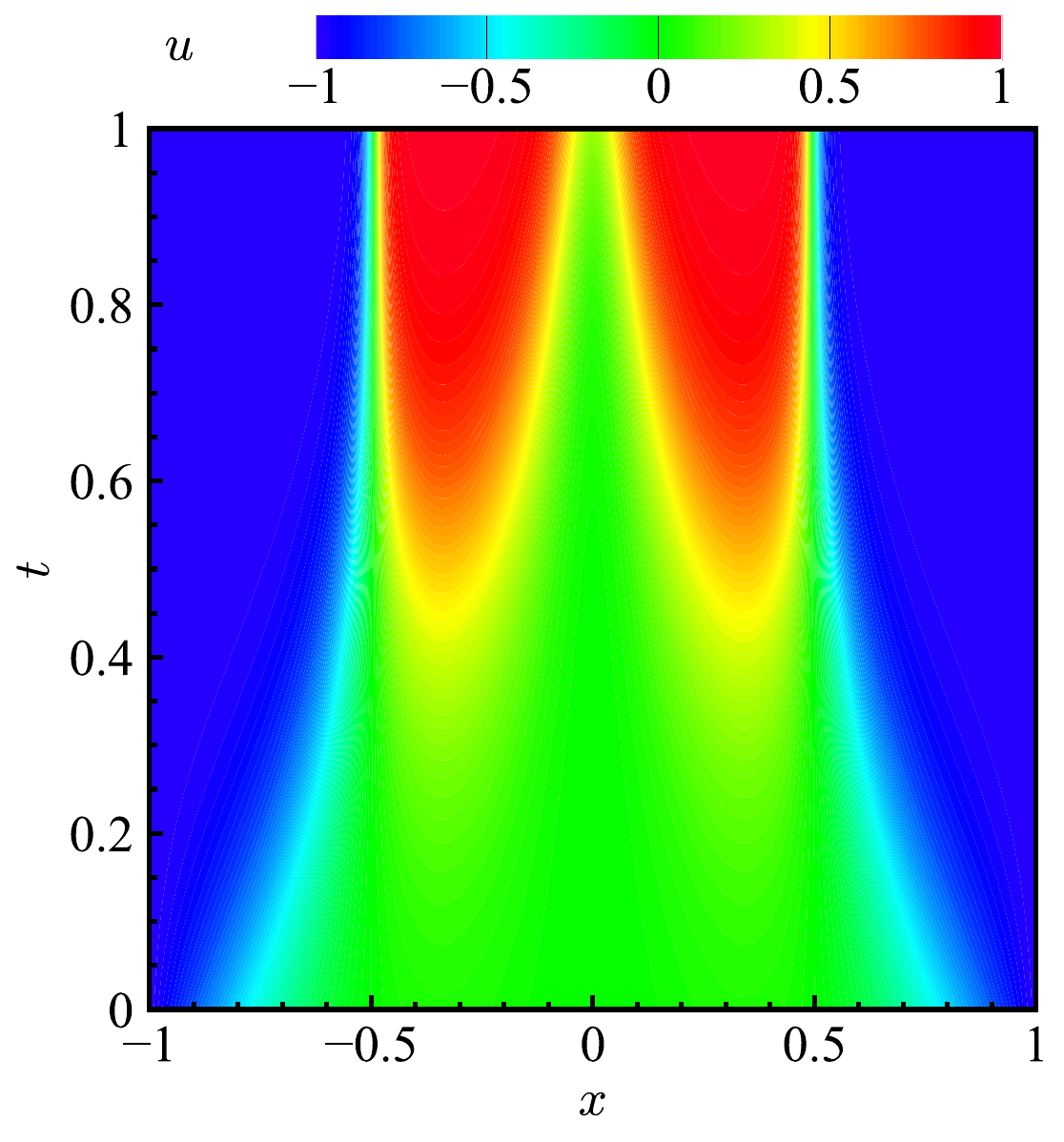}
	}
	\subfigure[PIFBNN prediction]{
		\includegraphics[height=4cm]{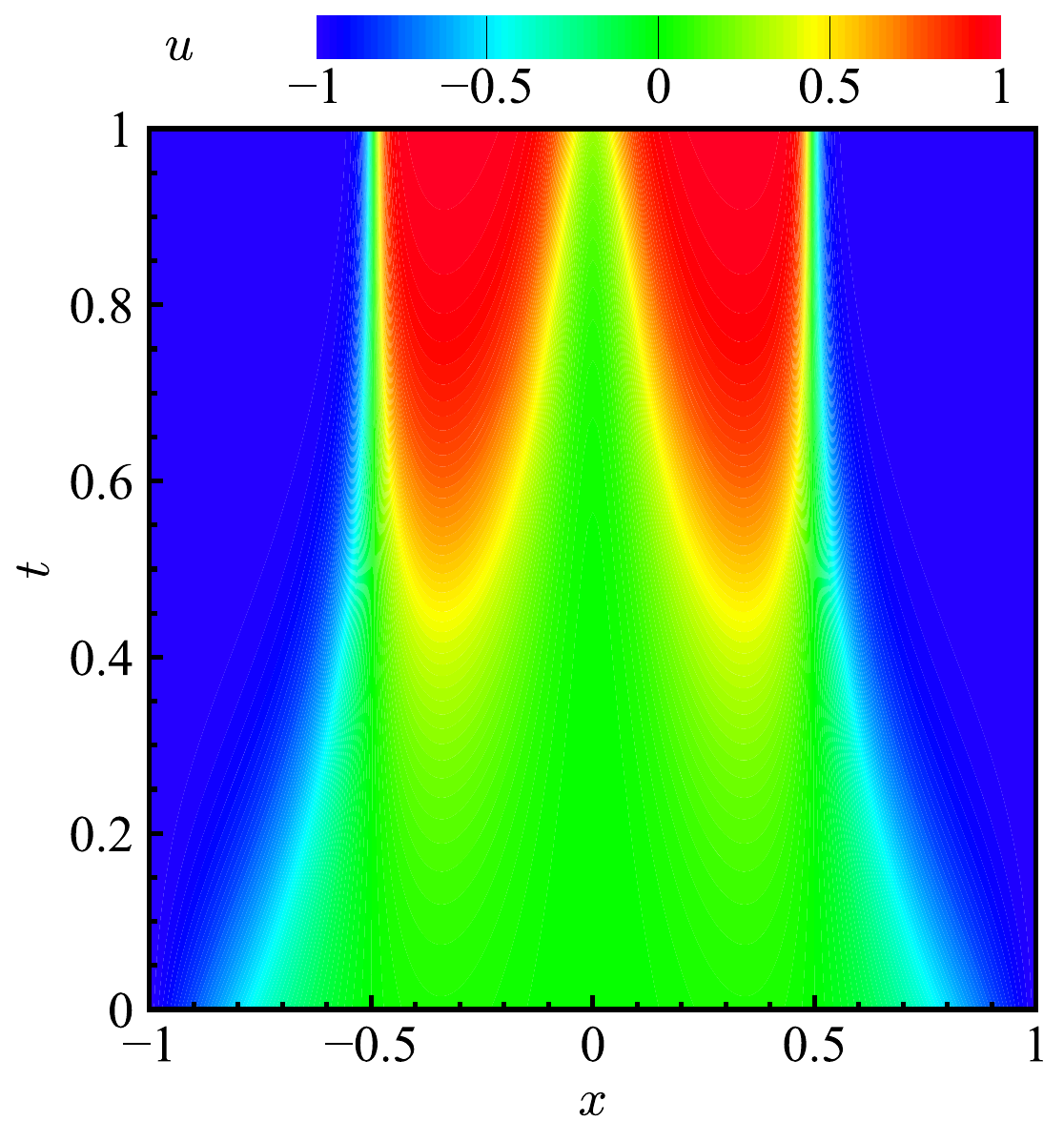}
	}
	\subfigure[PINN prediction]{
		\includegraphics[height=4cm]{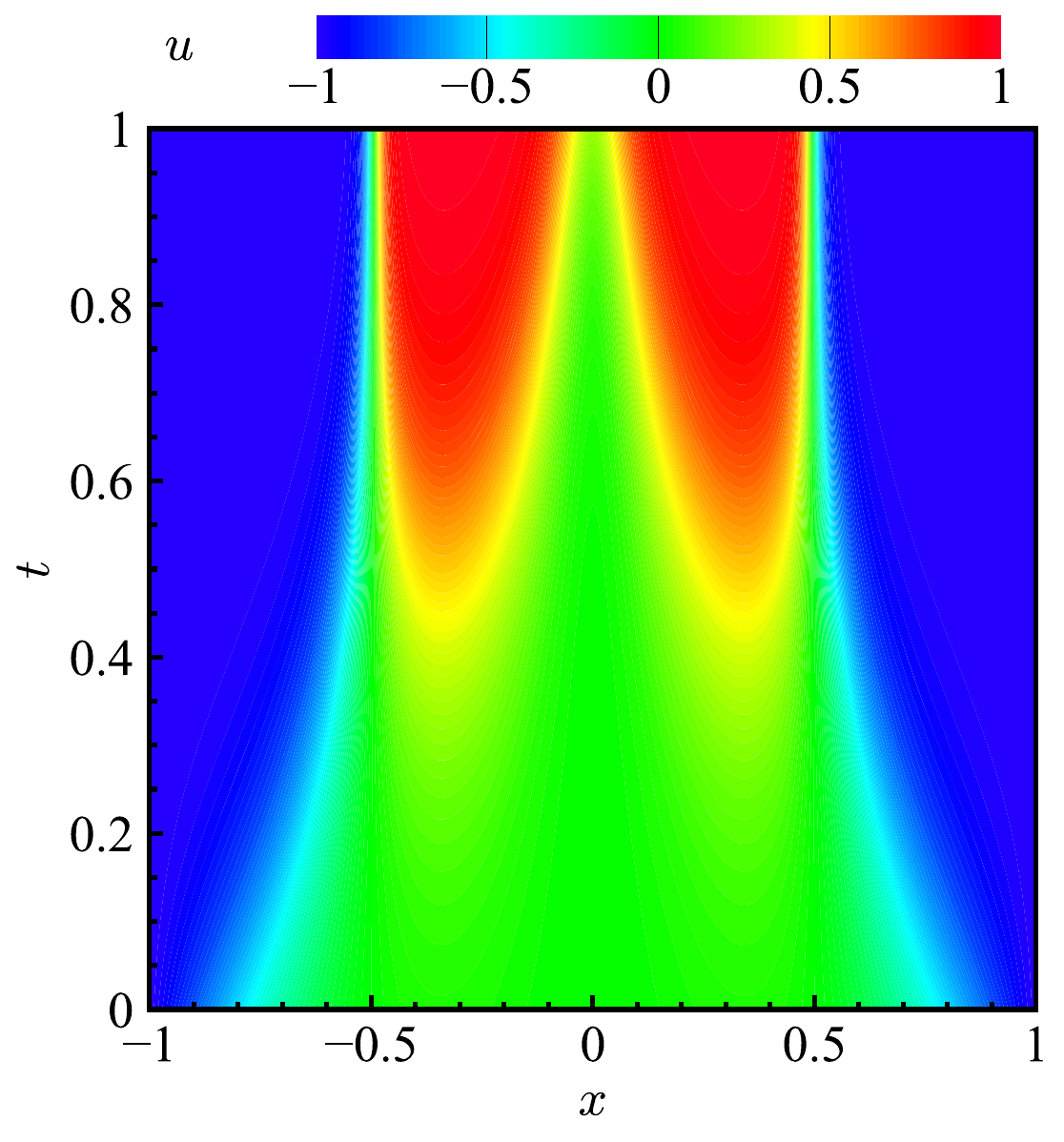}
	}
	
	\par %
	\subfigure[PIFBNN relative error]{
		\includegraphics[height=4cm]{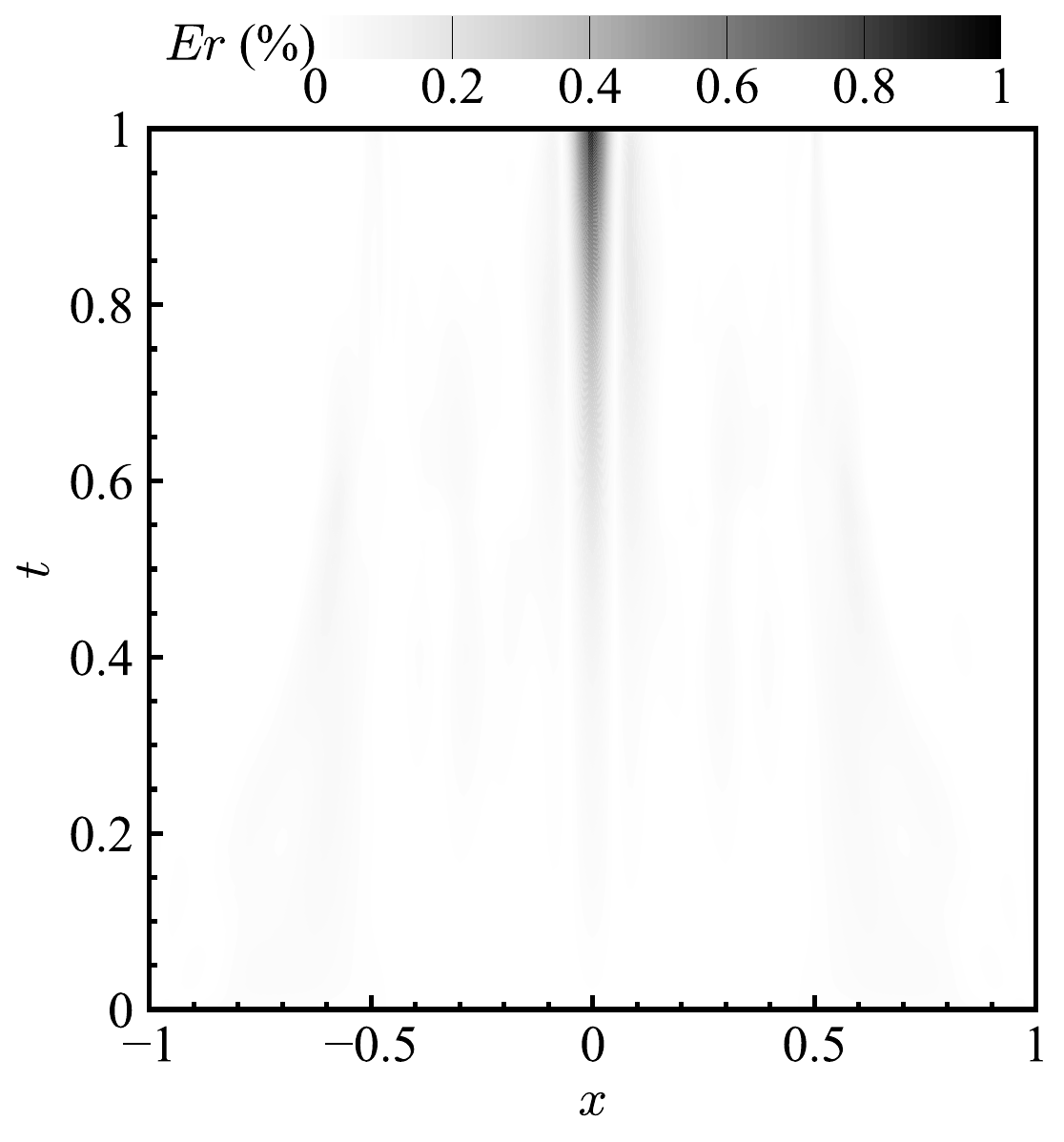}
	}
	\subfigure[PINN relative error]{
		\includegraphics[height=4cm]{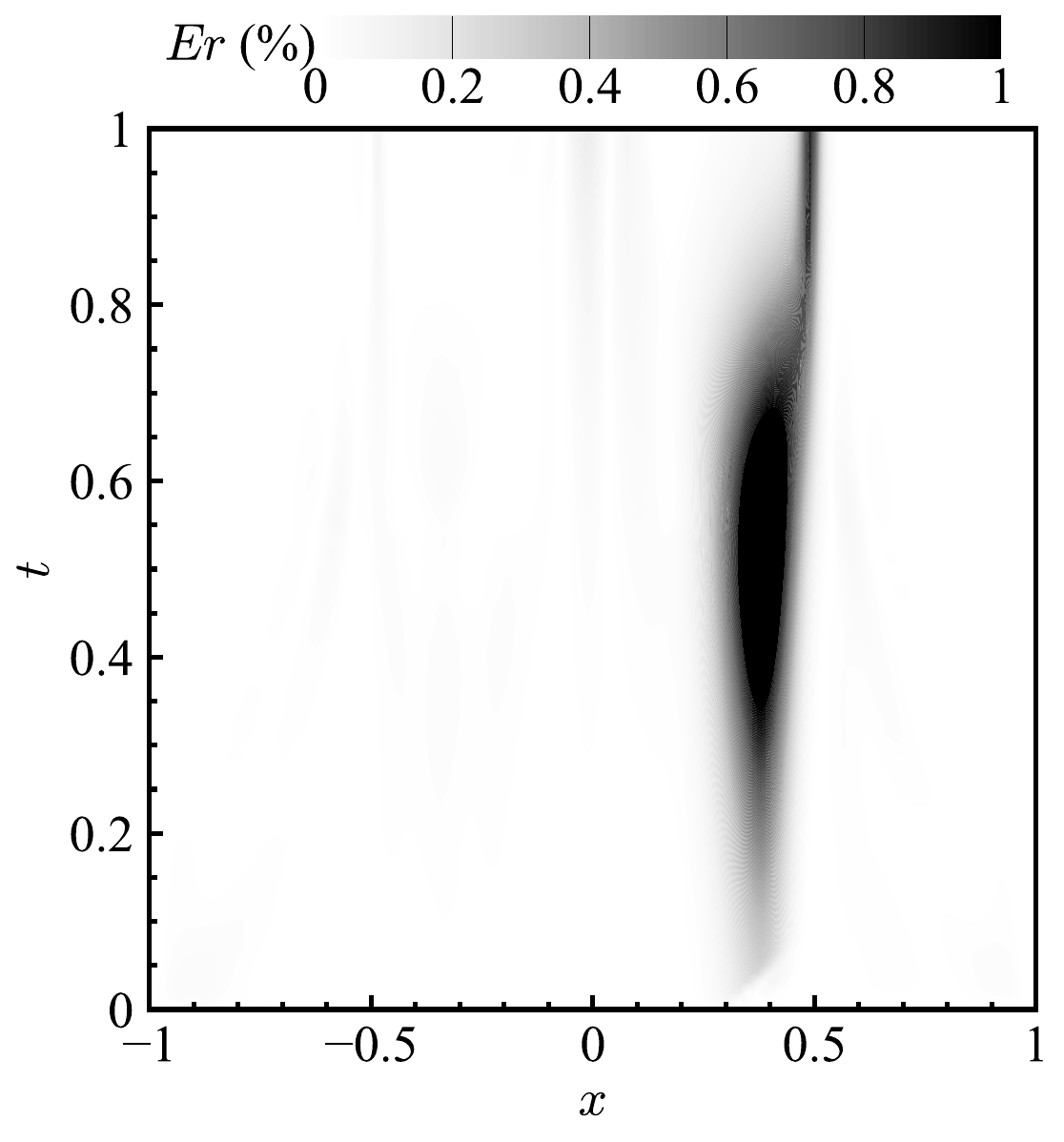}
	}
	\caption{Velocity and relative error contours predicted by PIFBNN and PINN models for Allen-Cahn equation. }
	\label{accontour}
\end{figure}

The predicted velocity and relative error contours are shown in Figure \ref{accontour}. Similar to the above results, both PINN and PIFBNN depict good prediction results for the AC equation. In the predicted velocity contour, the results predicted by PIFBNN and PINN models are consistent with the ground truth. However, significant differences can be observed in the relative error contour, where PIFBNN exhibits only a slight error in the strong convective region, whereas PINN shows a significant prediction error in the high-velocity region on the right side. Notably, the maximum value in the error contour is 1\%. Therefore, both PINN and PIFBNN show very small prediction errors, whereas the PIFBNN demonstrates superior prediction accuracy.

\subsection{Burgers equation}
The general form of the Burgers equation is 
\begin{equation}
	\frac{\partial u}{\partial t}+u \frac{\partial u}{\partial x}=\nu \frac{\partial^{2} u}{\partial x^{2}}.
\end{equation}

Here, $u$ represents the velocity of the fluid. $\nu=0.01/\pi$ is the viscosity kinematic.
In this study, the parameters are varied within the ranges 
\begin{equation}
	x \in[-1,1], \quad t \in[0,1].
\end{equation}

The Dirichlet boundary and initial conditions of Burgers equation are defined as
\begin{equation}
	u(-1, t)=u(1, t)=0, \quad u(x, 0)=-\sin (\pi x).
\end{equation}

\begin{figure}[h!] 
	\centering
	{
		\includegraphics[height=5.5cm]{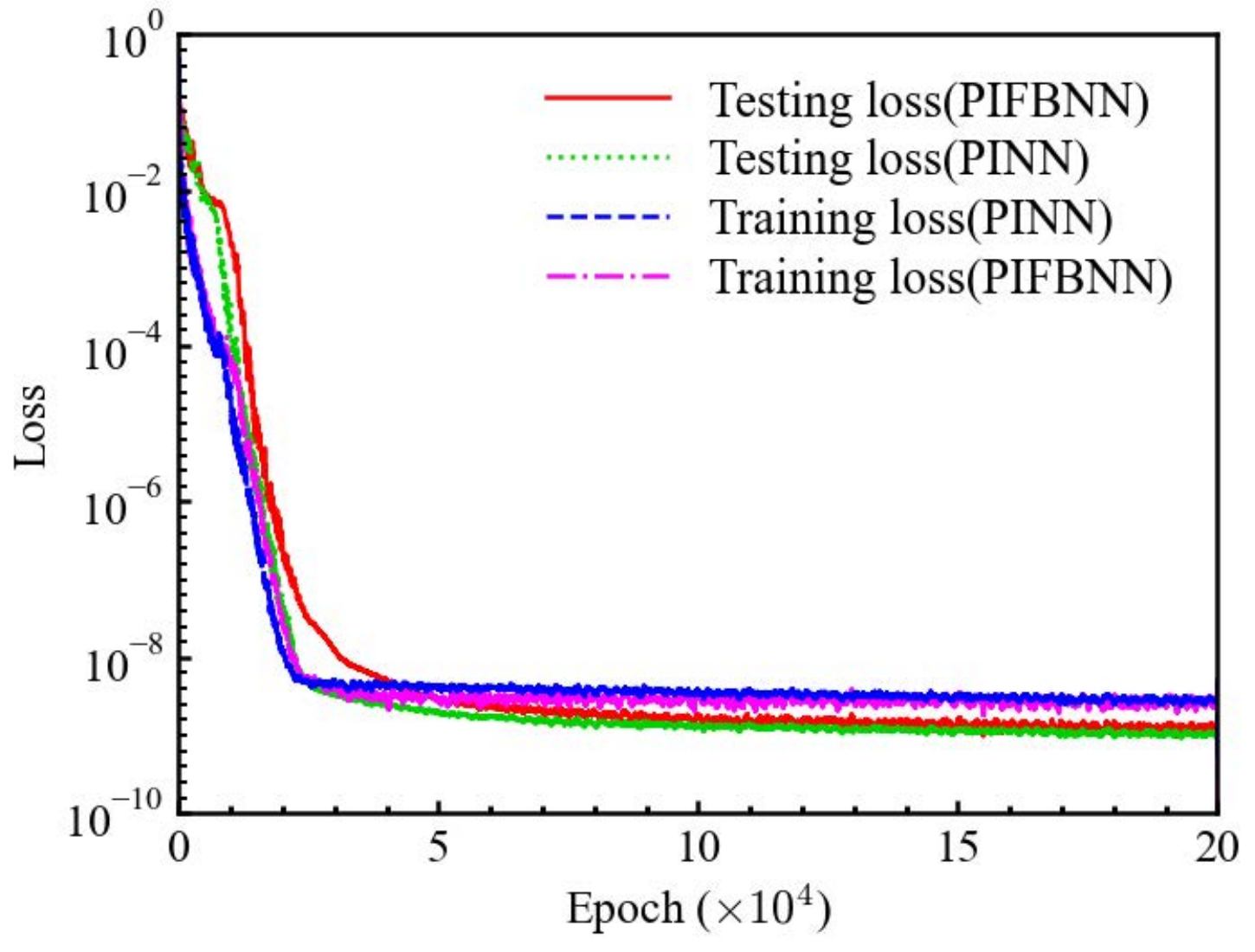}
	}
	
	\caption{Learning curves of PINN and PIFBNN models for Burgers equation. }
	\label{bgpredloss}
\end{figure}

Due to the existence of nonlinear convection terms, the solution to the Burgers equation can develop shock waves and finally become discontinuous. For this purpose, we use this case to validate the ability of PIFBNN for solving various scientific and engineering problems involving sharp features and discontinuities. We use the Burgers equation bencmark dataset of deepxde\cite{lu2021deepxde}, and 1000 test points are randomly sampled within the domain to evaluate the performance of the network.
We consistently use Latin hypercube sampling method to randomly sample 50 points at t = 0 as initial condition constraint and randomly sampled 50 points at the boundaries of $x = -1$ and $x = 1$ as the boundary condition constraints. By randomly sampling 10000 points within the domain to calculate the residual of the governing equation, we aim to predict the solution to Burgers equation of the entire spatiotemporal domain based solely on initial and boundary conditions, as well as governing equation constraints.
The training and testing losses of PIFBNN and PINN are shown in Figure \ref{bgpredloss}.

\begin{figure}[h!] 
	\centering
	{
		\includegraphics[height=5.5cm]{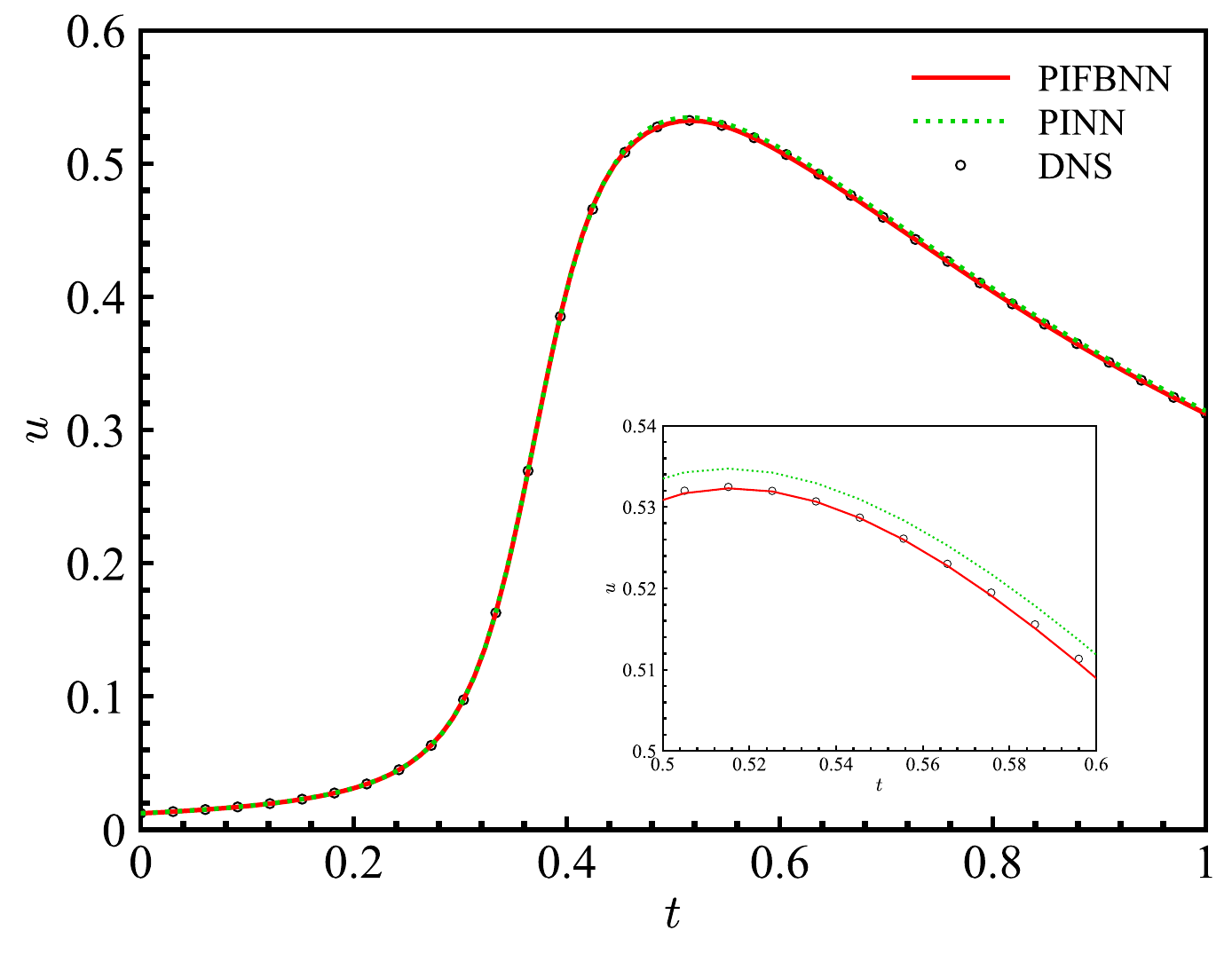}
	}
	
	\caption{Comparisons of velocity profiles using PINN and PIFBNN models for Burgers equation at x=0. }
	\label{bgpred0x}
\end{figure}

\begin{figure}[h!] 
	\centering
	{
		\includegraphics[height=5.5cm]{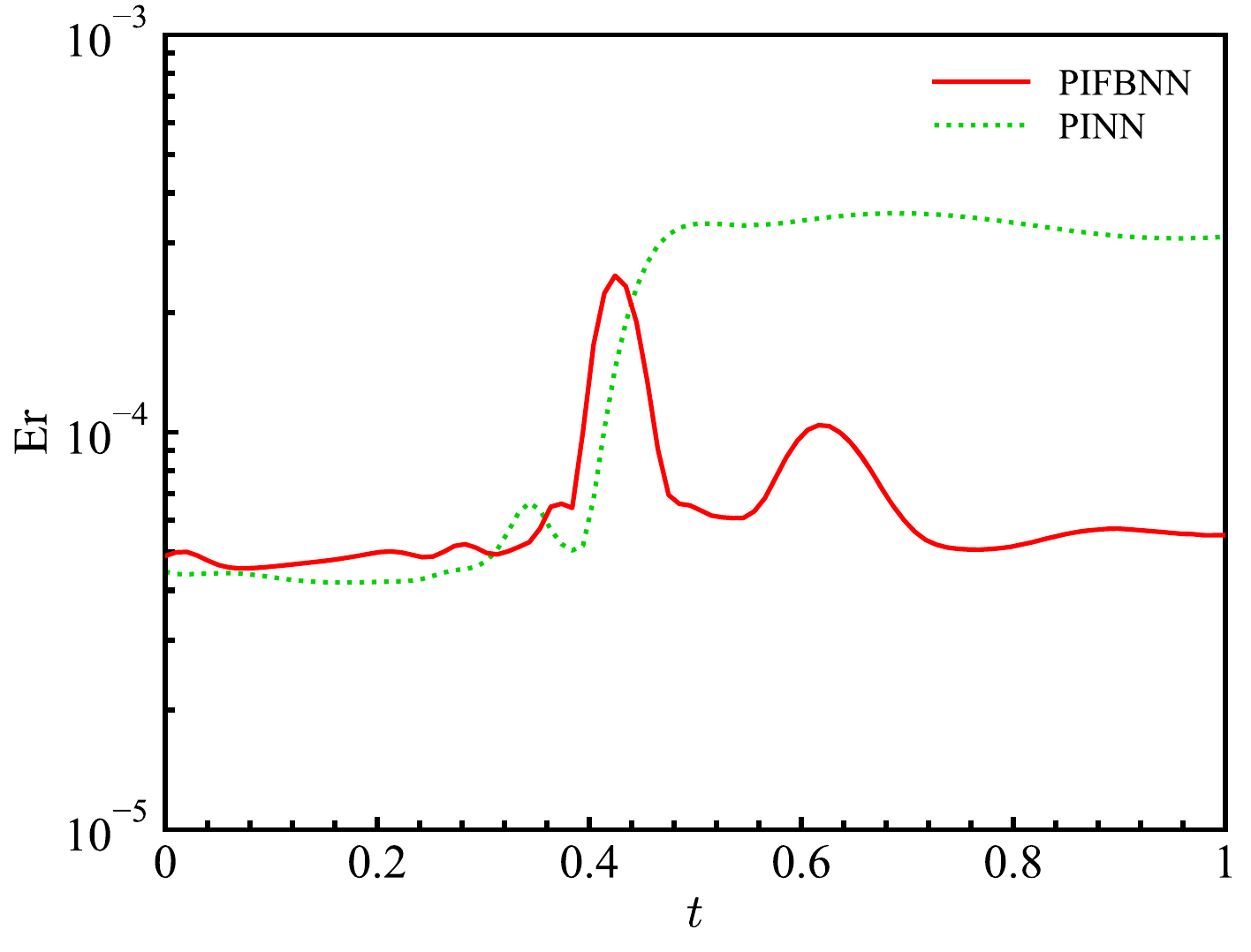}
	}
	
	\caption{Relative errors of PINN and PIFBNN models for Burgers equation. }
	\label{bgpreder}
\end{figure}

Both PINN and PIFBNN effectively learned the boundary and initial conditions, resulting in low testing losses for both models. This demonstrates that each model can successfully approximate  the solution to the Burgers equation.
Since the presence of discontinuous solutions in Burgers equation at $x = 0$, we focused on prediction performance at x = 0. The comparisons of velocity profiles with PINN and PIFBNN at $x = 0$ are shown in Figure \ref{bgpred0x}.
Prior to t = 0.4, wh, en there are no discontinuous solutions to the Burgers equation, i.e., when shock waves are not generated, the predicted solutions of PINN and PIFBNN are relatively consistent with the ground truth. However, as time progresses, shock waves emerge after t>0.45, and the Burgers equation exhibits discontinuous solutions, thereby enhancing nonlinearity and increasing the complexity of network-based predictions. Results predicted by the PINN are generally larger than the ground truth at this stage, whereas velocity predicted by PIFBNN model fits the ground truth well. This demonstrates that the PIFBNN has a better ability to predict discontinuous solutions compared to PINN.
The relative error curves of PINN and PIFBNN over time are shown in Figure \ref{bgpreder}.

\begin{figure}[!htbp]
	\centering
\subfigure[PIFBNN relative error]{
	\includegraphics[height=4cm]{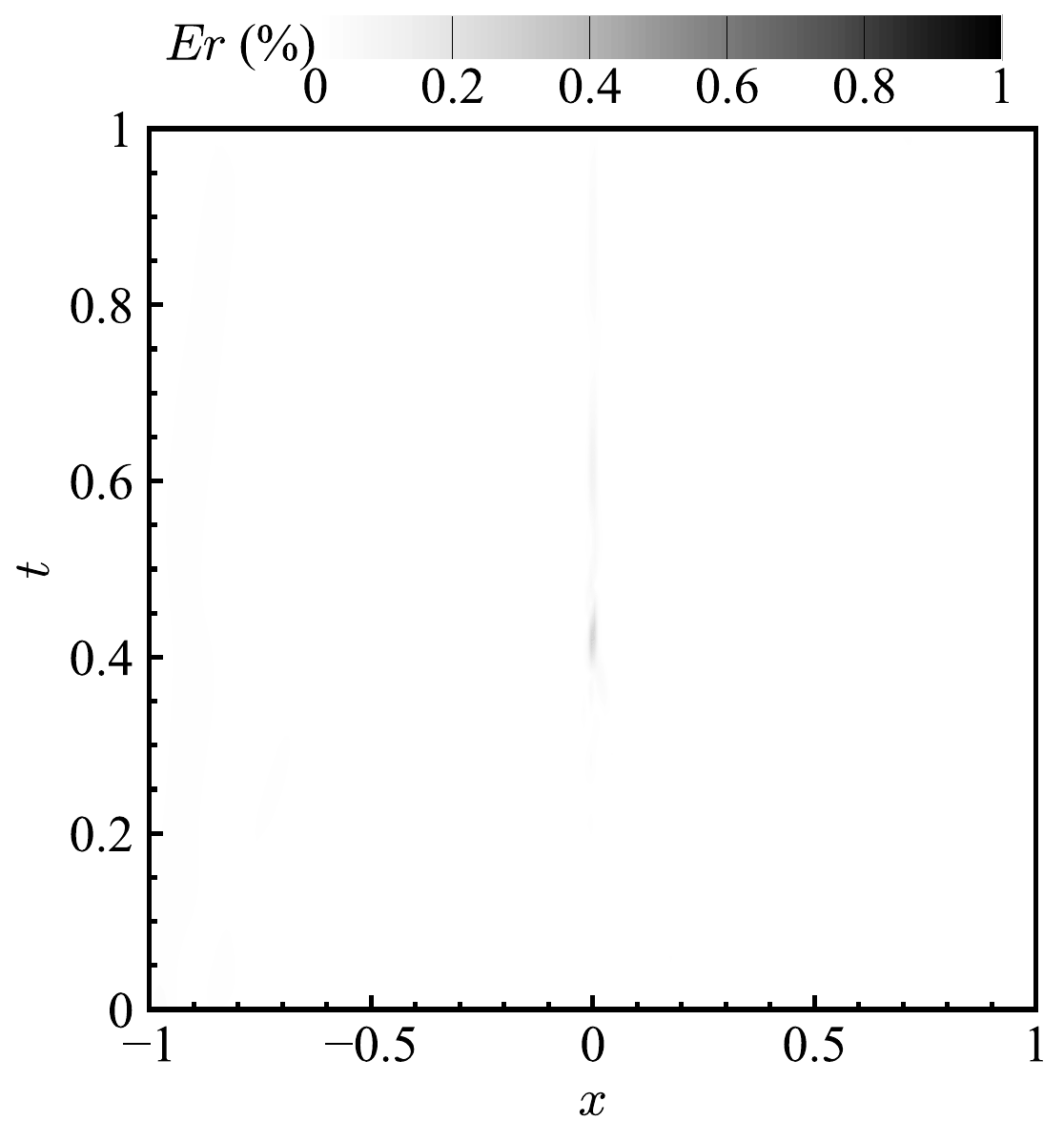}
}
\subfigure[PINN relative error]{
	\includegraphics[height=4cm]{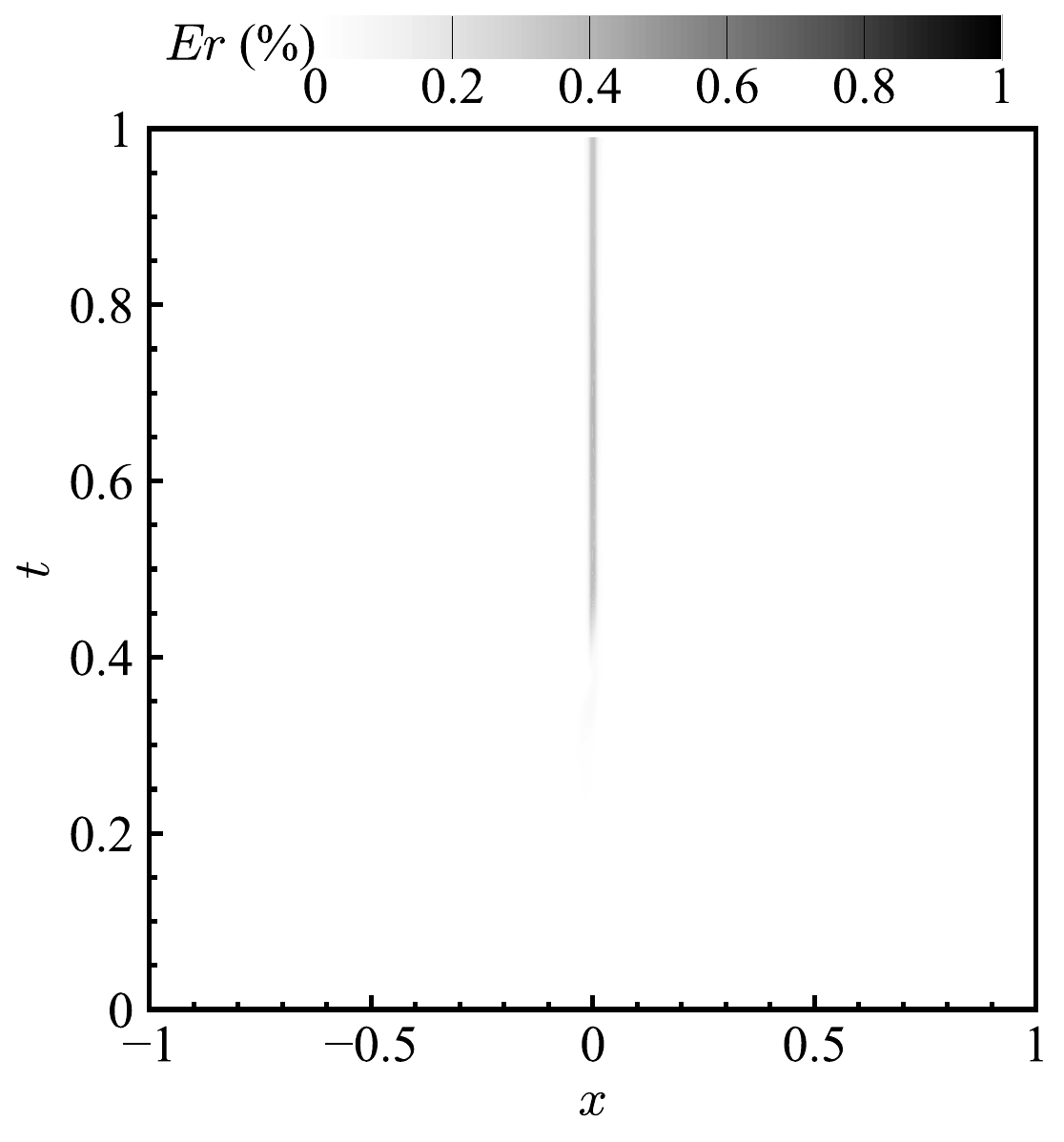}
}
	
	\caption{ Relative error contours predicted by PIFBNN and PINN models for the Burgers equation. }
	\label{bgcontour}
\end{figure}

Similar to the conclusion drawn above, when t<0.4 and no discontinuous solutions are generated, relative errors of PINN and PIFBNN are relatively low, demonstrating good predictive ability. However, when t>0.4 and discontinuous solutions are generated, PINN exhibits larger prediction errors, while PIFBNN maintains excellent prediction accuracy. Notably, both PINN and PIFBNN exhibited high relative error peaks around t = 0.45, due to the sudden appearance of discontinuous solutions, resulting in the generation of error peaks. However, during the later stages of training, PIFBNN seems to effectively learn the pattern of discontinuity. As a result, predicted error of PIFBNN quickly decreases to a stable average level. By contrast, PINN appears unable to capture the correct discontinuous solution, with its error remaining relatively high thereafter.
The relative error contour predicted by the PIFBNN and PINN models are presented in Figure \ref{bgcontour}.

It is evident that these two physics informed neural networks, except for slight errors in the presence of discontinuous solutions, correctly predict the solution of the Burgers equation in other regions. However, PINN has a relatively large relative error of prediction throughout the discontinuous region, whereas PIFBNN has only a minor error over the entire domain.
\subsection{Helmholtz equation}
The two-dimensional Helmholtz equation is widely used in physics and engineering, particularly in fields such as waves, acoustics, electromagnetics, and quantum mechanics. The form of the Helmholtz equation is closely related to the wave equation. However, it describes a steady-state or time-harmonic situation, representing the distribution of a certain wave in space.
We restrict our study to a two-dimensional case, specifically with the wave number $k_{0}=2 n \pi $, where $n=2$. Under this condition, the equation can be simplified as
\begin{equation}
	-u_{x x}-u_{y y}-k_{0}^{2} u=f, \quad \Omega=[0,1]^{2}.
\end{equation}

The Dirichlet boundary conditions and source terms are expressed as
\begin{equation}
	u(x, y)=0, \quad(x, y) \in \partial \Omega,
\end{equation}

\begin{equation}
	f(x, y)=k_{0}^{2} \sin \left(k_{0} x\right) \sin \left(k_{0} y\right).
\end{equation}

\begin{figure}[h!] 
	\centering
	{
		\includegraphics[height=5.5cm]{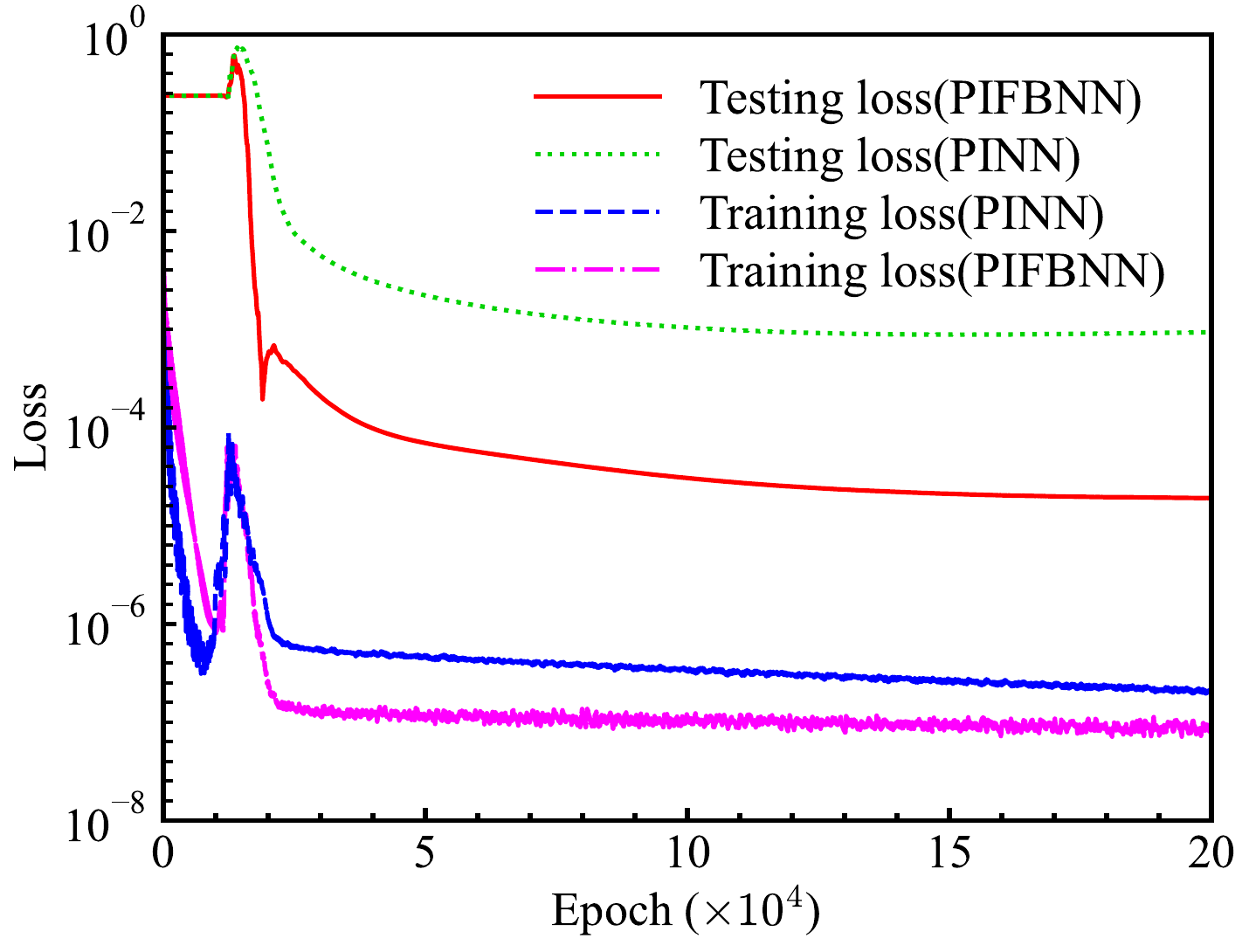}
	}
	
	\caption{Learning curves of  PINN and PIFBNN models for Helmholtz equation. }
	\label{hmpredloss}
\end{figure}

In this case, the exact solution to Helmholtz equation is
\begin{equation}
	u(x, y)=\sin \left(k_{0} x\right) \sin \left(k_{0} y\right),
\end{equation}
where $u_ {xx} $ and $ u_ {yy}$ represent the bending or curvature in the $x$- and $y$-directions, respectively, reflecting the local wave response in those directions. $k_{0}^{2}$ is related to the propagation characteristics of waves, reflecting the relationship between the wave amplitude and wave number. $f$ represents an external source applied to the system that affects the distribution of fluctuations or physical quantities.
It is difficult for general neural networks to accurately predict the Helmholtz equation due to its strong periodicity and repeated oscillation characteristics. This case can examine the ability of the proposed FBNN architecture to learn the periodicity and the repetitive oscillation behavior. 10 points per wavelength are selected along each of the four boundaries served as boundary conditions to train the neural network in predicting the solution of the entire domain. Additionally, 30 points per wavelength are sampled in each direction within the domain as test points to evaluate the prediction performance of the neural network. PINN and PIFBNN calculate residuals using 1600 randomly sampled points within the domain, and the labeled data are directly calculated from the analytical solution. Owing to the strong periodicity and nonlinearity of the Helmholtz equation, the ratio of Fourier nodes is set to 0.6.
The training and testing losses of PINN and PIFBNN are shown in Figure \ref{hmpredloss}.

\begin{figure}[h!] 
	\centering
	\subfigure[$y$=0.2m]{
		\includegraphics[height=5.5cm]{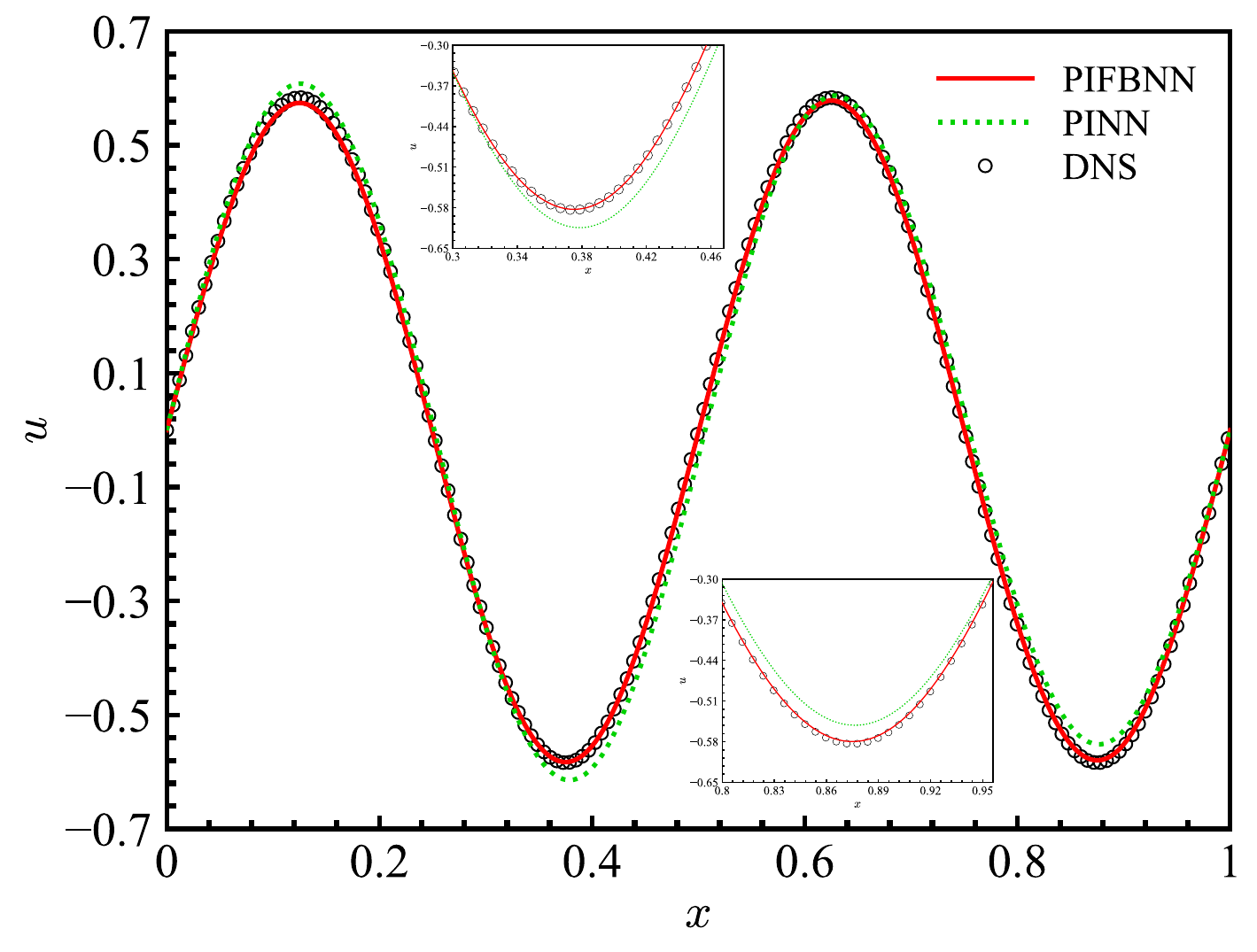}
	}
	\subfigure[$y$=0.8m]{
		\includegraphics[height=5.5cm]{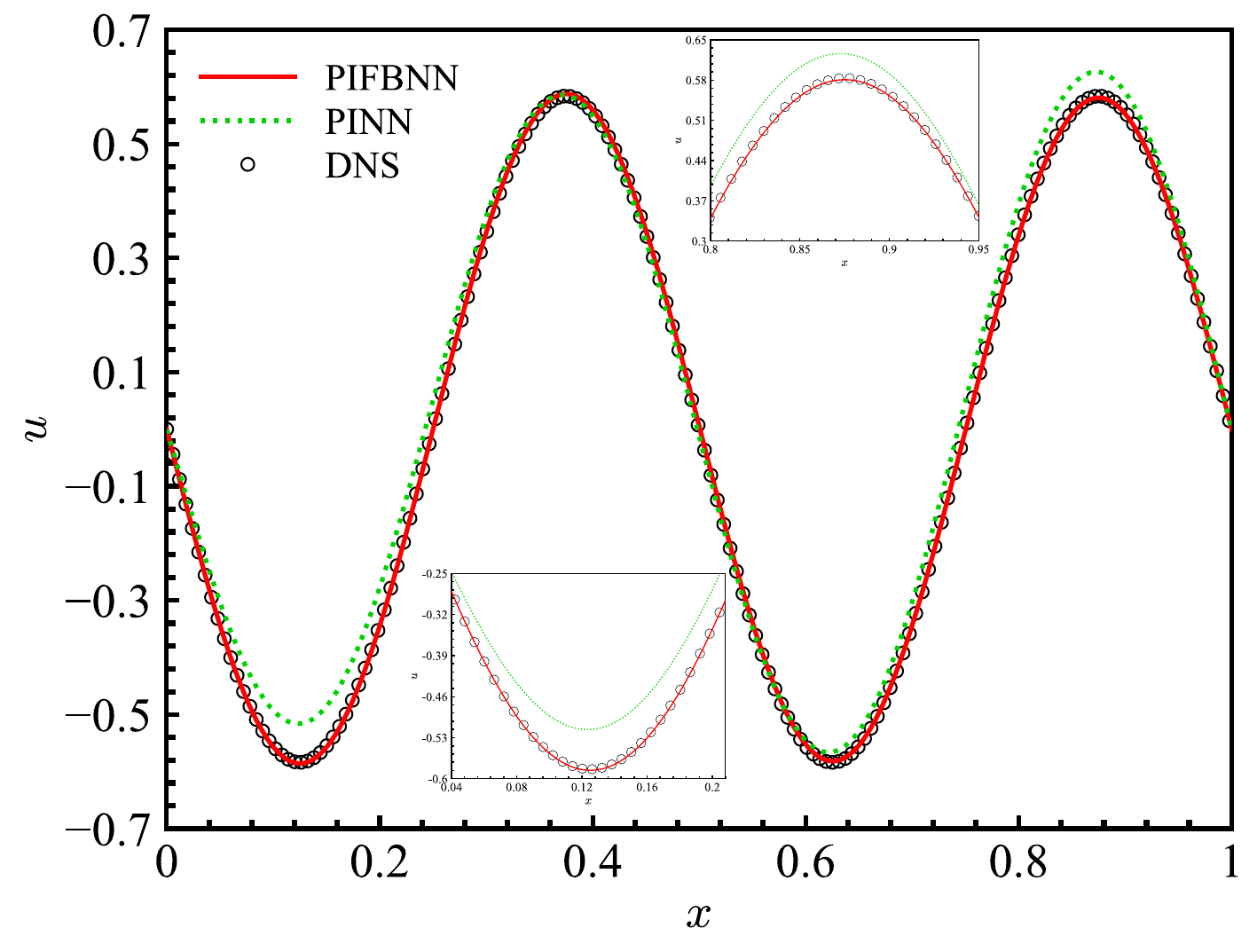}
	}
	
	\caption{Comparisons of velocity profiles using PINN and PIFBNN models for Helmholtz equation at $y$=0.2 and $y$=0.8. }
	\label{hmpred0.20.8y}
\end{figure}

\begin{figure}[h!] 
	\centering
	{
		\includegraphics[height=5.5cm]{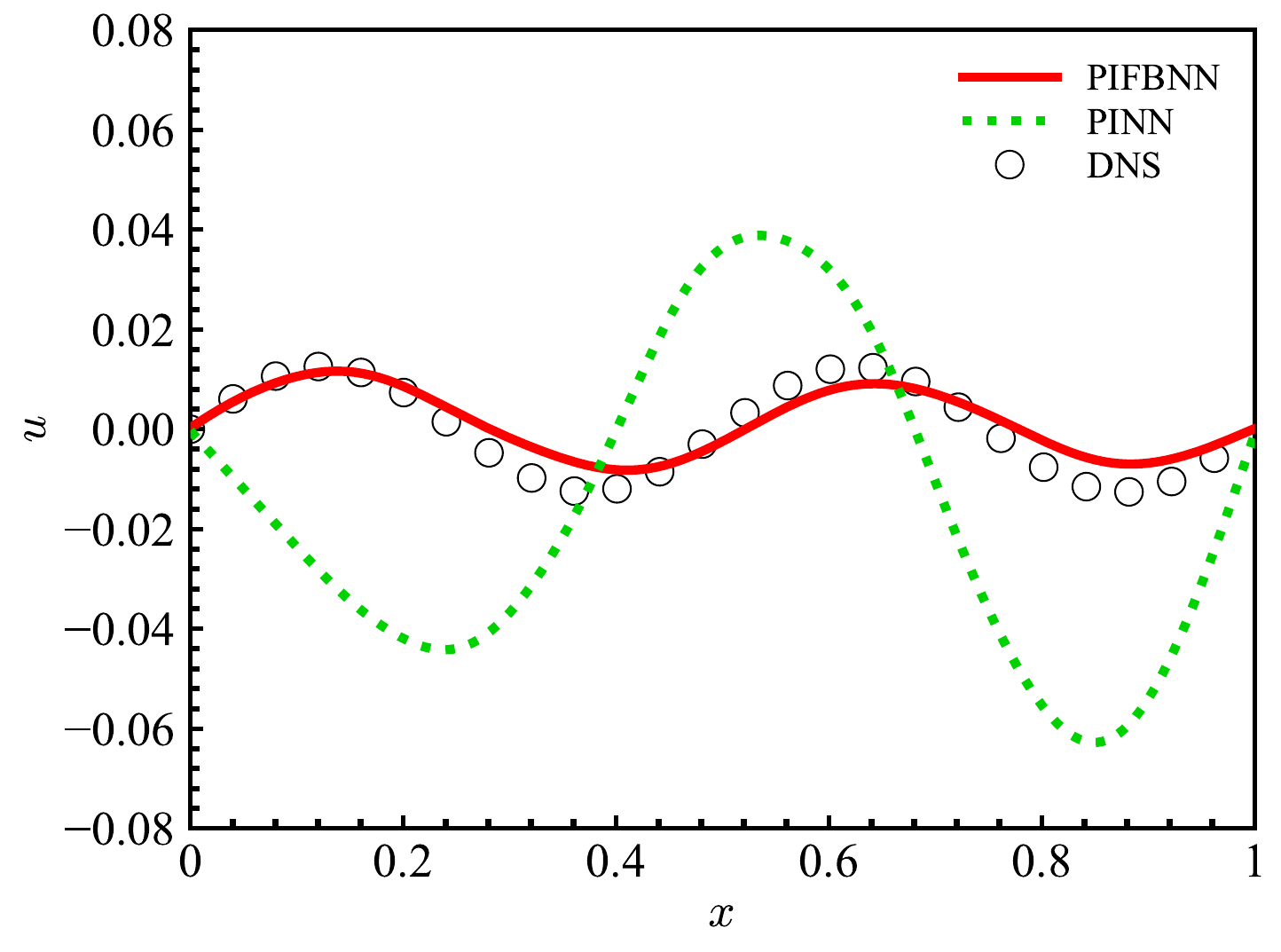}
	}
	
	\caption{Comparisons of velocity profiles using PINN and PIFBNN models for Helmholtz equation at  $y$=0.5 m  }
	\label{hmpred0.5y}
\end{figure}

There is a certain gap in the learning effects of PINN and PIFBNN on the boundary conditions. PIFBNN can better capture the periodicity of the boundary conditions, and thus accurately predict the solution for the entire domain. This indicates that PIFBNN has significantly better generalization than PINN in periodic modeling, and that PIFBNN is more likely to capture the periodic information of physical information residuals, achieving better prediction of periodic features.
The velocity profiles predicted by PINN and PIFBNN for Helmholtz equation at $y$=0.2 m and $y$=0.8 m are compared in Figure \ref{hmpred0.20.8y}.
It is evident that PINN has significant prediction errors in both peaks and valleys, with irregularity in its prediction performance. By contrast, the entire prediction curve of PIFBNN is consistent with the analytical solution.
The comparisons of velocity profiles using PINN and PIFBNN for the Helmholtz equation at $y$=0.5 m are shown in Figure \ref{hmpred0.5y}.  Note that the upper and lower bounds of the image are only $\pm$0.08 m.

\begin{figure}[h!] 
	\centering
	{
		\includegraphics[height=5.5cm]{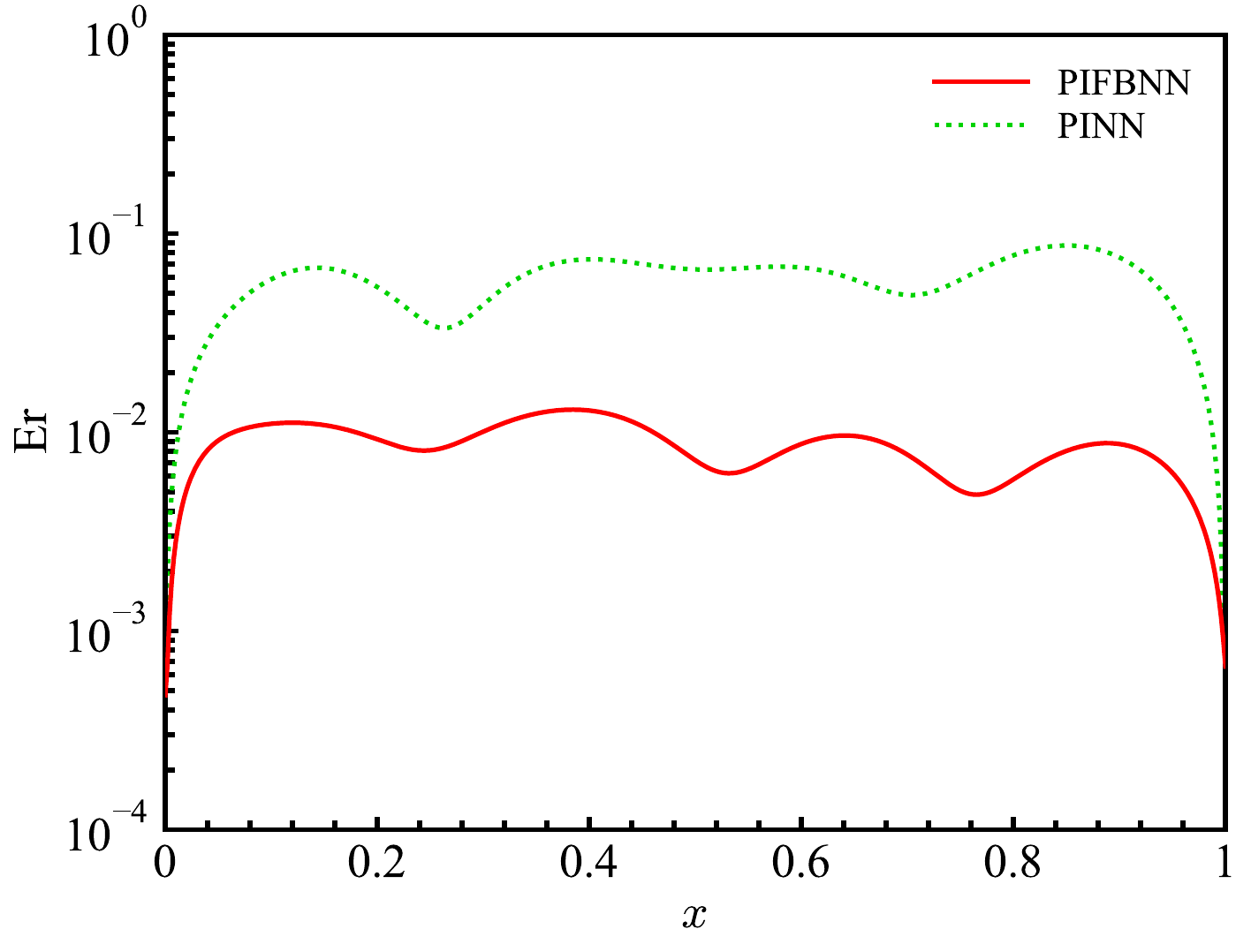}
	}
	
	\caption{Relative errors of PINN and PIFBNN models for Helmholtz equation. }
	\label{hmpreder}
\end{figure}

From Figure \ref{hmpred0.5y}, it is evident that PINN cannot accurately predict the periodicity when $y = 0.5$. PINN wrongly predicts valleys for occurrences where peaks exist. By contrast, PIFBNN yields pretty results that closely agree with the analytical solution. The relative error curves of PINN and PIFBNN predictions of the Helmholtz equation along $x$ are illustrate in Figure \ref{hmpreder}.

\begin{figure}[!htbp]
	\centering
\subfigure[PIFBNN relative error]{
	\includegraphics[height=4cm]{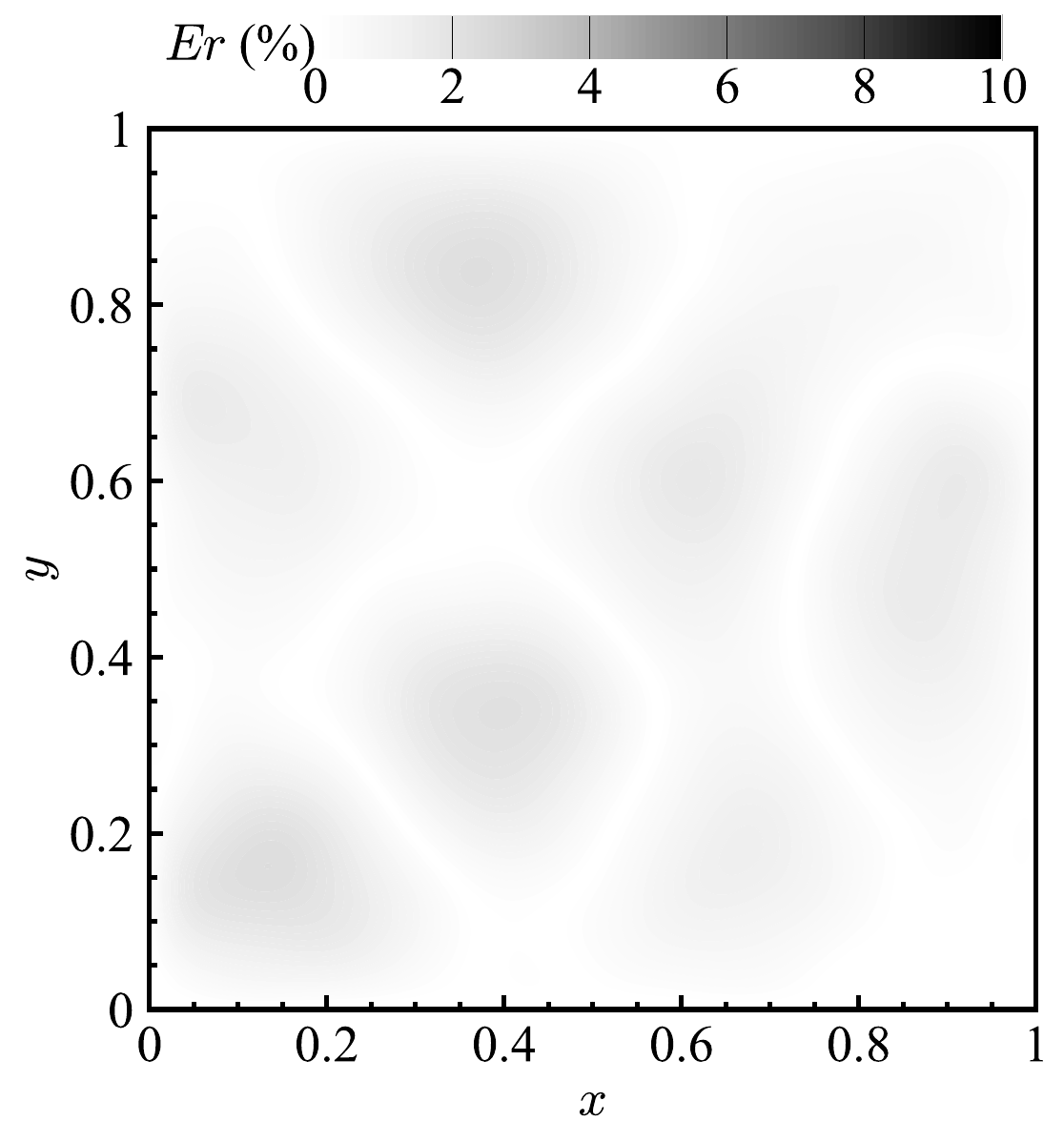}
}
\subfigure[PINN relative error]{
	\includegraphics[height=4cm]{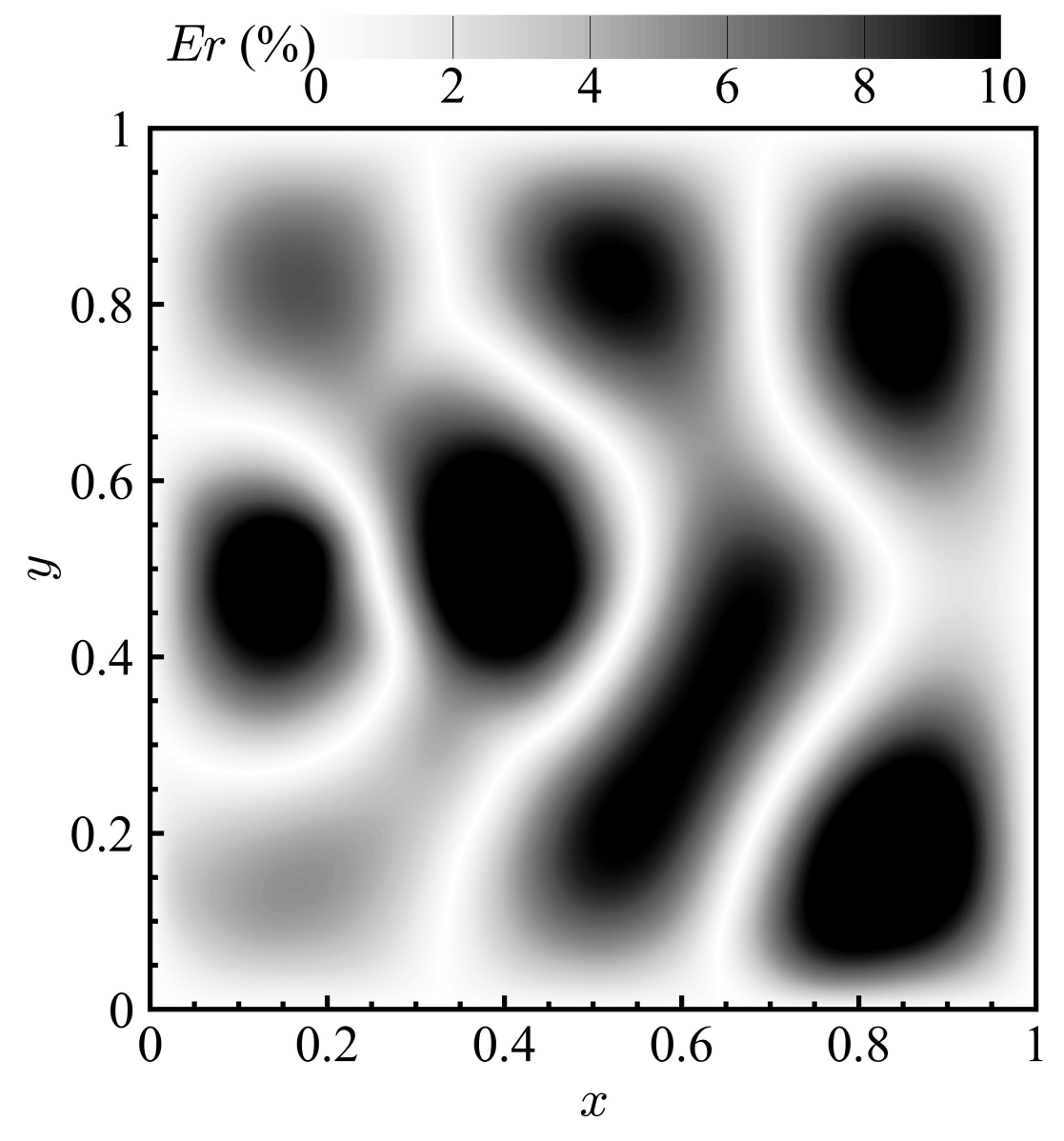}
}
	\caption{Relative error contours predicted by PIFBNN and PINN models for Helmholtz equation. }
	\label{hmpredres}
\end{figure}

Both PINN and PIFBNN have relatively low prediction errors owing to boundary constraints at $x = 0$ and 1. As the prediction gradually approaches the domain, both errors begin to increase slowly, and periodic changes occur with the periodic variation of the function. When the nonlinear characteristics are enhanced within a period, the relative errors generally grow up. The error of PIFBNN is always significantly smaller than that of PINN.
The relative error contours predicted by the PIFBNN and PINN are shown in Figure \ref{hmpredres}.

Similar to the relative error curve, the prediction errors of both PINN and PIFBNN exhibited periodic variations with changes in the function period, with the most obvious errors occurring in areas with strong nonlinearity. However, the overall error of PIFBNN is significantly lower than that of PINN, with relative errors significantly lower than 10\%. PINN may exhibit an error exceeding 10\% in regions characterized by significant nonlinearity. The prediction of Helmholtz equation serves as a representative to highlight the ability of PIFBNN to model periodicity and capture nonlinear characteristics.
\subsection{Kovasznay flow}
Kovasznay flow equation is an idealized model used to describe the steady-state flow of a single-phase incompressible fluid within an infinite domain. It is used to study the transition from laminar to turbulent flows. Kovasznay flow is usually expressed as a two-dimensional solution to the Navier–Stokes equations, assuming that the fluid at infinity is uniform and stationary. The governing equations can generally be expressed as
\begin{equation}
	u \frac{\partial u}{\partial x}+v \frac{\partial u}{\partial y}=-\frac{\partial p}{\partial x}+\frac{1}{R e}\left(\frac{\partial^{2} u}{\partial x^{2}}+\frac{\partial^{2} u}{\partial y^{2}}\right),
\end{equation}

\begin{equation}
	u \frac{\partial v}{\partial x}+v \frac{\partial v}{\partial y}=-\frac{\partial p}{\partial y}+\frac{1}{R e}\left(\frac{\partial^{2} v}{\partial x^{2}}+\frac{\partial^{2} v}{\partial y^{2}}\right),
\end{equation}
with Dirichlet boundary conditions defined as 
\begin{equation}
	u(x, y)=0, \quad(x, y) \in \partial \Omega,
\end{equation}
where $u$ and $v$ are respectively streamwise and transverse velocities, and $p$ is pressure. The dimensionless quantity $Re$ denotes Reynolds number.The exact solutions for $u$, $v$, and $p$ in this equation are given by
\begin{equation}
	u=1-e^{\lambda x} \cos (2 \pi y),
\end{equation}

\begin{equation}
	v=\frac{\lambda}{2 \pi} e^{\lambda x} \sin (2 \pi x),
\end{equation}

\begin{equation}
	p=\frac{1}{2}\left(1-e^{2 \lambda x}\right).
\end{equation}

Here, the parameter is $\lambda=\frac{Re}{2 }-\sqrt{\frac{Re^2}{4 }+4 \pi^{2}}$.
This equation represents a simplified version of the Navier--Stokes equations in fluid dynamics, focusing on the conservation of momentum in the fluid flow. 

\begin{figure}[h!] 
	\centering
	\subfigure[Training loss]{
		\includegraphics[height=5.5cm]{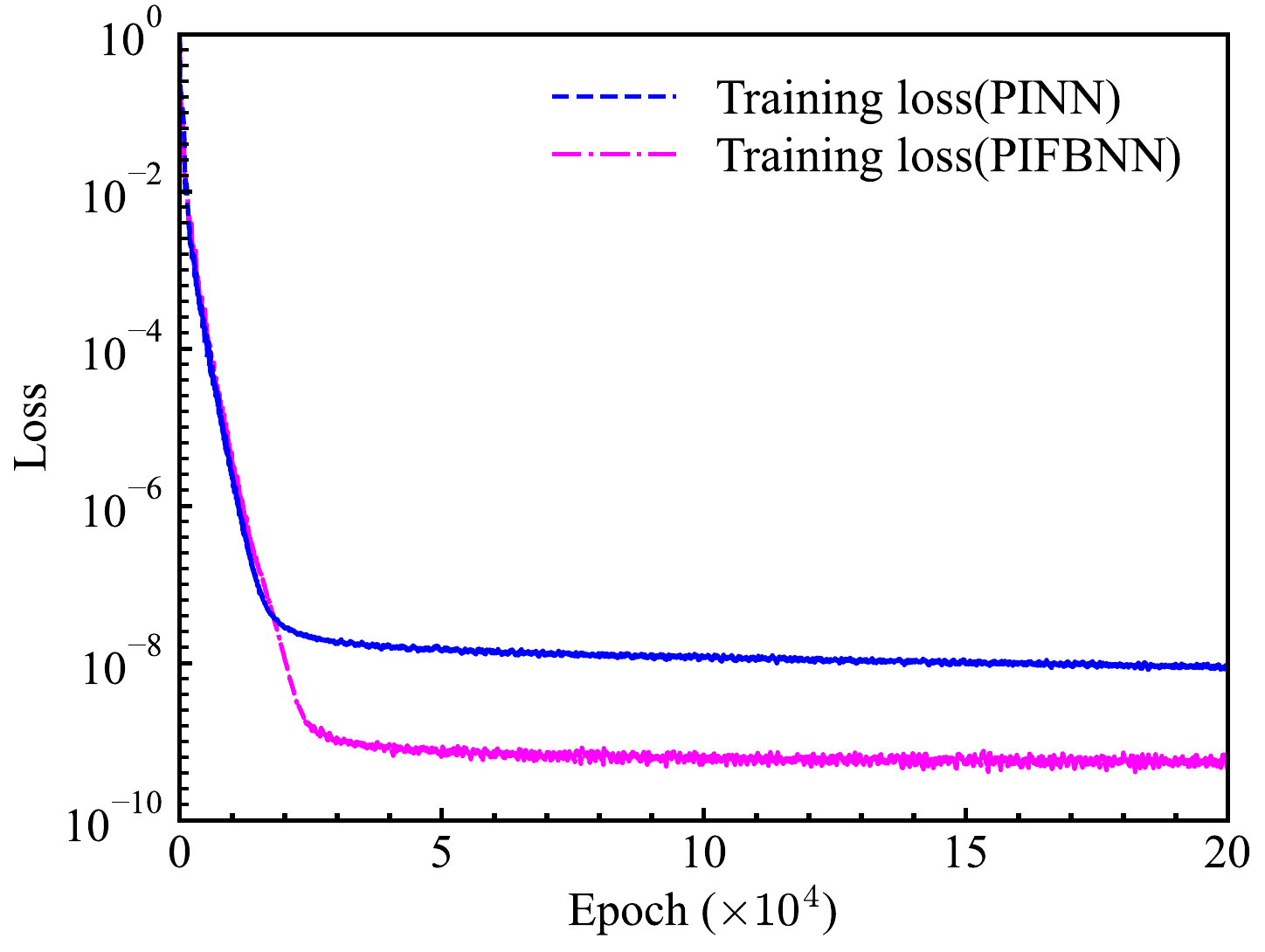}
	}
	\subfigure[Testing loss]{
		\includegraphics[height=5.5cm]{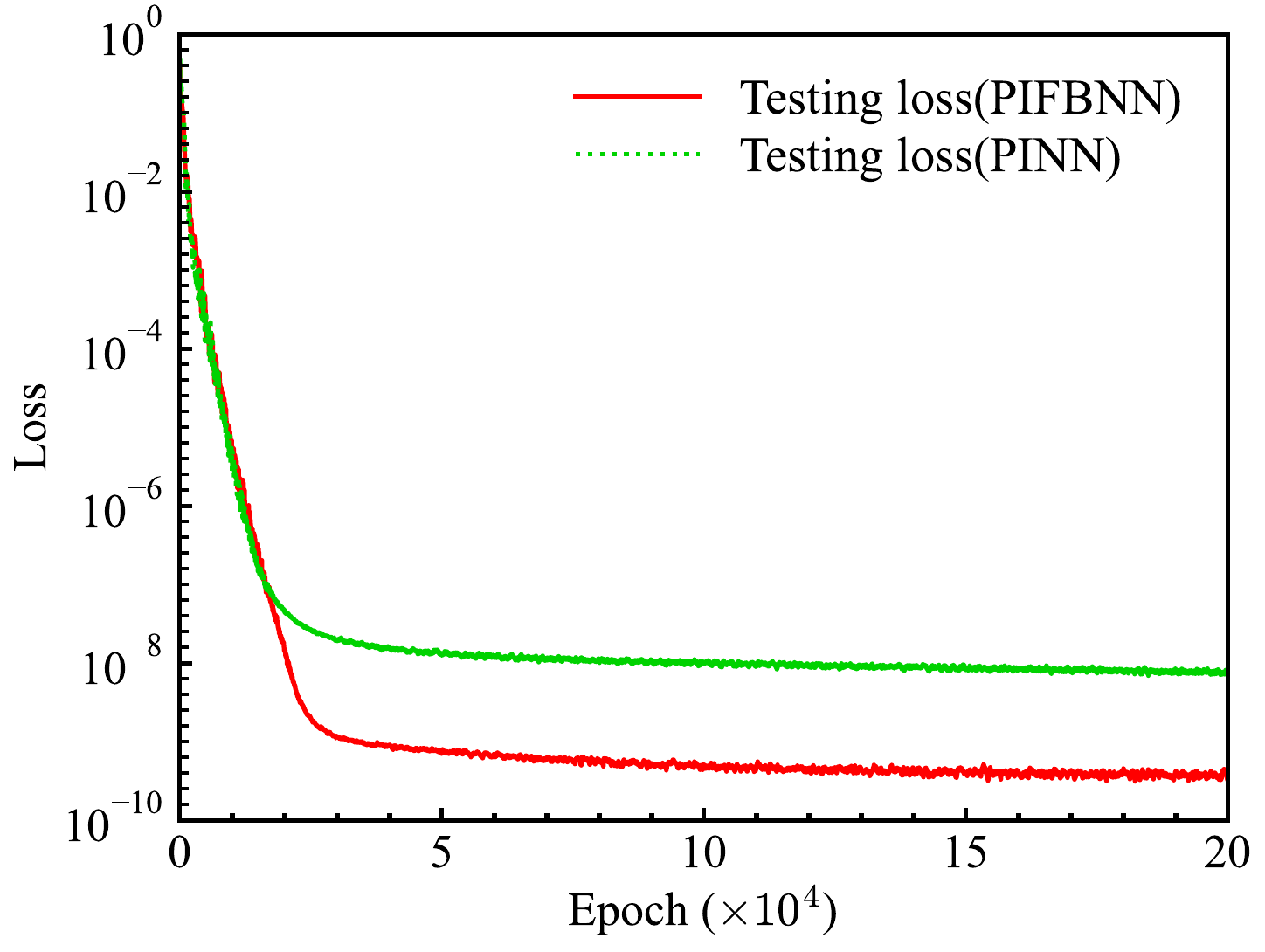}
	}
	
	\caption{Learning curves of PINN and PIFBNN models for Kovasznay flow. }
	\label{kovpredloss}
\end{figure}

The domain of Kovasznay flow is $x\in[-0.5,1]$ and $y\in[-0.5,1.5]$. A total of 101 points are randomly selected as the boundary condition constraints for each boundary, and 2601 internal points are randomly sampled within the domain to calculate the physical information residuals. This evaluation targets to train neural networks to predict the entire computational domain using only boundary conditions and governing equation constraints. A total of 400 points are randomly sampled within the domain as test points to evaluate the prediction accuracy of the neural network, with their corresponding labels obtained directly from the analytical solution.
The training and testing losses of PIFBNN and PINN are displayed in Figure \ref{kovpredloss}.

\begin{figure}[h!] 
	\centering
	\subfigure[Streamwise flow velocity: $u$]{
		\includegraphics[height=3.5cm]{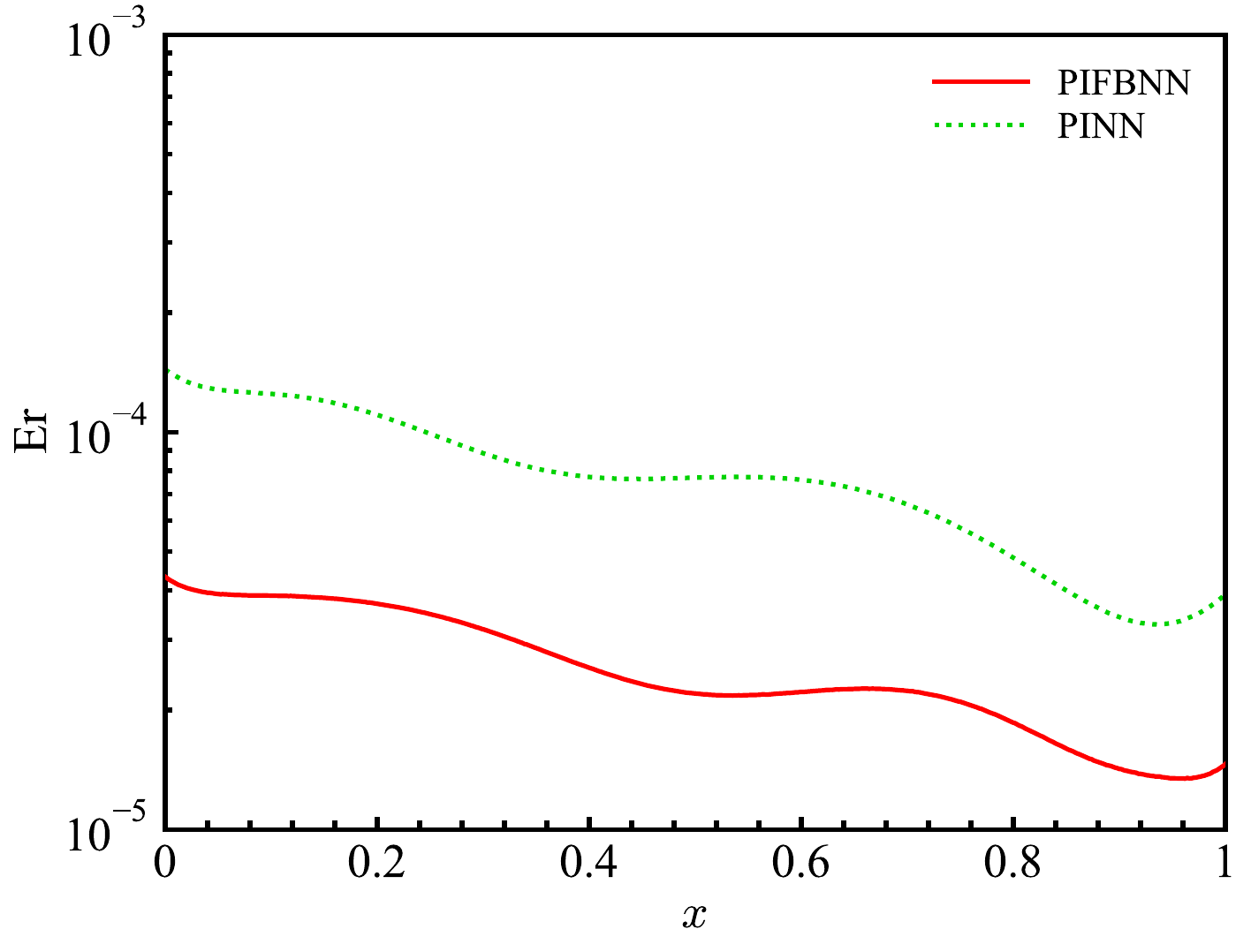}
	}
	\subfigure[Transverse velocity: $v$]{
		\includegraphics[height=3.5cm]{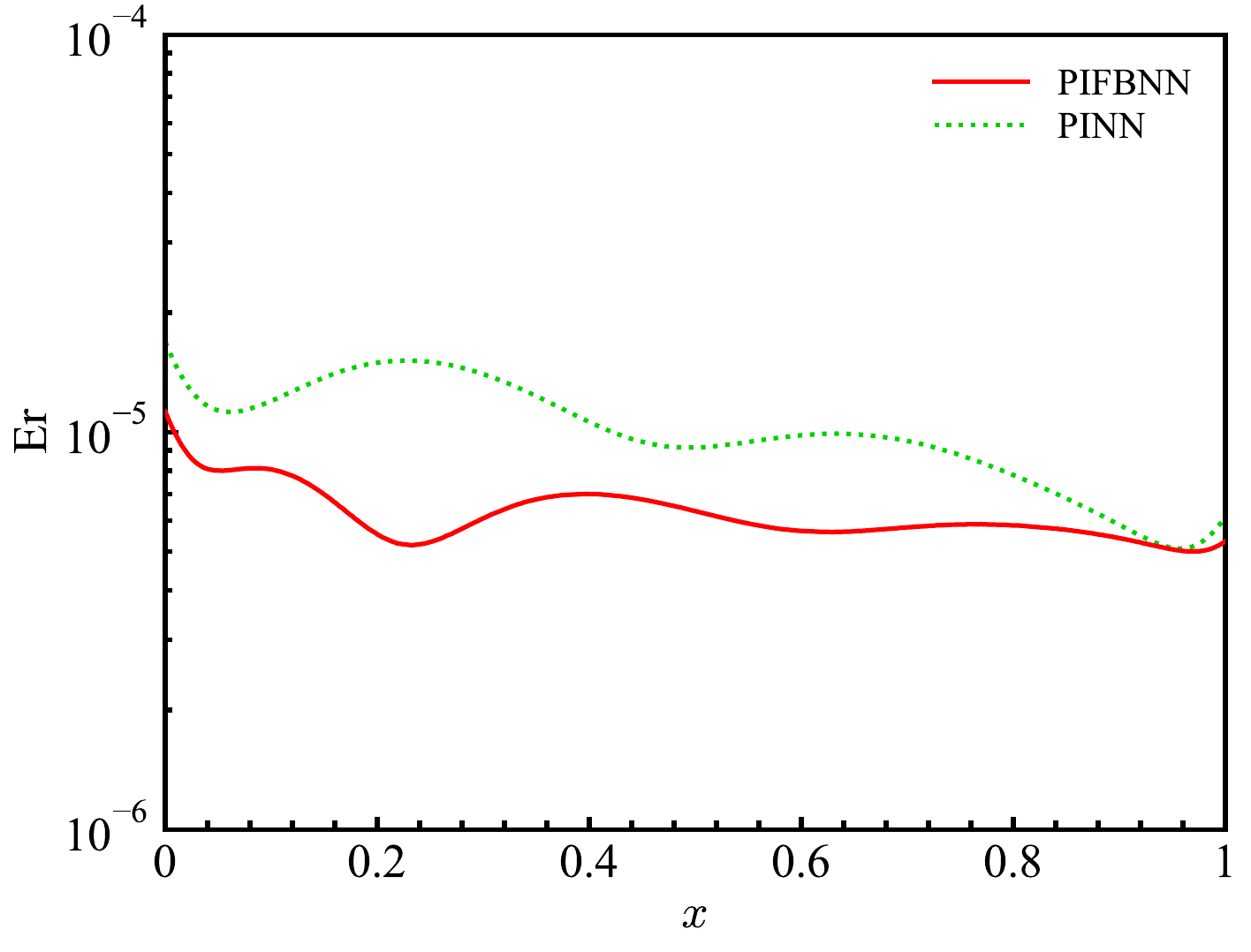}
	}
	\subfigure[Pressure: $p$]{
		\includegraphics[height=3.5cm]{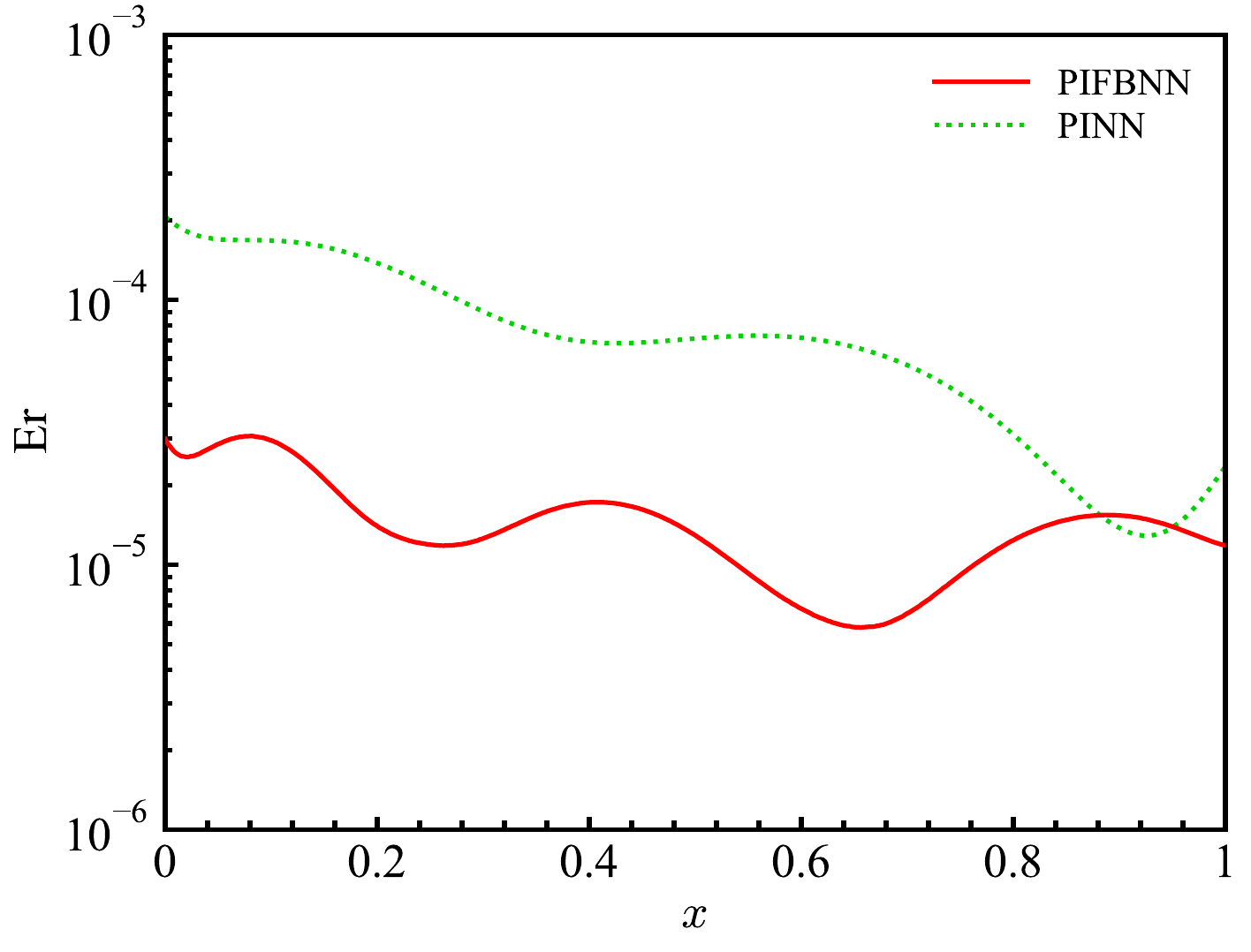}
	}
	
	\caption{Relative errors of PINN and PIFBNN models for Kovasznay flow. }
	\label{kovpreder}
\end{figure}

It can be seen that the learning efficiency of boundary conditions directly affects the prediction performance of the entire region without overfitting, and both depict good generalization performance. The training loss of PIFBNN is significantly lower than that of PINN, indicating that it learns the boundary conditions effectively. Consequently, the testing loss of PIFBNN is lower than that of PINN.
The relative errors of PINN and PIFBNN's predictions for Kovasznay flow along $x$ direction are demonstrated in Figure \ref{kovpreder}.

\begin{figure}[!htbp]
	\centering
	
	\includegraphics[height=4cm]{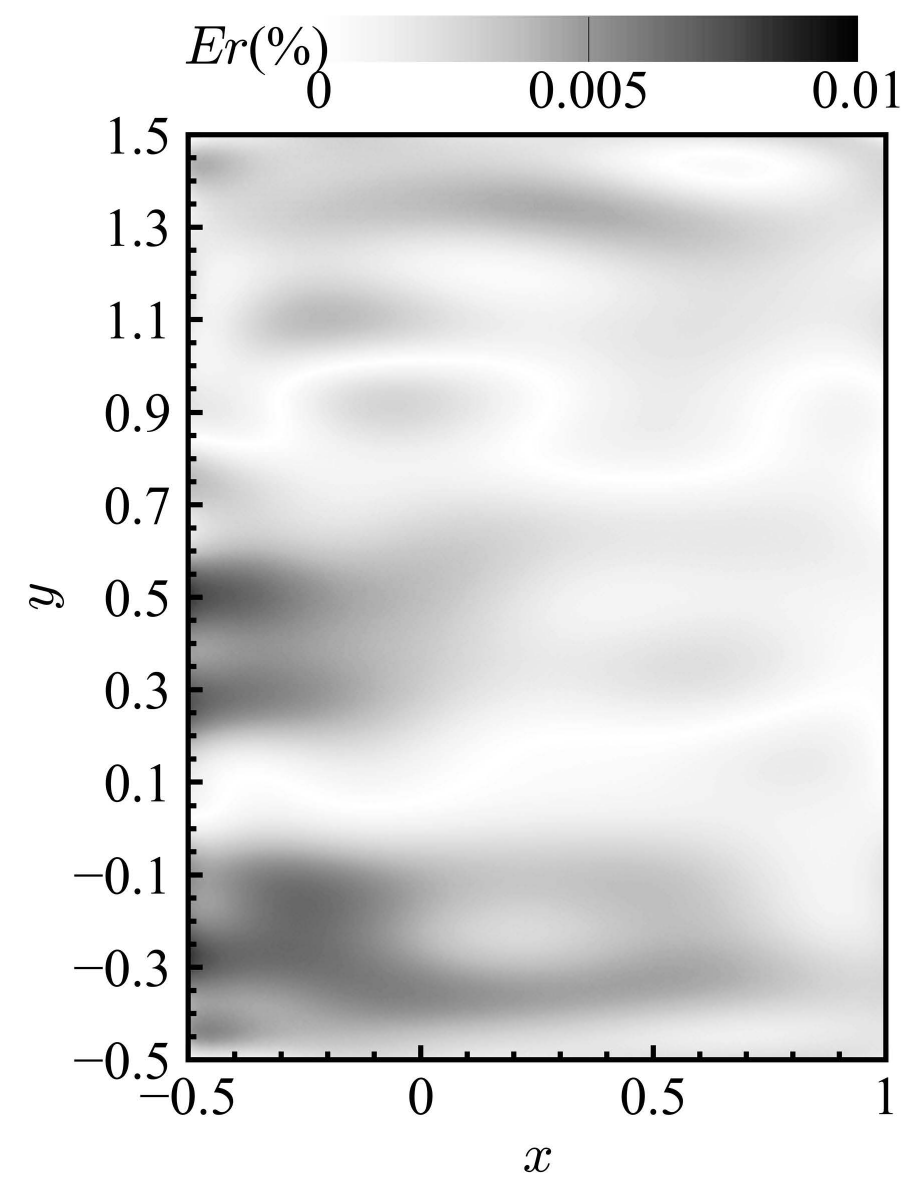}
	\hspace{1.2cm}
	\includegraphics[height=4cm]{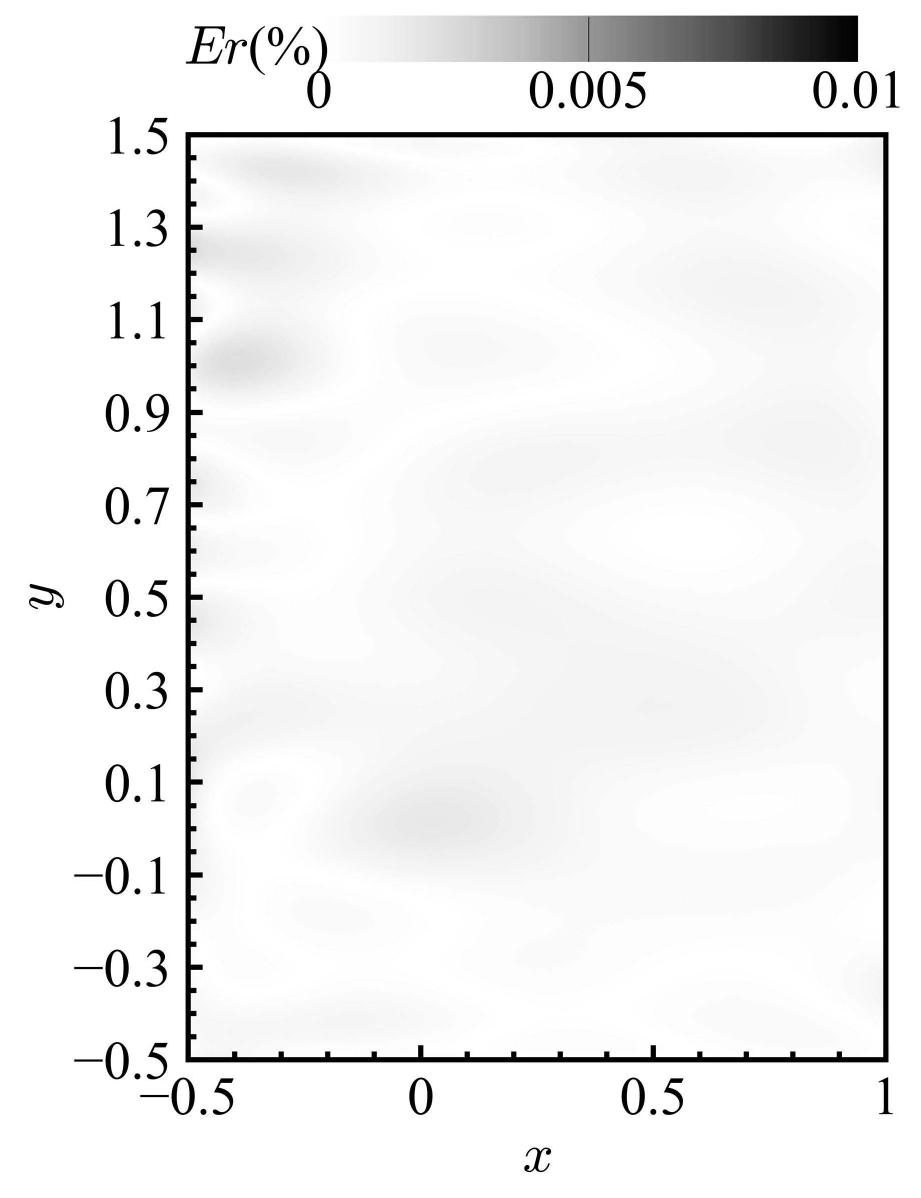}
	\hspace{1.2cm}
	\includegraphics[height=4cm]{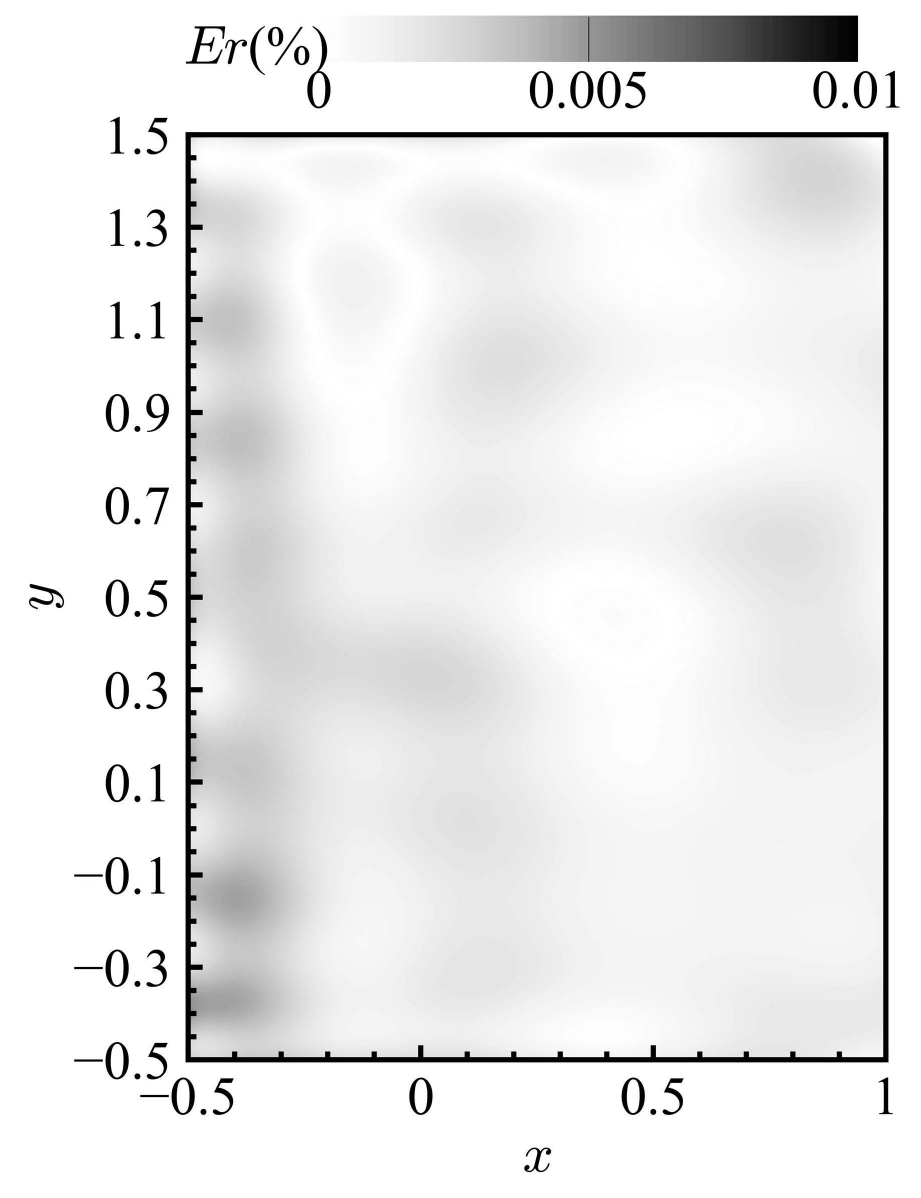}
	
	\par %
	\subfigure[Streamwise flow velocity: $u$]{
		\includegraphics[height=4cm]{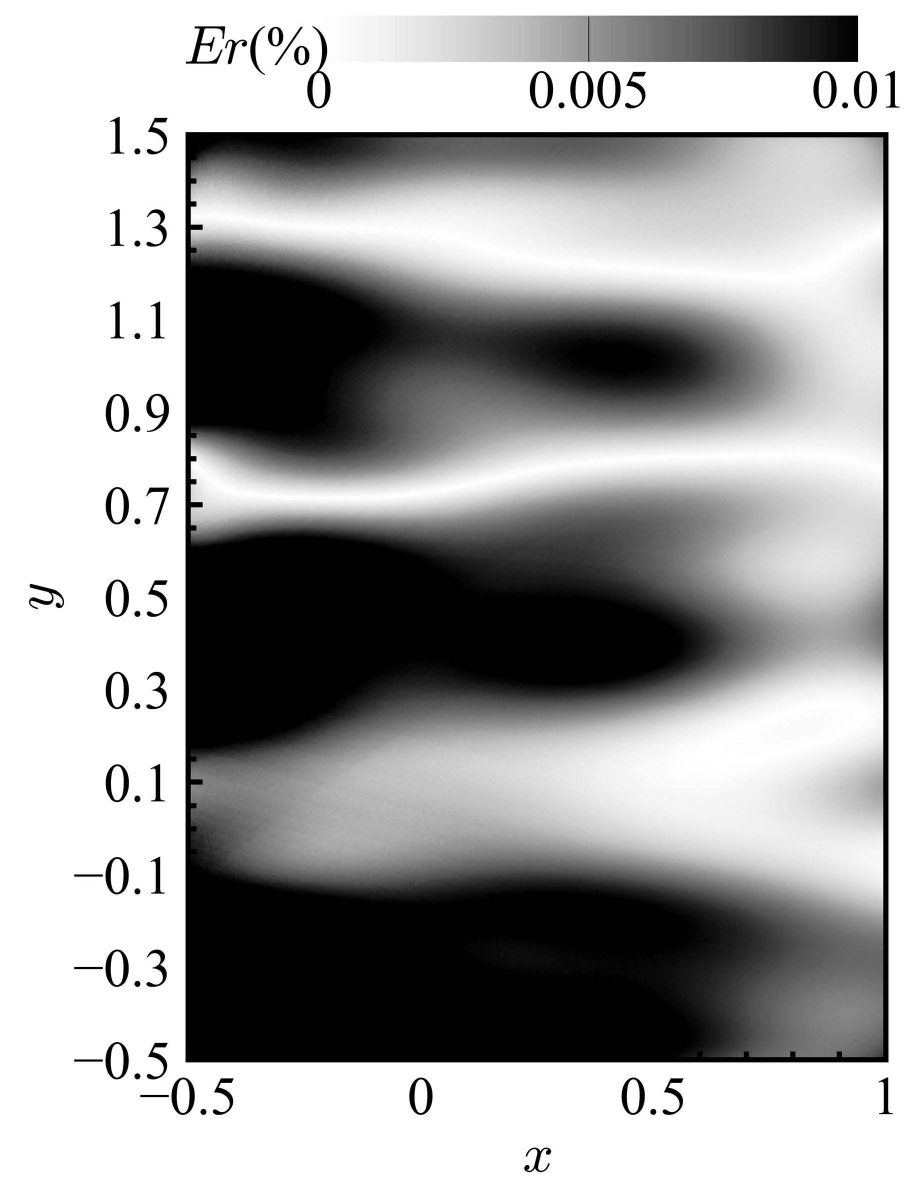}
	}
	\hspace{1cm}
	\subfigure[Transverse velocity: $v$]{
		\includegraphics[height=4cm]{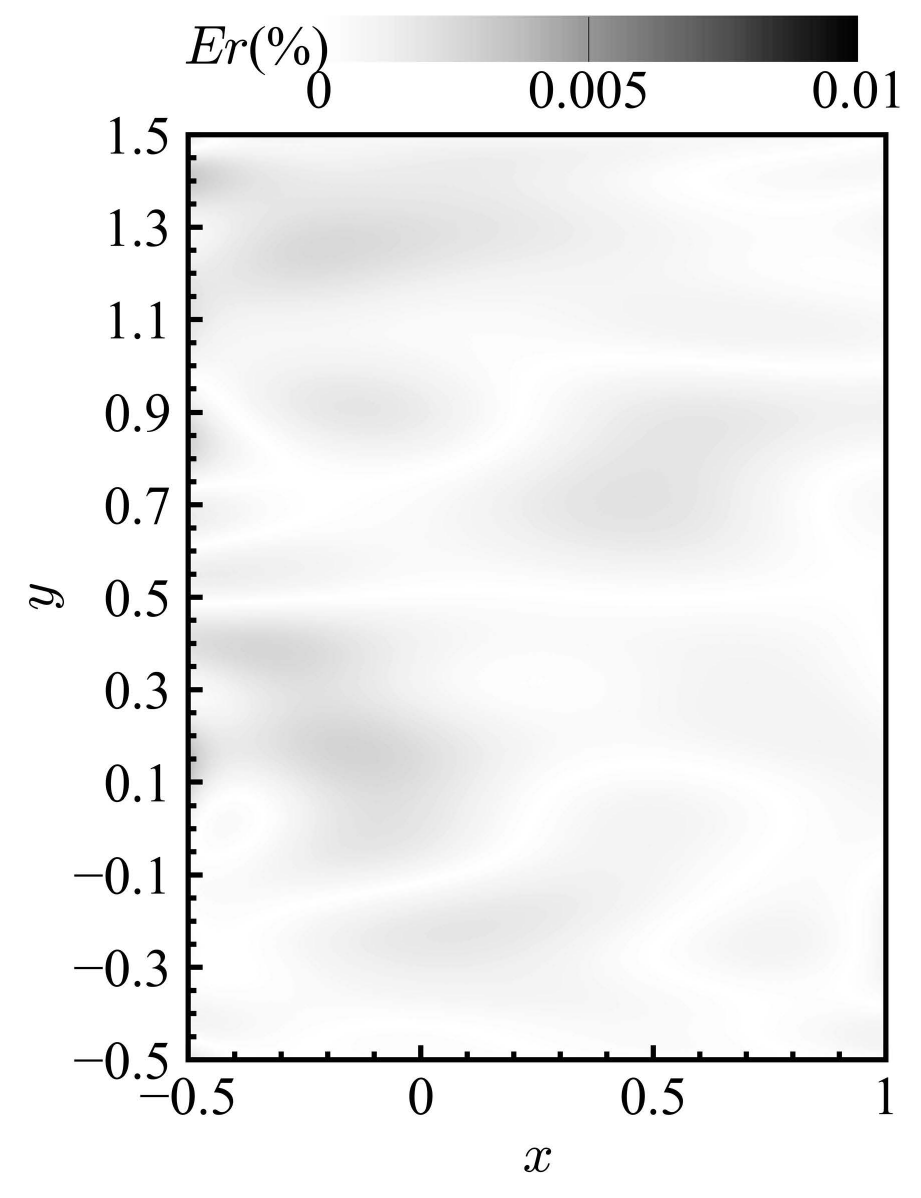}
	}
	\hspace{1cm}
	\subfigure[Pressure: $p$]{
		\includegraphics[height=4cm]{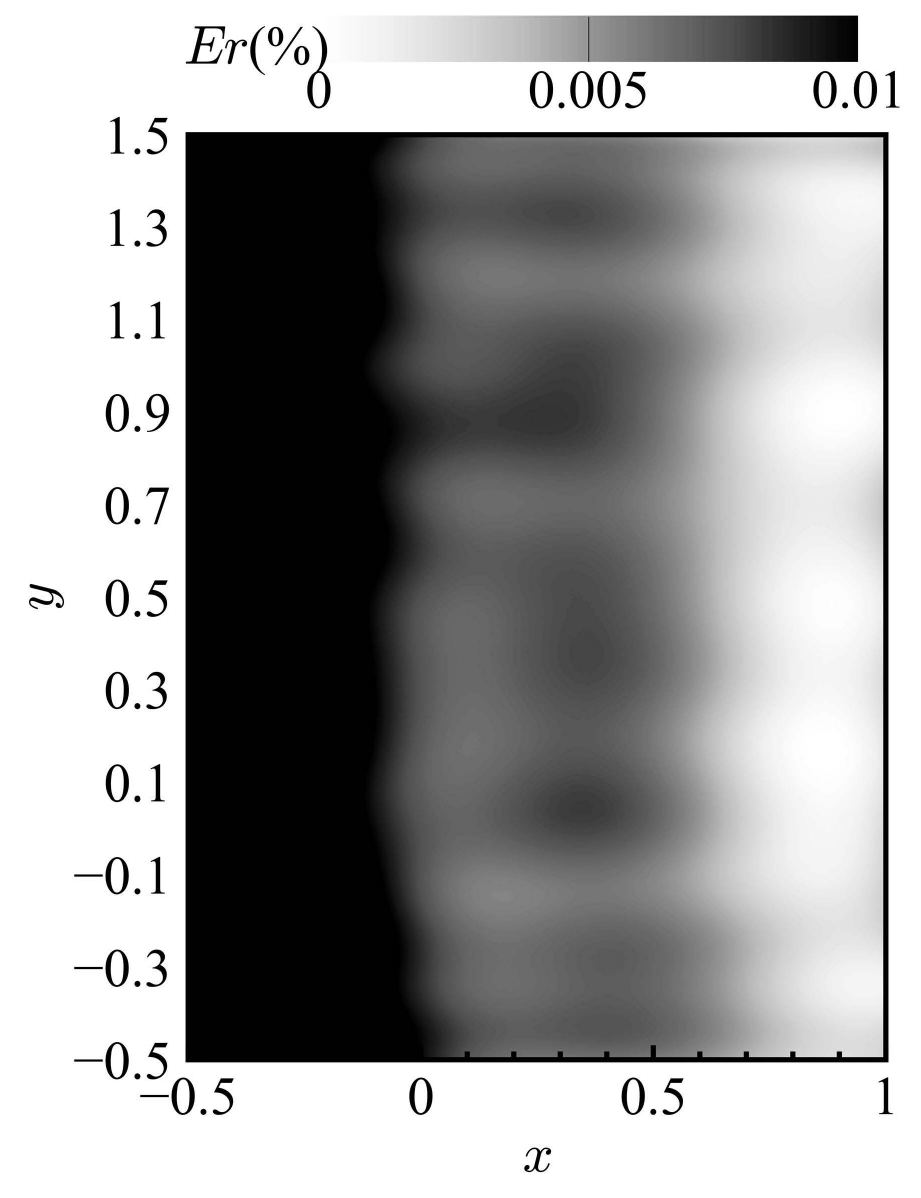}
	}
	
	\caption{Relative error contours using PIFBNN and PINN models for the Kovasznay flow: the first line is predicted by PIFBNN, the second line is predicted by PINN.}
	\label{kovpredres}
\end{figure}

It can be observed that relative errors predicted by PIFBNN for all variables:$ u, v$ and $p$ are significantly smaller than that by PINN. When $x$ approaches to zero, the Kovasznay flow has a large velocity gradient, hence the prediction error is generally large in the early stage of $x$. As the value of $x$ increases, the shock wave gradually subsides, and the error gradually decreases. Among all predicted results, $v$ had the lowest prediction error. Within the computational domain, the region and the value of high-velocity gradient of $v$ is less compared with u. When $x$ is larger than zero, the shock wave ends, making it easier for neural networks to capture the fluid characteristics. The shock wave of $u$ occupies the entire $x$-direction, and the velocity gradient is huge, therefore, the overall prediction error is relatively large, and the error significantly decreases with a decrease of velocity gradient in the later stage of $x$. The prediction of pressure $p$ is generally close to or equal to zero when $x<0.2$, which can lead to a large relative error. As the $x$-region moves backward, $p$ gradually increases, and the relative error gradually decreases.
The relative error contours predicted by PINN and PIFBNN for the Kovasznay flow are demonstrated in Figure \ref{kovpredres}.

Consistent with the previous results, the relative error of $u$ had a denser distribution in the high-speed region. The relative error of $v$ is lower in the entire region owing to the presence of small velocity gradients and high-speed area distributions. The relative error of $p$ is mostly concentrated in the low-pressure and zero-pressure regions when $x$ is low. It can be seen that among all the results, the relatively poor predictions of PIFBNN are significantly less severe than those of PINN.

\subsection{Two-dimensional cylinder wake flow}
The cylinder wake flow is a classic incompressible flow governed by the Navier--Stokes equation. There are many representative physical phenomena in cylindrical wake flows, such as the shedding of wake vortices (Kármán vortex streets) when $Re>47$. Studying the shedding and flow phenomena of two-dimensional cylindrical vortices is important for determining the flow separation laws and fluid viscosity phenomena under boundary conditions. The governing equation for the two-dimensional cylinder wake flow is the incompressible two-dimensional Navier--Stokes equation, defined as
\begin{equation}
	\frac{\partial u}{\partial t}+u \frac{\partial u}{\partial x}+v \frac{\partial u}{\partial y}=-\frac{\partial p}{\partial x}+\nu\left(\frac{\partial^{2} u}{\partial x^{2}}+\frac{\partial^{2} u}{\partial y^{2}}\right),
\end{equation}
\begin{equation}
	\frac{\partial v}{\partial t}+u \frac{\partial v}{\partial x}+v \frac{\partial v}{\partial y}=-\frac{\partial p}{\partial y}+\nu\left(\frac{\partial^{2} v}{\partial x^{2}}+\frac{\partial^{2} v}{\partial y^{2}}\right),
\end{equation}
\begin{equation}
	\frac{\partial u}{\partial x} + \frac{\partial v}{\partial y} = 0,
\end{equation}
where $u$ and $v$ are respectively streamwise and transverse velocities, and $p$ is pressure. The dimensionless quantity $\nu$ denotes kinematic viscosity. In this case, the following values are considered: a free-flow velocity of $u=1, v=0$, cylinder diameter $D=1$, and kinematic viscosity $\nu=0.01$. The fluid flow develops into a periodic steady state, and a Kármán vortex street appears in the wake. We refer to Raissi et al.\cite{raissi2019physics} and used the spectral/HP element solver Nektar++\cite{karniadakis2005spectral} to obtain the dataset. A uniform free-flow velocity distribution is applied to the left boundary, and a zero-pressure outflow condition is applied to the right boundary at a diameter of 25 mm downstream of the cylinder. The computational domain is $[-15,25] \times[-8,8]$, and periodic boundary conditions are applied at the top and bottom. For the convenience of neural network prediction, we only consider spatial domain $x\in[1,8] $, $y\in[-2,2]$ and temporal domain $t\in[0,7]$.

To achieve prediction within this small computational domain, Dirichlet boundary conditions are used as training data at every 0.1 $t$ interval. Specifically, 100 points on the upper and lower boundaries and 50 points on the left and right boundaries are sampled as Dirichlet boundary conditions at every interval. A total of 14000 collection points are sampled throughout the entire computational domain to calculate the physical residuals, and 8000 labeled points are selected to evaluate the performance of neural networks. The proposed neural networks use ten hidden layers, each with 100 neurons and employ $tanh$ as the activation function. The target of neural networks is to predict the velocity $(u, v)$ and pressure $p$ across the entire domain using only Dirichlet boundary conditions and governing N--S equation, without using any labeled data. Due to the time domain $t\in[0,7]$ only contains approximately two cycles, the periodicity is not evident in this case. Therefore, a FBNN framework with a Fourier nodes ratio of 0.3 is used.
\begin{figure}[h!] 
	\centering
	{
		\includegraphics[height=5.5cm]{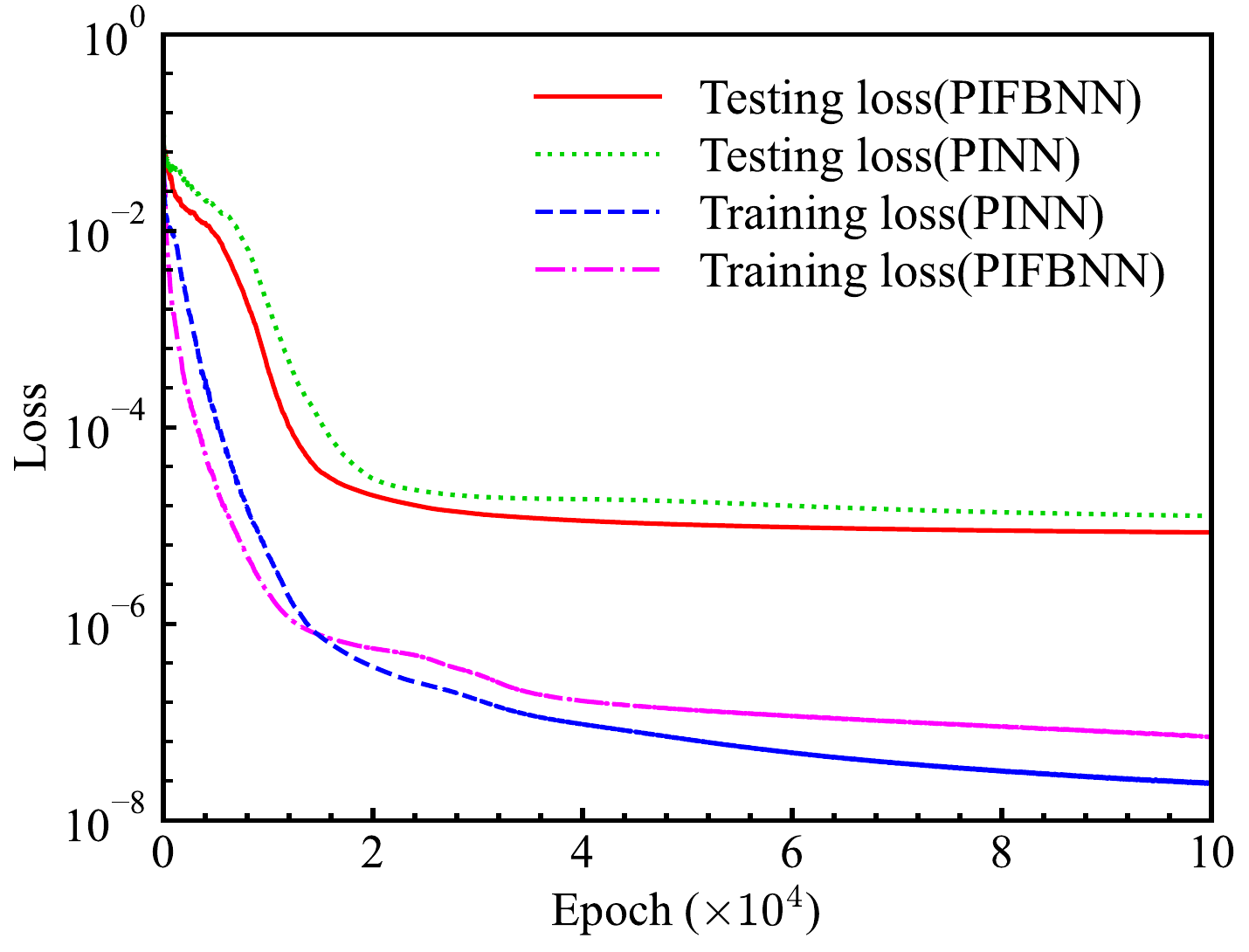}
	}
	
	\caption{Learning curves of PINN and PIFBNN models for cylinder wake flow. }
	\label{cypredloss}
\end{figure}
\begin{figure}[h!] 
	\centering
	{
		\includegraphics[height=5.5cm]{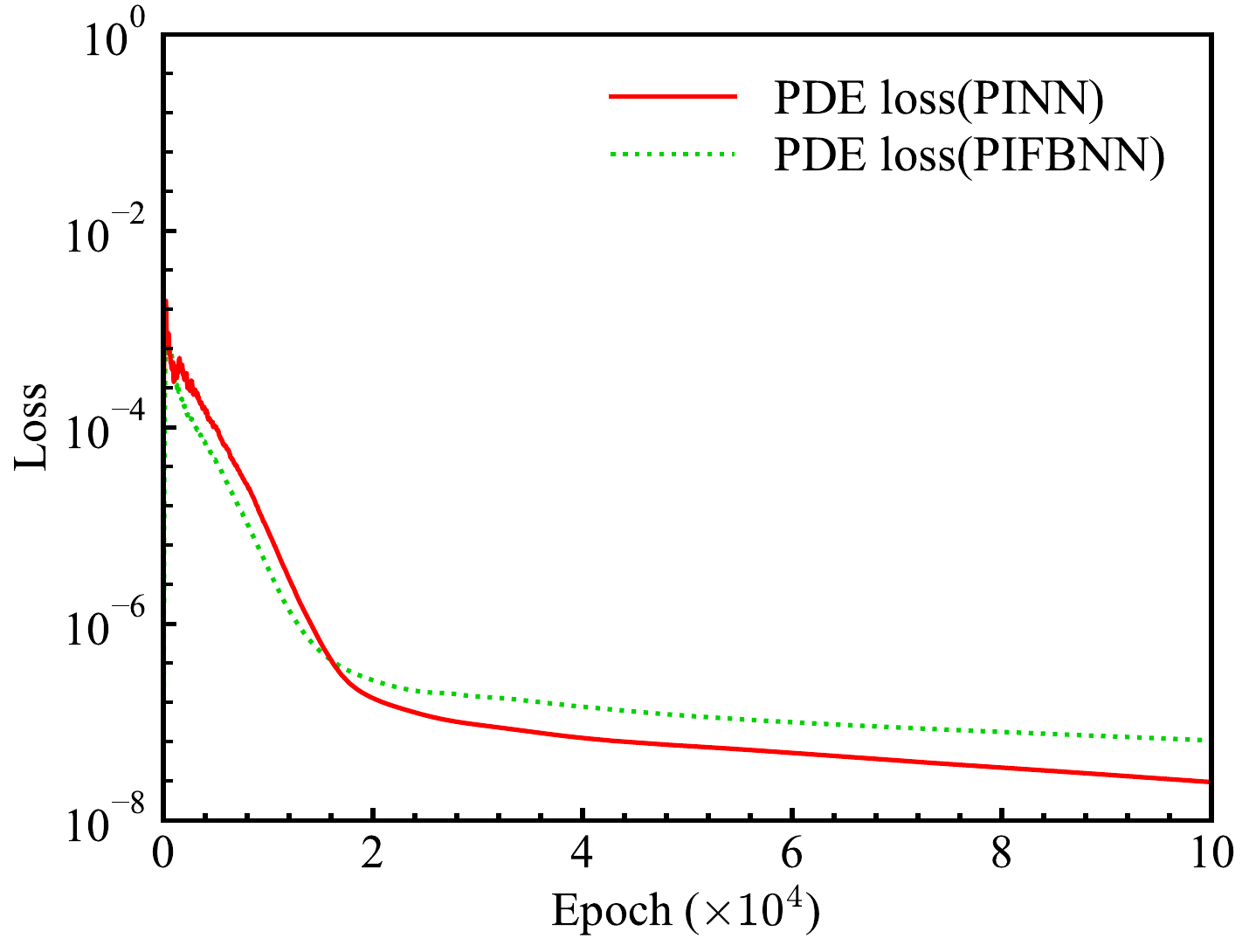}
	}
	
	\caption{PDE loss of PINN and PIFBNN models for cylinder wake flow. }
	\label{cypredpdeloss}
\end{figure}
The learning curves of PINN and PIFBNN for the cylindrical wake flow are shown in Figure \ref{cypredloss}. It can be observed that both PINN and PIFBNN achieve relatively low training losses overall. PIFBNN exhibits lower training loss in the initial phase, whereas PINN demonstrates superior convergence with reduced training loss in later stages, indicating that PINN has effective learning of the boundary conditions at the end. Notably, the testing loss of PIFBNN is always lower than PINN. This phenomenon can be attributed to the lower physical residuals of PIFBNN compared to PINN, indicating that PIFBNN more effectively leverages the PDE residuals to enhance prediction accuracy. To validate this, Figure \ref{cypredpdeloss} presents the PDE loss curves. Notably, PINN achieves both lower training loss and lower PDE loss than PIFBNN, yet its prediction performance is inferior. This discrepancy suggests that PINN exhibits overfitting, it minimizes the training and PDE losses excessively but fails to generalize well to the actual problem, whereas PIFBNN maintains a better balance between optimization and generalization.
\begin{figure}[h!] 
	\centering
	\subfigure[Streanwise flow velocity: $u$]{
		\includegraphics[height=3.5cm]{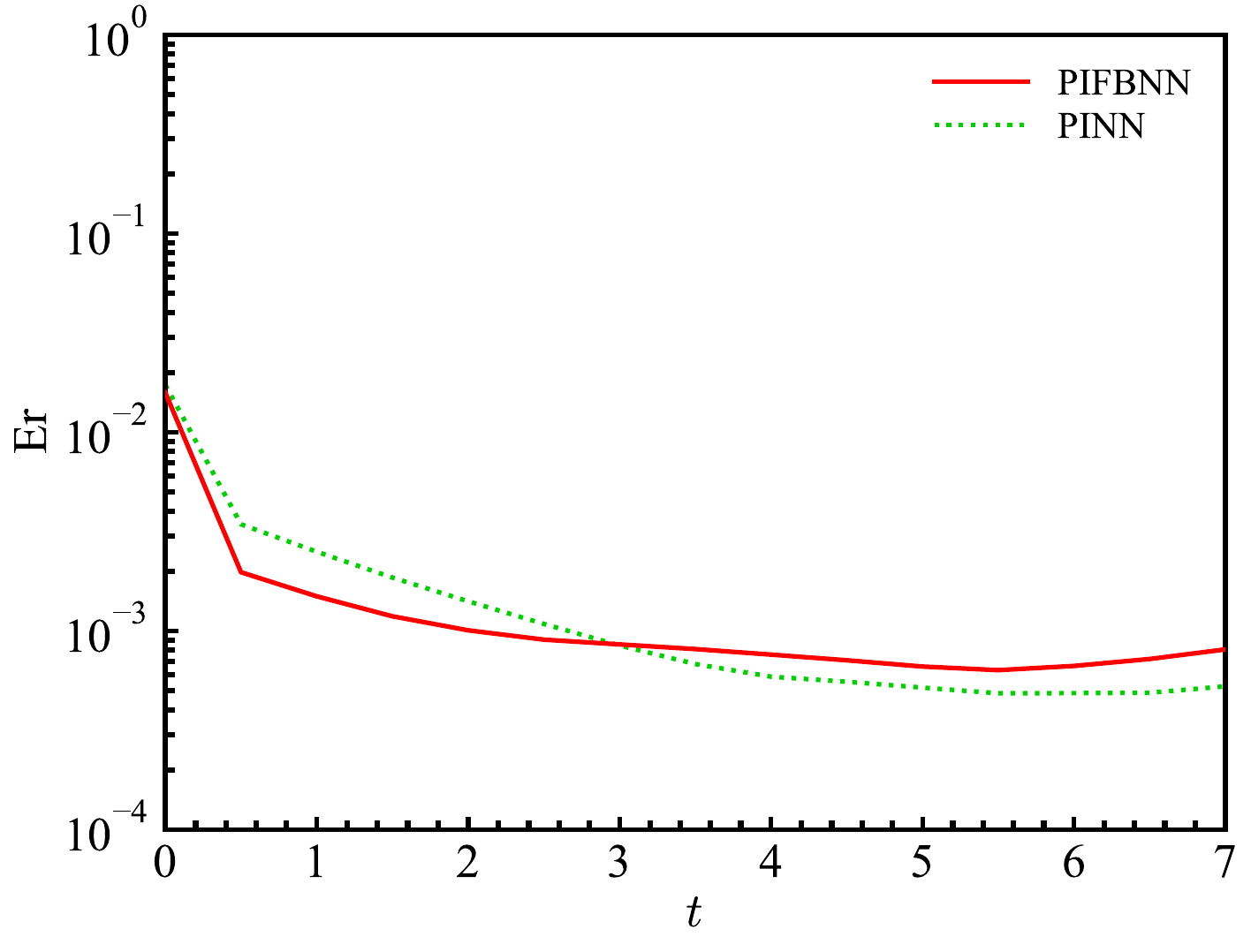}
	}
	\subfigure[Transverse velocity: $v$]{
		\includegraphics[height=3.5cm]{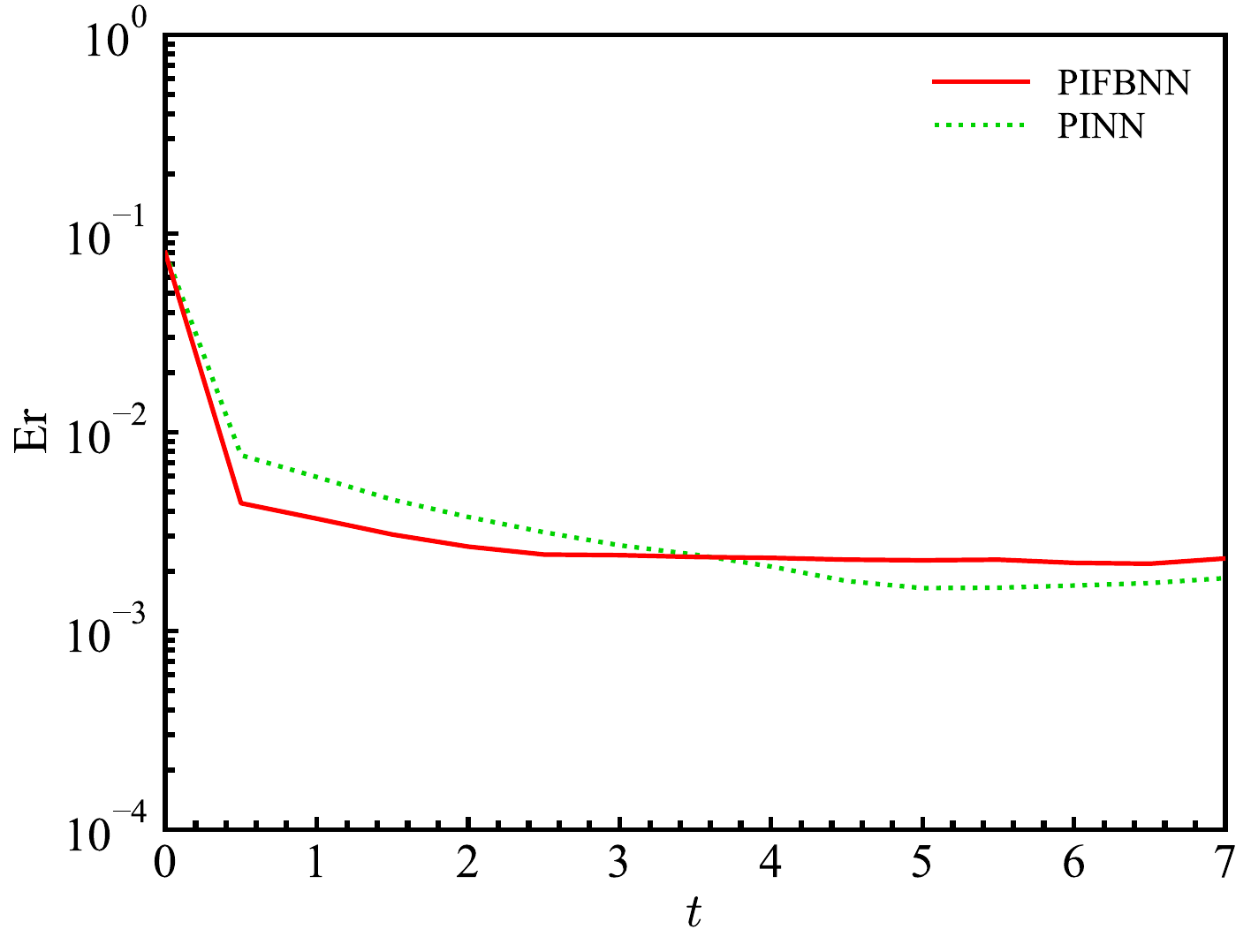}
	}
	\subfigure[Pressure: $p$]{
		\includegraphics[height=3.5cm]{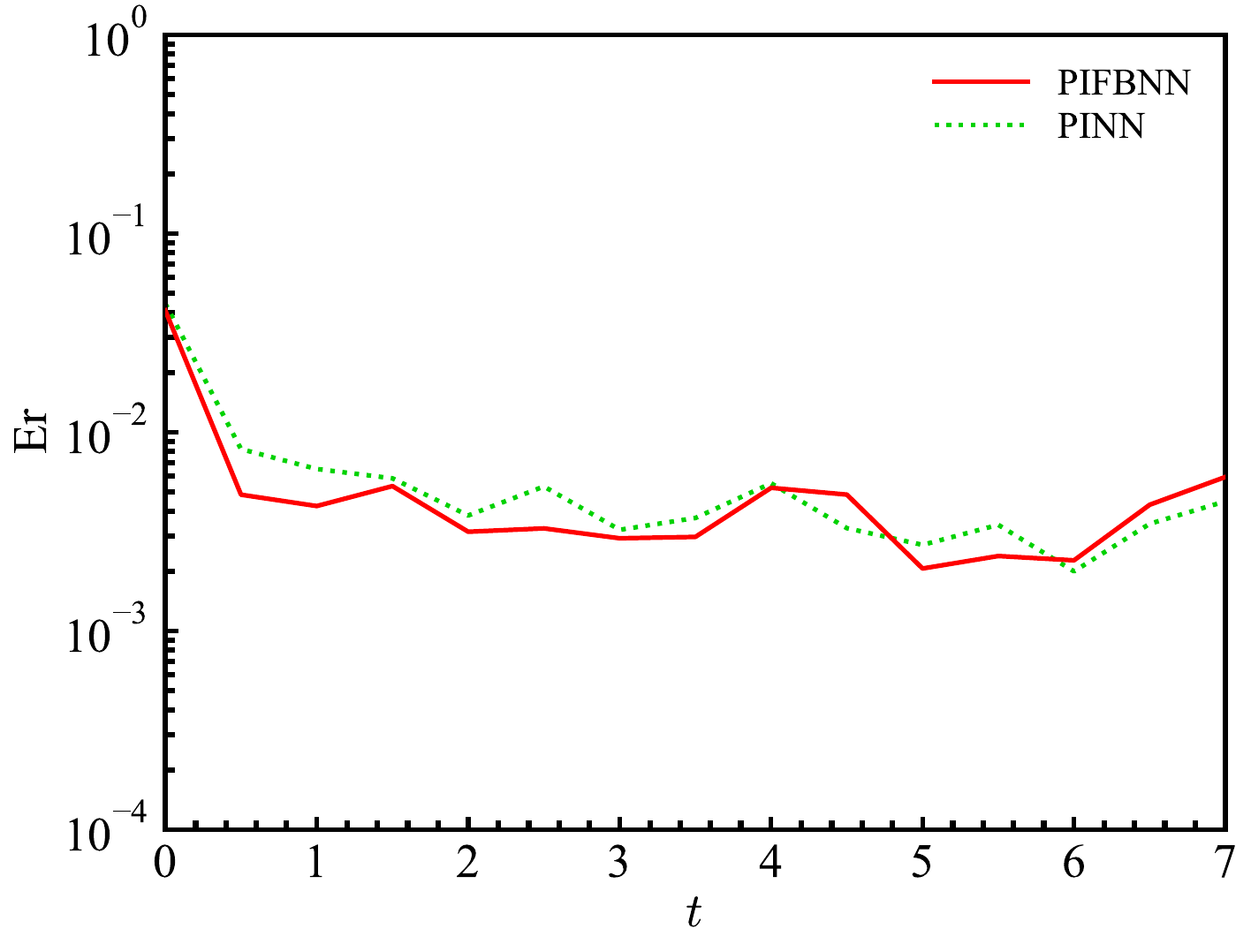}
	}
	
	\caption{Relative error curves of PIFBNN and PINN models for cylinder wake flow.}
	\label{cypreder}
\end{figure}
The relative errors of cylinder wake flow predicted by the PINN and PIFBNN are shown in Figure \ref{cypreder}. It can be seen that all output results have large prediction errors at $t=0$. This occurs because the neural networks predicting the unsteady periodic flow do not incorporate the initial conditions, hence there are multiple solutions for the $\frac{\partial\bm{u}} {\partial t} $ term of the PDE. This problem can be solved by appropriately sampling the labeled data initially as the initial training condition. As time passed, the multiple-solution problem is solved, and the output errors rapidly decrease. The amplitude reduction in PIFBNN is significantly more pronounced than in PINN, with PIFBNN achieving convergence stability more rapidly. This is also why the PIFBNN has a smaller overall error in predicting the cylinder flow wake than PINN. Over time, the relative error of PINN gradually decreased and is lower than PIFBNN. The prediction error of $u$ is the lowest compared to $v$ and $p$ because $u$ represents the main stream direction of the free flow, and a larger value results in a smaller relative error. For the prediction of pressure $p$, the relative error of PIFBNN is almost always lower than that of PINN. The pressure variable $p$ is subject to the weakest physical constraints, with its only constraint appearing in the pressure gradient term $\frac{\partial p}{\partial \bm{x}}$ of the Navier-Stokes equations. Therefore, predicting $p$ is the most challenging task, and the advantage of PIFBNN in this aspect further highlights its strong ability to learn physical laws. Specifically, PIFBNN is effective at capturing information from weakly constrained terms in the equations and can represent the physical state of the wake with greater fidelity. This also suggests that PIFBNN is more suitable for predicting challenging tasks with limited physical information and provides excellent results.

\begin{figure}[!htbp]
	\centering
	{
		\includegraphics[height=3cm]{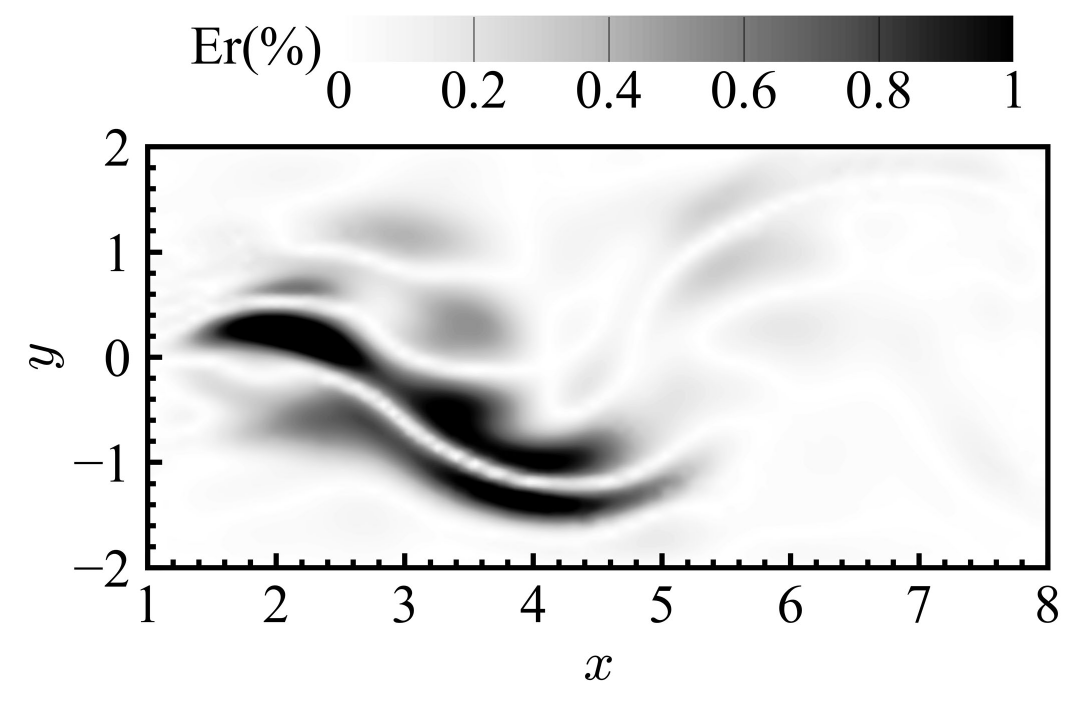}
	}
	{
		\includegraphics[height=3cm]{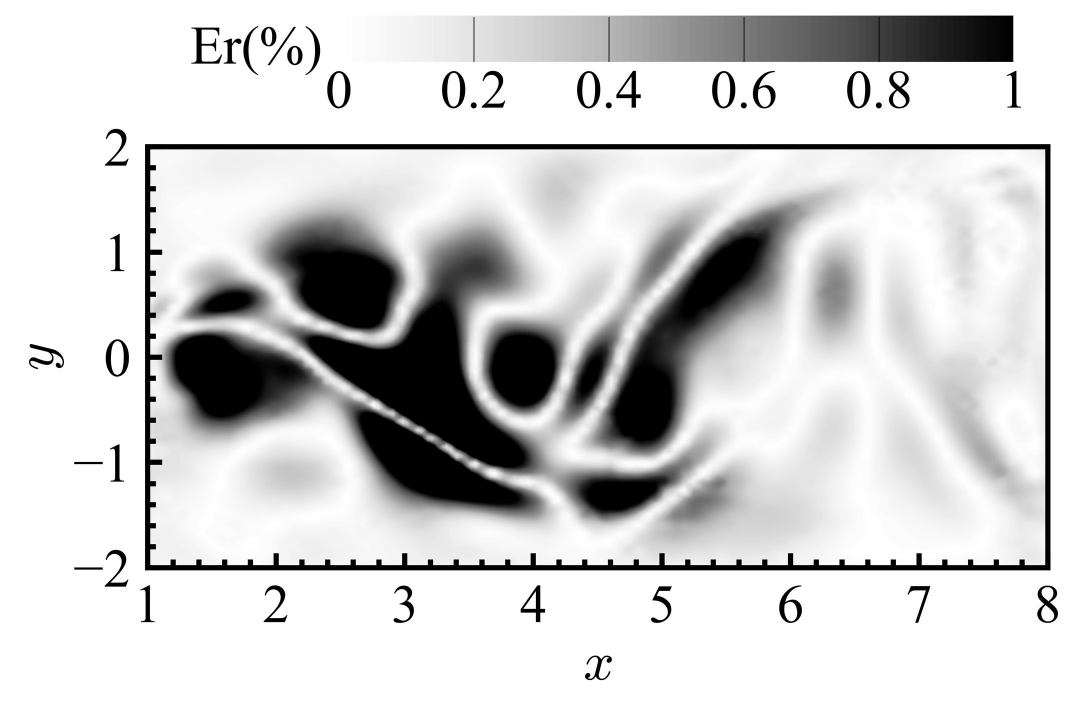}
	}
	{
		\includegraphics[height=3cm]{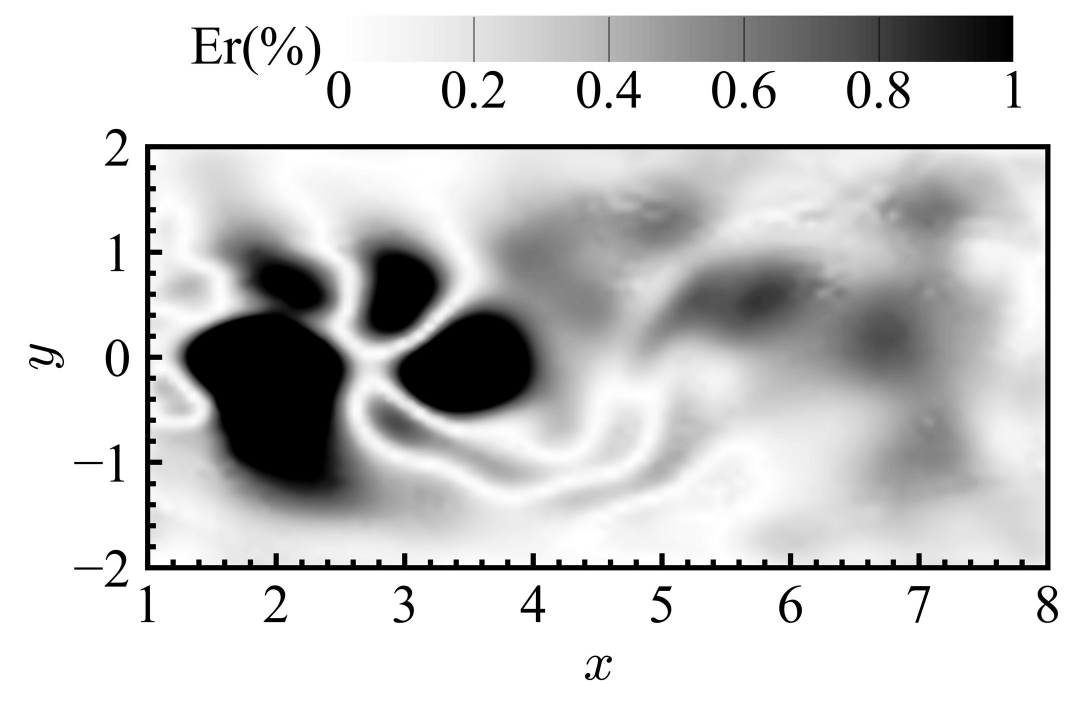}
	}
	{
		\includegraphics[height=3cm]{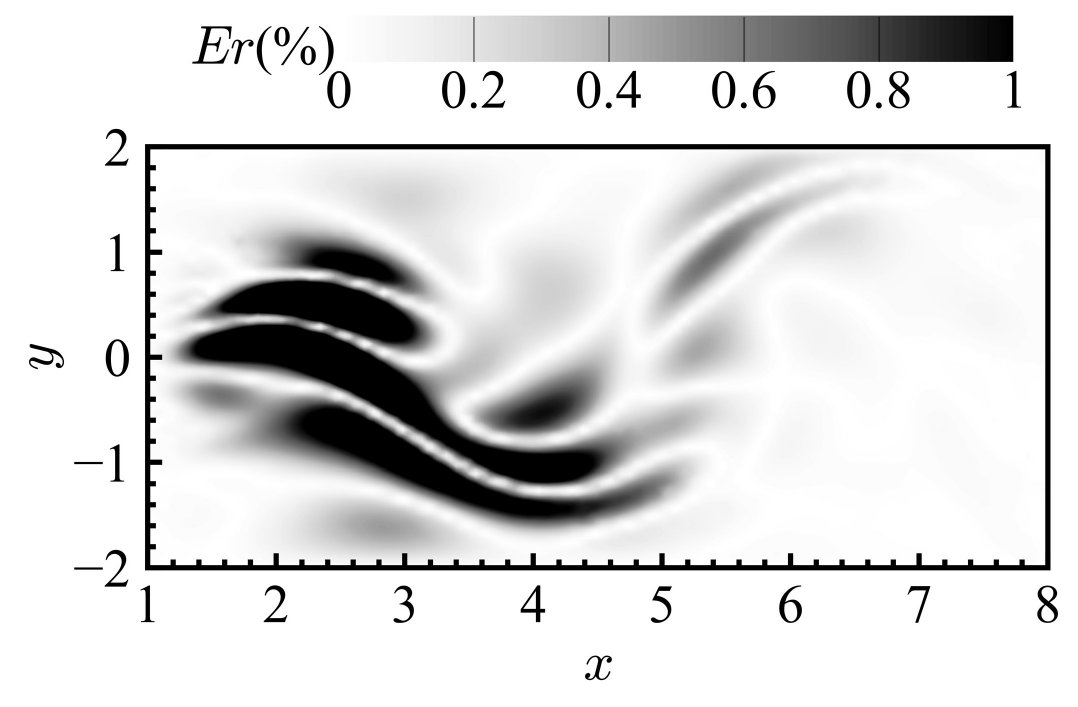}
	}
	\subfigure[$T=1$]
	{
		\includegraphics[height=3cm]{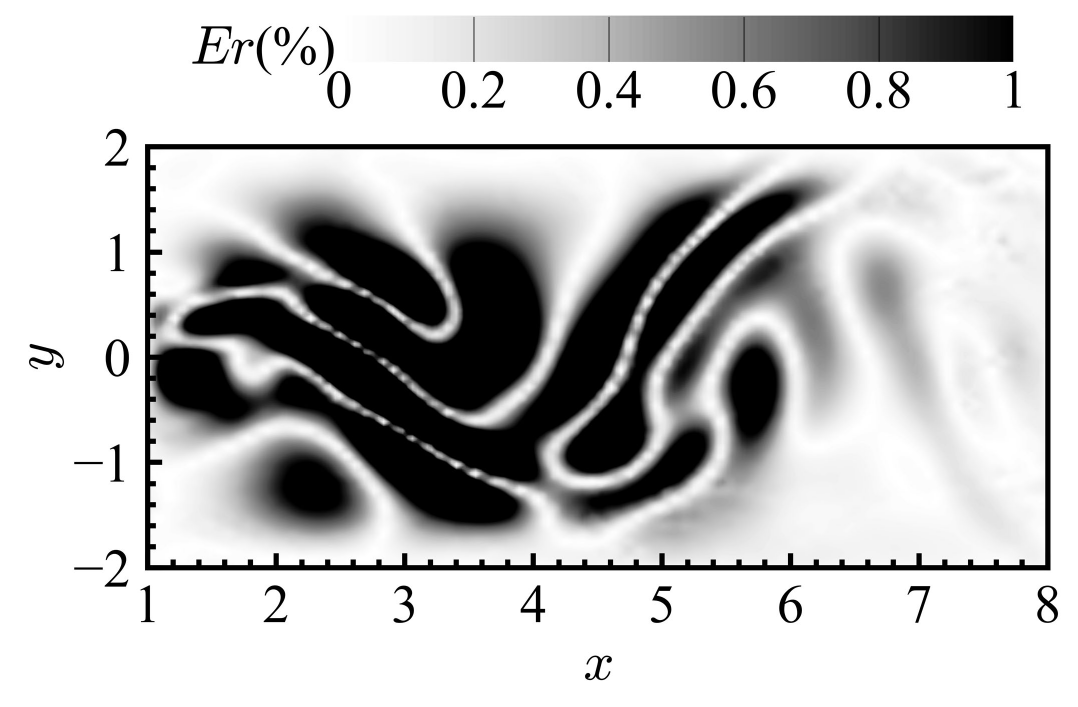}
	}
	{
		\includegraphics[height=3cm]{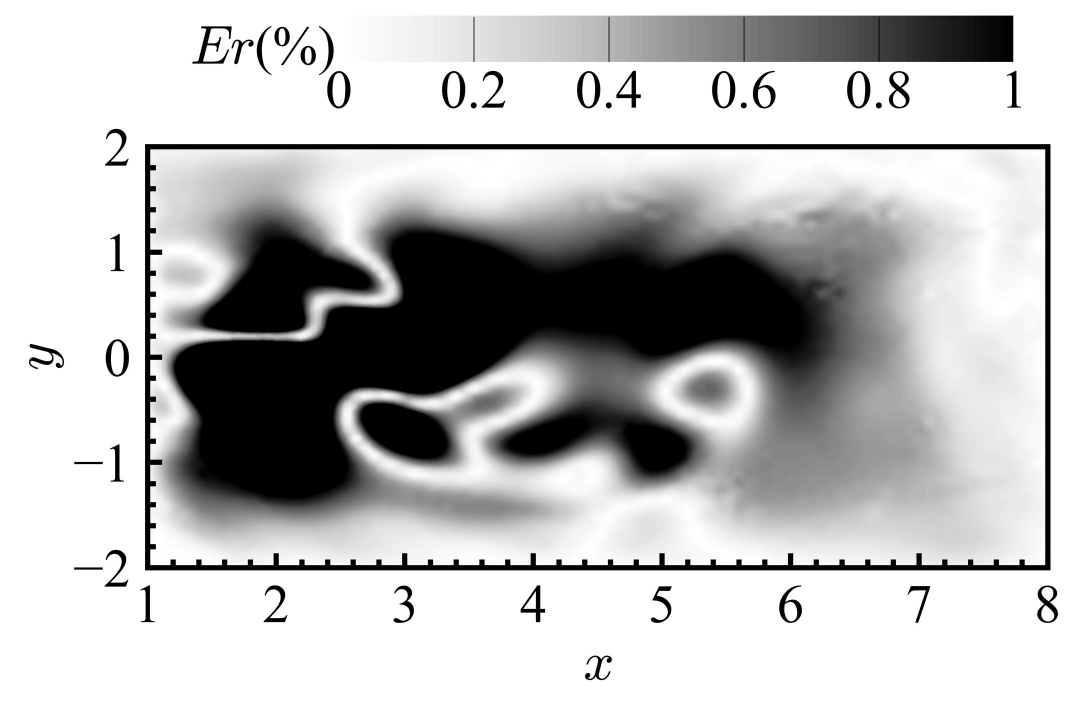}
	}
	
	{
		\includegraphics[height=3cm]{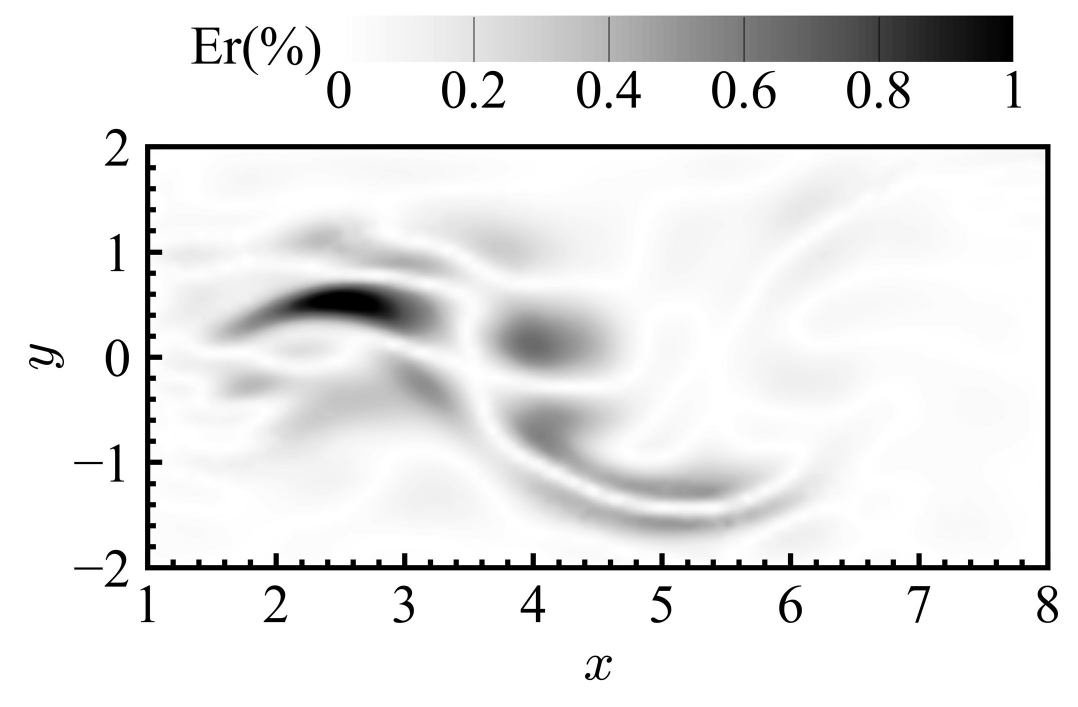}
	}
	{
		\includegraphics[height=3cm]{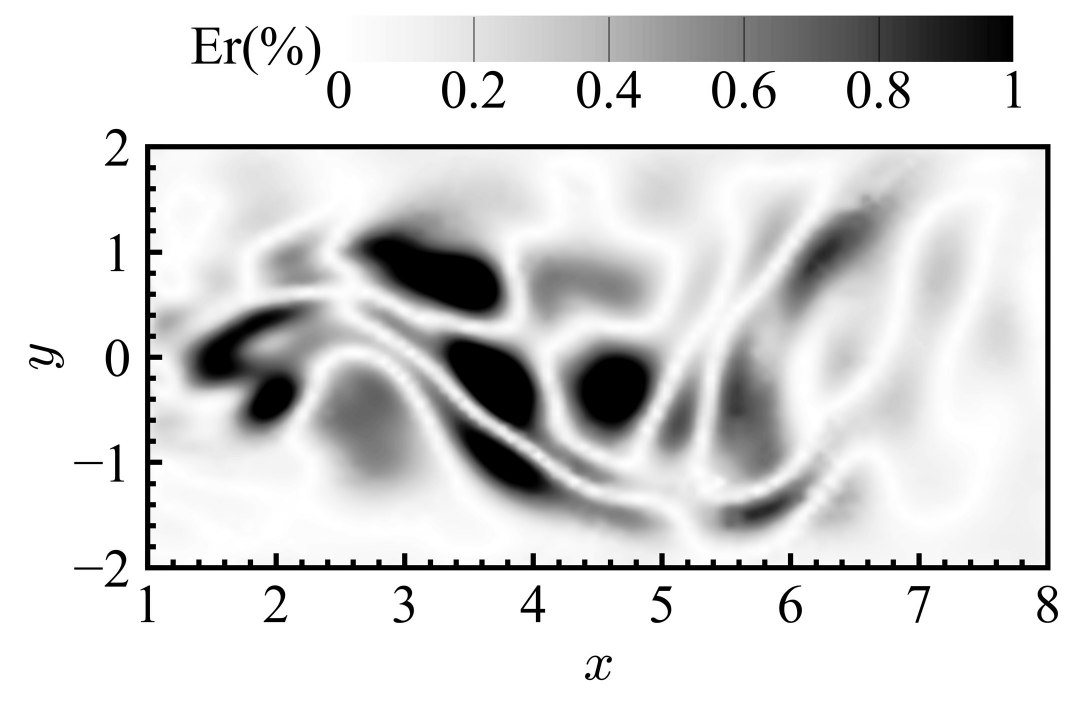}
	}
	{
		\includegraphics[height=3cm]{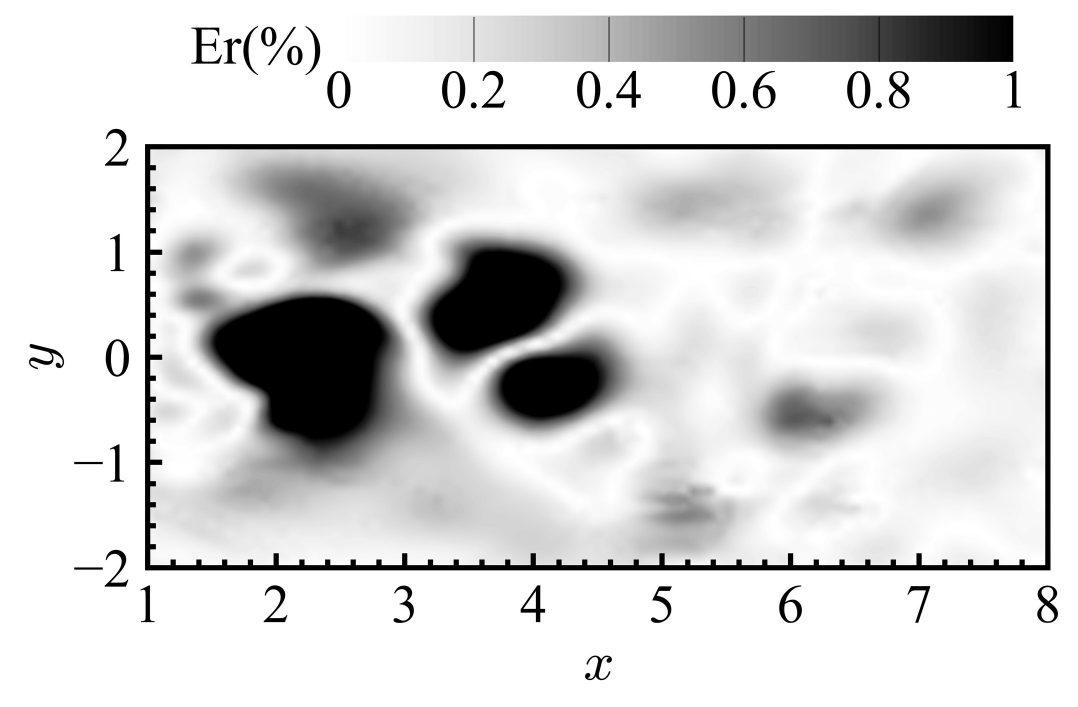}
	}
	
	{
		\includegraphics[height=3cm]{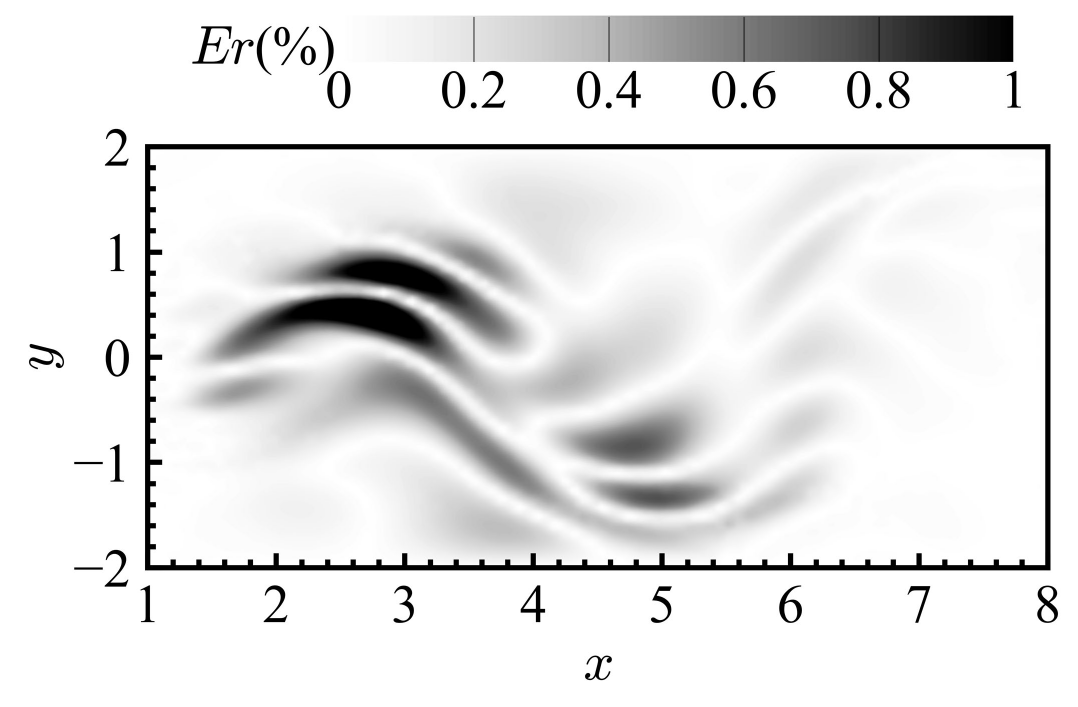}
	}
	\subfigure[$T=2$]
	{
		\includegraphics[height=3cm]{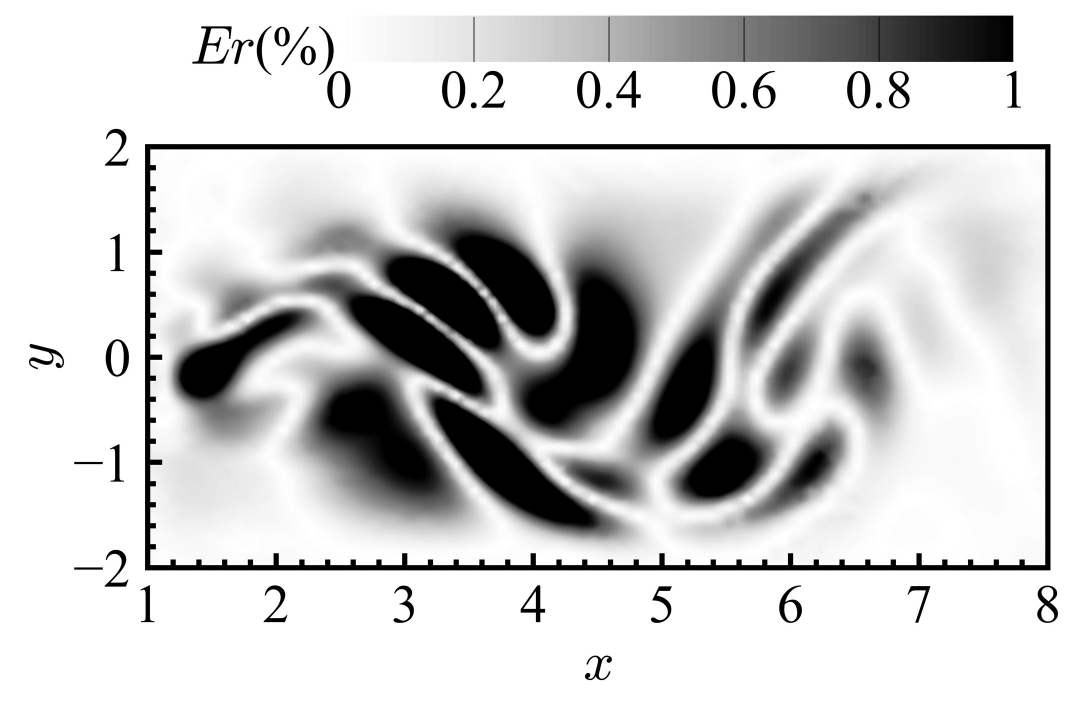}
	}
	{
		\includegraphics[height=3cm]{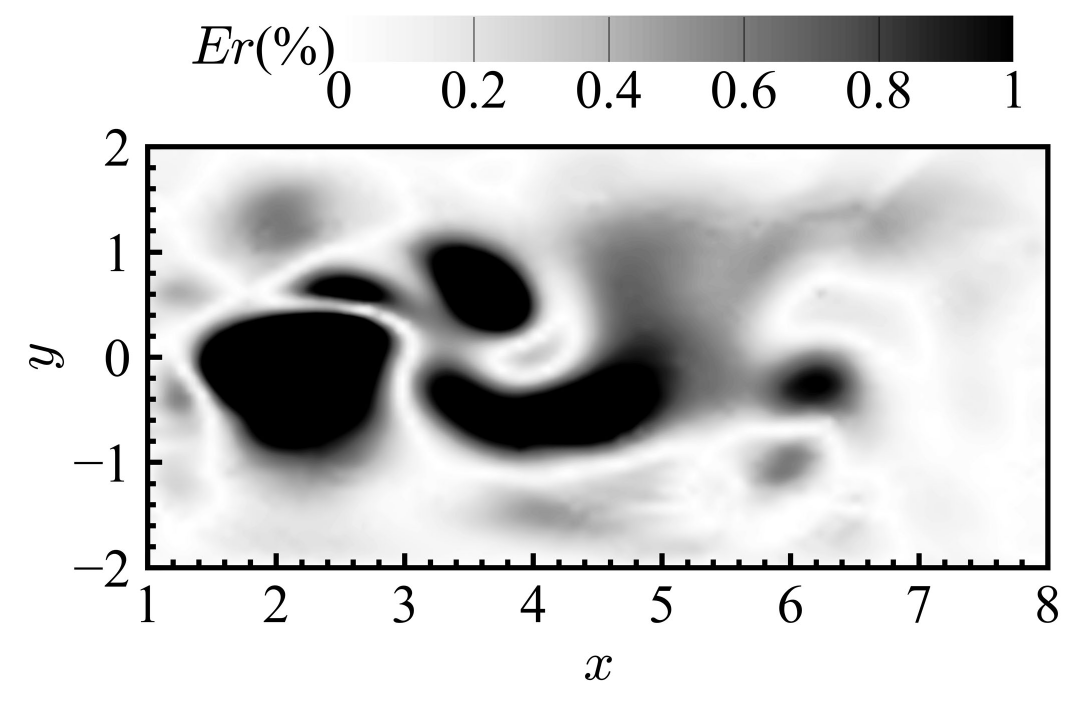}
	}
	{
		\includegraphics[height=3cm]{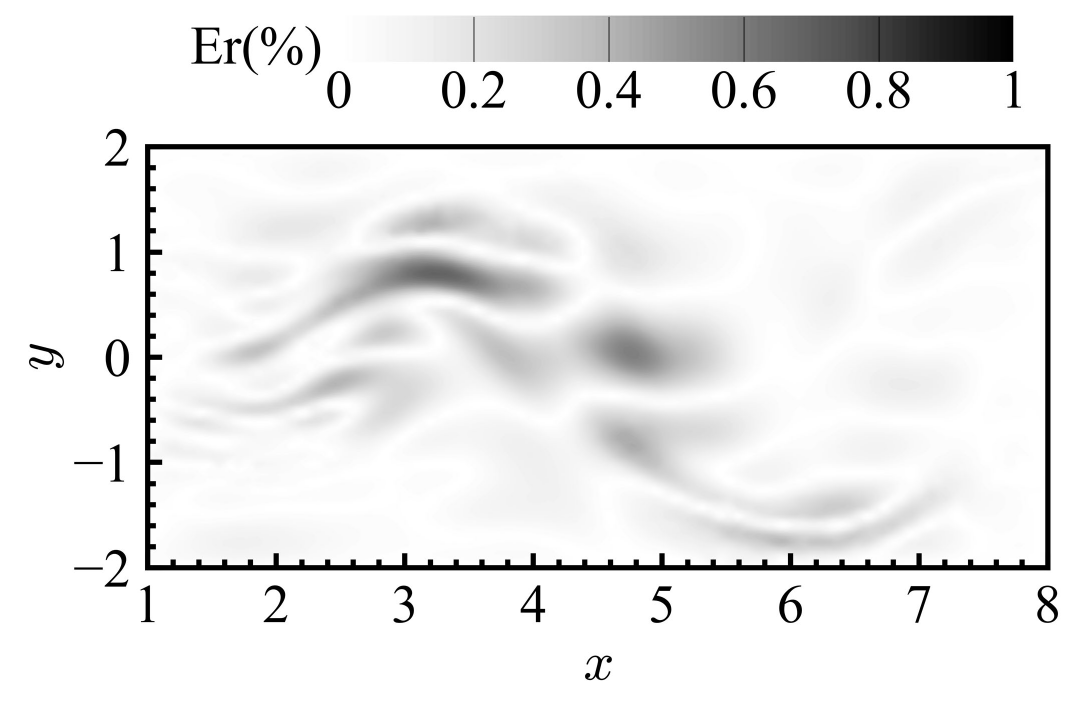}
	}
	{
		\includegraphics[height=3cm]{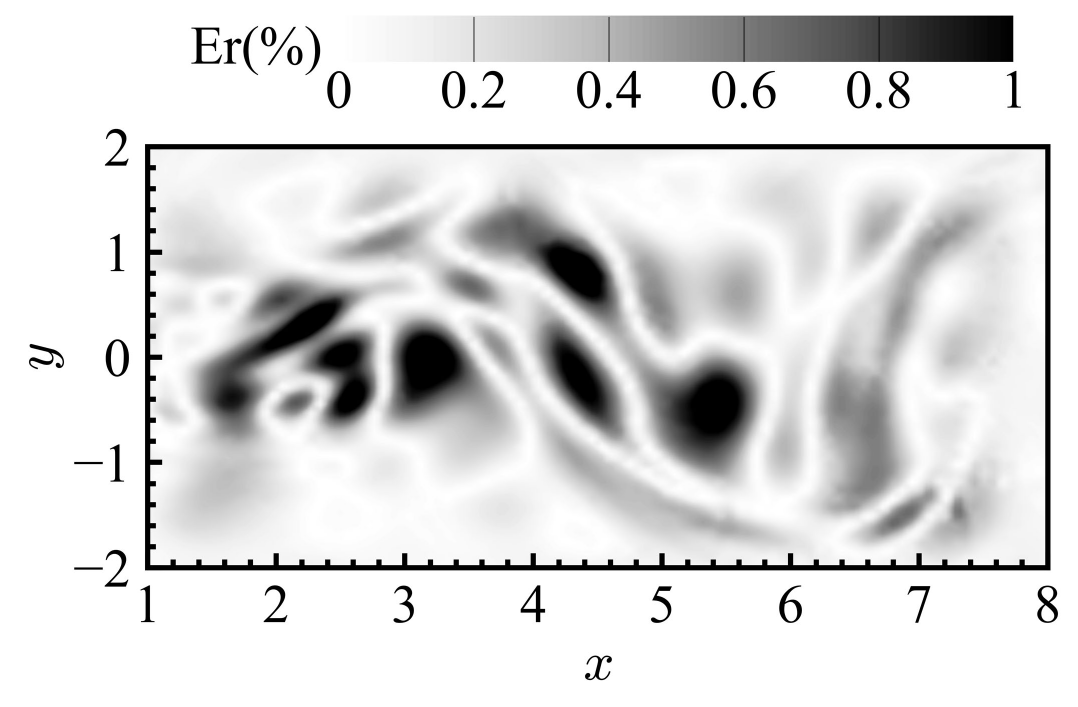}
	}
	{
		\includegraphics[height=3cm]{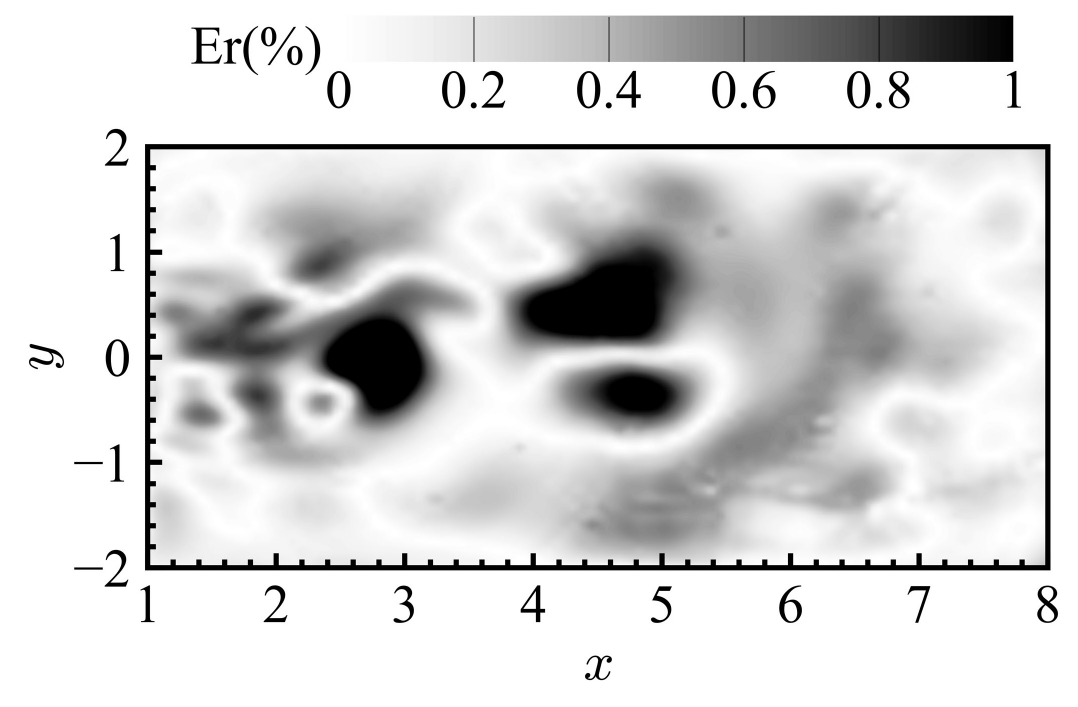}
	}
	{
		\includegraphics[height=3cm]{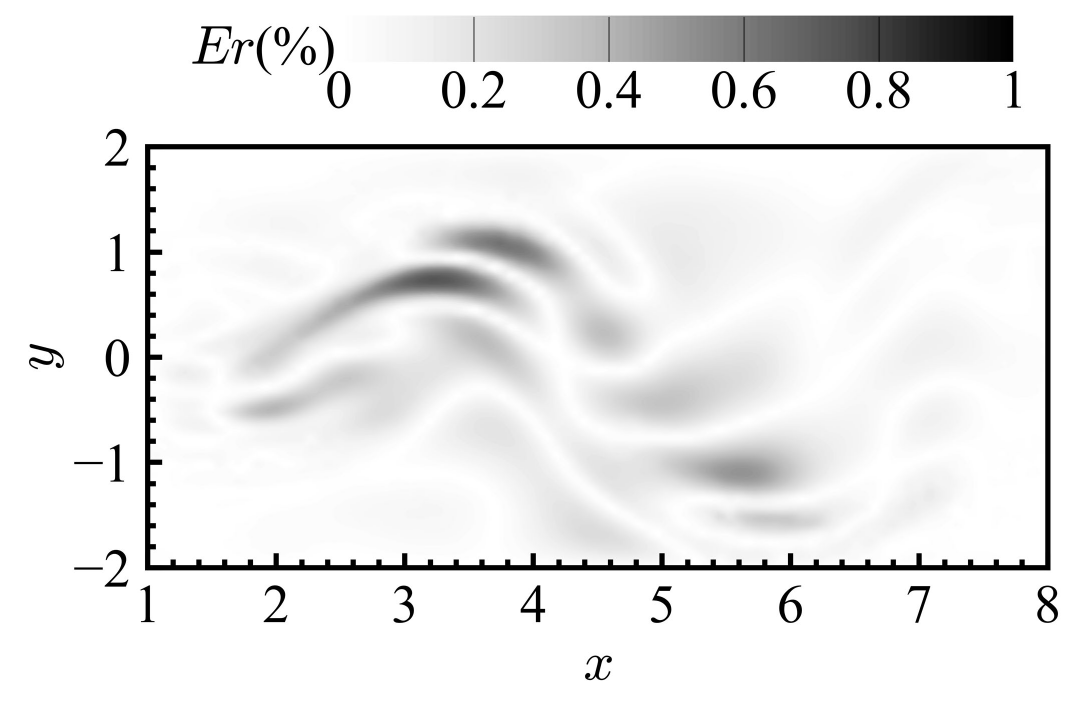}
	}
	\subfigure[$T=3$]
	{
		\includegraphics[height=3cm]{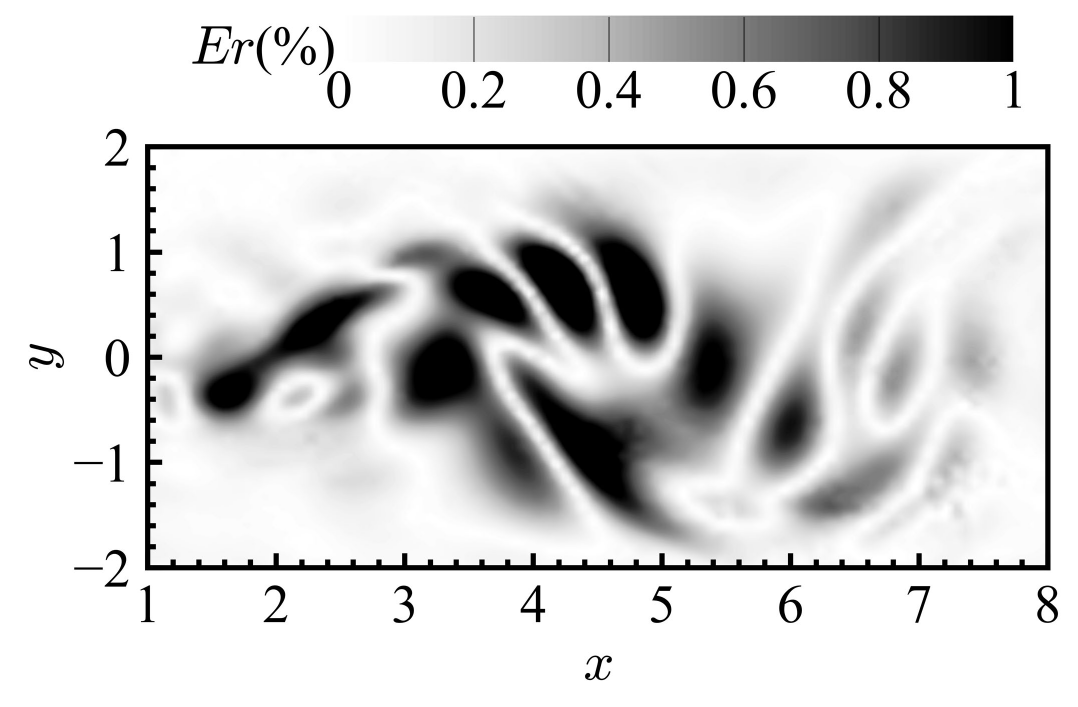}
	}
	{
		\includegraphics[height=3cm]{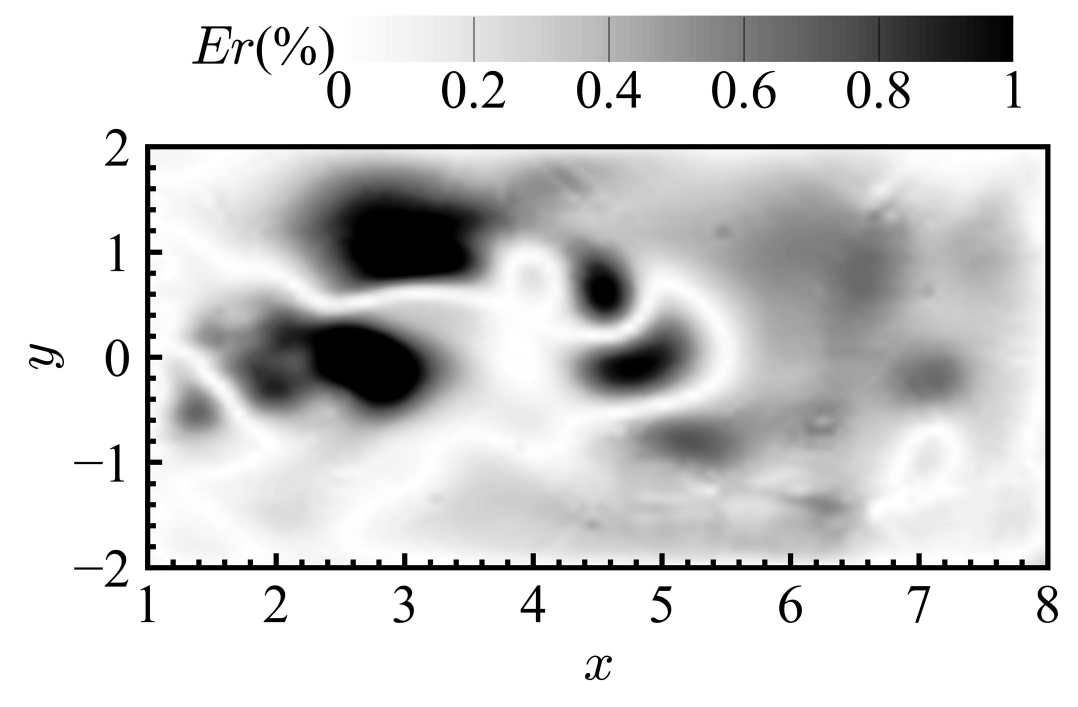}
	}
	
	\caption{Relative error contours for cylinder wake flow fields at T=1 (a), T=2 (b), and T=3 (c). For each time instance: top row shows PINN results, bottom row shows PIFBNN results, with velocity components $(u, v) $and pressure $(p)$ displayed left-to-right. }
	\label{cylinderres}
\end{figure}
The relative error contours for cylinder wake flow predicted by PINN and PIFBNN models are shown in Figure \ref{cylinderres}. Consistent with the observation described earlier, all networks has the smallest prediction error for $u$, whereas $v$ and $p$ are relatively difficult to predict. The prediction error distributions exhibit similar patterns across all neural network models, with errors predominantly concentrated in regions characterized by high velocity or pressure gradients. Across all temporal snapshots, PIFBNN consistently demonstrated both sparser error distributions and lower error magnitudes compared to PINN.

\subsection{Lid-driven cavity flow }
\begin{figure}[h!] 
	\centering
	{
		\includegraphics[height=5.5cm]{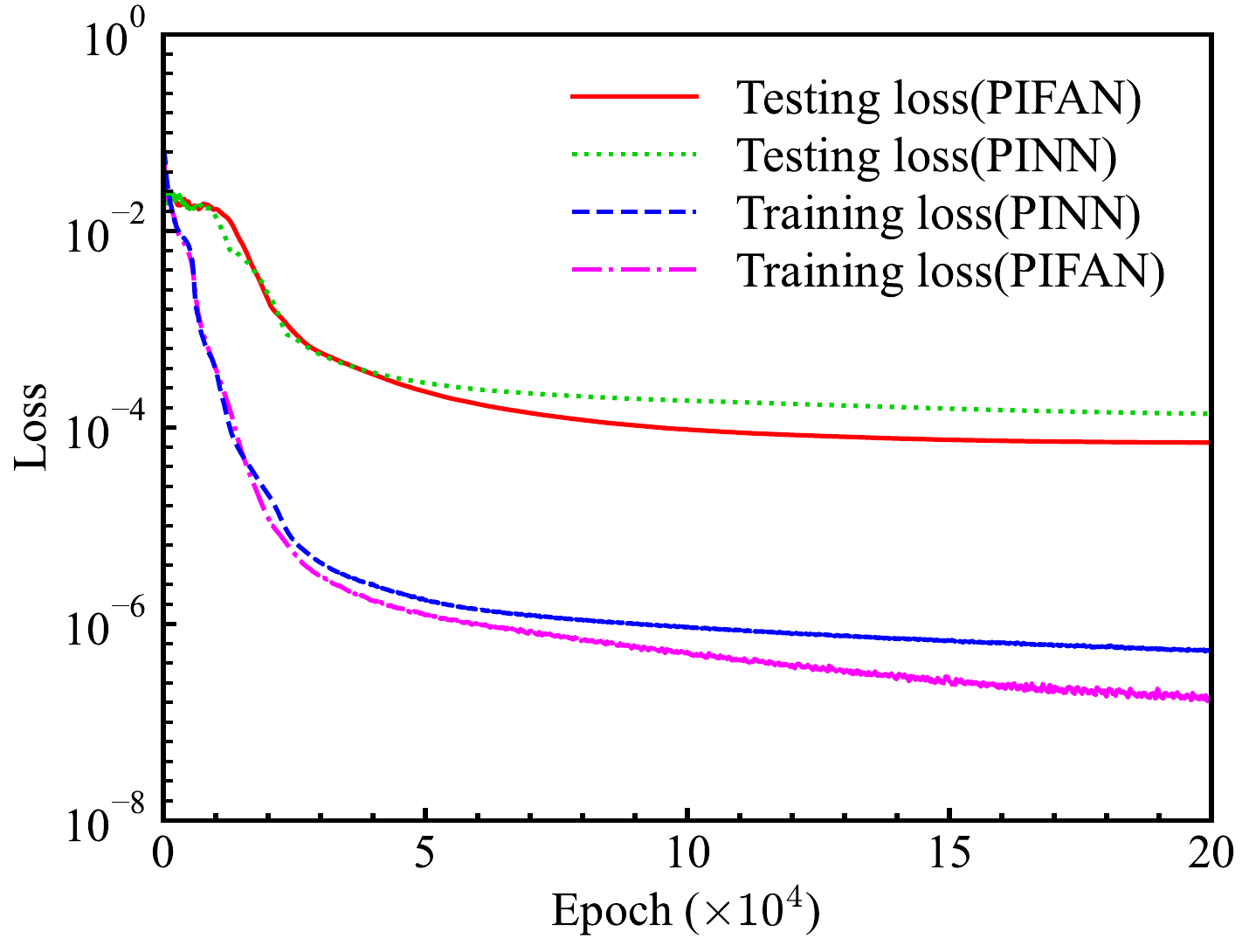}
	}
	
	\caption{Learning curves of PINN and PIFBNN models for lid-driven cavity flow. }
	\label{cavpredloss}
\end{figure}

Lid-driven cavity flow describes a canonical benchmark problem in fluid dynamics where incompressible viscous flow within a closed rectangular cavity is generated solely by the tangential motion of the top boundary wall. This flow configuration holds significant research importance and finds extensive applications across multiple disciplines, particularly in fluid dynamics, mechanical engineering, and physical sciences. The Lid-driven cavity flow is governed by the Navier--Stokes and continuity equations. For computational simplicity, we restrict our analysis to a two-dimensional lid-driven cavity flow configuration and simulate its evolution until steady-state conditions are achieved, which serves as the basis for our training dataset. The length of square cavity is 1, and the speed at which the top cover moves is $\mathbf{u}=(1,0)$.
The governing equations are represented as
\begin{equation}
	\nabla \cdot \mathbf{u} = 0,
\end{equation}
\begin{equation}
	\rho \left( \frac{\partial \mathbf{u}}{\partial t} + \mathbf{u} \cdot \nabla \mathbf{u} \right) = -\nabla p + \mu \nabla^2 \mathbf{u} + \mathbf{f},
\end{equation}
where $\mathbf{u}=(u,v)$ are velocity, and $p$ is pressure, $\mathbf{f}$ is source term. The dimensionless quantity $\mu$ denotes dynamic viscosity.The wall follows nonslip boundary condition. The bottom-left corner of the square cavity is fixed as the coordinate origin and expressed as
\begin{equation}
	\mathbf{u}(x,1) = (1, 0),
\end{equation}
\begin{equation}
	\mathbf{u}(0, y) = (0, 0), \quad \mathbf{u}(x,0) = (0, 0), \quad \mathbf{u}(1,y) = (0, 0).
\end{equation}
Lid-driven cavity flow provides a relatively simple model for studying unsteady flow and turbulence in fluid mechanics. Through the steady-state square cavity flow benchmark case, we demonstrate the capability of neural networks to effectively learn the governing Navier-Stokes equations and accurately capture key viscous vortex dynamics characteristics.
A Reynolds number of 1000 is considered for this case. To comply with previous baseline model studies, the neural network architecture is set to six layers with 50 neurons per layer. The dataset is derived from the Zeta dataset, which includes steady-state two-dimensional square cavity flow data with a Reynolds number of 1000 and a grid resolution of $51 * 51$. A total of 51 data points on each boundary are randomly sampled using Latin hypercube method as the boundary condition. Owing to the meshless nature of the physical information neural network, 10201 points are randomly sampled within the domain to calculate more detailed physical residuals. This evaluation aim to predict the entire field using only boundary conditions and physical residuals, without using any labeled data. The predictive performance of the network is evaluated by 1200 testing points randomly sampling from the Zeta dataset across the entire domain.
The learning curves of PINN and PIFBNN are shown in Figure \ref{cavpredloss}.

\begin{figure}[h!] 
	\centering
	\subfigure[Streamwise flow velocity: $u$]{
		\includegraphics[height=5.5cm]{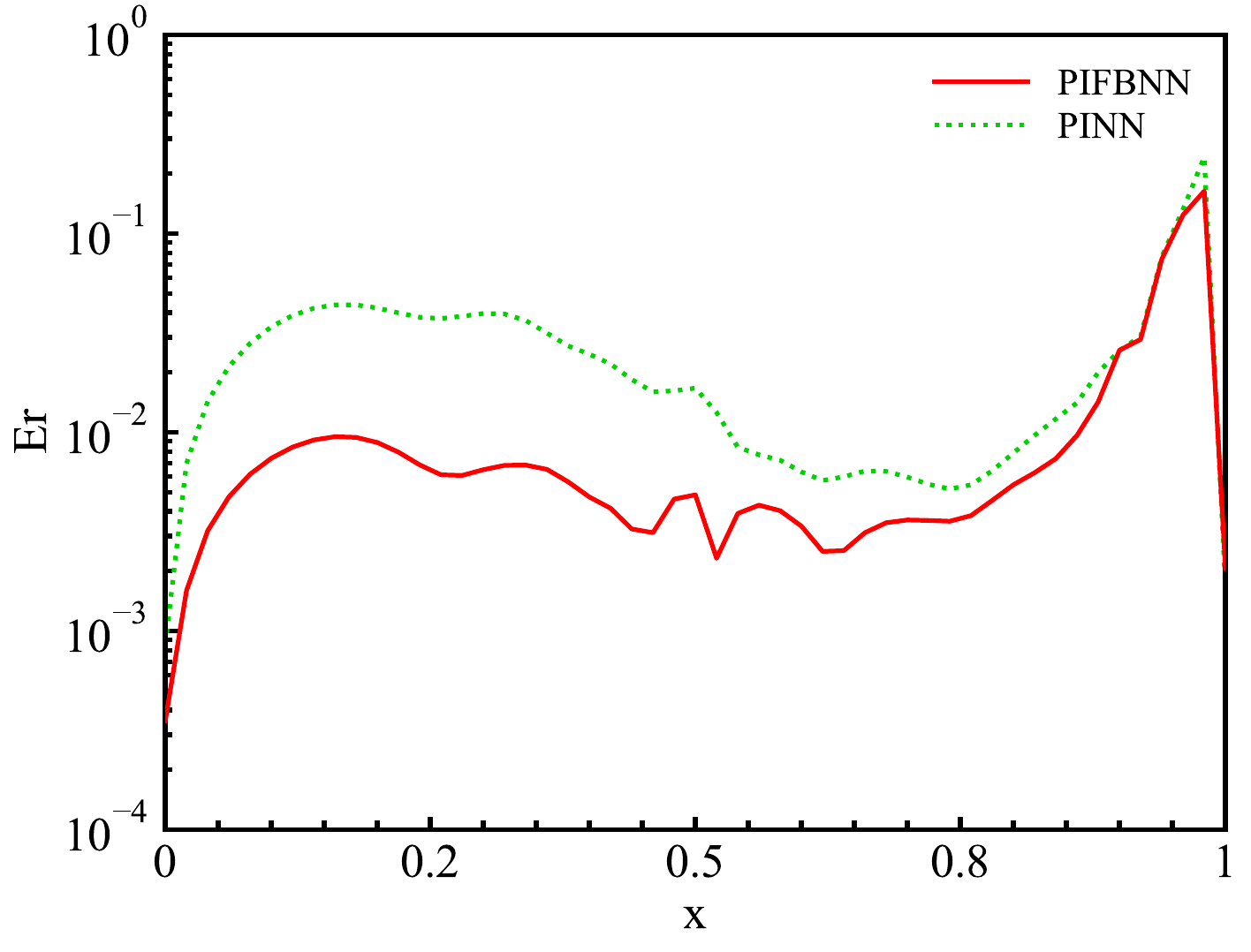}
	}
	\subfigure[Transverse velocity: $v$]{
		\includegraphics[height=5.5cm]{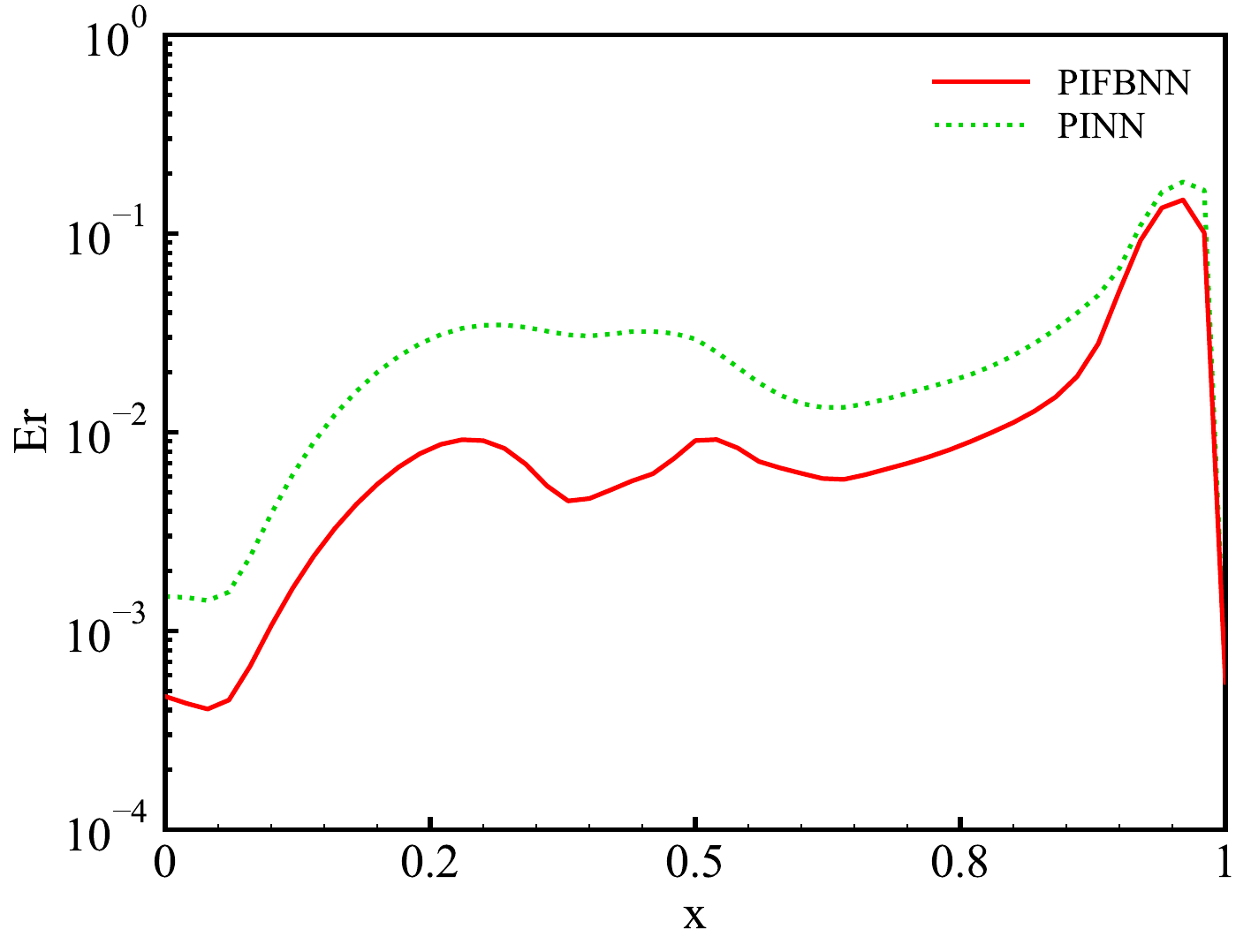}
	}
	
	\caption{The relative error curves of lid driven cavity flow using PINN and PIFBNN models.}
	\label{cavpreder}
\end{figure}

It can be observed that for boundary conditions, PIFBNN has a stronger ability to continue learning than PINN. As the number of iterations increases, PIFBNN can more accurately capture the boundary conditions, and the training loss shows a significant downward trend at the end of iteration. While PIFBNN has not yet reached full convergence and remains suboptimal, whereas PINN has converged and achieved the optimal result,which proves that PIFBNN still has the potential to achieve better results. Correspondingly, from the perspective of testing loss, PIFBNN achieves better results than PINN, indicating that it not only learns boundary conditions more accurately than PINN, but also achieves better prediction performance throughout the entire domain.
The relative errors of PINN and PIFBNN along $x$ are shown in Figure \ref{cavpreder}.

\begin{figure}[!htbp]
	\centering
	\raisebox{3.8cm}{(a)}
	{
		\includegraphics[height=4cm]{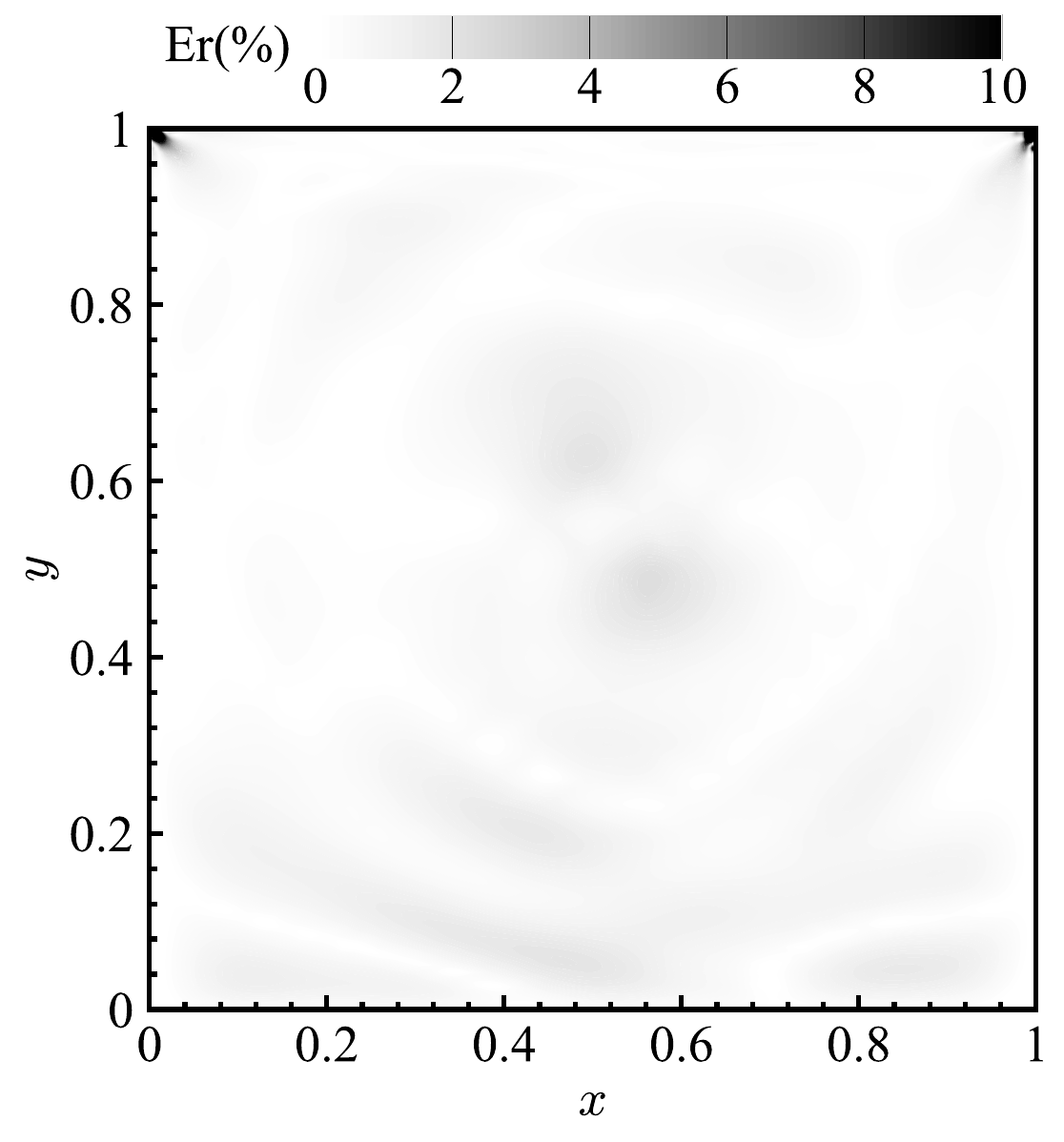}
	}
	{
		\includegraphics[height=4cm]{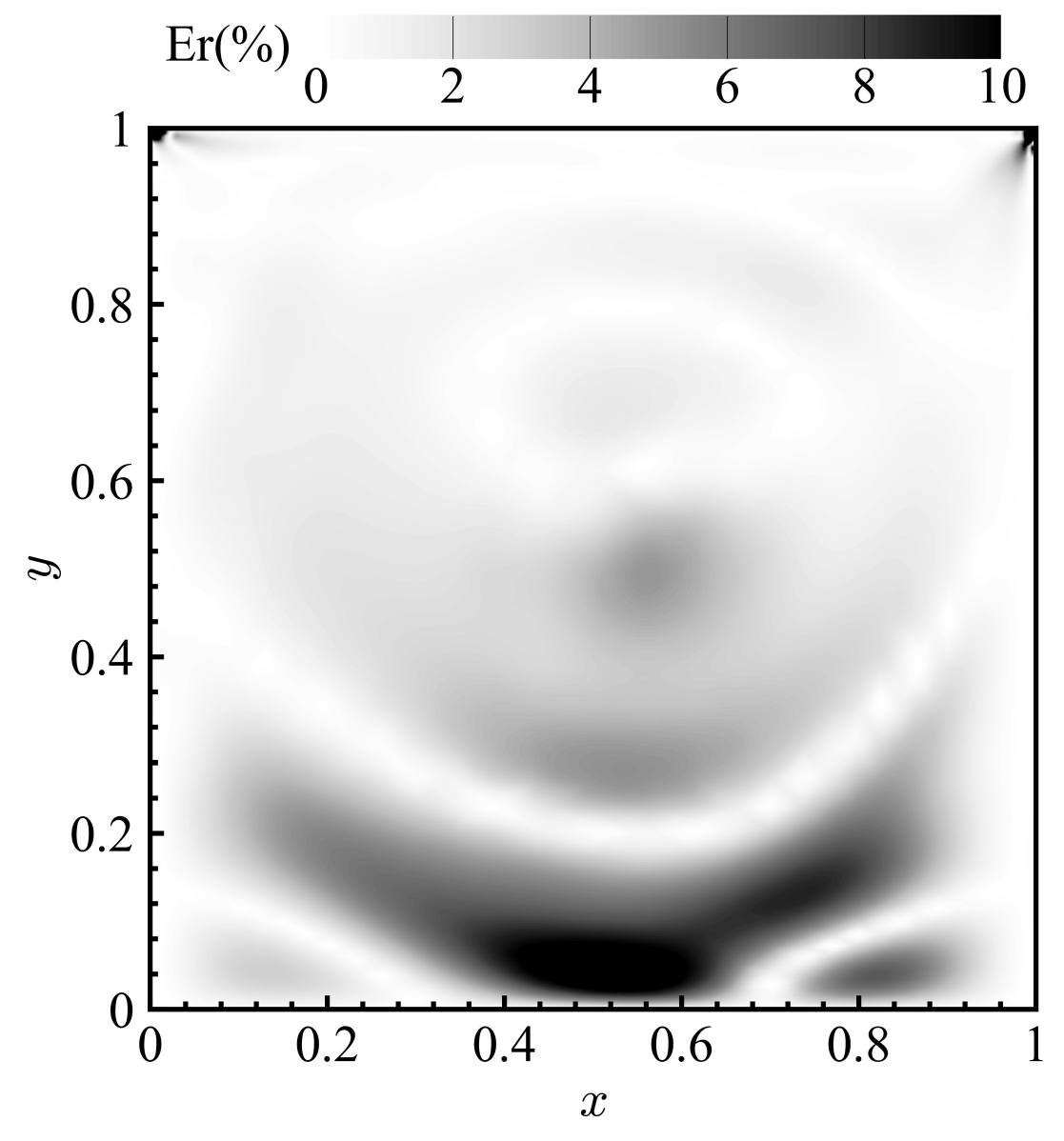}
	}
	\par %
	\raisebox{3.8cm}{(b)}
	{
		\includegraphics[height=4cm]{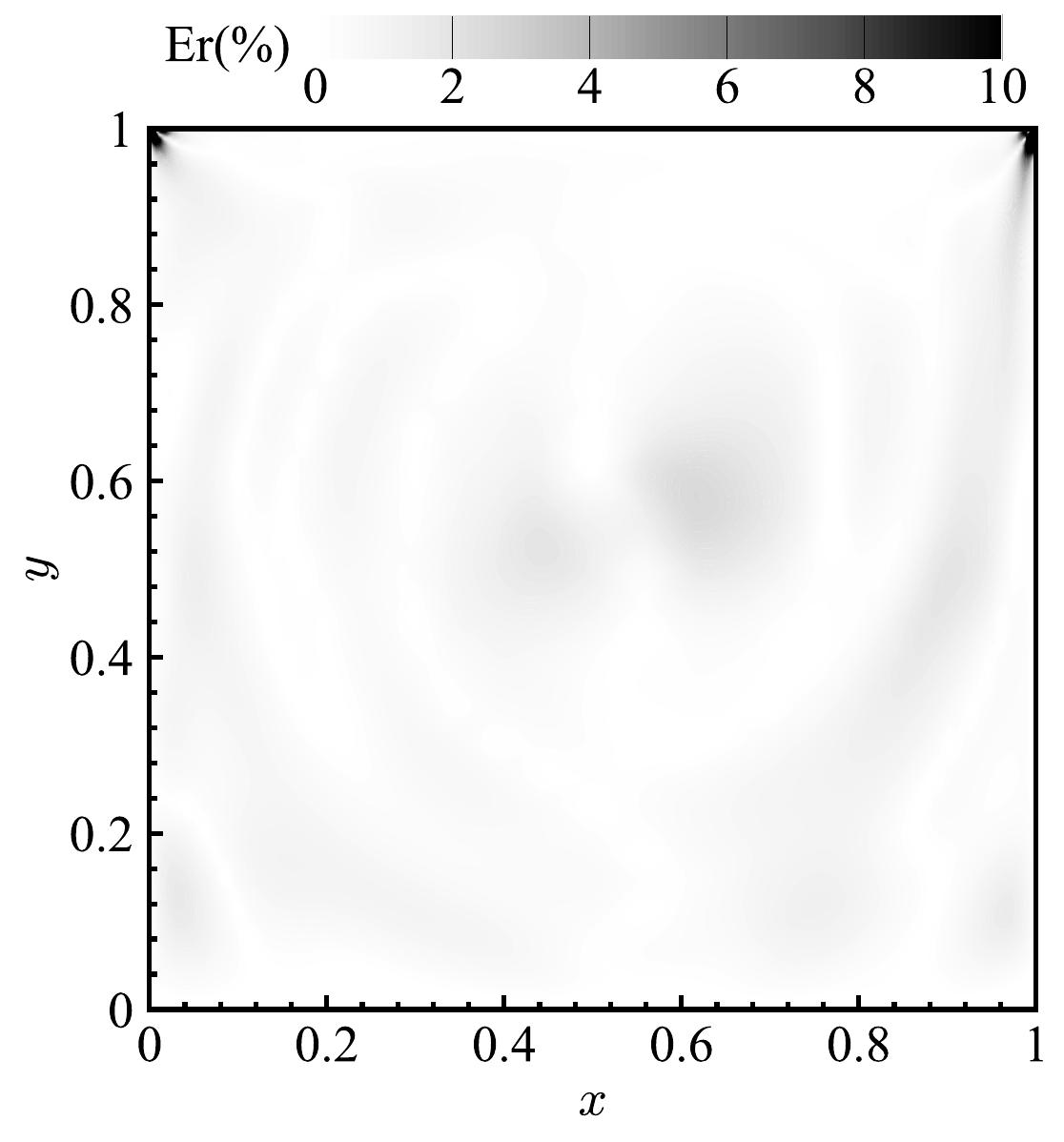}
	}
	{
		\includegraphics[height=4cm]{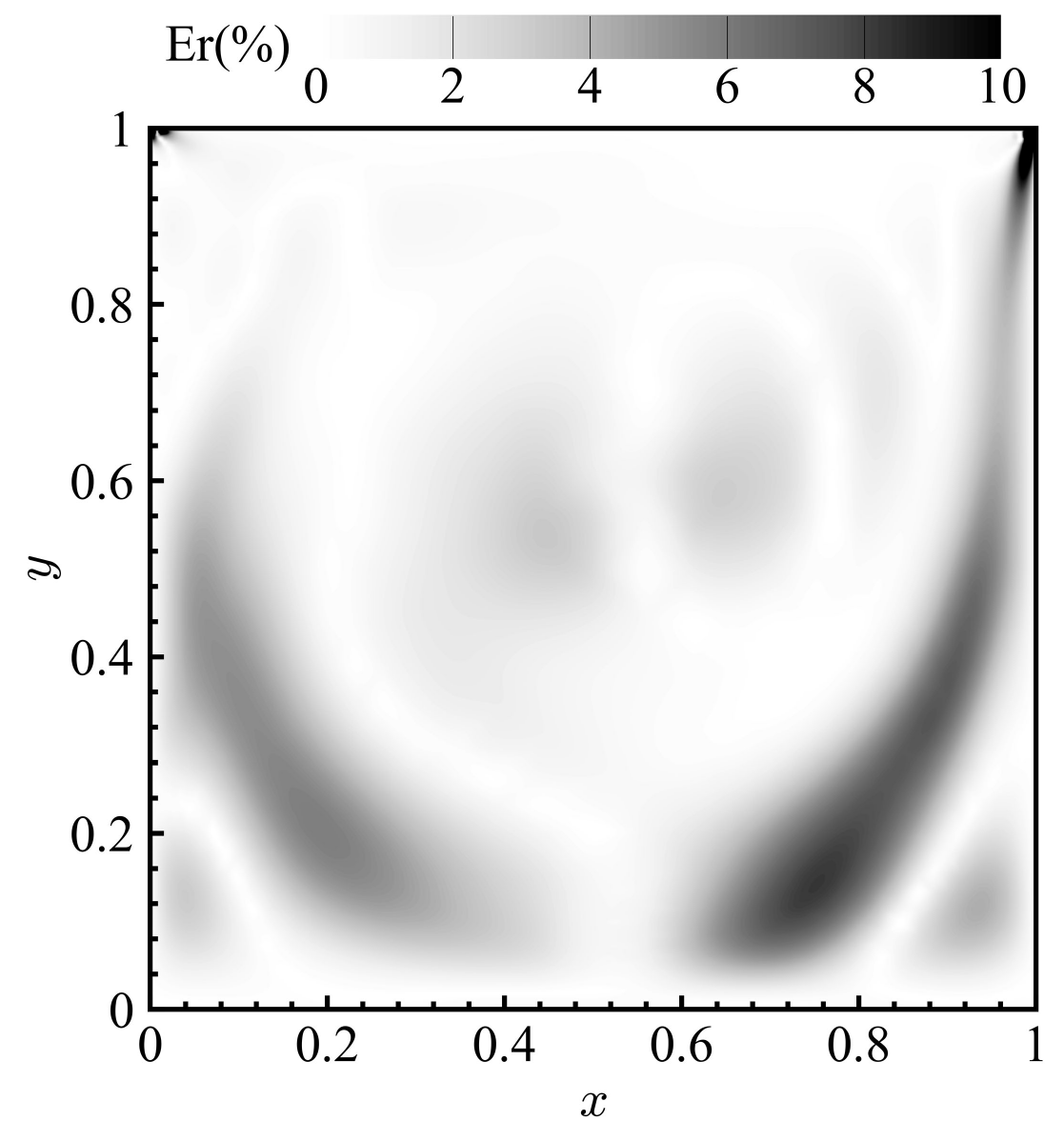}
	}
	\caption{The relative error contours for lid driven cavity flow predicted by the PIFBNN model (left column) and the PINN model (right column). (a):Streamwise flow velocity: $u$,(b):Transverse velocity: $v$}
	\label{cavres}
\end{figure}

It can be observed that the relative error of PIFBNN is lower than that of PINN along the entire $x$-direction. Notably, the error curves of $u$ and $v$ exhibit similar trends. At points $x = 0$ and $x = 1$, the errors are limited to a lower level owing to the boundary conditions, whereas the errors away form the boundary points tends to increase, especially when approaching $x = 1$, where error peaks appear. This may be due to conflicting boundary conditions in the top-cover area. The boundary condition at $x = 1$ and $x = 0$ in the top cover is $\mathbf{u} = (1,0)$, whereas for the non-slip wall at the same coordinate point, the boundary condition is $\mathbf{u} = (0,0)$. The discontinuity of the boundary conditions led to inaccurate predictions in a small region of the upper-left and upper-right corners of the cavity.
The relative error contours predicted by the PINN and PIFBNN are shown in Figure\ref{cavres}.

\begin{figure}[!htbp]
	\centering
	\subfigure[Ground truth]{
		\includegraphics[height=4cm]{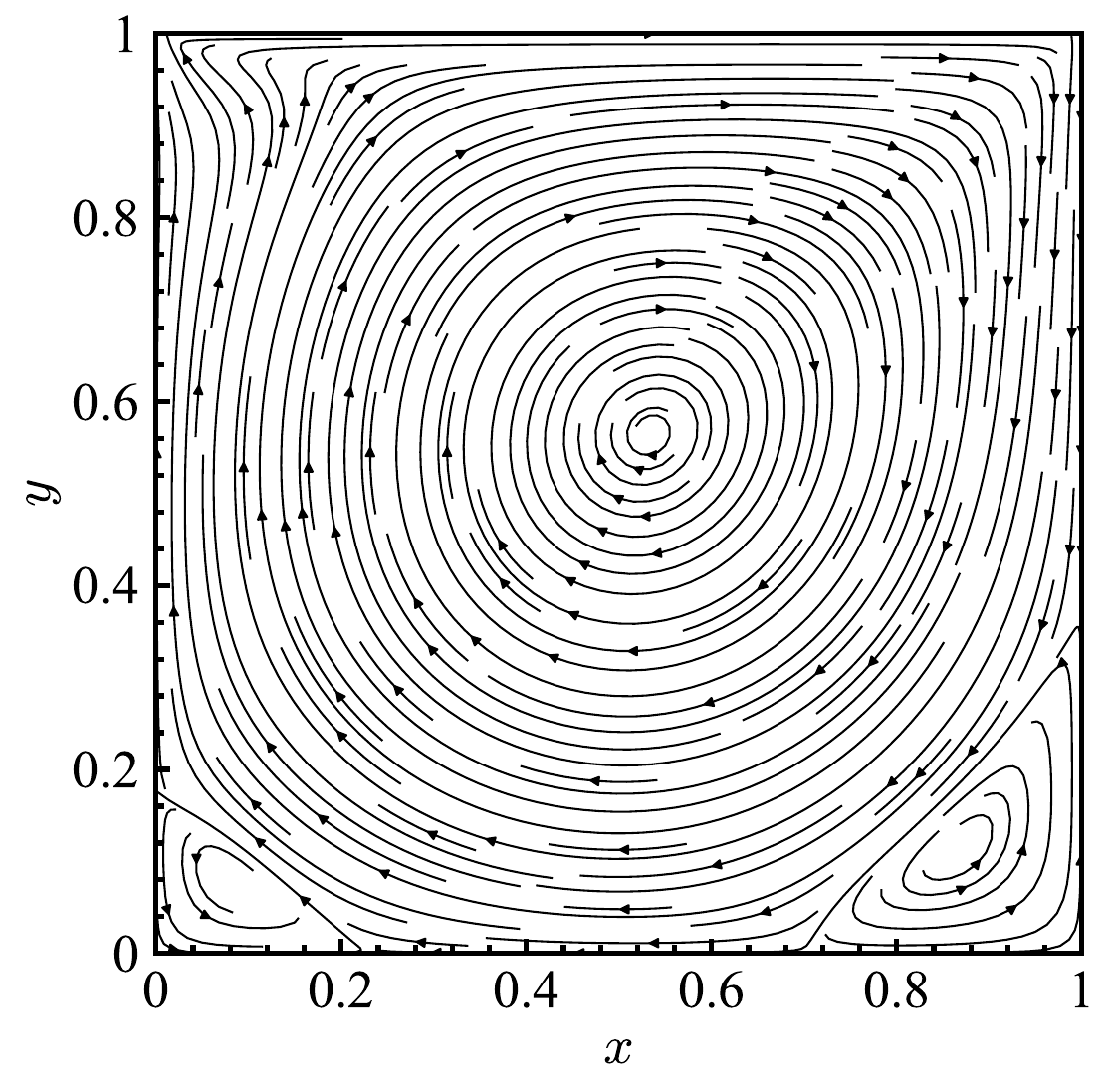}
	}
	\subfigure[PIFBNN]{
		\includegraphics[height=4cm]{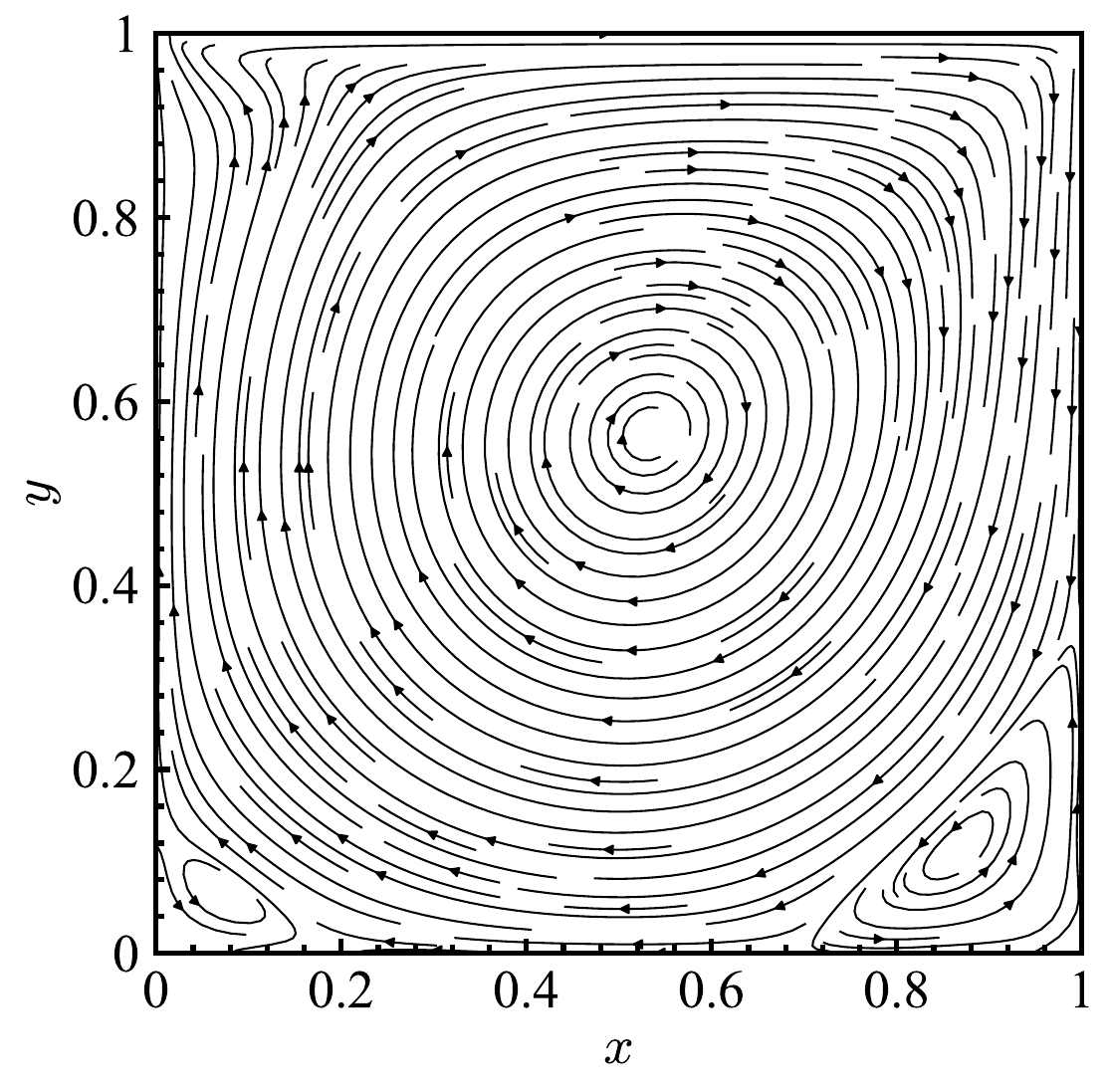}
	}
	\subfigure[PINN]{
		\includegraphics[height=4cm]{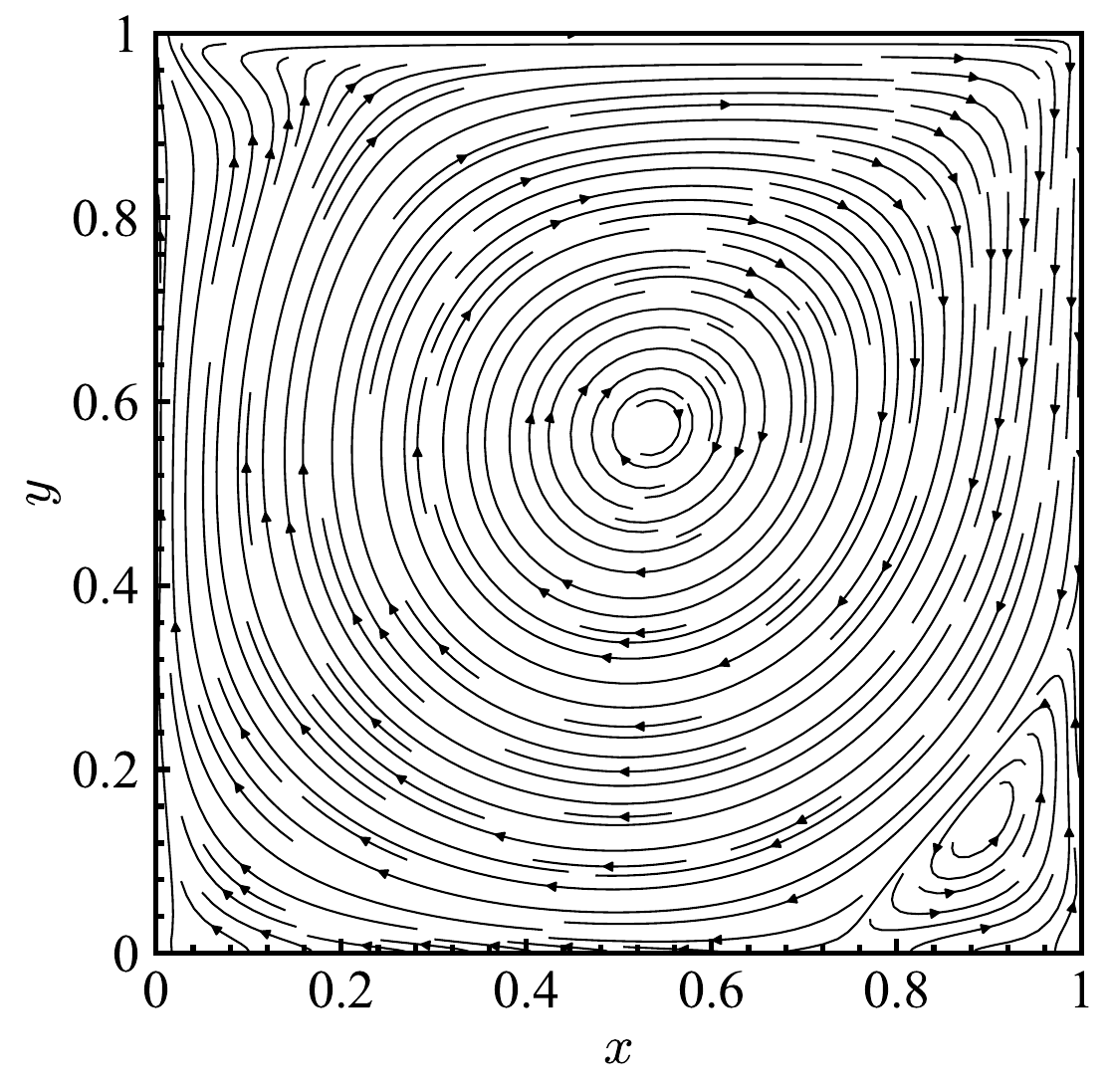}
	}
	
	\caption{Streamlines of lid driven cavity flow. }
	\label{cavstream}
\end{figure}

From the contours, it is evident that the errors of neural networks are concentrated in areas where vortices exist, that is, where the viscous effect is more pronounced. Similar to earlier observations, the relative error of PIFBNN in all regions is significantly smaller than that of PINN, indicating the stronger ability of PIFBNN to capture viscous features. Significant prediction errors occur in the upper-left and upper-right corners of the cavity, with the most pronounced discrepancies appearing in the upper-right region. These errors correspond to the numerical artifacts generated by the discontinuous boundary conditions described previously. Although the error is relatively large, only a small region near the boundary is occupied.
The streamlines of lid-driven cavity wake predicted by PIFBNN and PINN are illustrated in Figure \ref{cavstream}.

It can be learned that both neural networks can well predict the streamline trend of the large vortex in the middle of cavity. PIFBNN successfully predict each vortex, including the bottom-left and bottom-right corner vortices, whereas PINN does not. At the wall surfaces of the left and right corner vortices, PINN shows streamlines perpendicular to the wall, starting from the wall surface. This result violates the boundary conditions, which is a serious non-physical error, yet PIFBNN maintains full compliance with physical laws at all wall boundaries in its predictions. This indicates that PIFBNN strictly enforces boundary conditions and demonstrates its ability to capture the intrinsic physical properties of fluid vortex viscous motion more accurately and learn physical laws more effectively.

\section{\label{sec:level1/4} Influence of activation function selection on networks}

Conventional fully-connected artificial neural network (ANN) inherently possess limited nonlinear approximation capabilities due to their linear compositional structure. Consequently, the incorporation of suitable nonlinear activation functions across network layers becomes imperative to enable effective learning of nonlinear relationships. This consideration is particularly critical for physics-informed neural networks (PINNs) employing ANN architectures, as the activation function selection fundamentally determines the network's capacity to capture and represent nonlinear physical phenomena.
Two PDEs using ANN,FBNN,PINN and PIFBNN with $leaky-relu$ and $tanh$ as activation functions are predicted, with the exception of the activation function, all the neural network architectures and settings used are same as in Section \ref{sec:level1/3}. Owing to the lack of physical information embedded in the ANN and FBNN, these neural networks cannot predict the entire domain base solely on boundary conditions without labeled data. For this reason, in each case, we randomly sample labeled data points within the domain for neural network training, thus replacing the boundary or initial conditions. All the neural networks are trained using same strategy.

\subsection{Helmholtz equation}
\begin{figure}[h!] 
	\centering
	\subfigure[Training loss]{
		\includegraphics[height=5.5cm]{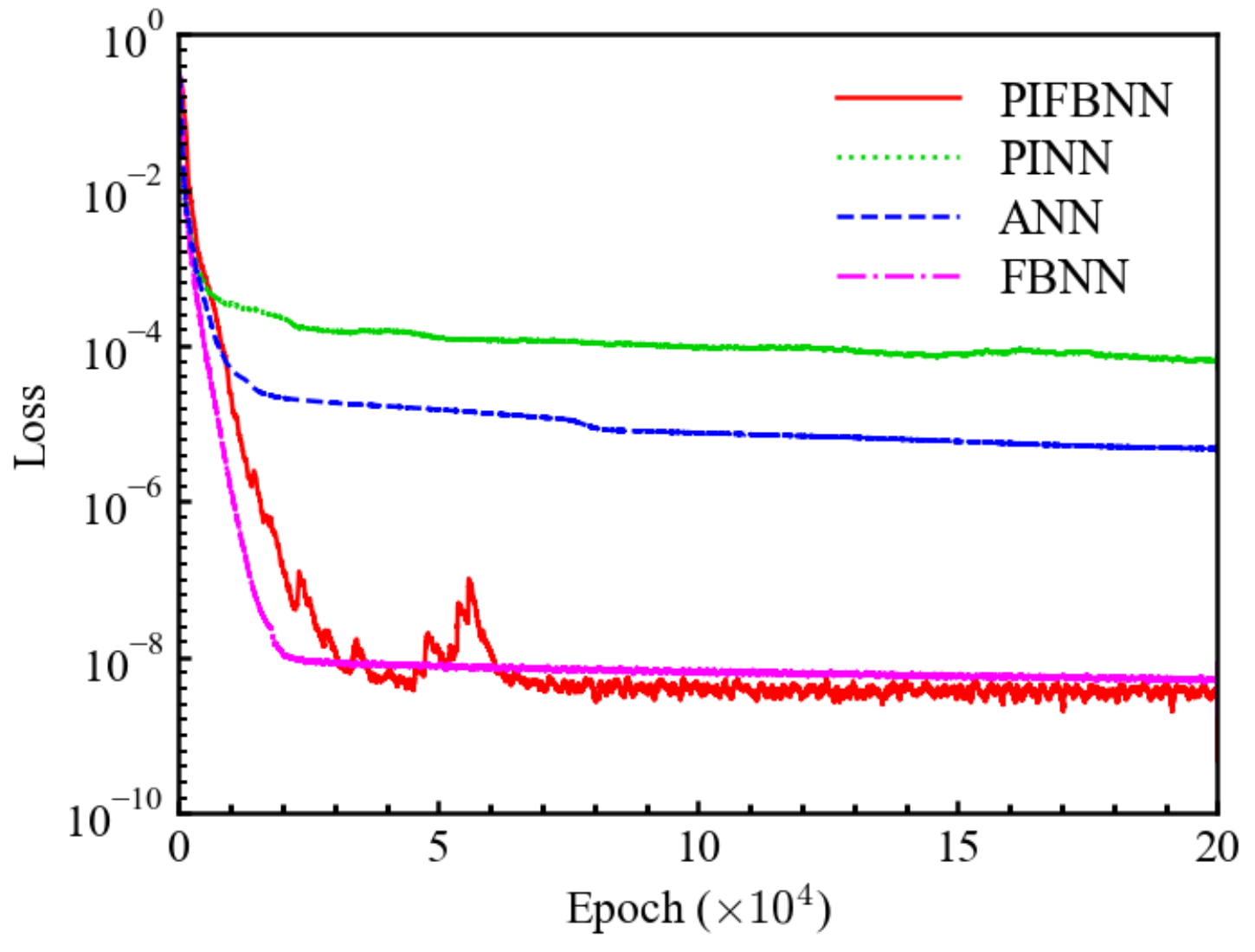}
	}
	\subfigure[Testing loss]{
		\includegraphics[height=5.5cm]{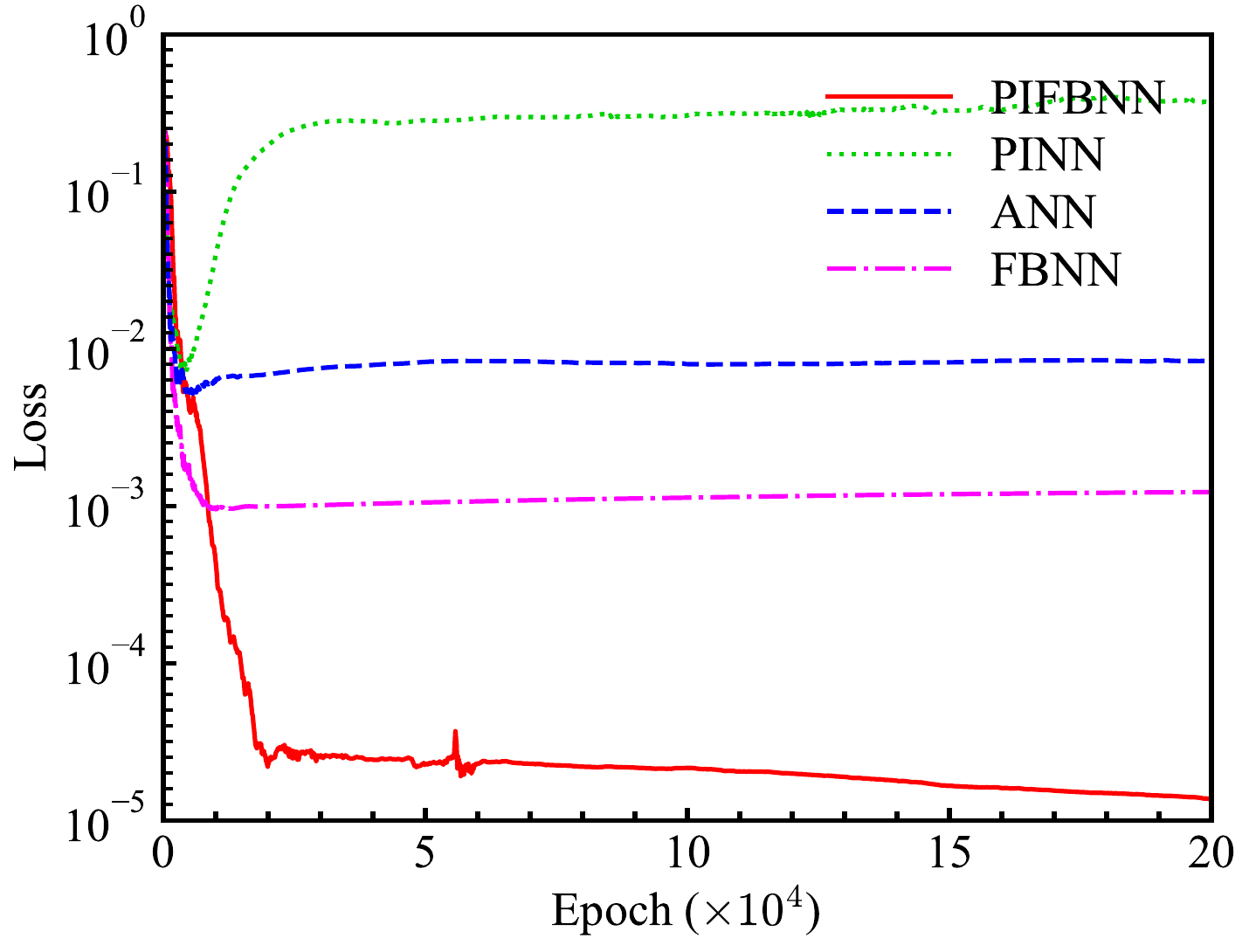}
	}
	\caption{Learning curves forHelmholtz equation reconstruction using different neural networks with $leaky-relu$ as the activation function.  }
	\label{hmfuncloss}
\end{figure}

In this example, 4*128 neural network architecture is used with a ratio of Fourier nodes is set to 0.6. A total of 225 points are randomly sampled within the same computational domain (as in the previous section) as labeled data for training and 1000 points as testing data to evaluate the performance of neural networks. Owing to the presence of labeled data, only 800 random sampling points are used for the residual calculation within the domain.
The learning curves of the Helmholtz equation reconstructed by different neural networks with $leaky-relu$ as the activation function are demonstrated in Figure \ref{hmfuncloss}.

Owing to the addition of a small number of points for calculating physical information residuals, the training loss of PIFBNN is slightly lower than that of FBNN. Perhaps due to the conflict between physical residual term and training loss term during neural network optimization, the training process of PIFBNN exhibits greater instability. However, due to optimization conflicts between competing loss terms, PINN employing $leaky-relu$ activation function fails to achieve successful training convergence, exhibiting significantly higher training loss values compared to ANN. Comparing FBNN and ANN separately, the FBNN depicts substantial advantages, including deeper learning of labeled data and lower testing losses. This indicates that FBNN architecture has stronger nonlinear learning and periodic modeling capabilities than ANN architecture. Notably, the training loss of PIFBNN is not much lower than FBNN because of optimization conflicts, but the test loss of PIFBNN is significantly lower than that of FBNN. These results demonstrate that physical residuals play a significant role in training process and prediction of neural networks.
The velocity profile of Helmholtz equation reconstructed by different neural networks are shown in Figure {hmfunclr0.20.50.8x}.

\begin{figure}[h!] 
	\centering
	\subfigure[$x=0.2$]{
		\includegraphics[height=5.5cm]{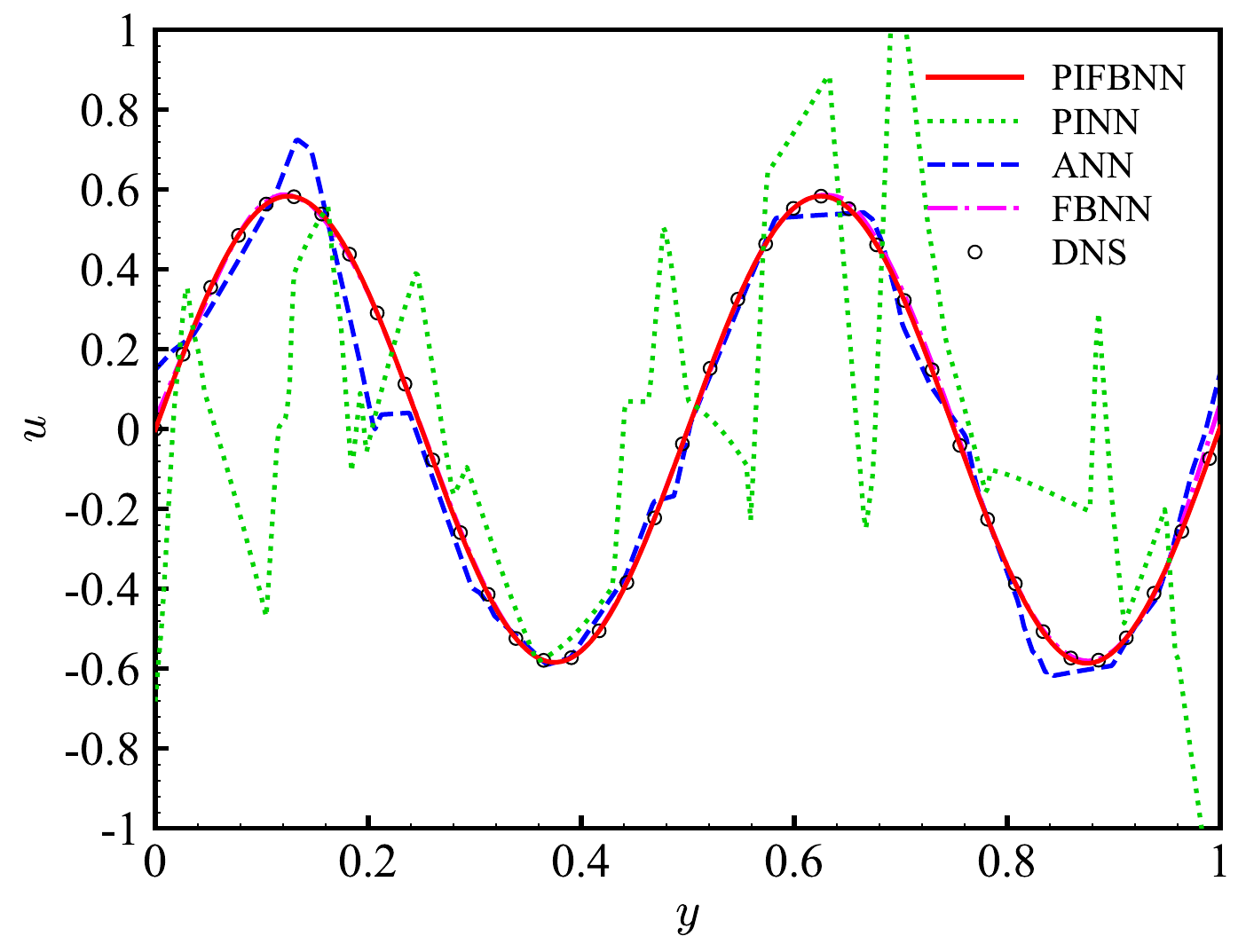}
	}
	\subfigure[$x=0.8$]{
		\includegraphics[height=5.5cm]{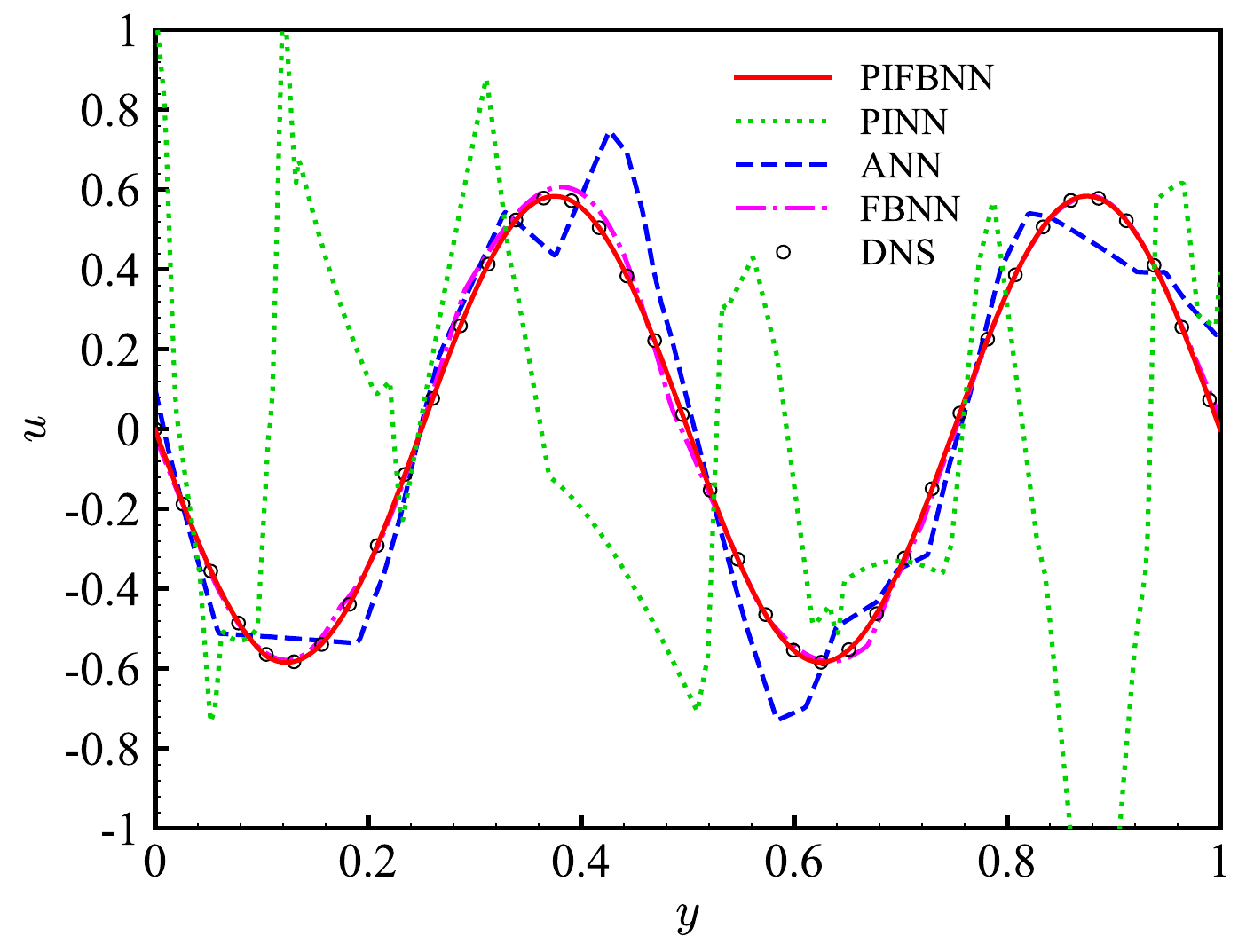}
	}
	\par%
	\subfigure[$x=0.5$]{
		\includegraphics[height=5.5cm]{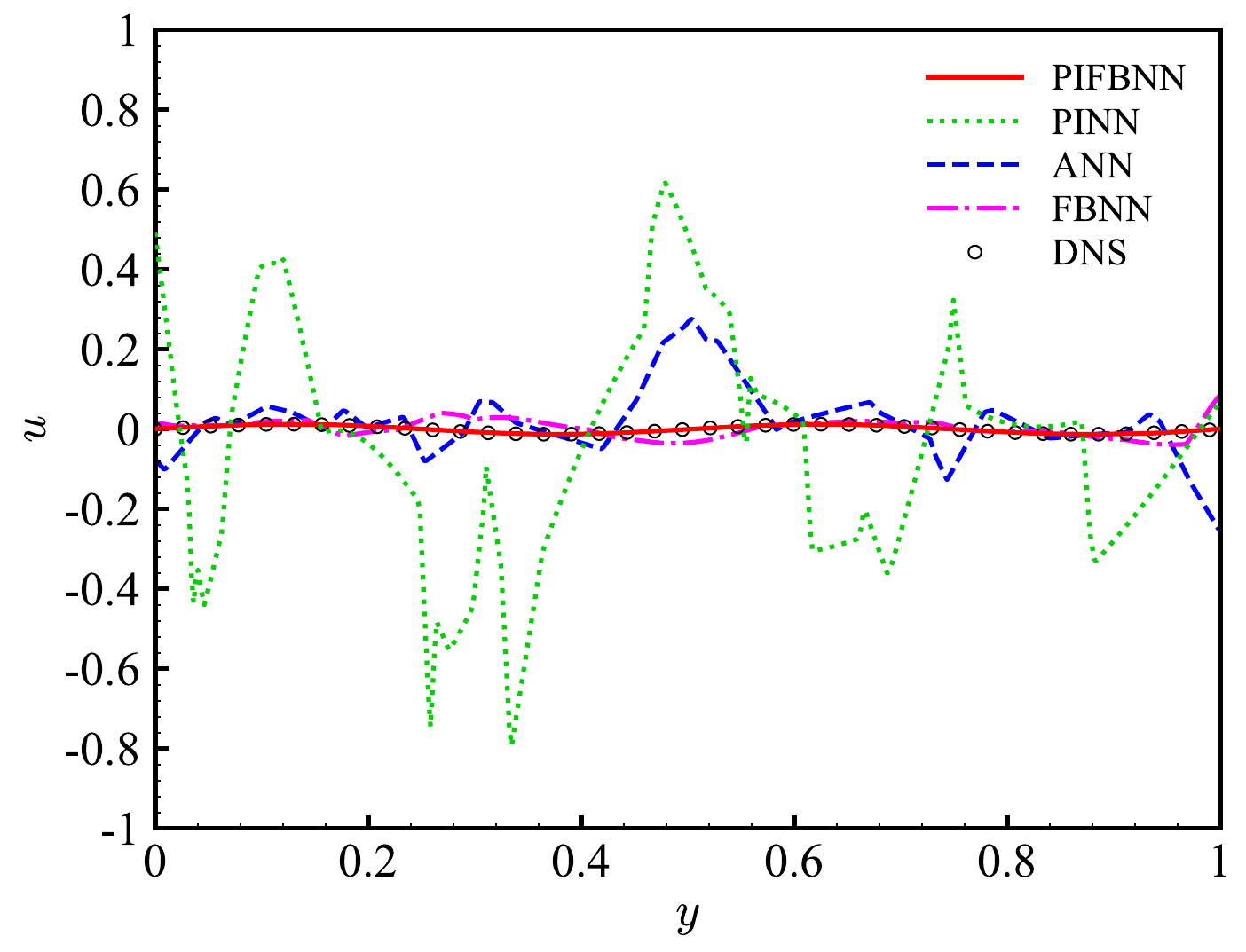}
	}
	\caption{Comparisons of velocity profiles for the reconstruction of the Helmholtz equation using different neural networks with $leaky-relu$ as the activation function.}
	\label{hmfunclr0.20.50.8x}
\end{figure}
\begin{figure}[h!] 
	\centering
	{
		\includegraphics[height=5.5cm]{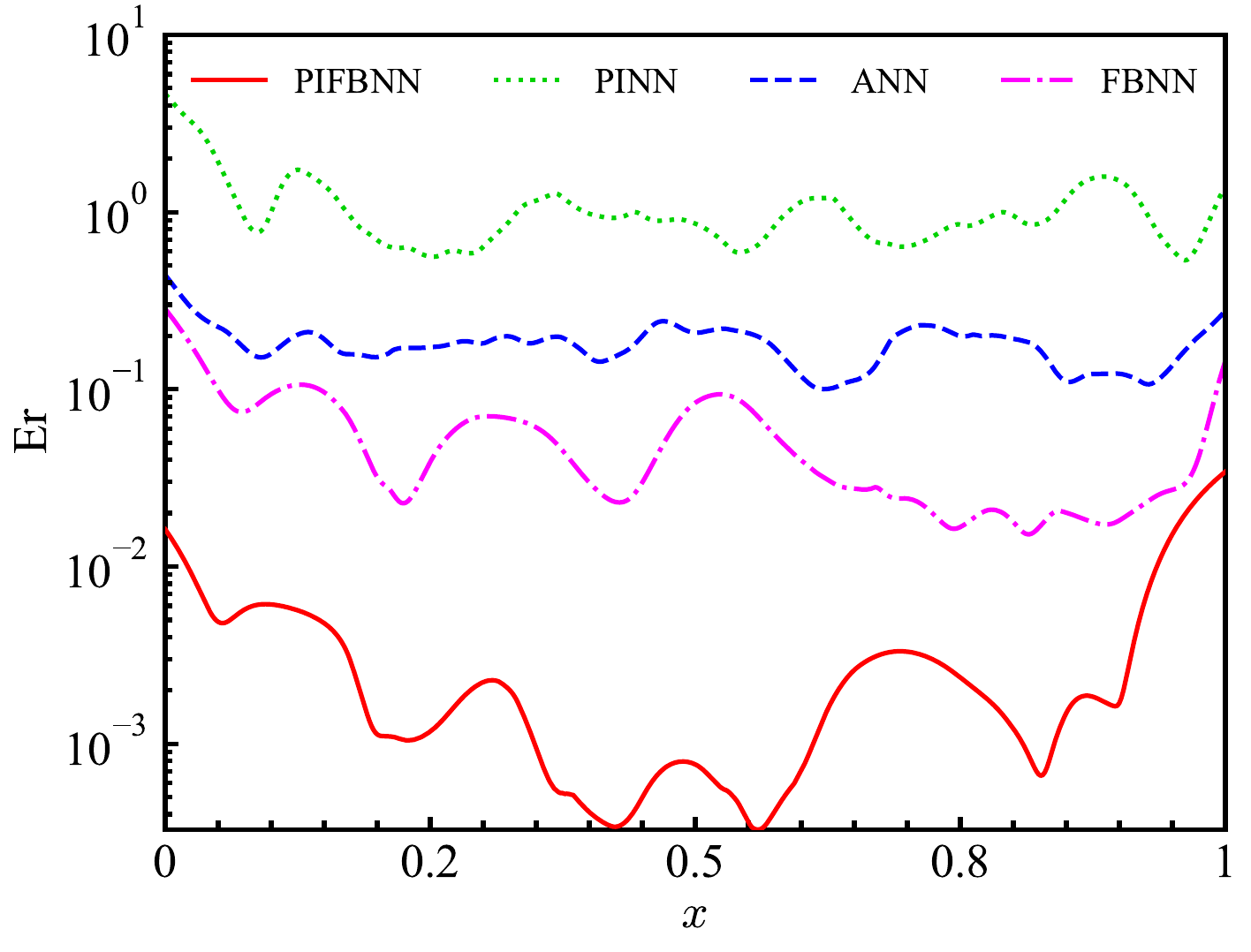}
	}
	\caption{Relative errors of different neural networks using $leaky-relu$ activation function for Helmholtz equation reconstruction. }
	\label{hmfunclrer}
\end{figure}

Regarding curve fitting performance, both PIFBNN and FBNN demonstrate satisfactory approximation of the ground truth. While FBNN exhibits localized fitting errors in regions exhibiting strong nonlinear characteristics, PIFBNN achieves complete agreement with all reference curves across the entire domain. ANN can predict the general trend of the curve, whereas the prediction by PINN  is completely incorrect. Notably, in the $x = 0.5$ curves, all neural network architectures demonstrate reduced prediction accuracy, which can be attributed to the diminished magnitude of ground truth values in this region. In this case, only PIFBNN can perfectly fit the ground truth curve. Althought FBNN has larger errors compare to PIFBNN, it is still significantly better than ANN and PINN.
The relative errors of Helmholtz equation predictions along $x$ reconstructed by neural networks using $leakyrelu()$ activation functions are shown in Figure \ref{hmfunclrer}.

\begin{figure}[!htbp]
	\centering
	\subfigure[ANN reconstruction]{
		\includegraphics[height=4cm]{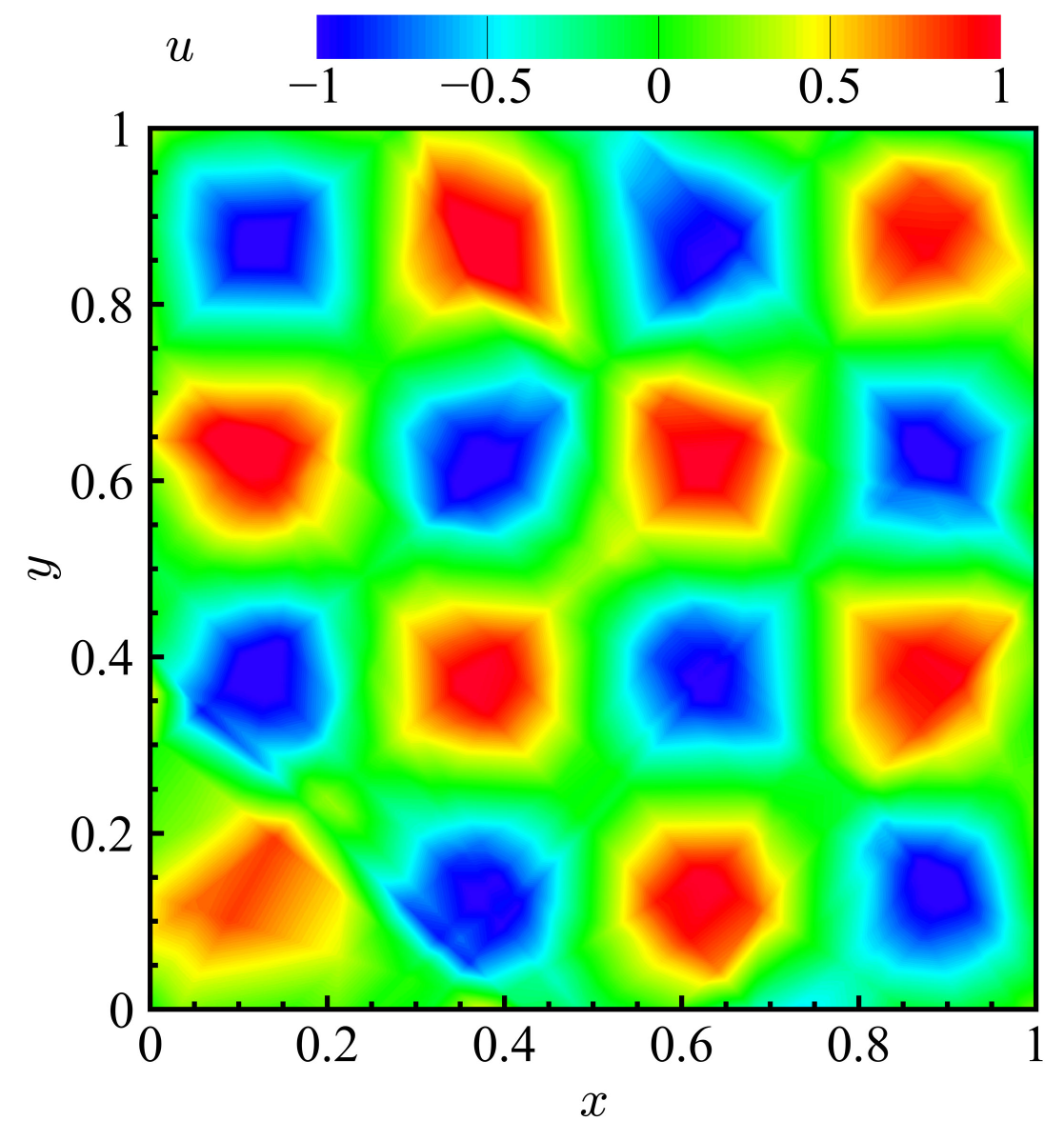}
	}
	\subfigure[FBNN reconstruction]{
		\includegraphics[height=4cm]{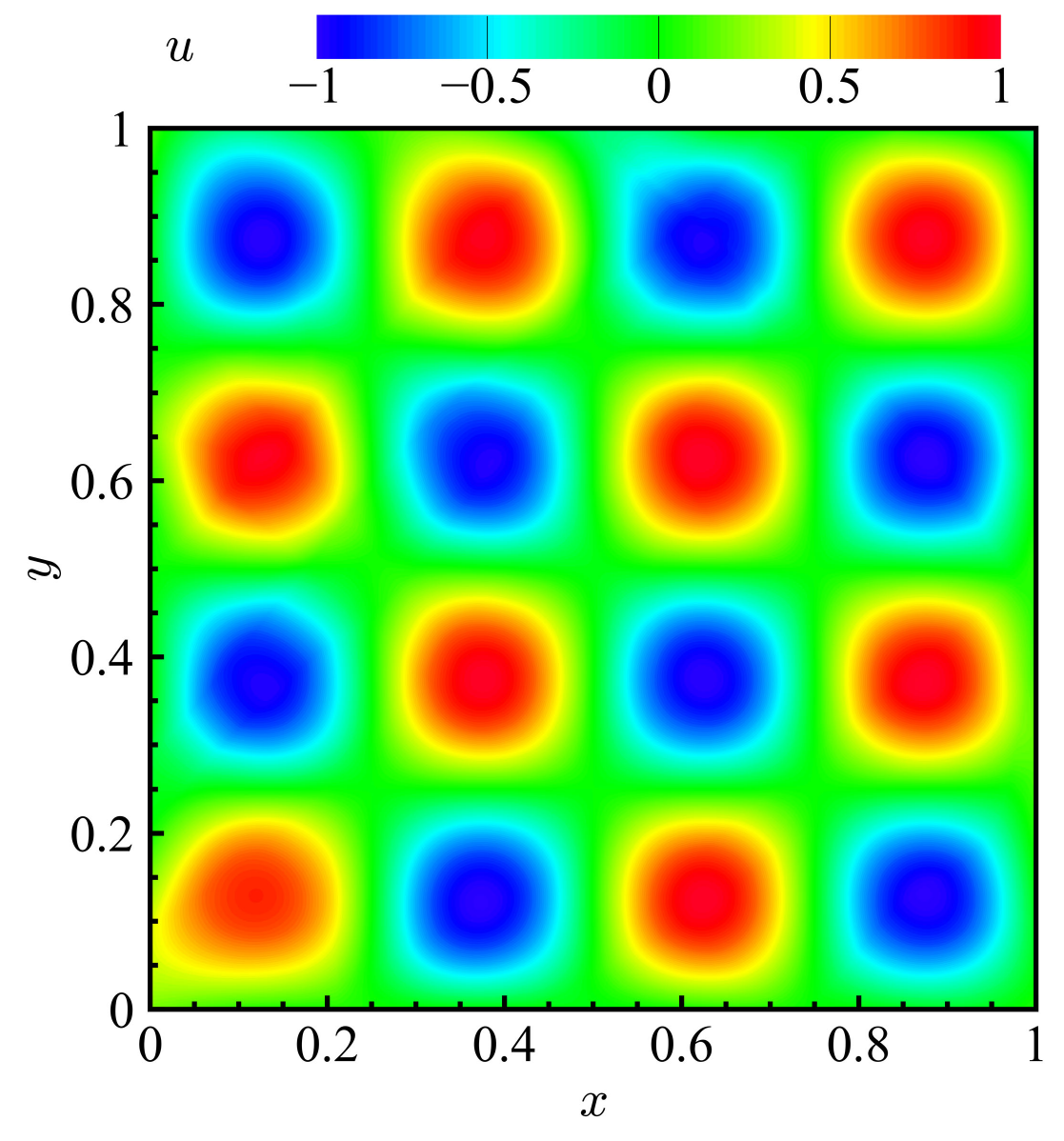}
	}
	\subfigure[PINN reconstruction]{
		\includegraphics[height=4cm]{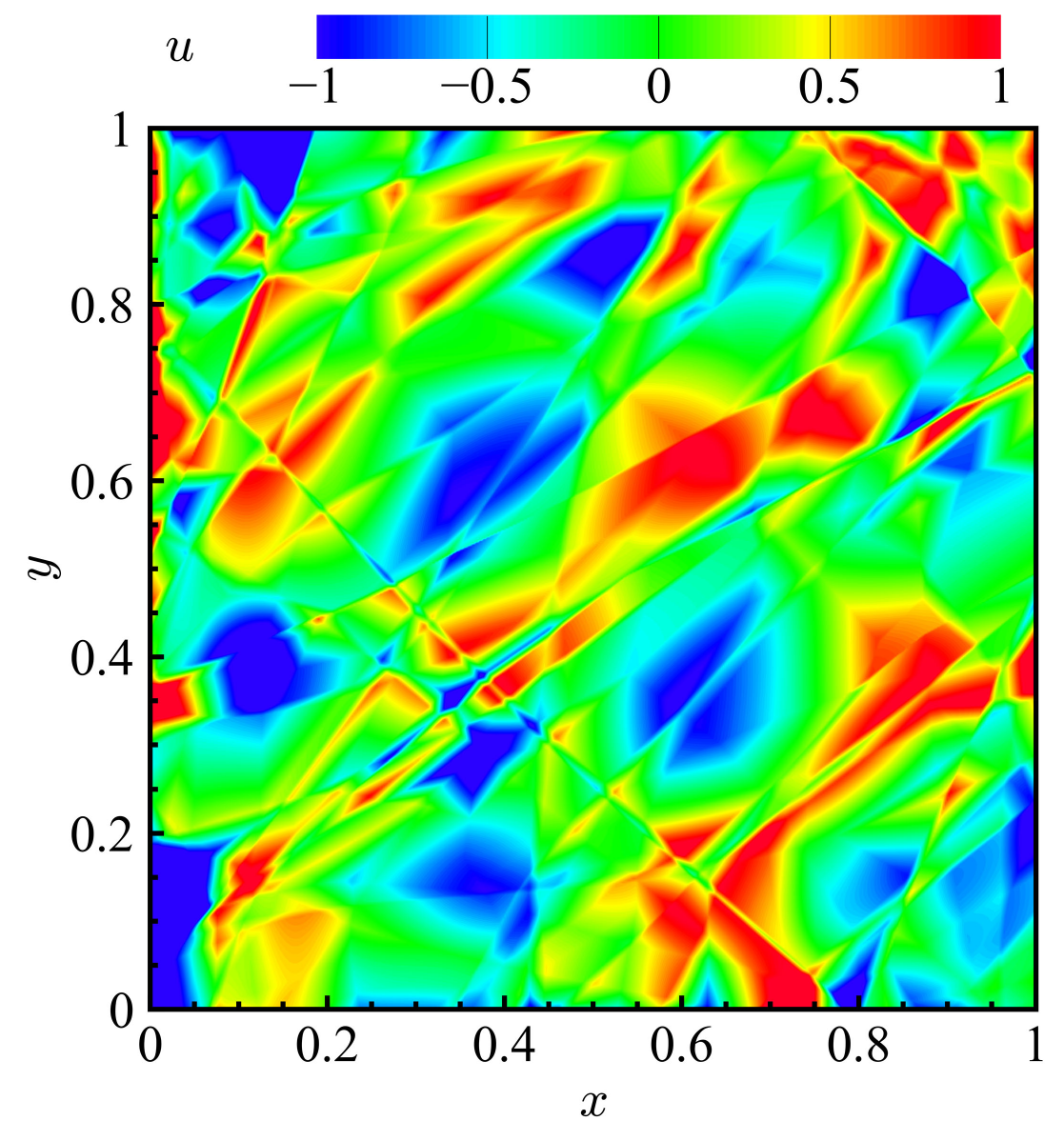}
	}
	\subfigure[PIFBNN reconstruction]{
		\includegraphics[height=4cm]{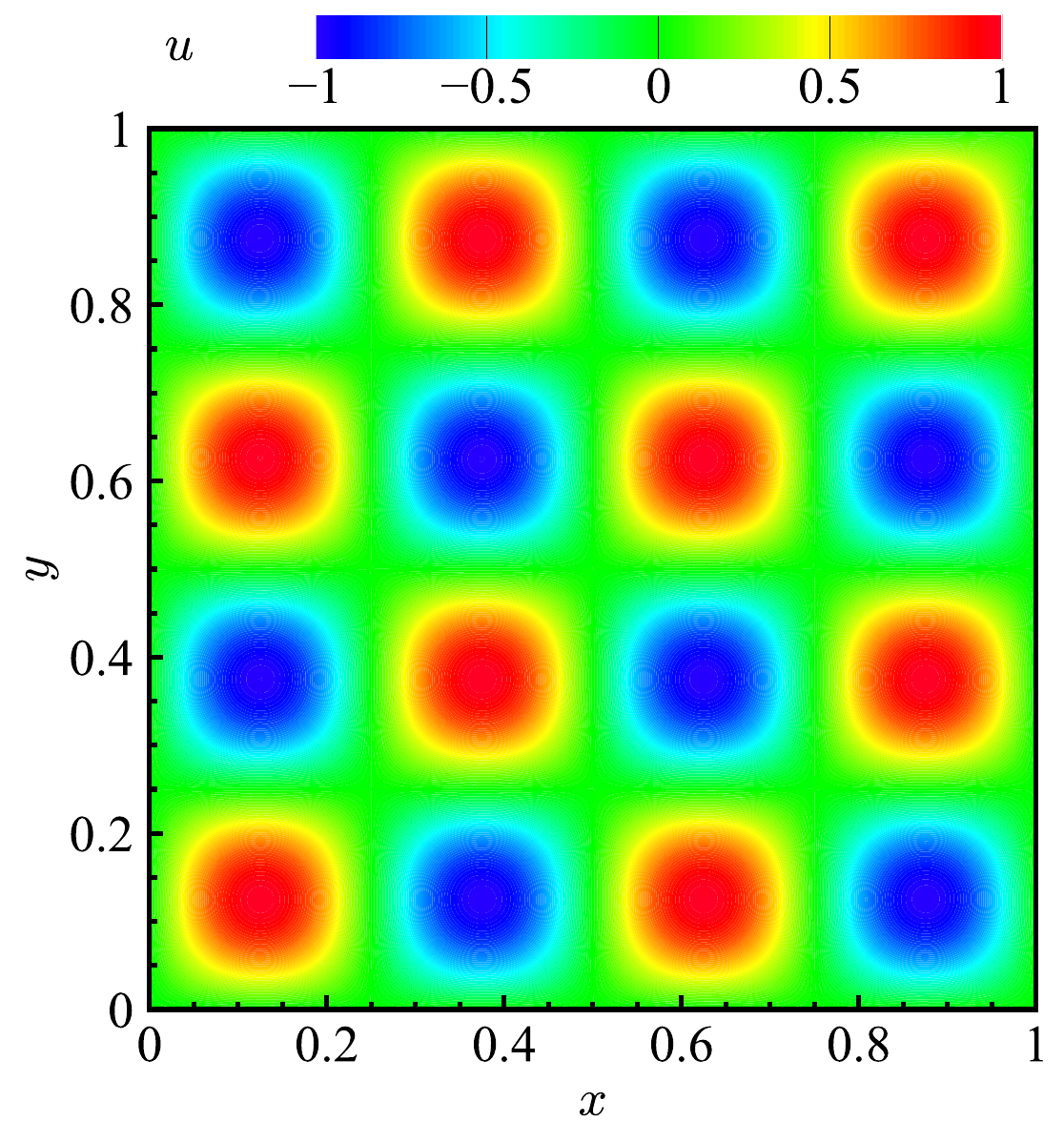}
	}
	
	\par %
	\subfigure[ANN relative error]{
		\includegraphics[height=4cm]{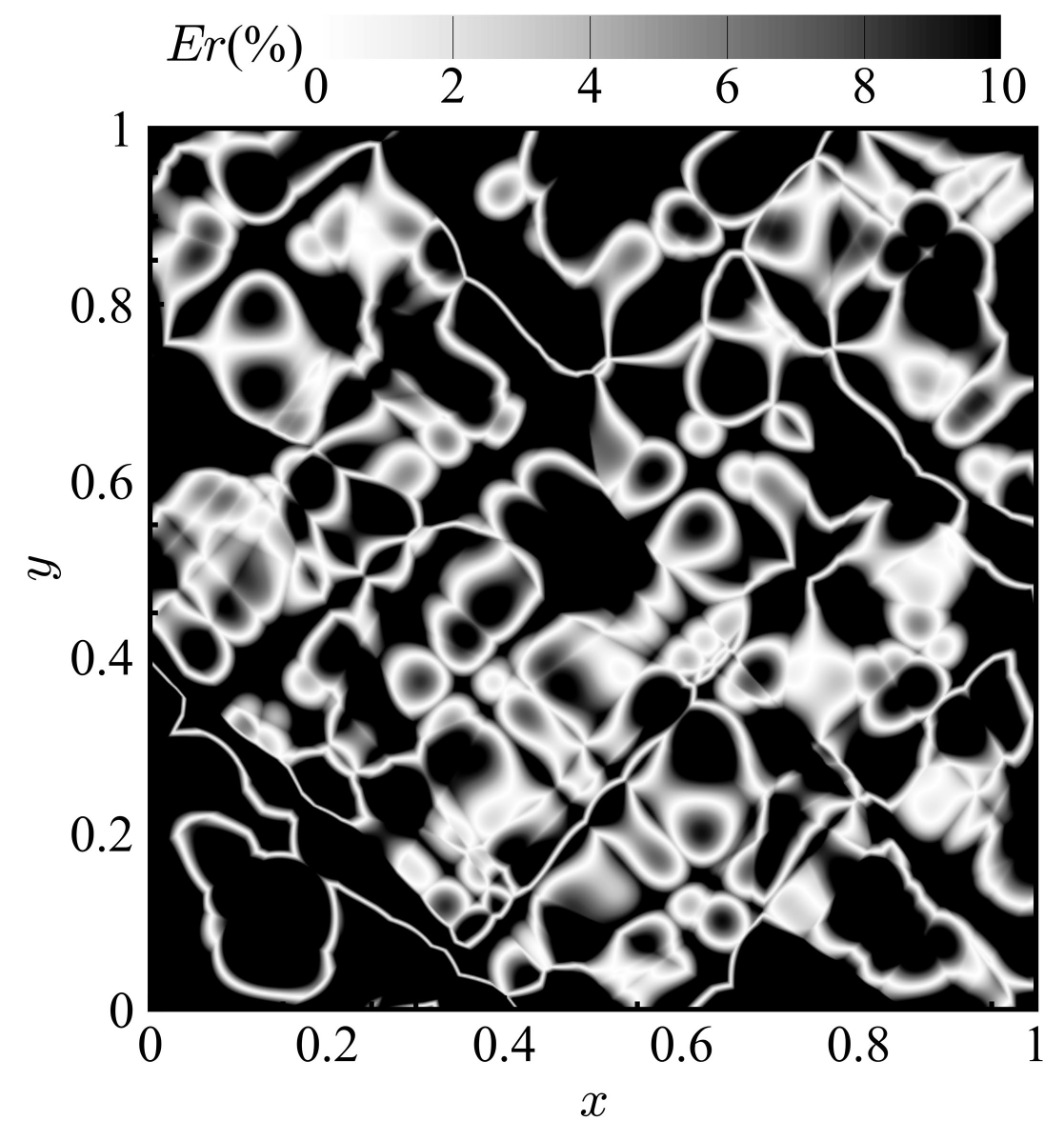}
	}
	\subfigure[FBNN relative error]{
		\includegraphics[height=4cm]{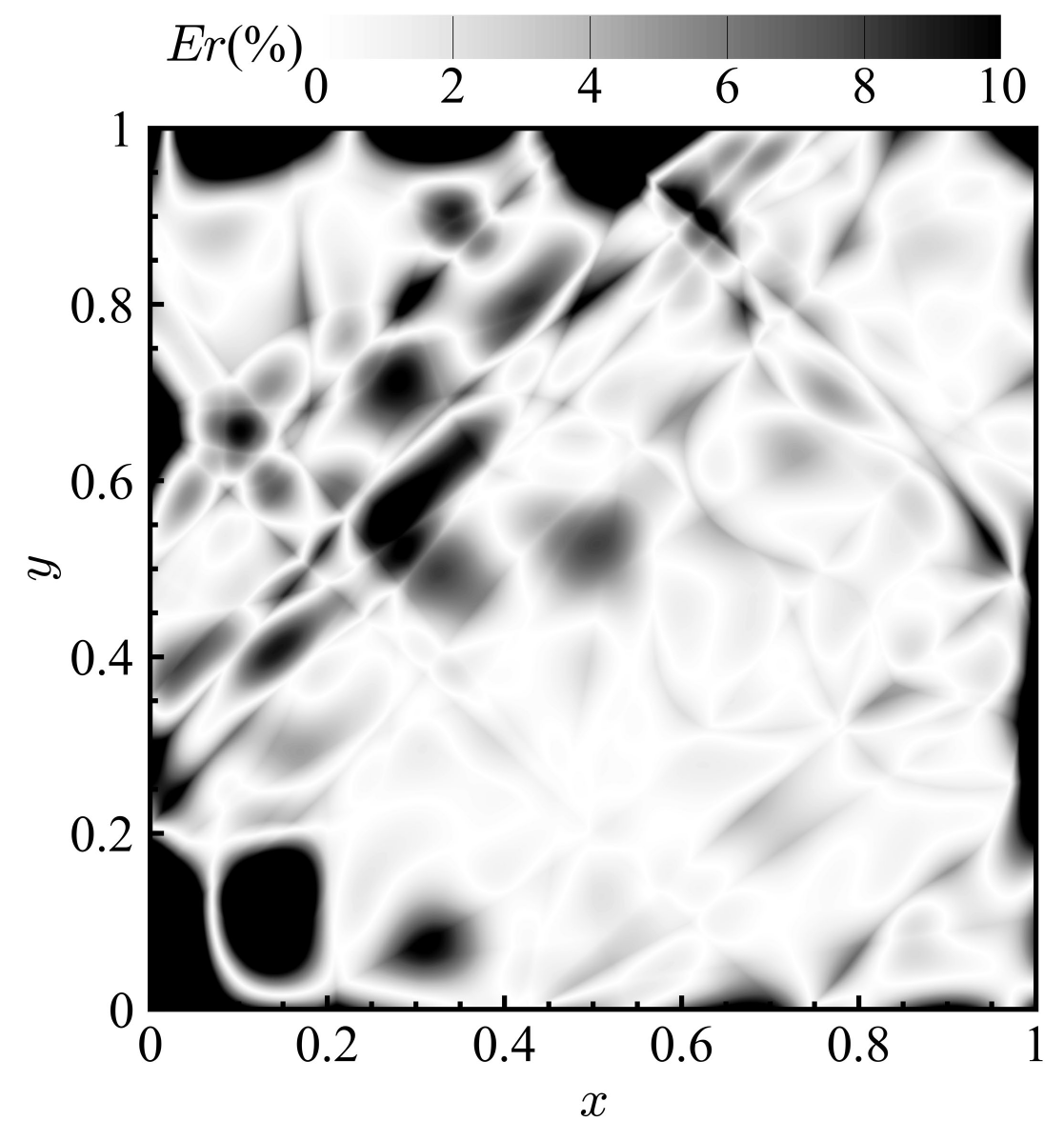}
	}
	\subfigure[PINN relative error]{
		\includegraphics[height=4cm]{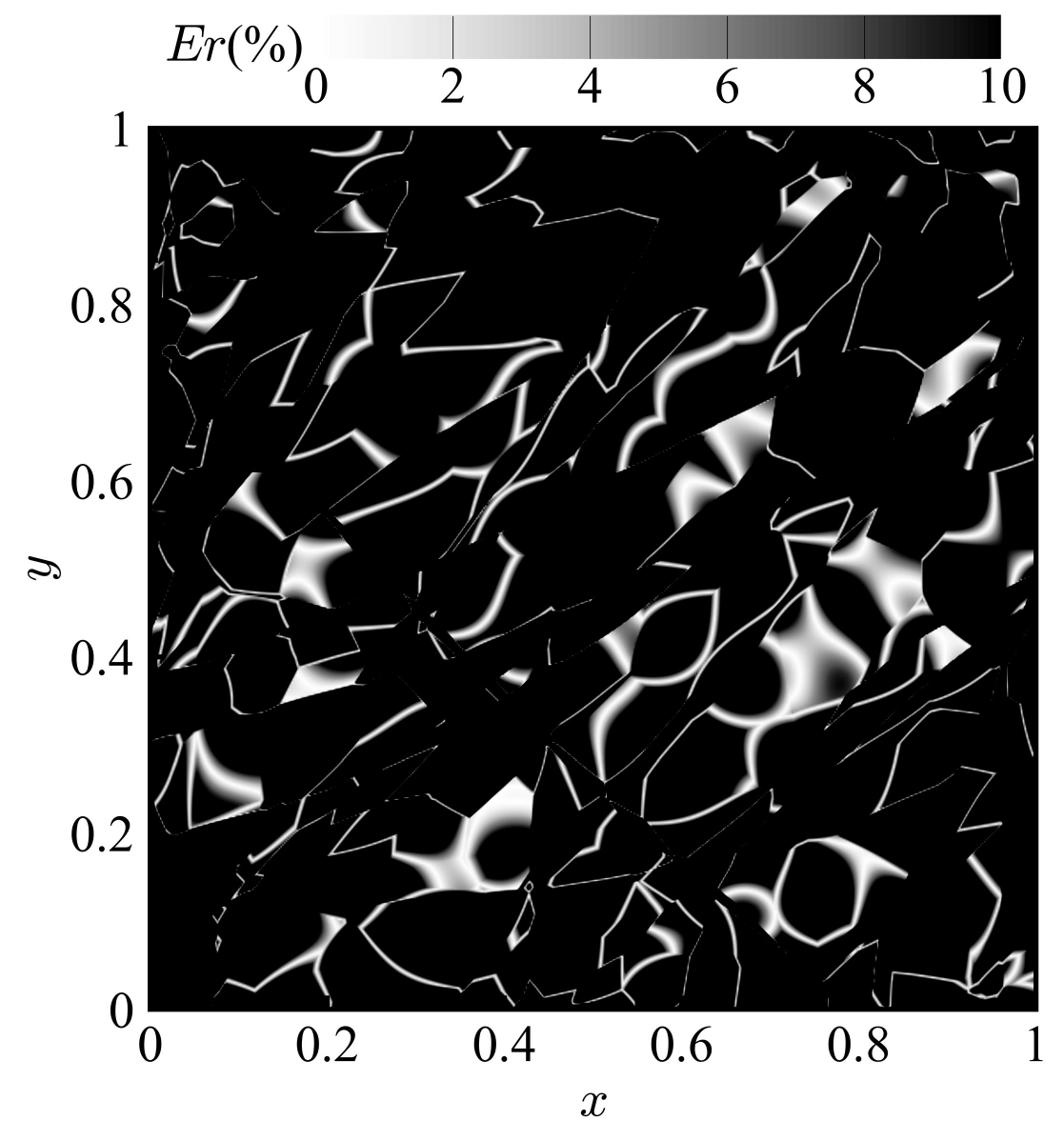}
	}
	\subfigure[PIFBNN relative error]{
		\includegraphics[height=4cm]{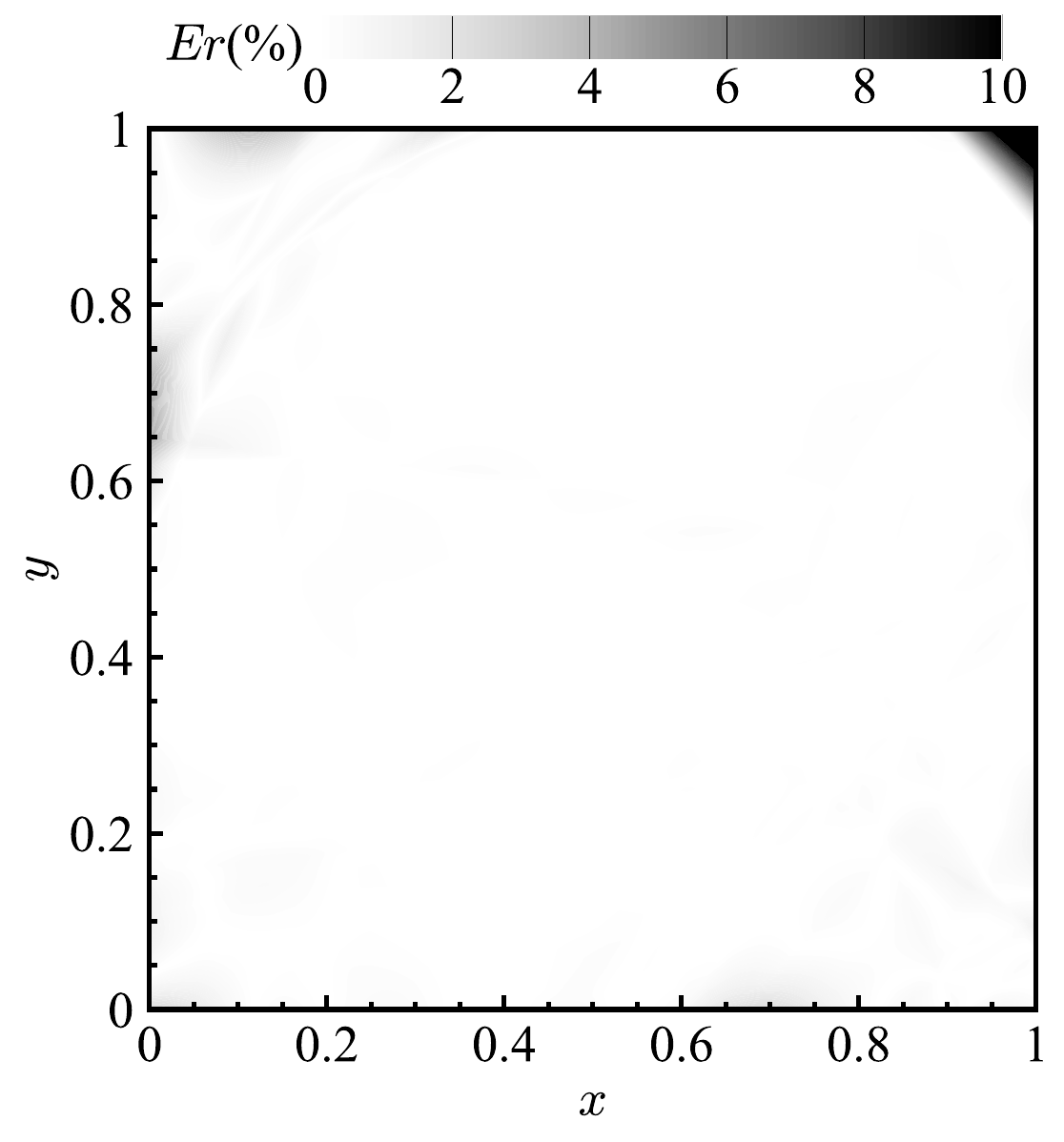}
	}

	\caption{Reconstruction velocity and relative error contours of Helmholtz equation with different neural networks using $leaky-relu$ as the activition function. }
	\label{hmfunclrcontour}
\end{figure}

From the relative error curves, we can see that PIFBNN exhibits superior performance with significantly reduced prediction errors compared to benchmark models. Furthermore, FBNN also shows considerable improvements in prediction accuracy over both conventional ANN and PINN approaches. Notably, as boundary conditions are not imposed, all networks exhibit relatively large errors near the boundary, consistent with expectations.
The reconstructed Helmholtz equation velocity profile using different neural networks with $leaky-relu$ as the activation function and the relative error contours are shown in Figure \ref{hmfunclrcontour}.

Similar to the previous observation, ANN is able to roughly capture the waveform of each cycle. However, the waveform is very rough and there are many sharp discontinuities. FBNN predicts relatively standardized waveforms in each cycle, exhibiting only minor prediction deviations near boundary transition regions. PINN confuses all waveforms and fails to capture the periodicity of the solution. PIFBNN perfectly predicts all the cycles and waveforms with almost no visible flaws. From the error contours, compared to the ANN, FBNN only has errors at the boundaries and in certain strongly nonlinear regions. By contrast, the errors of ANN and PINN cover almost the entire region. PIFBNN achieves significantly reduced prediction errors, with the only exception of localized discrepancies confined to the upper-right boundary region. Across the remaining computational domain, the model maintains high-accuracy predictions consistent with ground truth data.

In this case ,the use of $leaky-relu$ activation function in PINN led to an optimization conflict among the loss terms, ultimately resulting in training failure. Owing to the absence of single-interval in $leaky-relu$ activation functions, neural network architectures including PINN and ANN encounter significant challenges in accurately approximating highly nonlinear target functions and periodic oscillatory solutions. However, due to Fourier nodes in FBNN are equivalent to $cosine$ and $sine$ activations to some extent, FBNN still have a strong nonlinear learning ability when selecting linear-like activation functions.
 
\begin{figure}[h!] 
	\centering
	{
		\includegraphics[height=5.5cm]{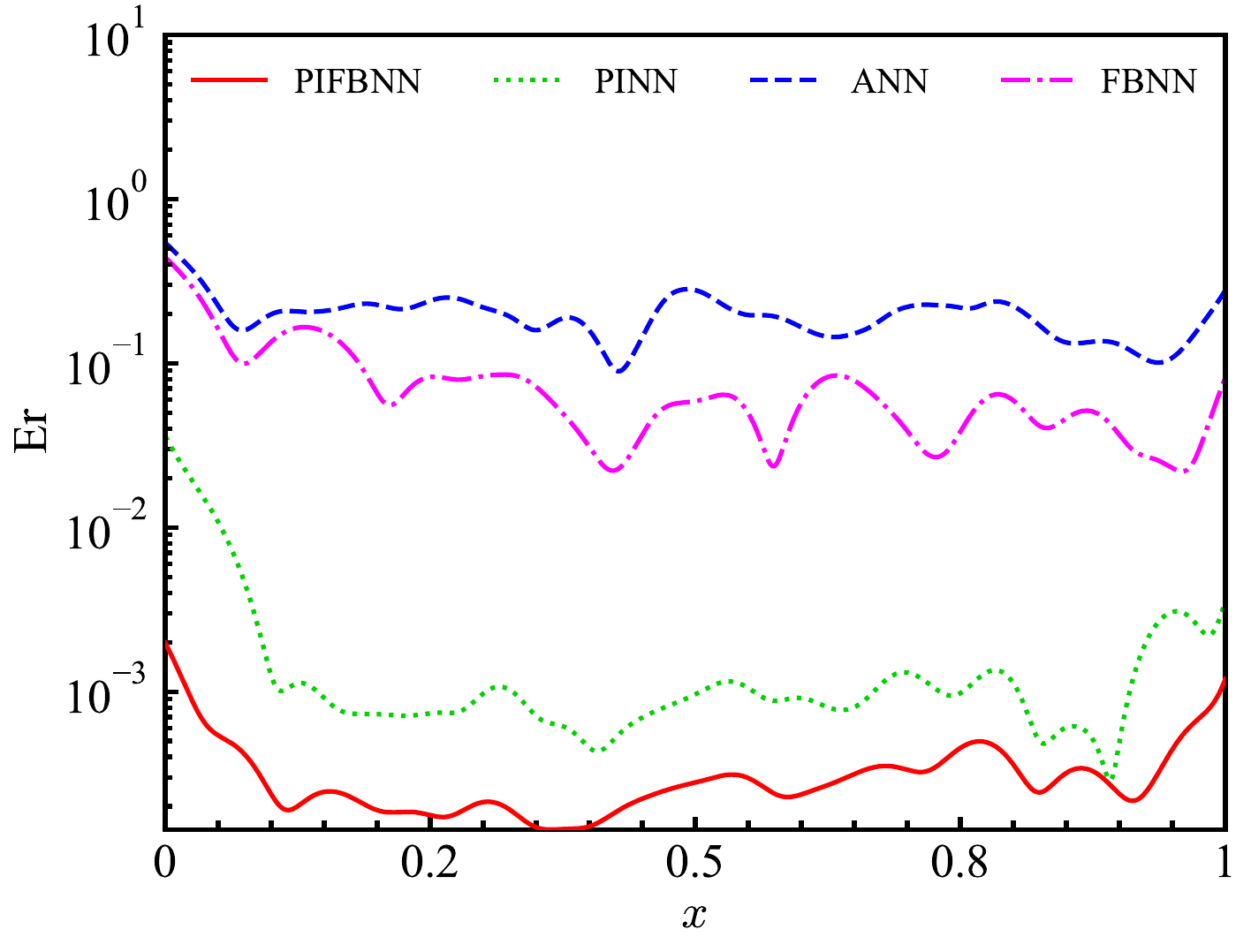}
	}
	
	\caption{Relative errors of different neural networks using $tanh$ as the activation function for Helmholtz equation reconstruction. }
	\label{hmfuncther}
\end{figure}

To control for a unique variable, all other settings and training strategies unchanged and the neural network is retrained using $tanh$ as the activation function. The relative error curves along $x$ reconstructed by different neural networks using $tanh$ activation function are illustrated in Figure\ref{hmfuncther}.

\begin{figure}[!htbp]
	\centering
	\subfigure[ANN]{
		\includegraphics[height=4cm]{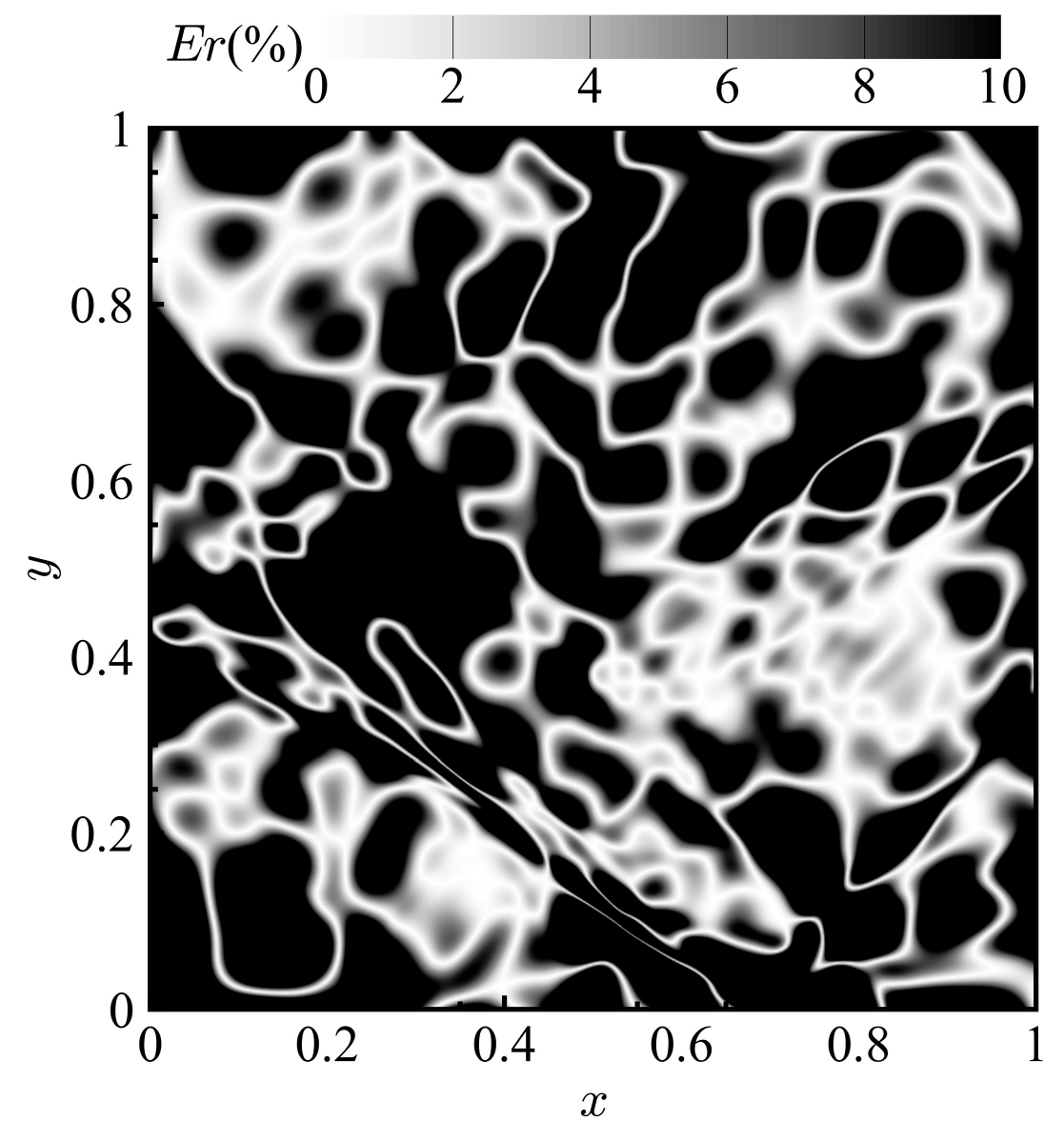}
	}
	\subfigure[FBNN]{
		\includegraphics[height=4cm]{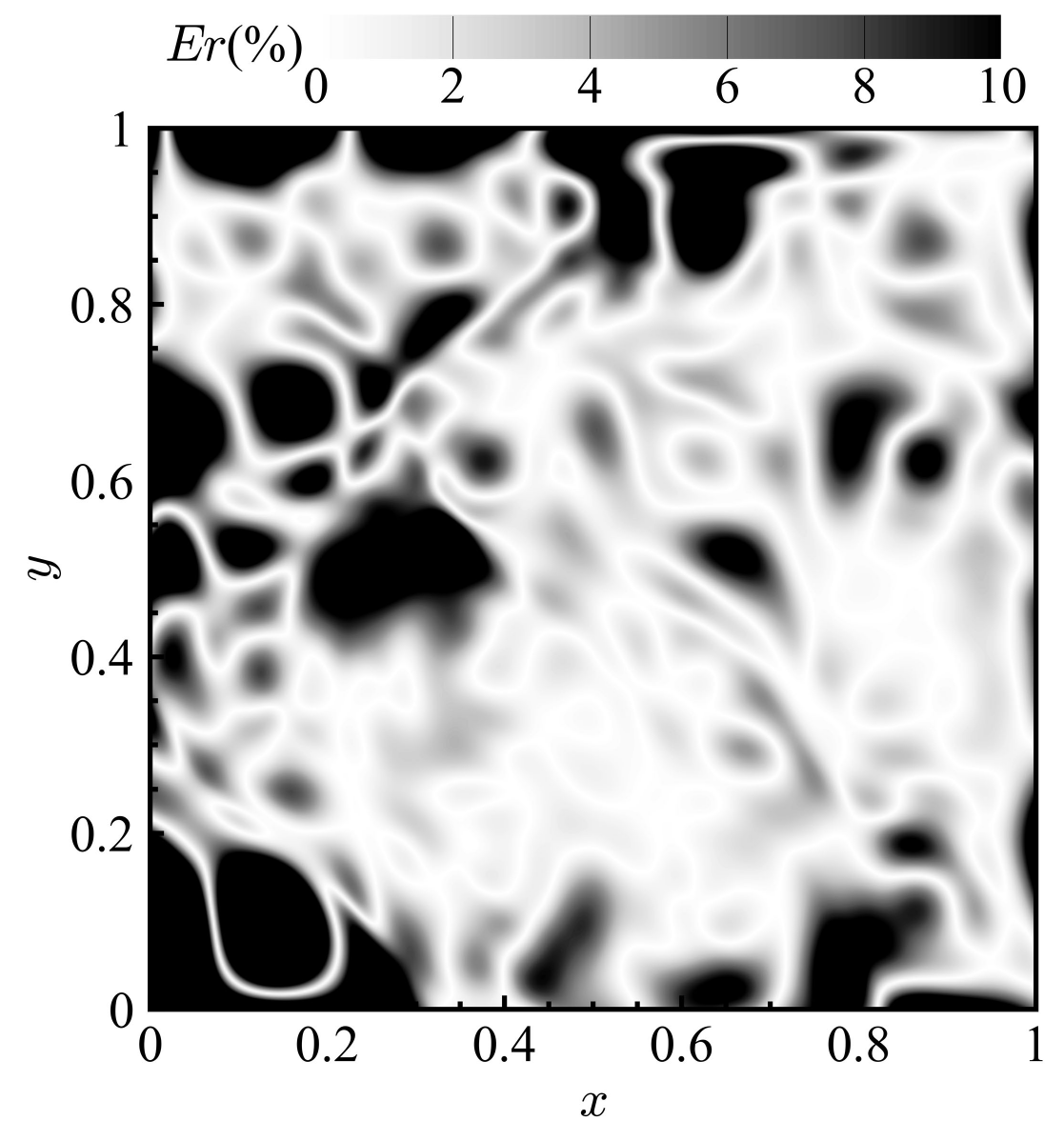}
	}
	\subfigure[PINN]{
		\includegraphics[height=4cm]{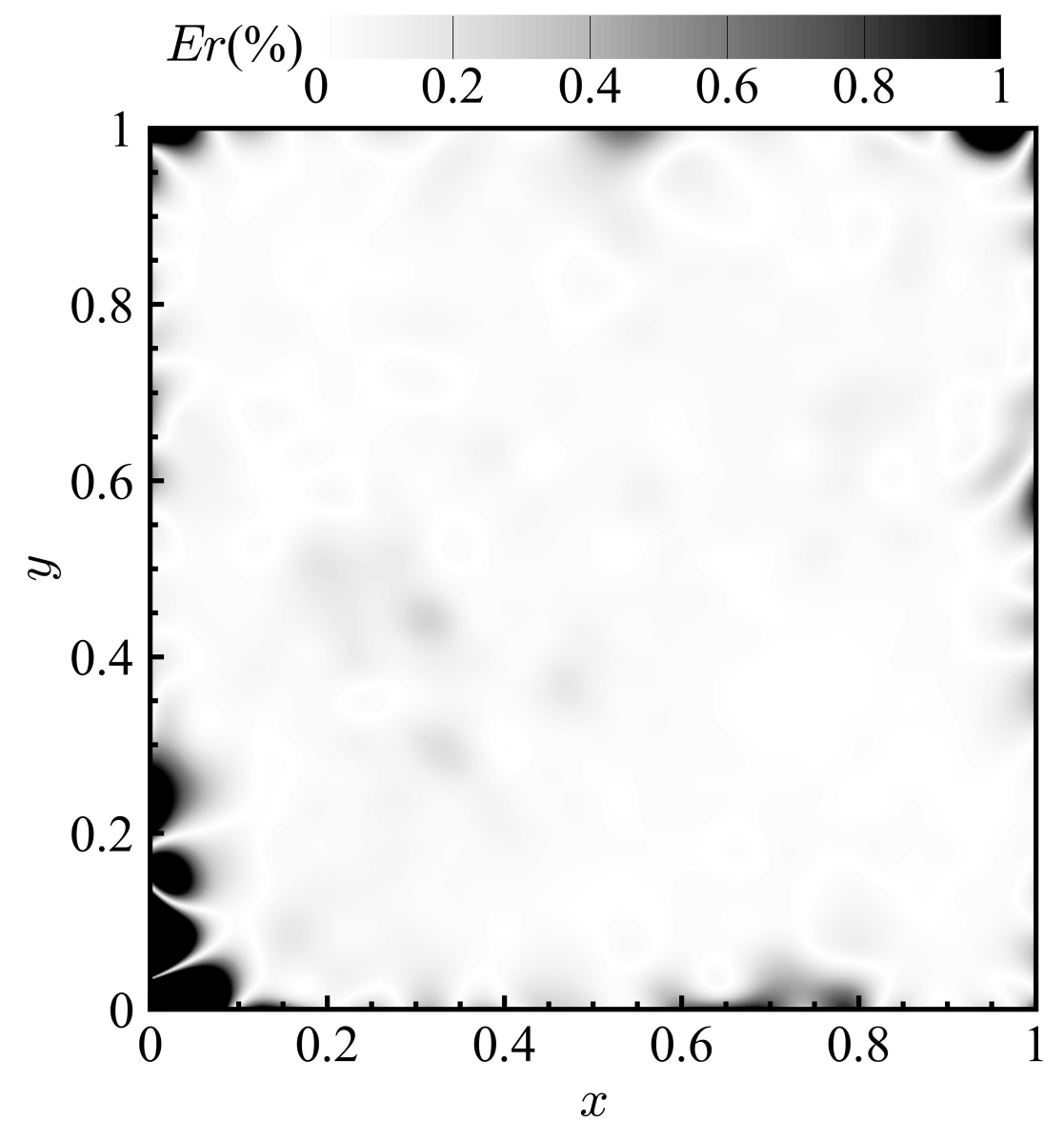}
	}
	\subfigure[PIFBNN]{
		\includegraphics[height=4cm]{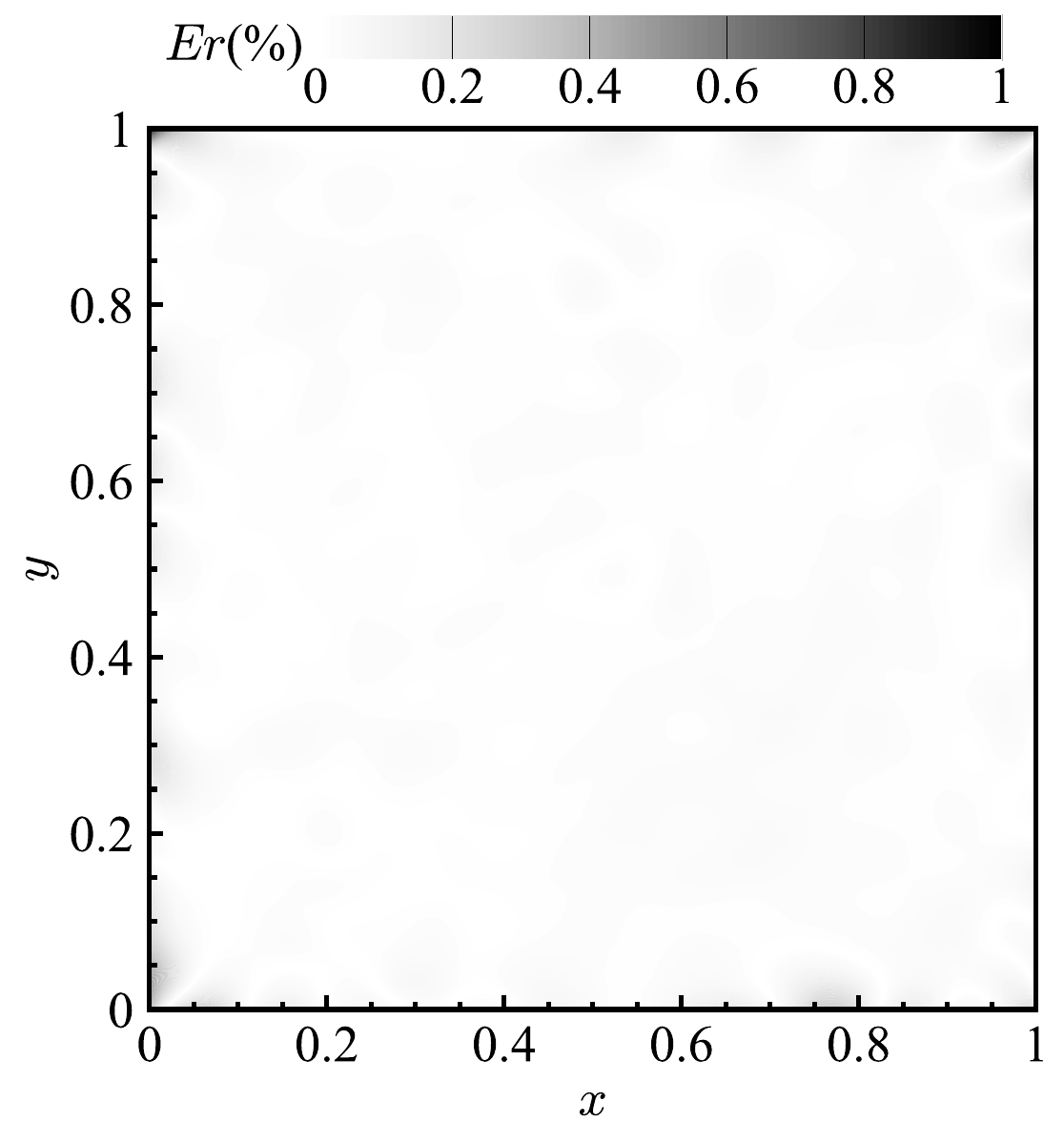}
	}
	\caption{Reconstruction relative error contours of different neural networks using $tanh$ as the activition function for Helmholtz equation.}
	\label{hmfuncthres}
\end{figure}

As observed, PINN shows significant improvement when using $tanh$ as the activation function owing to the superior nonlinear characteristics compared with $Leaky-relu$. With the addition of physical information residuals, the relative error is lower than that of FBNN, which is more consistent with expectations. Comparative analysis reveals that both FBNN and PIFBNN consistently outperform conventional ANN and PINN architectures, respectively. Notably, the performance enhancement of FBNN and PIFBNN exhibits limited sensitivity to activation function selection, indicating that the superior predictive capability stems primarily from their structural innovations rather than activation function optimization.
The relative error contours reconstructed by different neural networks with $tanh$ as the activation function are shown in Figure \ref{hmfuncthres}.

Analysis of relative error contours demonstrates measurable performance enhancements in both ANN and PINN when compared to architectures employing $leaky-relu$ activation functions. Particularly, the  performance of PINN is prominently good. Except for a slight error distribution at the boundary, low prediction error is maintained in the other regions. However, the performances of FBNN and PIFBNN do not show significant improvements. The prediction error of PIFBNN is optimized, while FBNN maintains an error distribution profile analogous to conventional neural networks utilizing  $leakyrelu()$ activation function. This indicates that the Fourier nodes of FBNN replace the nonlinear learning ability of the activation function to some extent, making it less affected by the choice of the activation function.
\subsection{Allen-Cahn equation}
\begin{figure}[h!] 
	\centering
	\subfigure[$leaky-relu$ models]{
		\includegraphics[height=5.5cm]{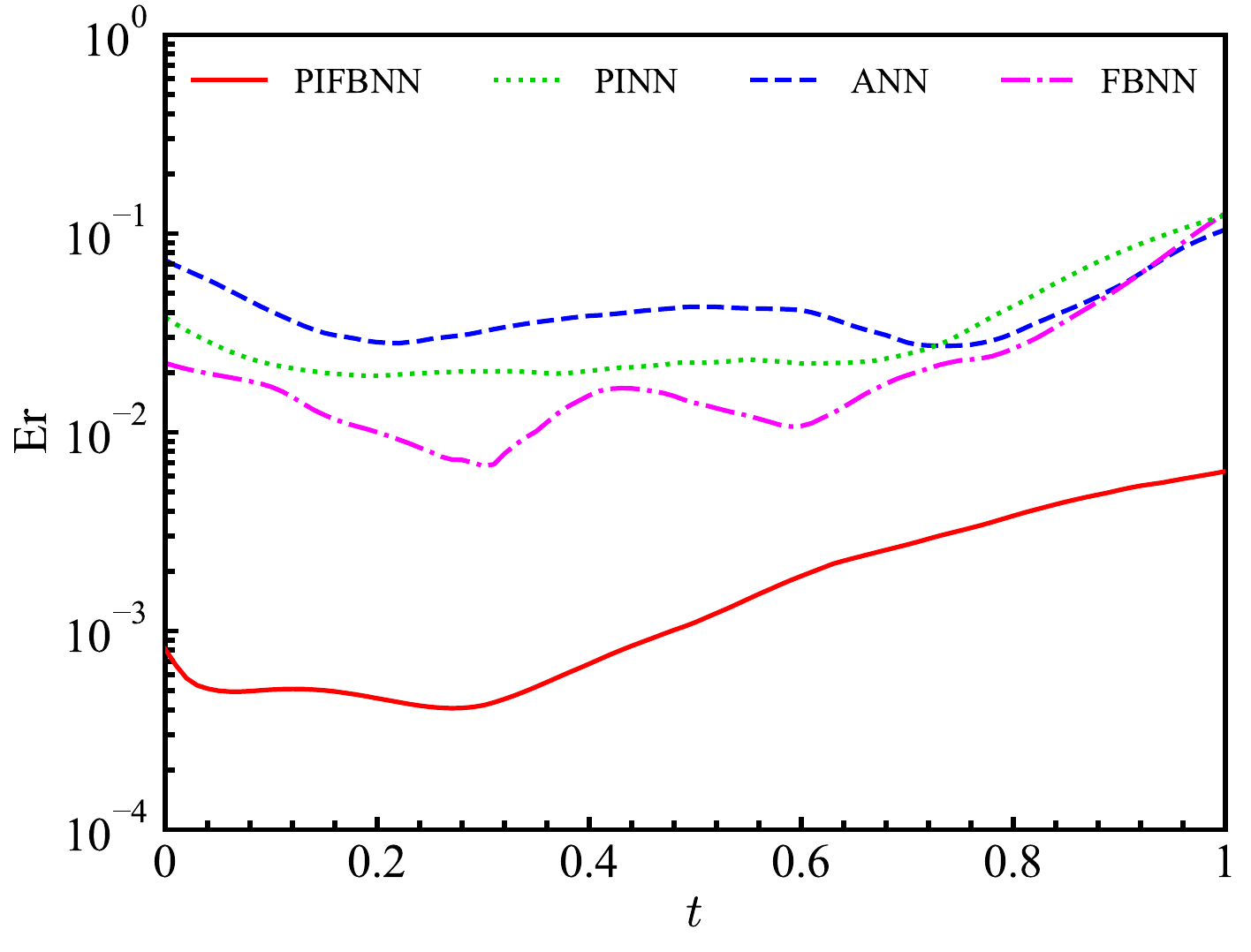}
	}
	\subfigure[$tanh$ models]{
		\includegraphics[height=5.5cm]{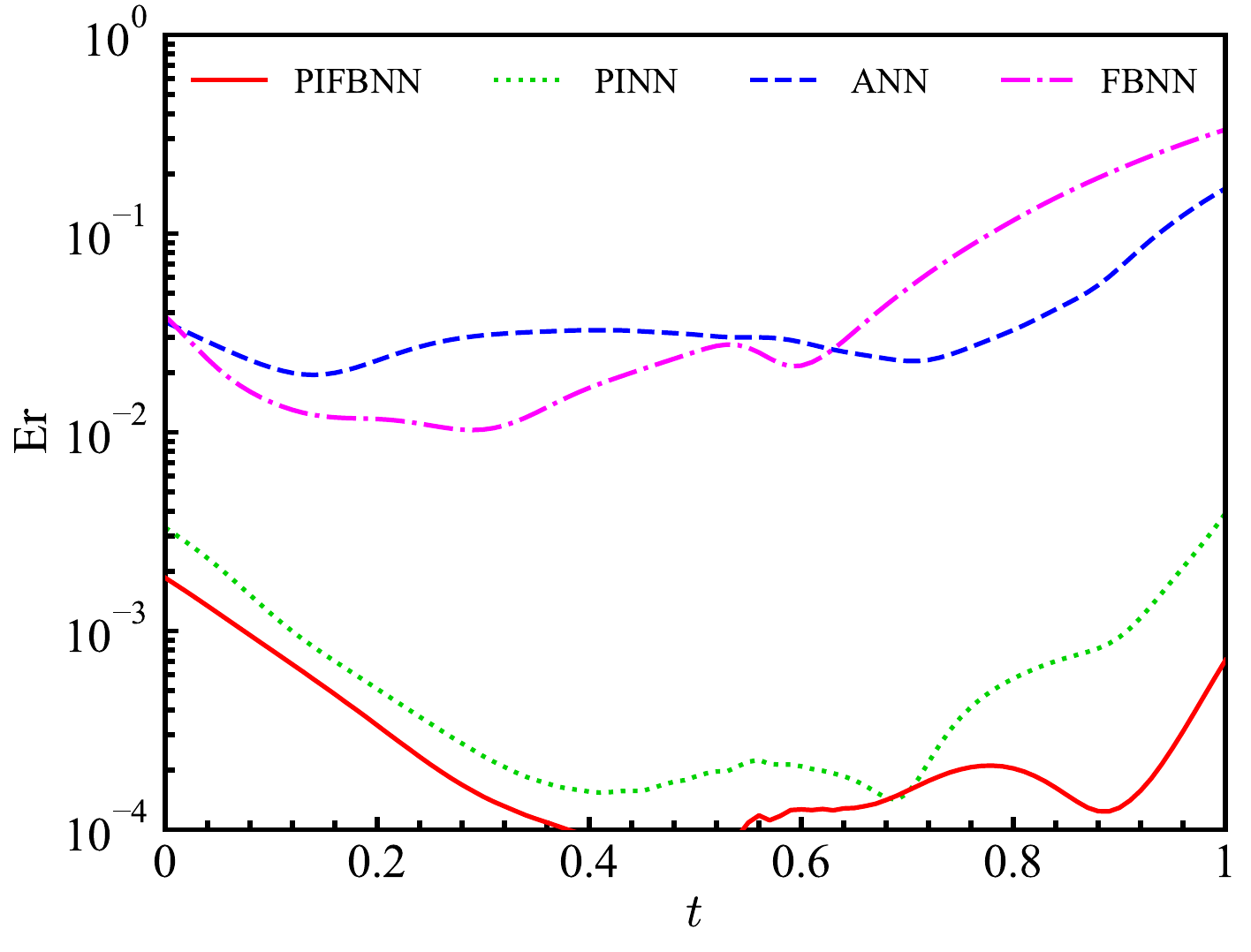}
	}
	\caption{Reconstruction relative errors using different neural networks for Allen-Cahn equation.  }
	\label{acfuncer}
\end{figure}

In this case, the neural network architecture used in this example has four hidden layers, each with 128 neurons. The ratio of Fourier nodes is 0.3. Similar to the previous case, a total of 225 points are randomly sampled within the domain as labeled data to train the neural network models and replace the boundary conditions. In addition, 1000 test points are randomly selected to evaluate the reconstruction performance. A total of 800 points within the domain are randomly sampled to calculate the physical information residuals.
For comparative analysis of reconstruction accuracy, the relative errors of neural network predictions are systematically evaluated across different activation functions. The relative errors along $t$ reconstructed by different neural networks using different activation functions  are shown in Figure \ref{acfuncer}.

\begin{figure}[!htbp]
	\centering
	
	{
		\includegraphics[height=4cm]{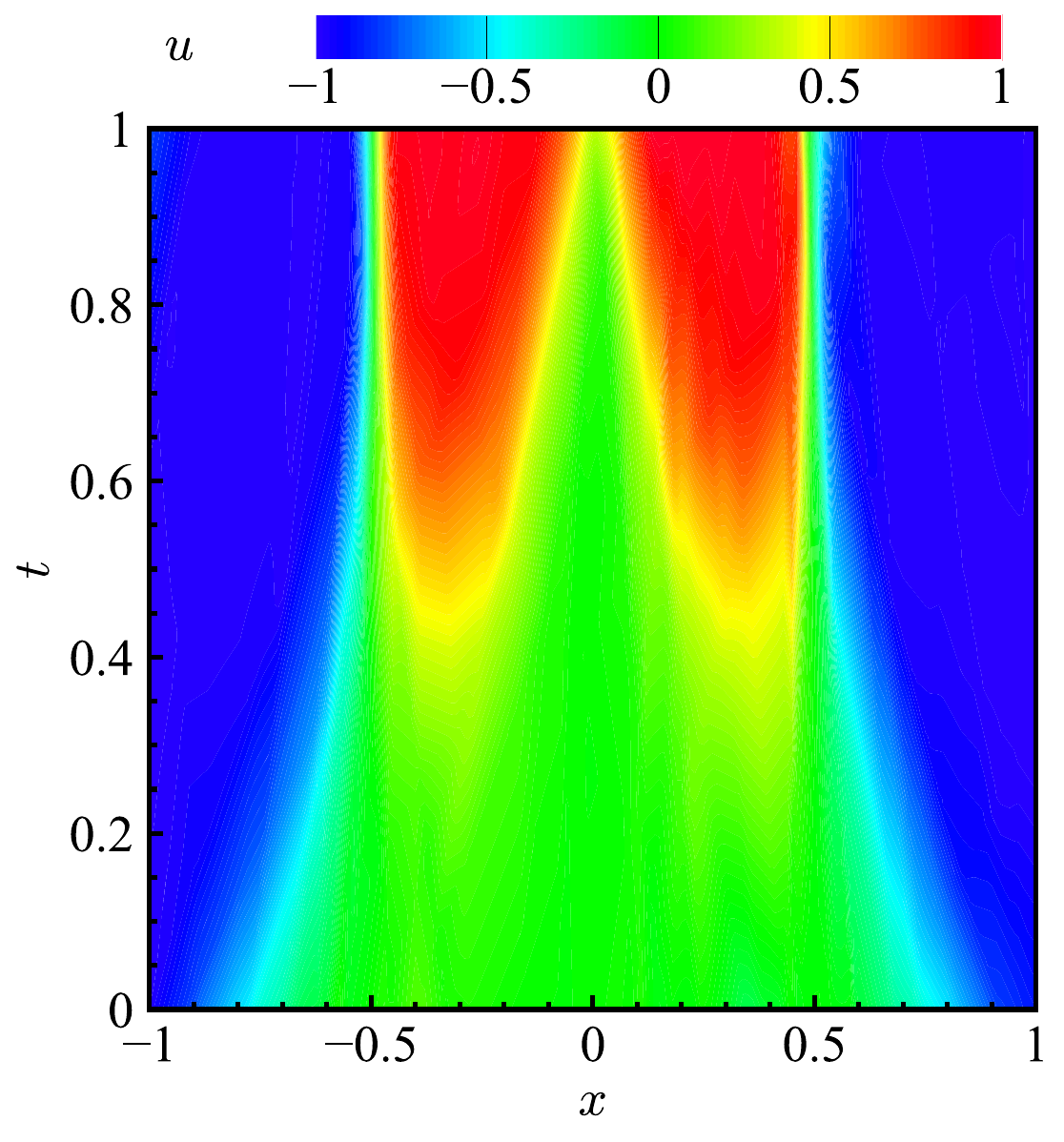}
	}
	{
		\includegraphics[height=4cm]{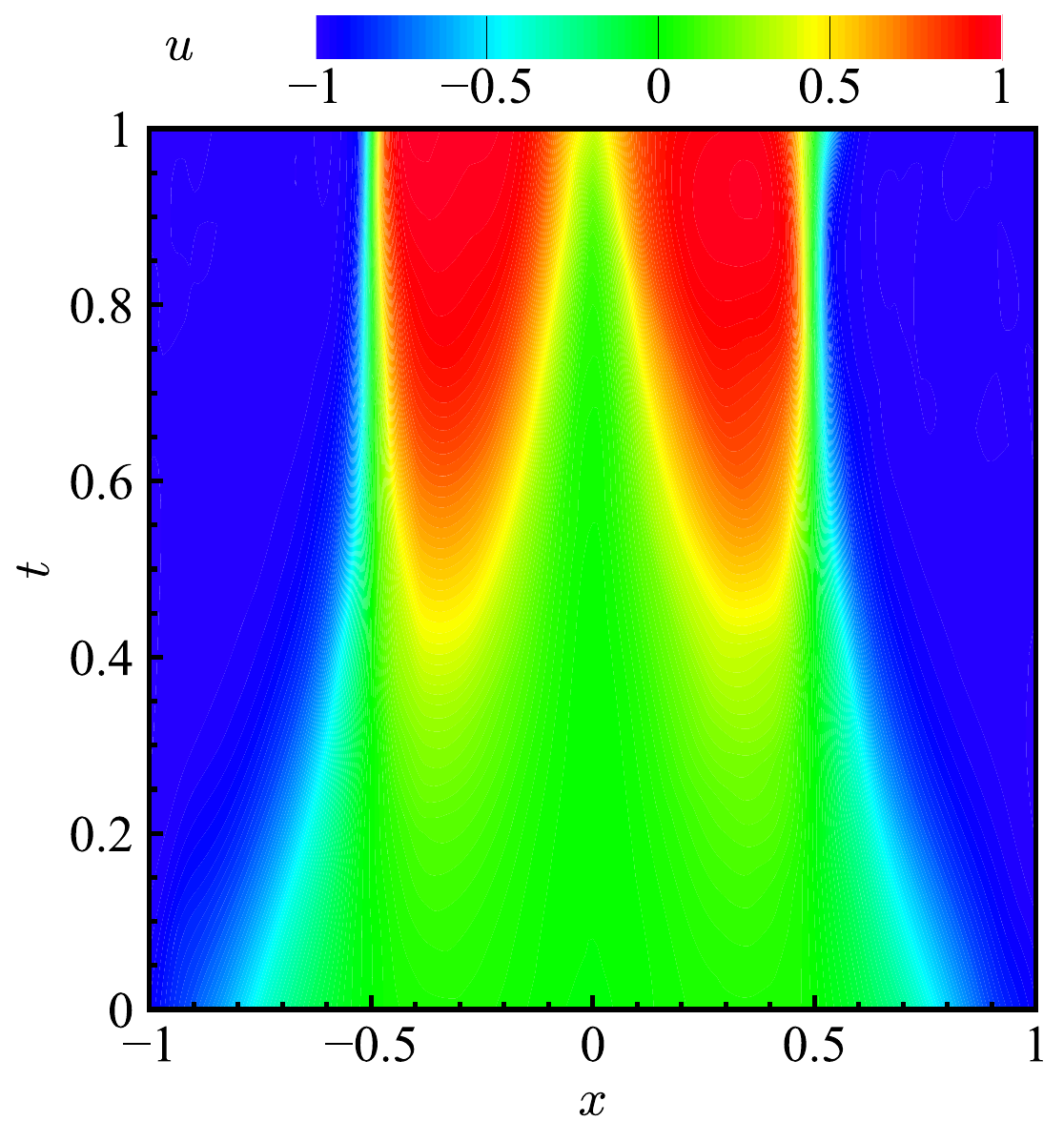}
	}
	{
		\includegraphics[height=4cm]{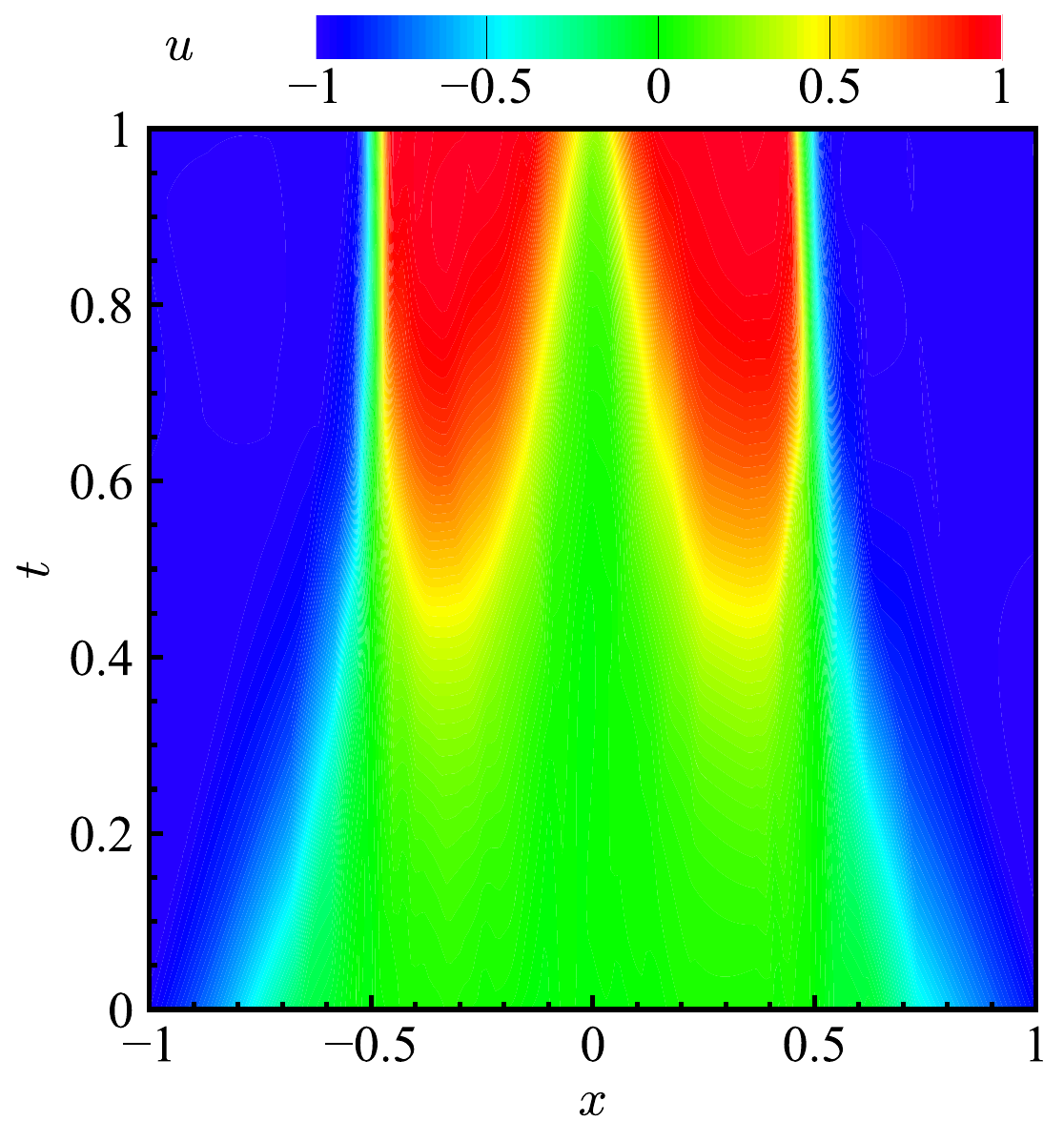}
	}
	{
		\includegraphics[height=4cm]{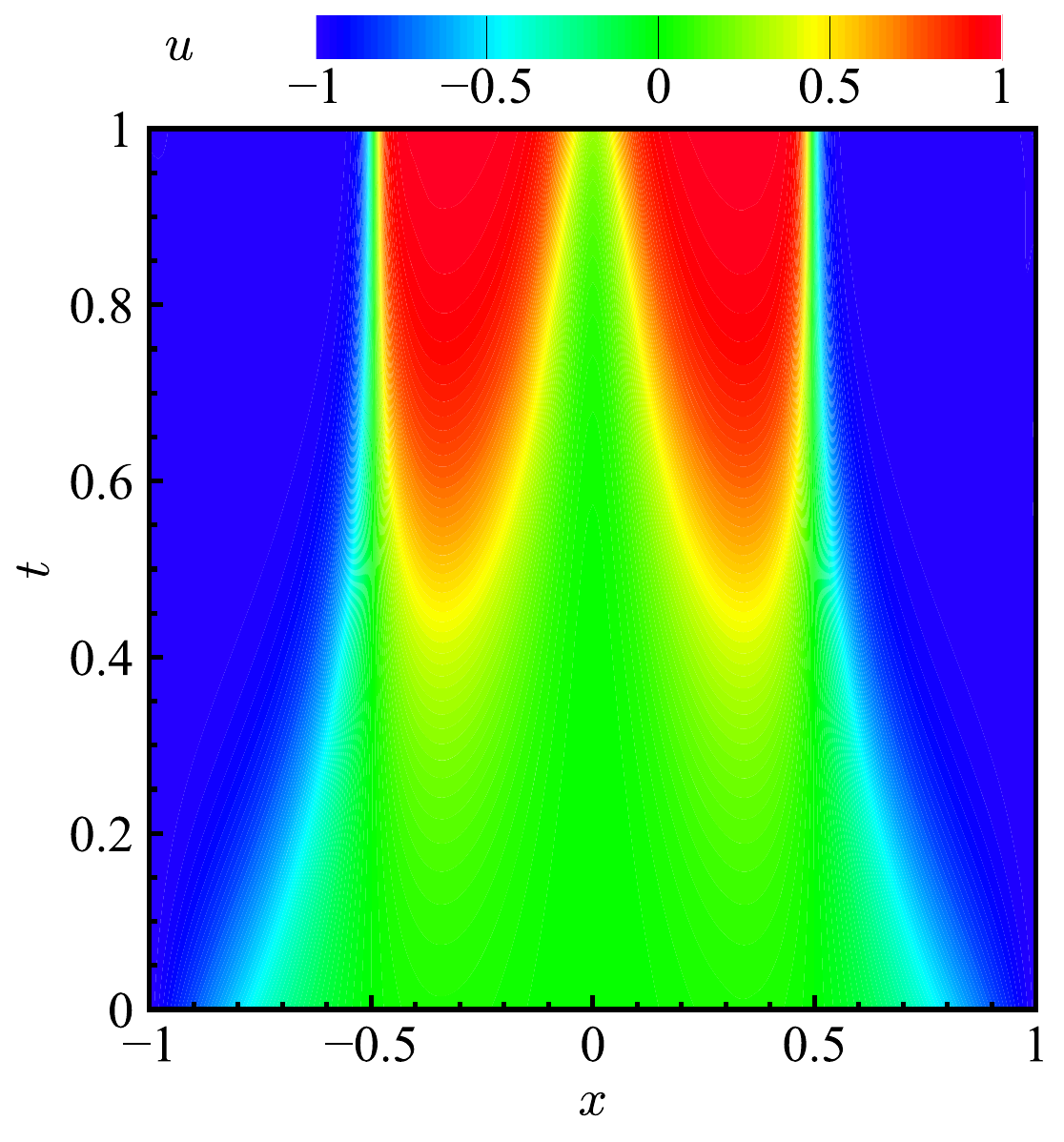}
	}
	
	\par %
	\subfigure[ANN]{
		\includegraphics[height=4cm]{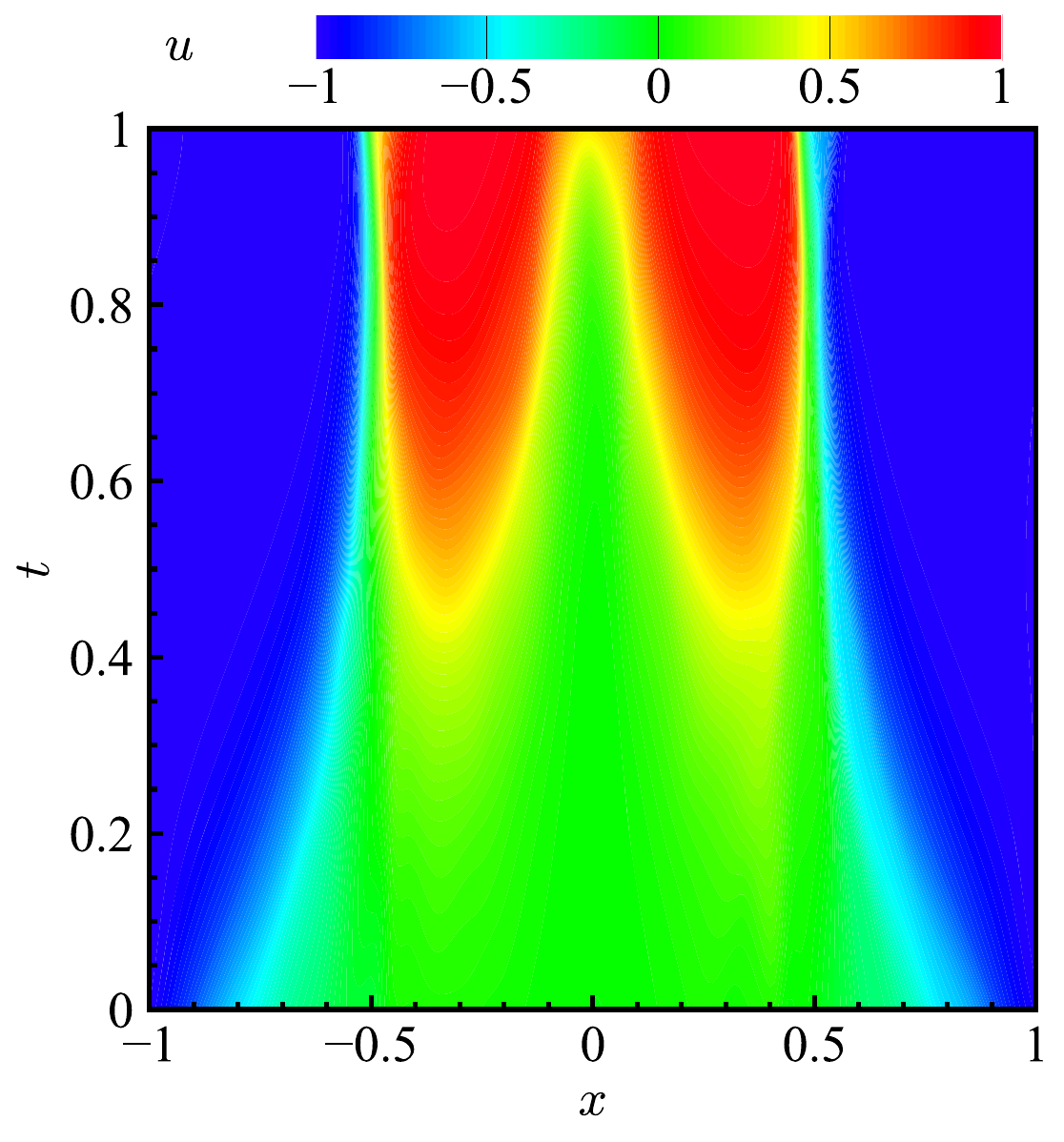}
	}
	\subfigure[FBNN]{
		\includegraphics[height=4cm]{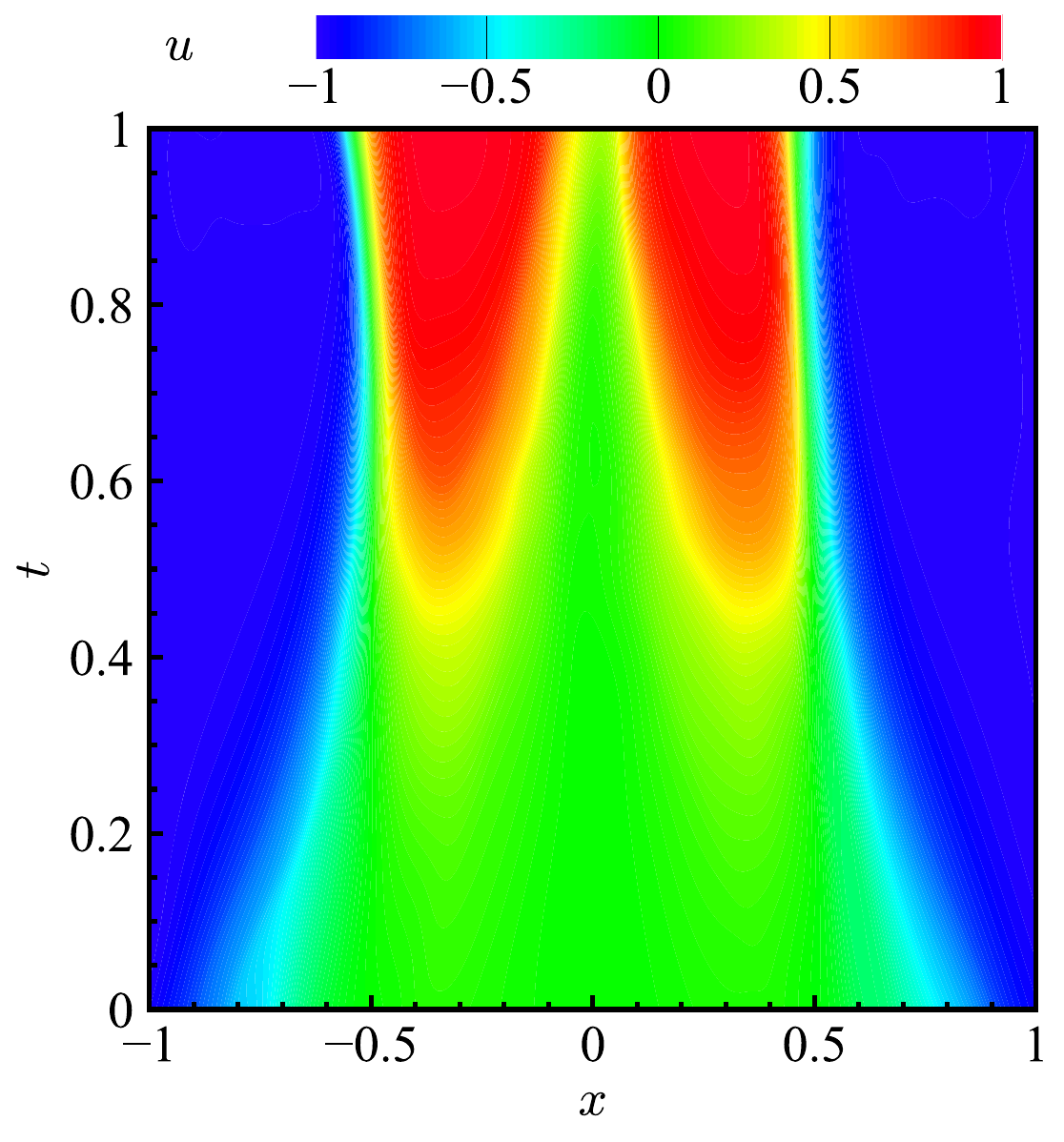}
	}
	\subfigure[PINN]{
		\includegraphics[height=4cm]{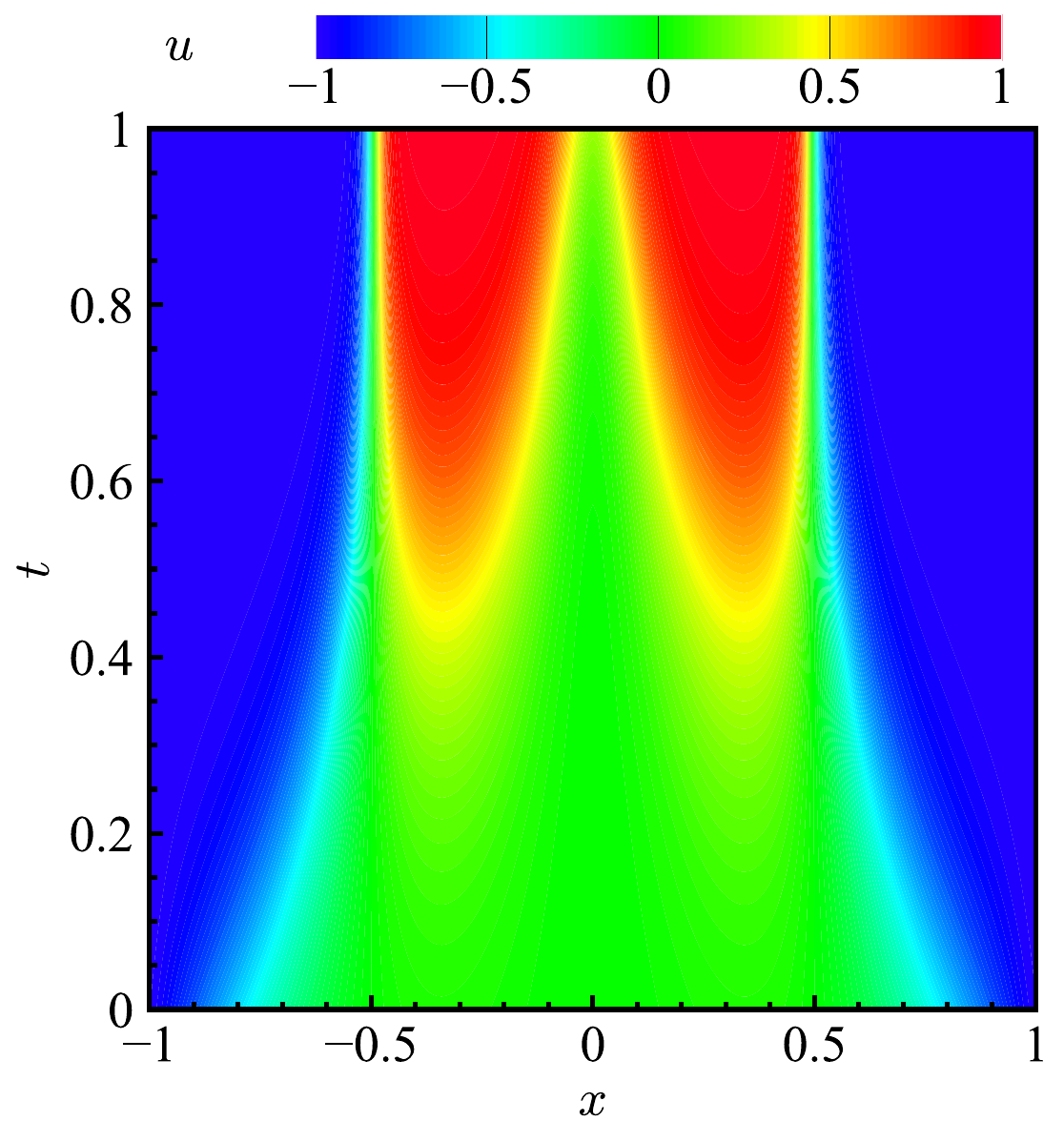}
	}
	\subfigure[PIFBNN]{
		\includegraphics[height=4cm]{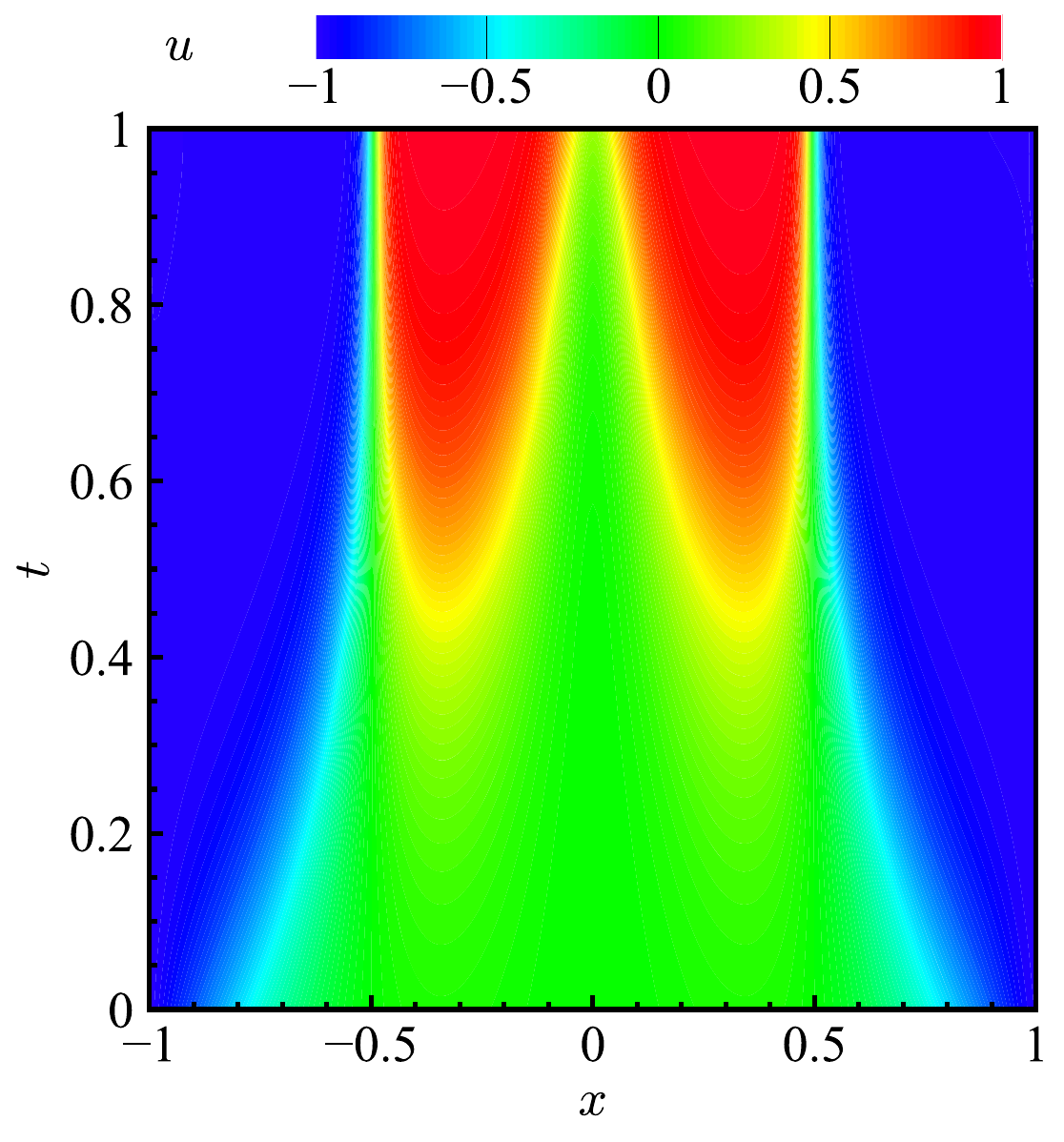}
	}
	
	\caption{Reconstruction velocity contours of different neural networks using $leaky-relu$ as the activition function for Allen-Cahn equation.The first line use the $leaky-relu$, and the second line use the $tanh$ activation function.}
	\label{acfuncpred}
\end{figure}

As shown in the previous case, PINN using $leaky-relu$ as the activation function exhibits a relatively large relative error when calculating the Allen--Cahn equation, which is almost same as the error of ANN. The reconstruction error of FBNN without adding physical information is always smaller than that of PINN. After changing $tanh$ as the activation function, both PINN and ANN improve to varying degrees. The relative error of PINN exhibits a reduction of multiple orders of magnitude, achieving comparable accuracy to the PIFBNN. However, there is no significant change in the order of magnitude between the errors of FBNN and PIFBNN, regardless of which activation function is used. The error of PIFBNN is always lower than that of PINN, indicating that the FBNN framework is minimally affected by the choice of activation function. Notably, the reconstruction error of the ANN using $tanh$ as the activation function is comparable to that of FBNN, and even lower than that of FBNN when $t>0.6$. However, the error of PIFBNN always maintains its advantage, which further confirms that the FBNN framework can more fully utilize the physical information and capture finer physical features.
The contours of Allen--Cahn equation for the neural networks using different activation functions are demonstrated in Figure \ref{acfuncpred}.

\begin{figure}[!htbp]
	\centering
	{
		\includegraphics[height=4cm]{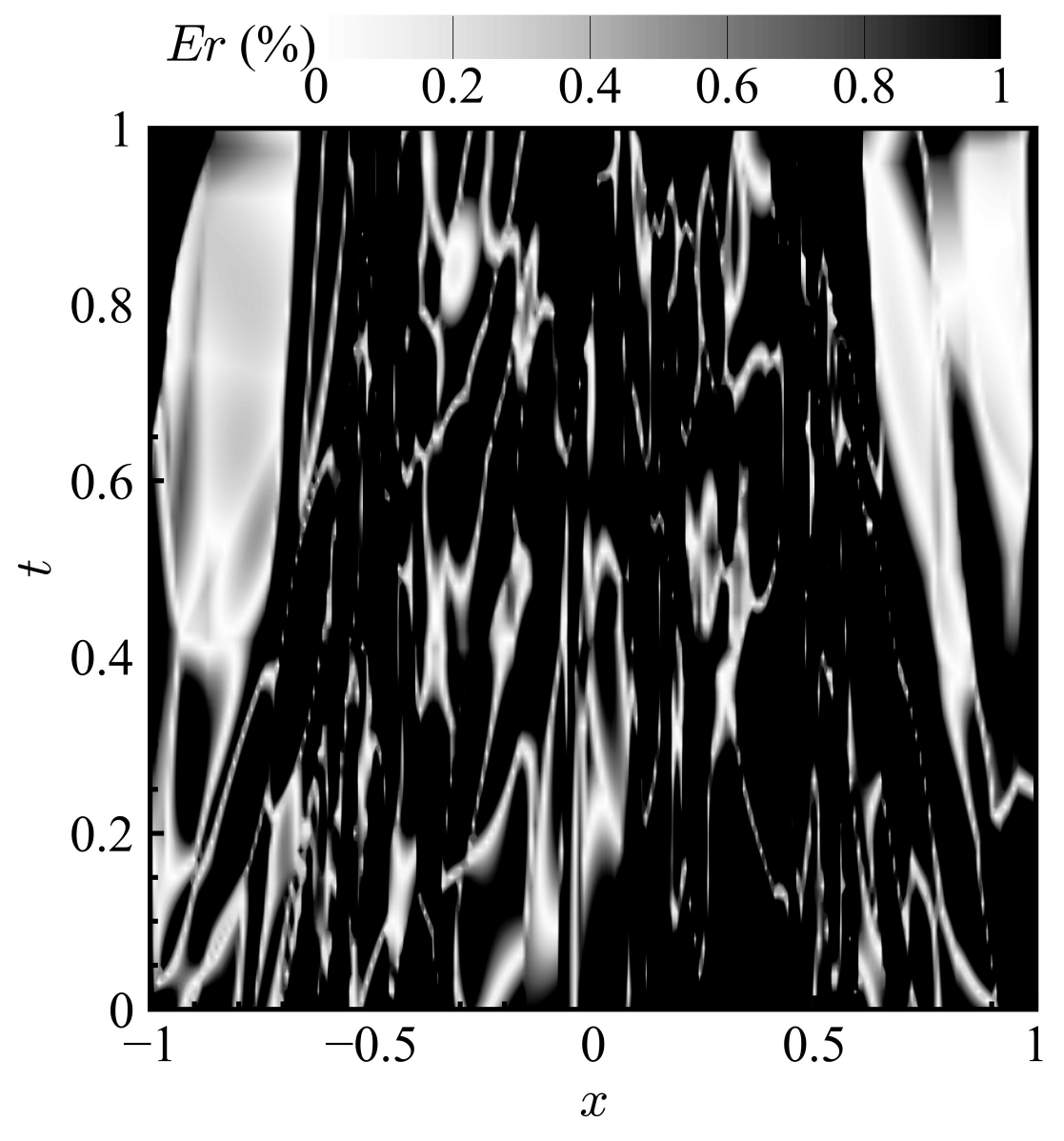}
	}
	{
		\includegraphics[height=4cm]{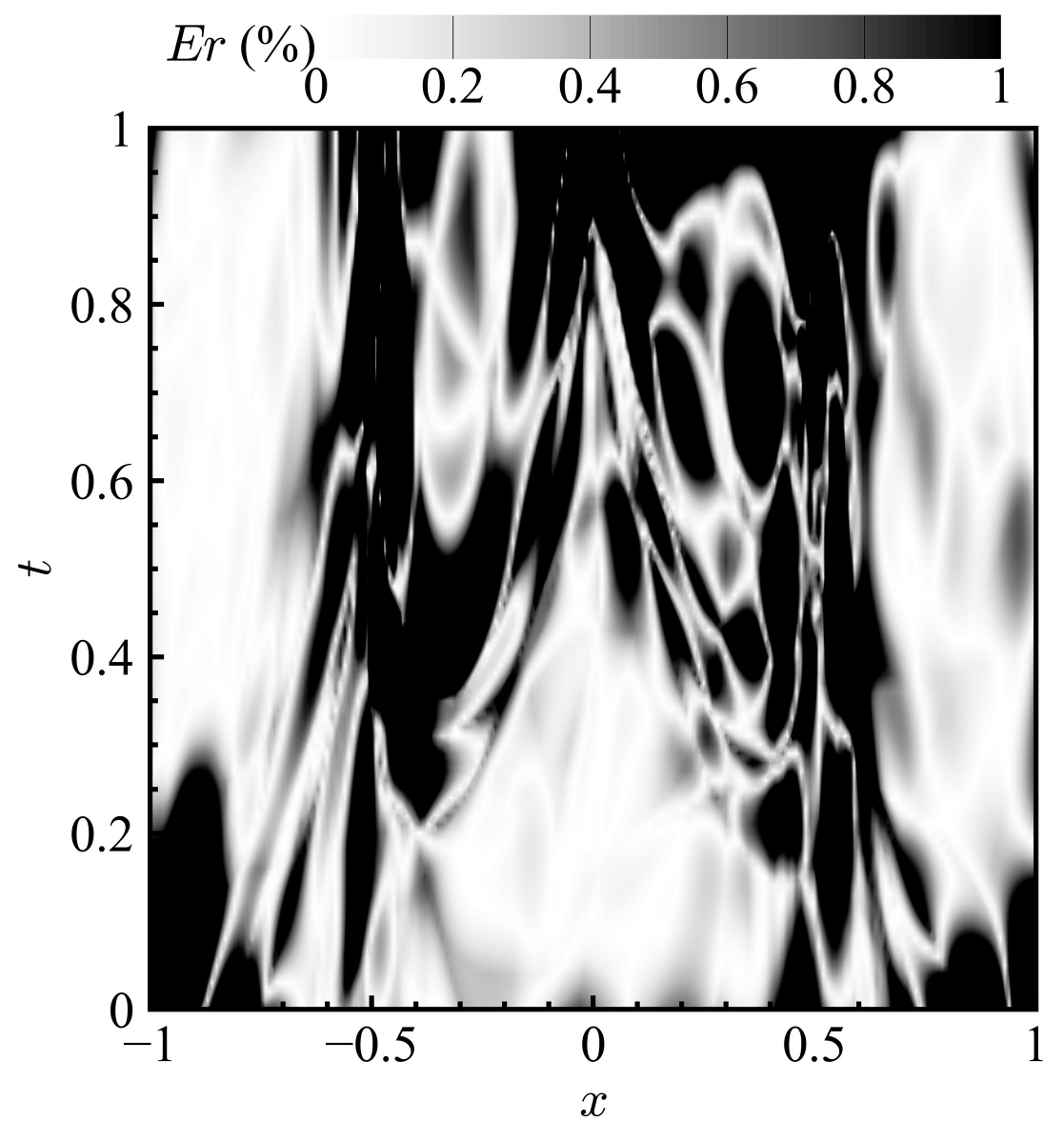}
	}
	{
		\includegraphics[height=4cm]{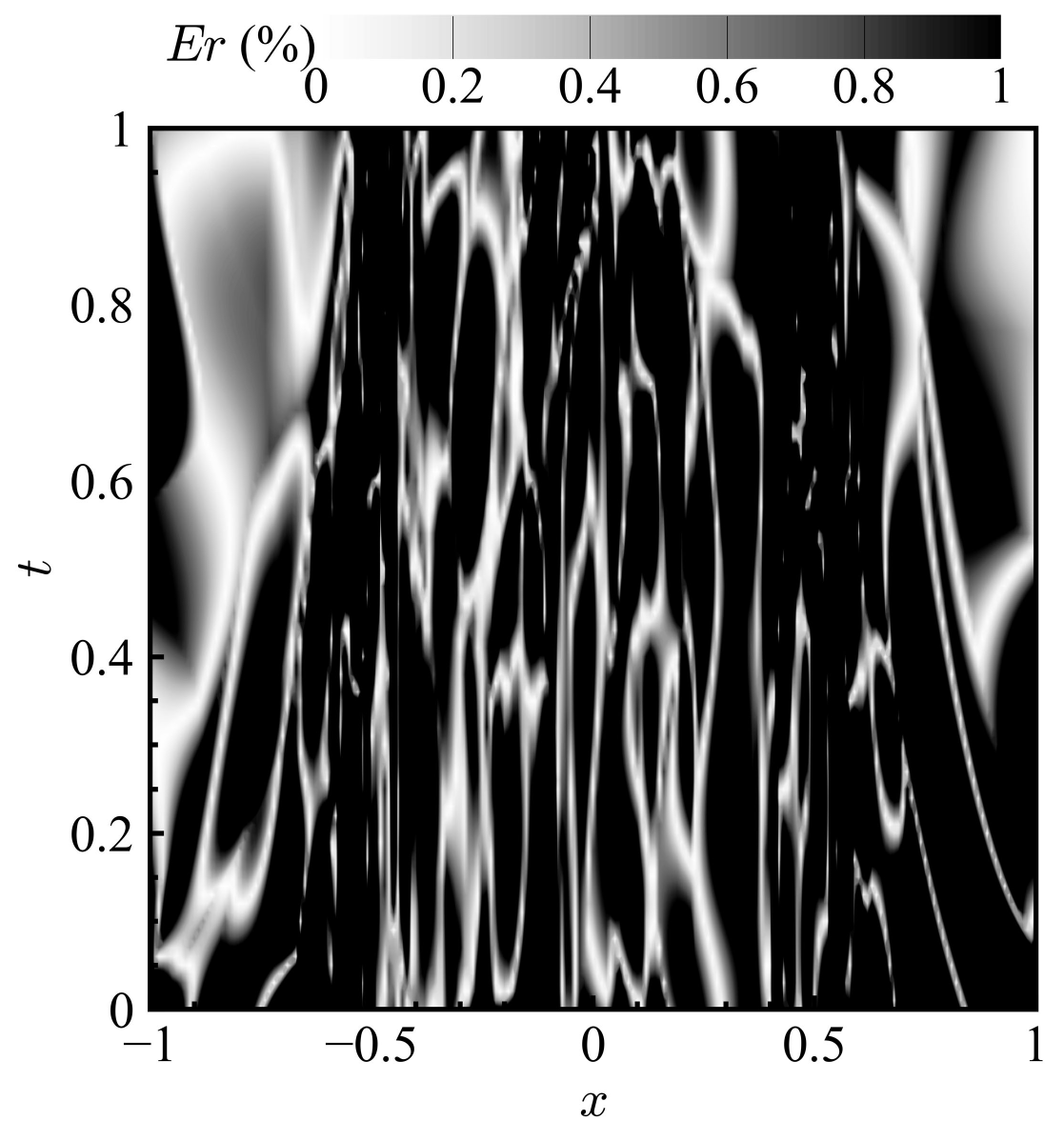}
	}
	{
		\includegraphics[height=4cm]{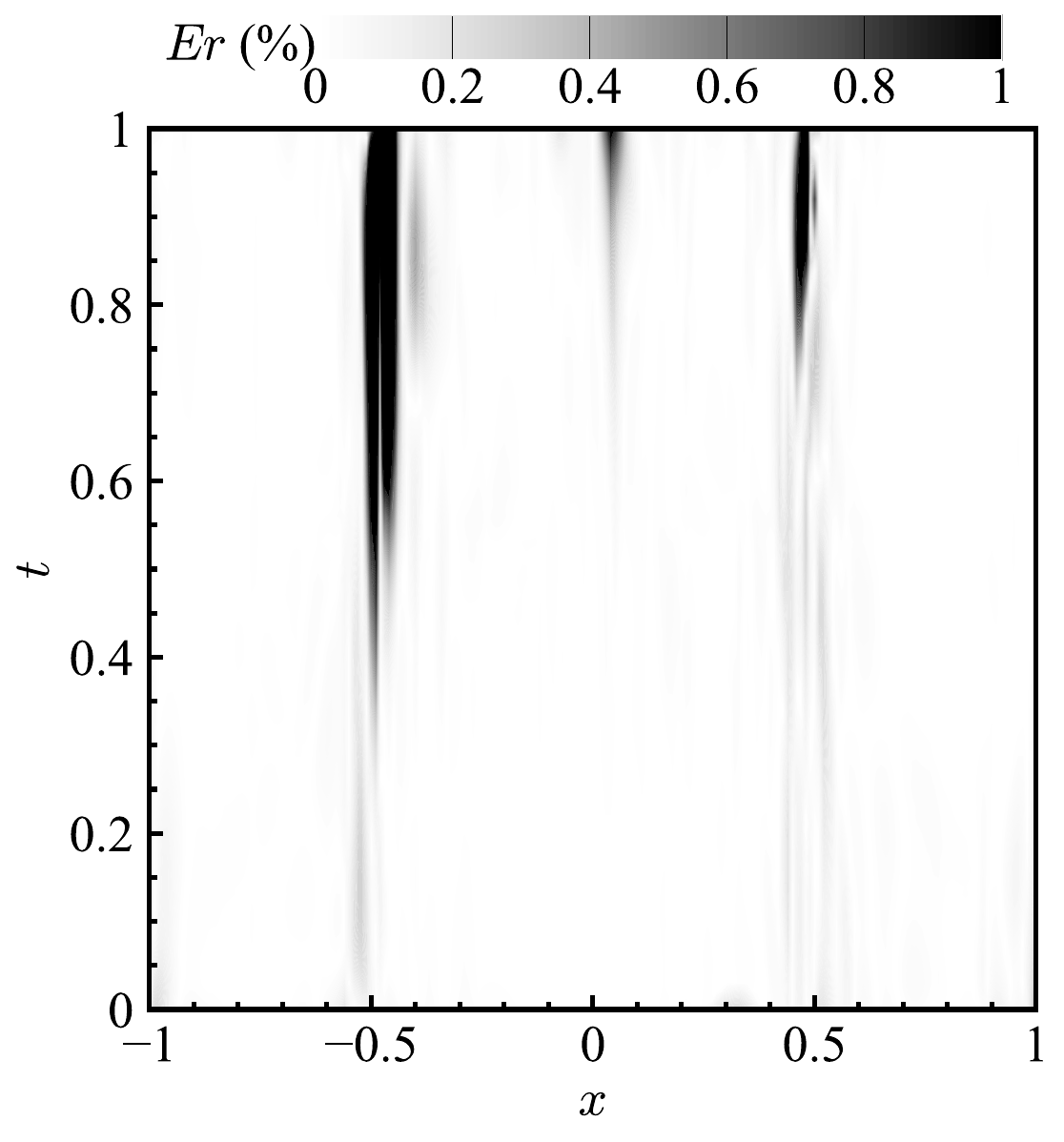}
	}
	
	\par %
	\subfigure[ANN]{
		\includegraphics[height=4cm]{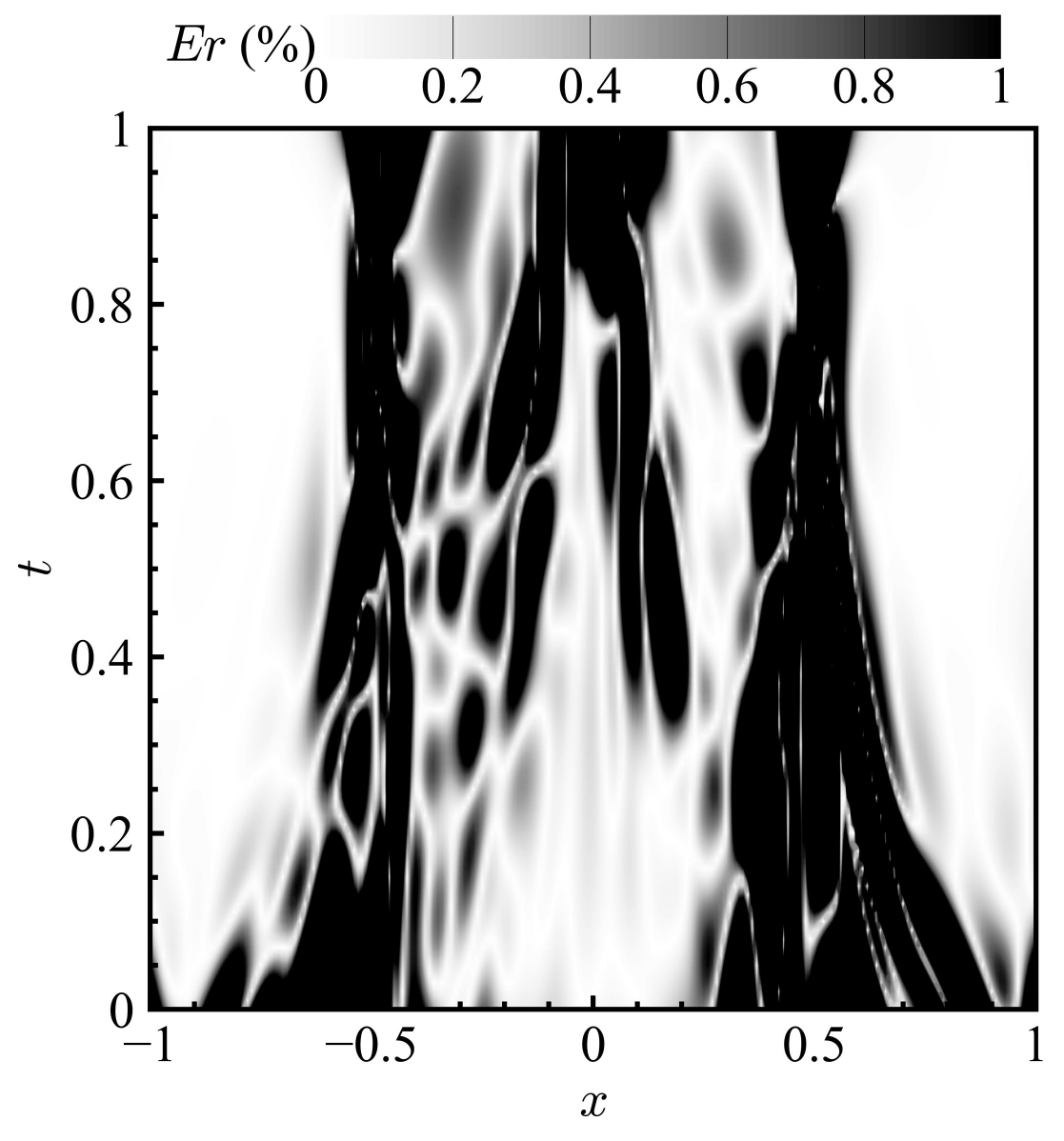}
	}
	\subfigure[FBNN]{
		\includegraphics[height=4cm]{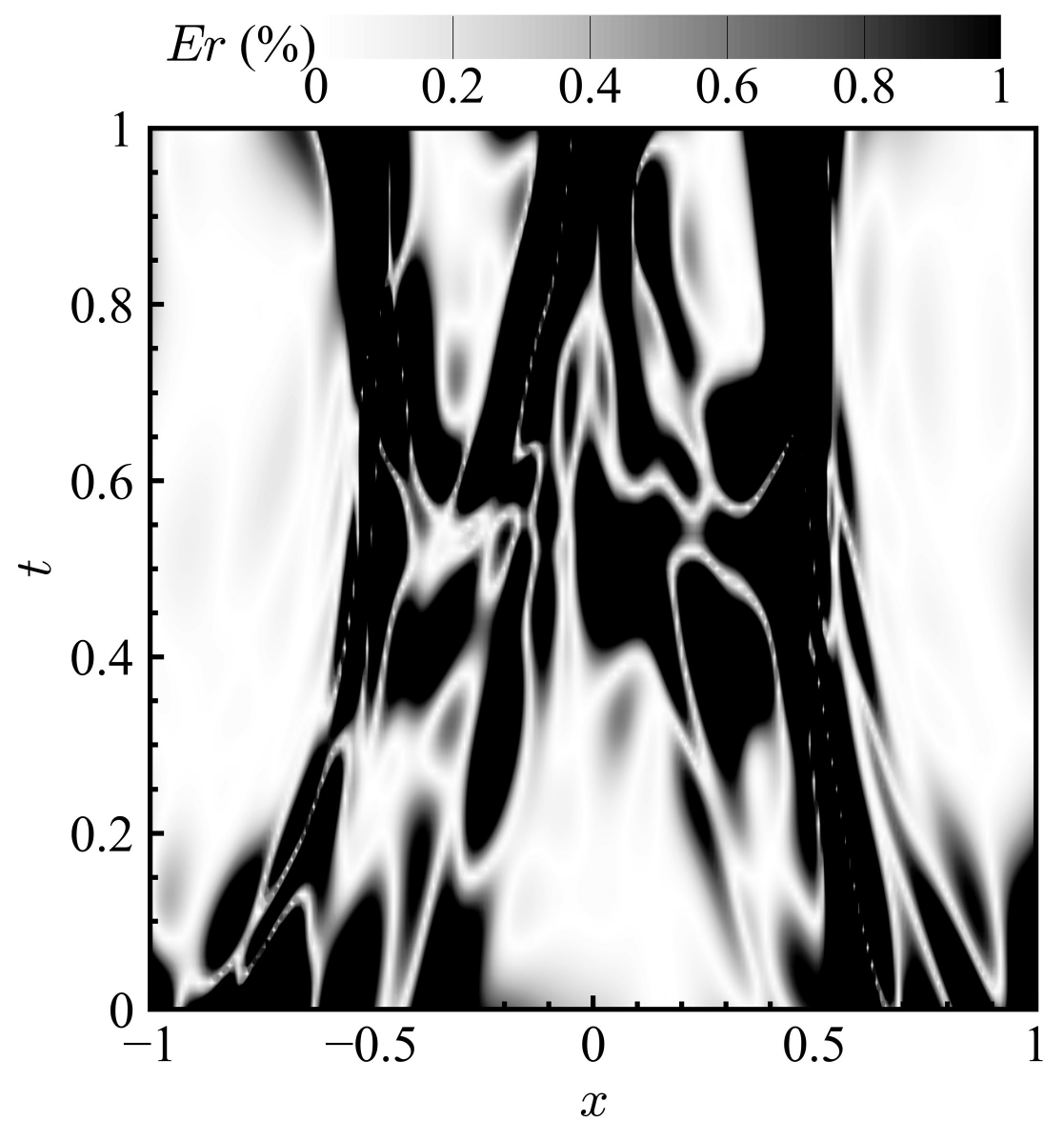}
	}
	\subfigure[PINN]{
		\includegraphics[height=4cm]{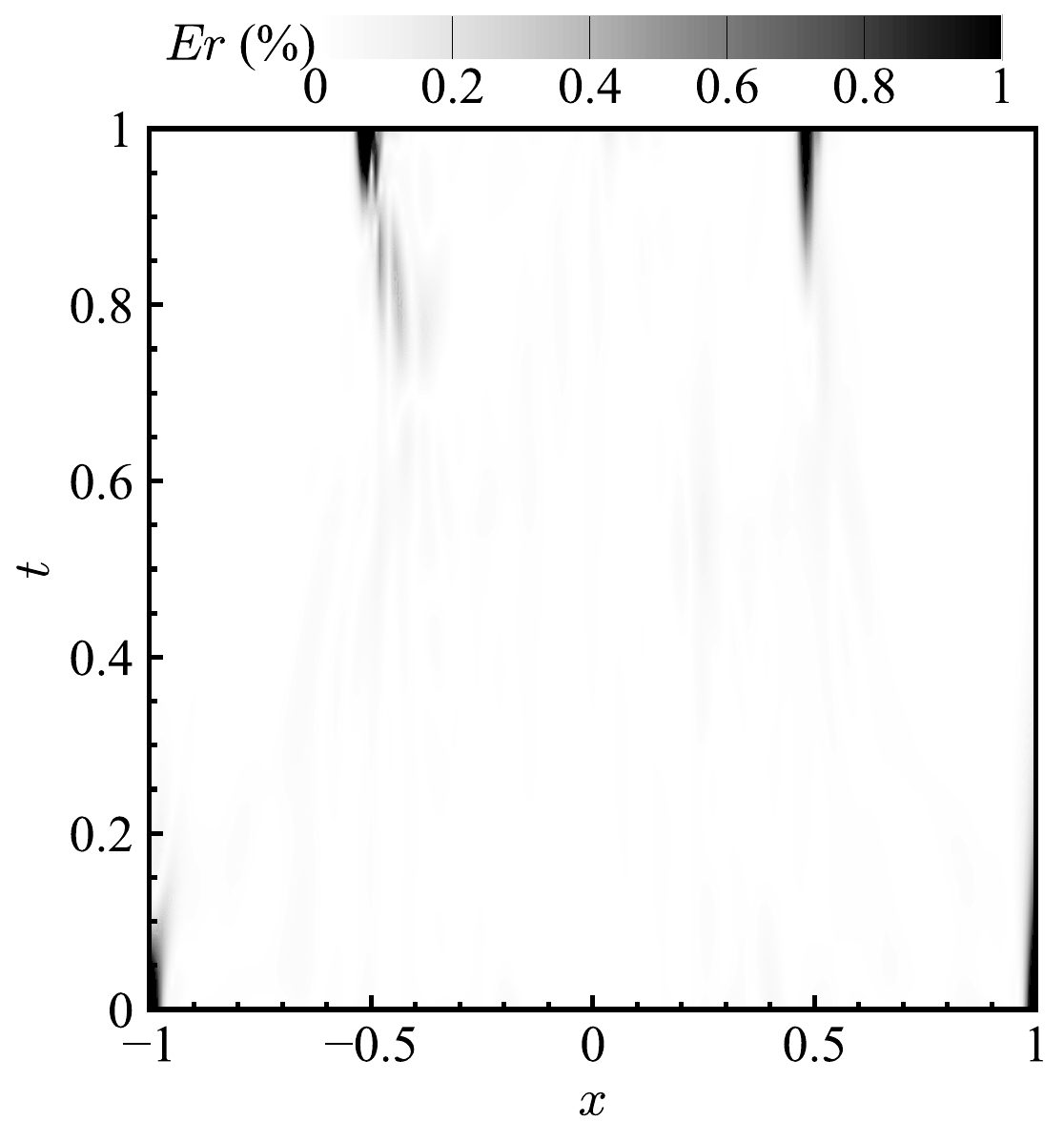}
	}
	\subfigure[PIFBNN]{
		\includegraphics[height=4cm]{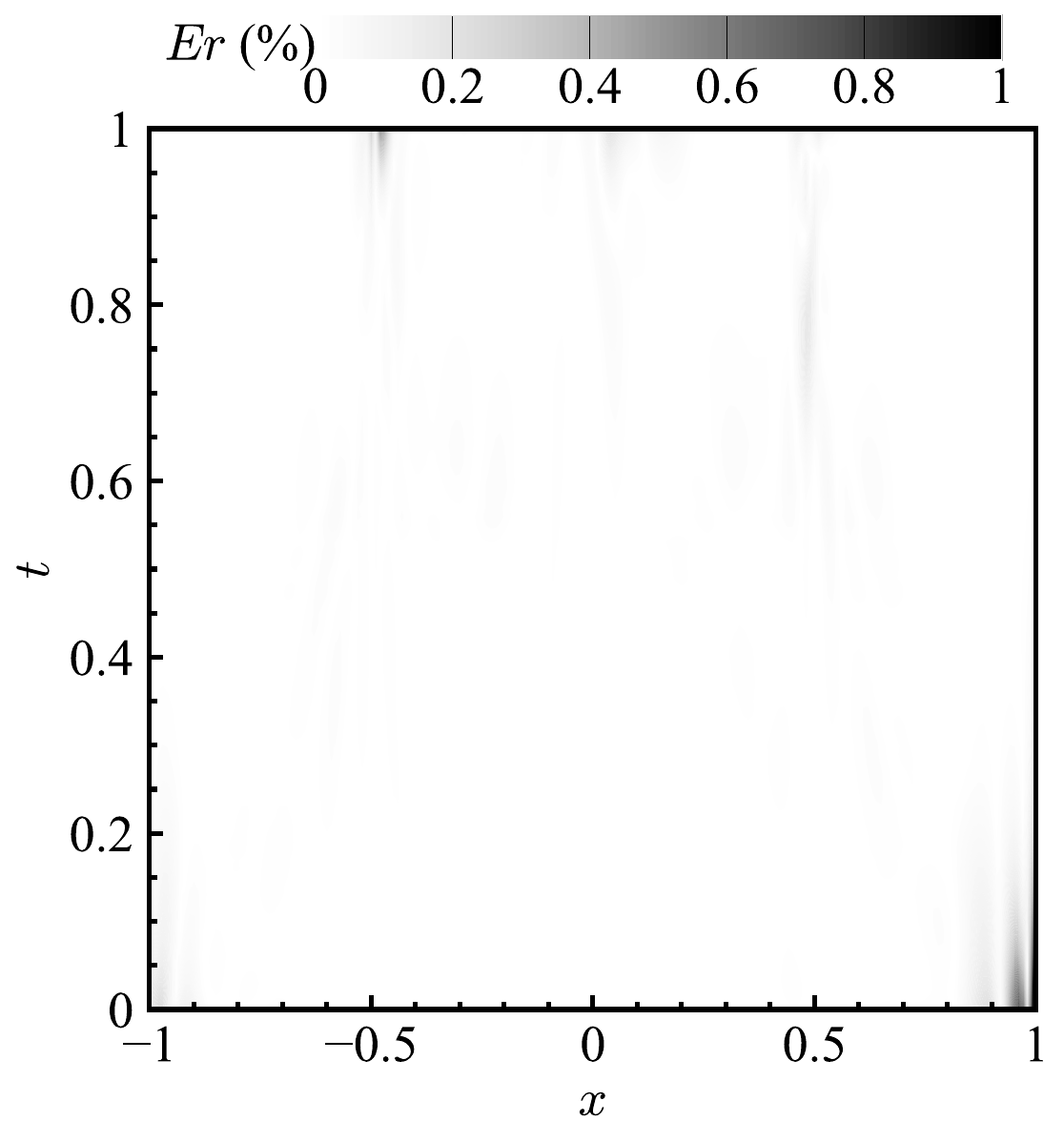}
	}
	
	\caption{Reconstruction relative error contours of different neural networks using $leaky-relu$ as the activition function for Allen-Cahn equation.The first line use the $leaky-relu$, and the second line use the $tanh$ activation function.}
	\label{acfuncres}
\end{figure}

The contours of relative error reconstructed by neural networks using different activation functions are displayed in Figure \ref{acfuncres}.
Longitudinal comparison of reconstruction performance demonstrates that both ANN and PINN employing $tanh$ activation functions achieve superior accuracy compared to $leaky-relu$, particularly in high-velocity flow regions. This can be better verified by relative error contours, with the highest improvement observed in PINN, where the error distribution across the entire field is reduced to only partial errors at boundary. However, FBNN and PIFBNN do not show significant changes in the reconstruction or error contours. The inference from horizontal comparison is similar to the above conclusion, in all cases the errors of FBNN and PIFBNN are lower than those of the ANN and PINN. Particularly, the PIFBNN and PINN with added physical information have more obvious advantages. 

Through comparative analysis of these representative cases with distinct activation functions, we demonstrate that the FBNN framework exhibits enhanced capability in both nonlinear feature extraction and physical information encoding compared to conventional architectures. Simultaneously, the presence of Fourier nodes further enhances the networks' generalization of activation function selection, solving the problem of sensitivity to activation function selection in ordinary ANN frameworks.

\section{\label{sec:level1/5}Field reconstruction of PDEs with sparse data points}

In experimental fluid dynamics and engineering applications, flow field datasets frequently exhibit partial sparsity due to measurement limitations or data acquisition constraints, presenting a persistent challenge for analysis. Concurrently, CFD simulations face substantial computational overhead when employing dense spatial discretization. Consequently, developing accurate flow field reconstruction methodologies from sparse measurements constitutes a critical requirement for practical engineering applications.
In this section, we use Burgers and Helmholtz equations to evaluate neural network performance in sparse flow field reconstruction. The sparse labeled data functionally emulate boundary condition constraints, thus eliminating the need for explicit boundary condition specification in the current neural network implementations. Owing to the use of labeled data, our nonphysical information neural network can also be trained. In this section, we use the ANN,FBNN,PINN and PIFBNN for training and testing. Through comparative analysis between ANN-FBNN and PINN-PIFBNN pairs, the enhanced capabilities of the FBNN framework are rigorously demonstrated. The training device and dataset sources used are the same as those described in Section \ref{sec:level1/3}.

\subsection{Burgers equation}
\begin{figure}[h!] 
	\centering
	\subfigure[Training loss]{
		\includegraphics[height=5.5cm]{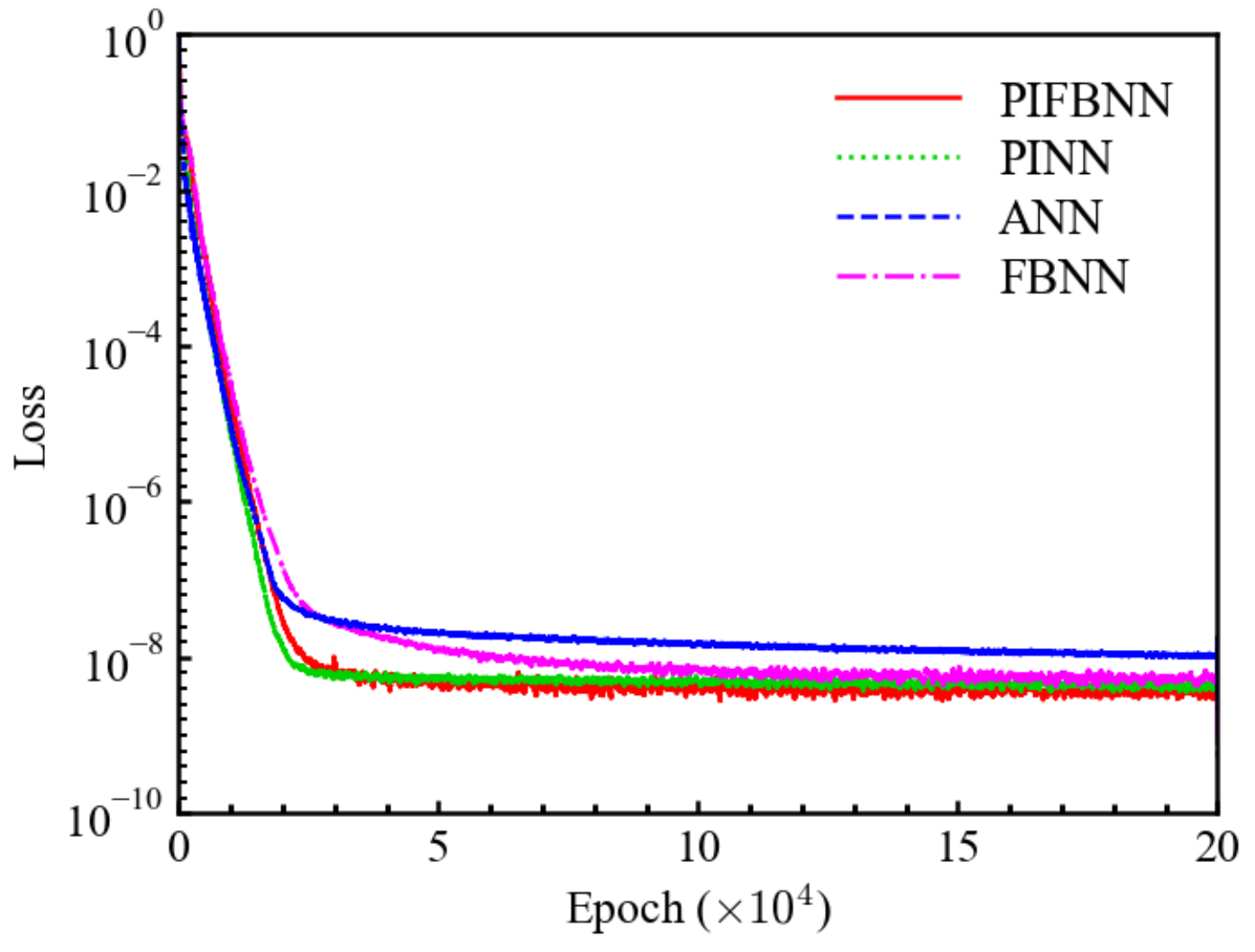}
	}
	\subfigure[Testing loss]{
		\includegraphics[height=5.5cm]{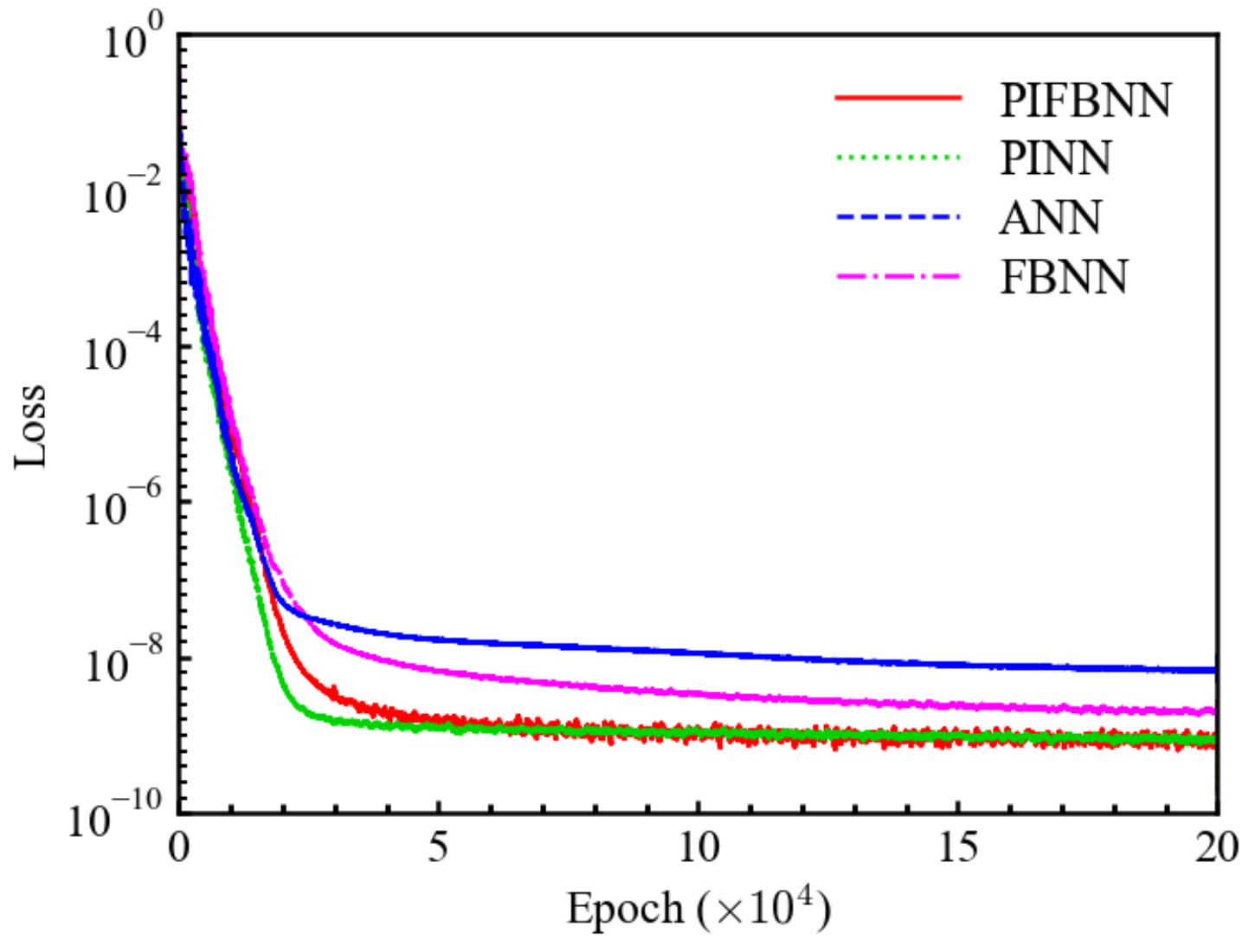}
	}
	\caption{Learning curves for Burgers equation reconstruction using different neural networks. }
	\label{bgrecoloss}
\end{figure}

The same 10 layers network, with 50 neurons per layer, is used. A total of 500 points are randomly sampled within the computational domain as labeled data to train the neural network models. From the above conclusion, we observe that owing to the existence of discontinuous solutions in Burgers equation, the difficulty of neural networks' prediction significantly increases at $x=0$ where shock waves exist, resulting in significant errors. Therefore, 150 points are abstracted in $x\in [-0.2, 0.2]$ region as labeled encrypted data to achieve better reconstruction results. Owing to the presence of labeled data, the number of initial and boundary condition points are reduced. At $t = 0$, 35 points are randomly sampled as initial condition constraints, and at the $x = 1$ and $x = -1$ boundaries, 35 points are selected as boundary conditions. Further, PINN and PIFBNN randomly sampled 10000 points within the domain to calculate residuals.
The learning curves are presented in Figure \ref{bgrecoloss}.

\begin{figure}[h!] 
	\centering
	{
		\includegraphics[height=5.5cm]{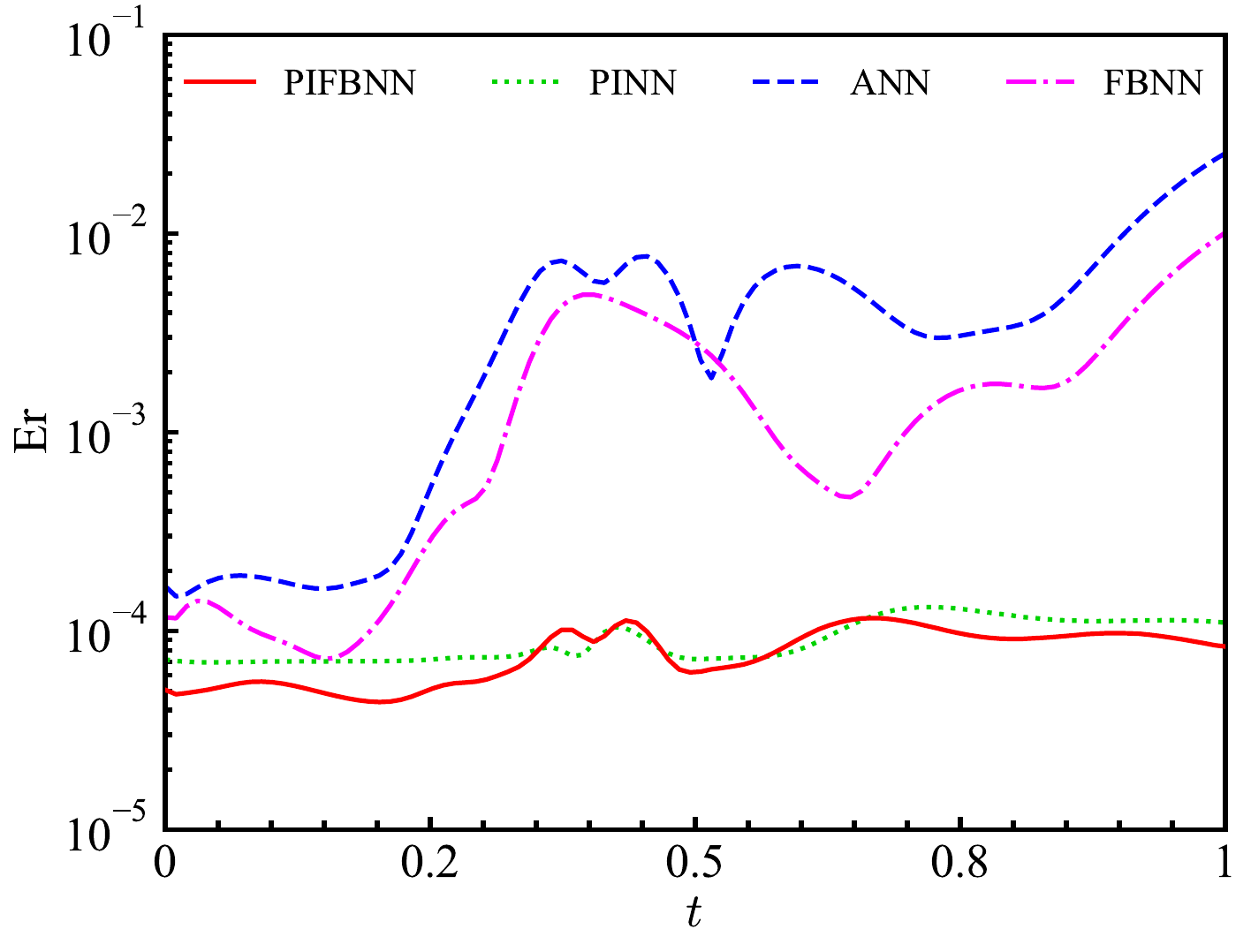}
	}
	
	\caption{Relative errors of different neural networks for Burgers equation reconstruciton.  }
	\label{bgrecoer}
\end{figure}

As observed, owing to the addition of labeled data and encrypted training data, the training and testing losses of all neural networks are not significantly different. This consistency confirms the adequacy of the provided training dataset for model optimization. FBNN has a slower convergence of training loss compared with PINN and PIFBNN because of the absence of physical information residuals. However, as the number of iterations increased, the training loss of FBNN gradually approached that of PINN, rendering it advantageous over the ANN. From the testing loss curves, it is evident that FBNN has a lower testing loss than ANN, PINN and PIFBNN have significantly lower losses than ANN and FBNN, respectively, which is in-line with expectations. The difference between PINN and PIFBNN seems insignificant because all networks have sufficient labeled data and physical residual points for training. PIFBNN appears to have a slower convergence speed, due to the presence of Fourier nodes increases the number of learnable weight coefficients. However, these parameters also gives the FBNN framework a stronger fitting ability. Furthermore, analysis of the learning curves reveals that the augmented learnable parameters do not adversely affect convergence rates. There is no significant difference in convergence speed between PINN and PIFBNN.
The relative errors along time $t$ of the Burgers equation reconstructed by the different neural networks are displayed in Figure \ref{bgrecoer}.

 From the relative error curves of the entire domain, it is evident that FBNN has significant advantages over ANN, PIFBNN over PINN, respectively. Moreover, owing to the addition of labeled data, the reconstruction velocity profile have significantly lower errors than the results predicted in Section \ref{sec:level1/3}. Although we encrypted labeled data in discontinuous solution region, the error curves show that the neural networks without physical residuals have an increased error when t>0.4, that is, when the shock wave appears. However, PINN and PIFBNN always maintain stable errors. The results demonstrate that physics-informed architectures (PINN and PIFBNN) can effectively predict discontinuous solutions when augmented with physical constraints, underscoring the critical role of physical information embedding. Furthermore, the feature-based framework of PIFBNN yields superior prediction accuracy compared to conventional PINN, establishing it as the optimal architecture for this class of problems.
\subsection{Helmholtz equation}
\begin{figure}[h!] 
	\centering
	\subfigure[Training loss]{
		\includegraphics[height=5.5cm]{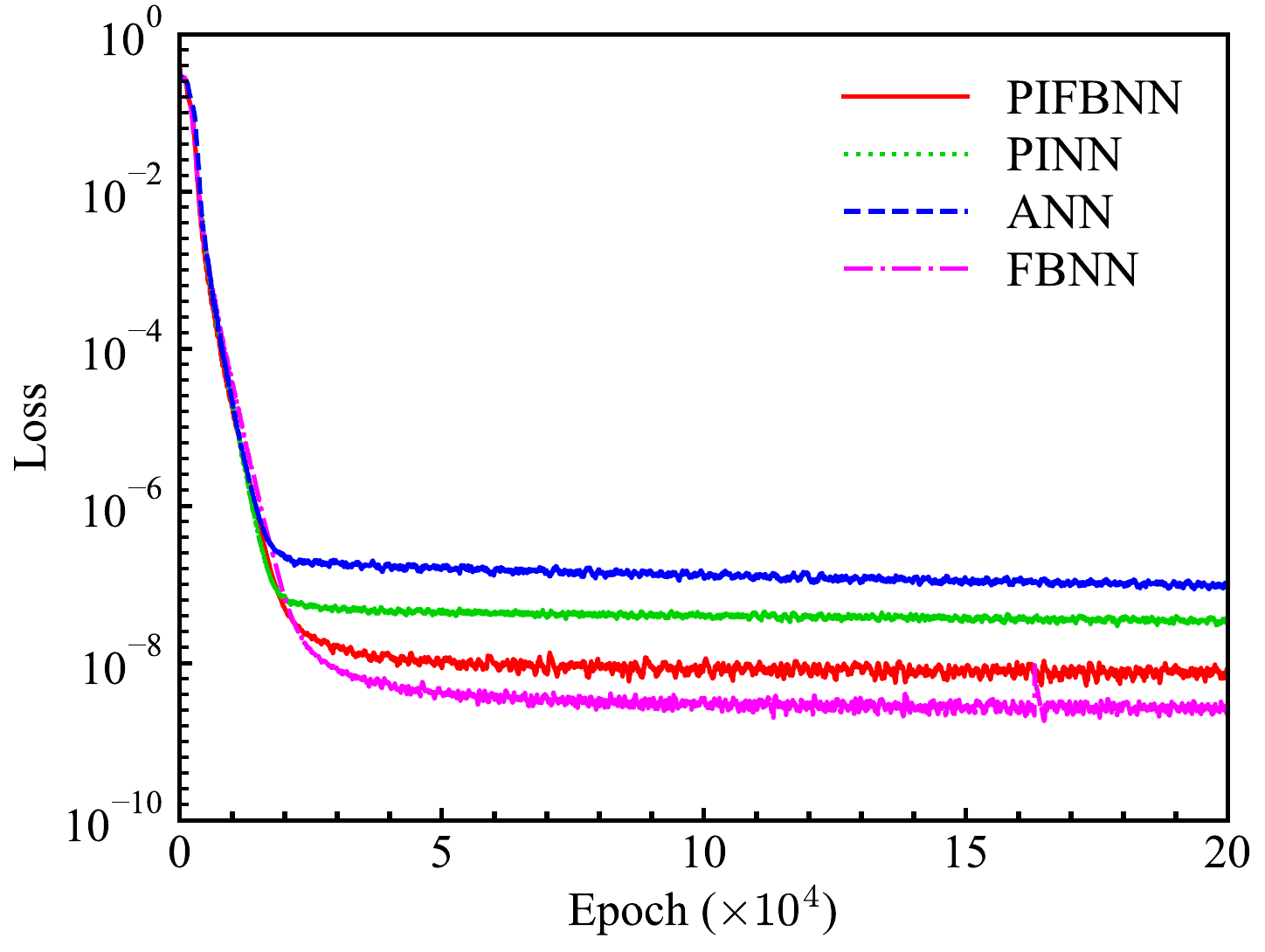}
	}
	\subfigure[Testing loss]{
		\includegraphics[height=5.5cm]{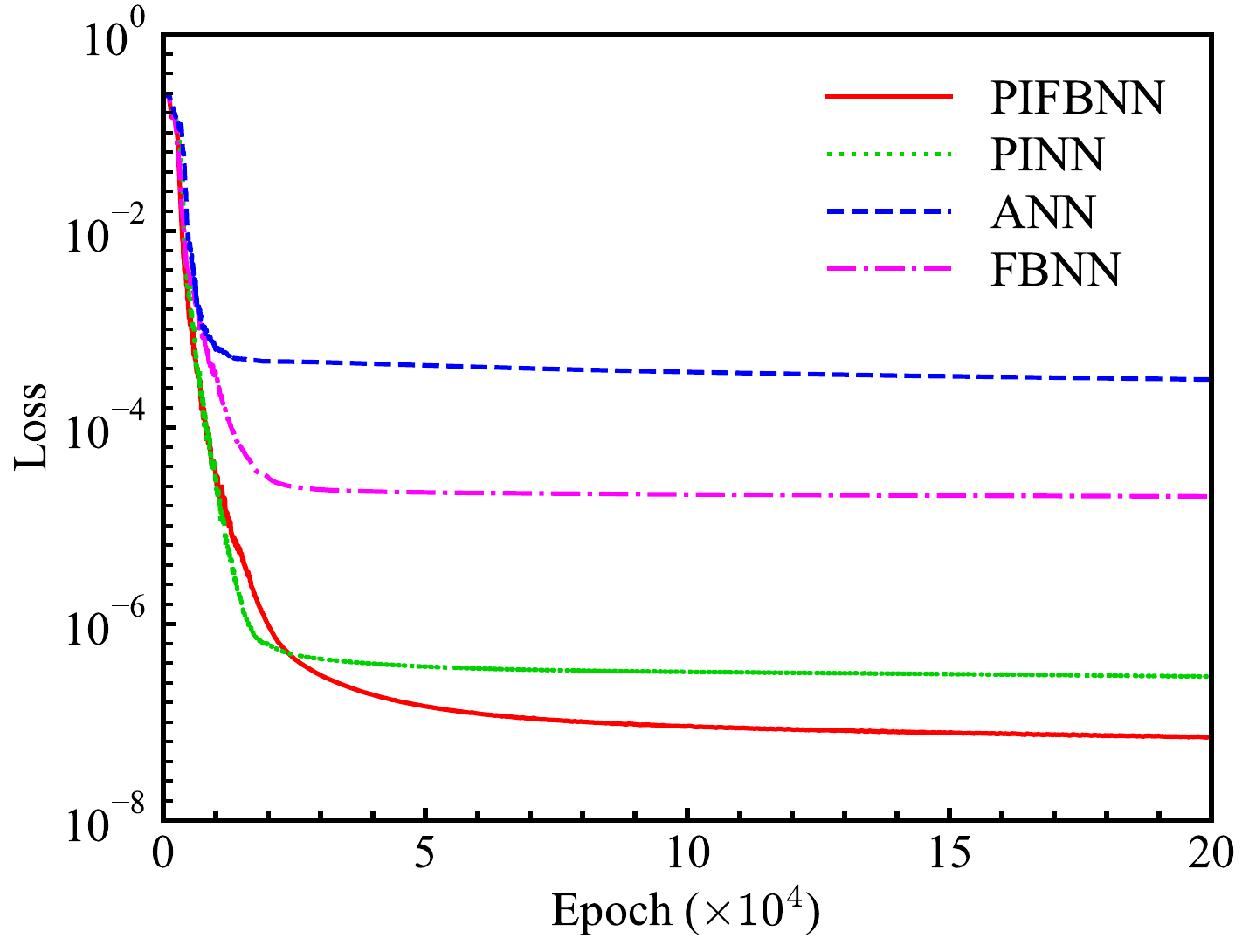}
	}
	\caption{Learning curves for the Helmholtz equation reconstruction using different neural networks. }
	\label{hmrecoloss}
\end{figure}

\begin{figure}[h!] 
	\centering
	{
		\includegraphics[height=5.5cm]{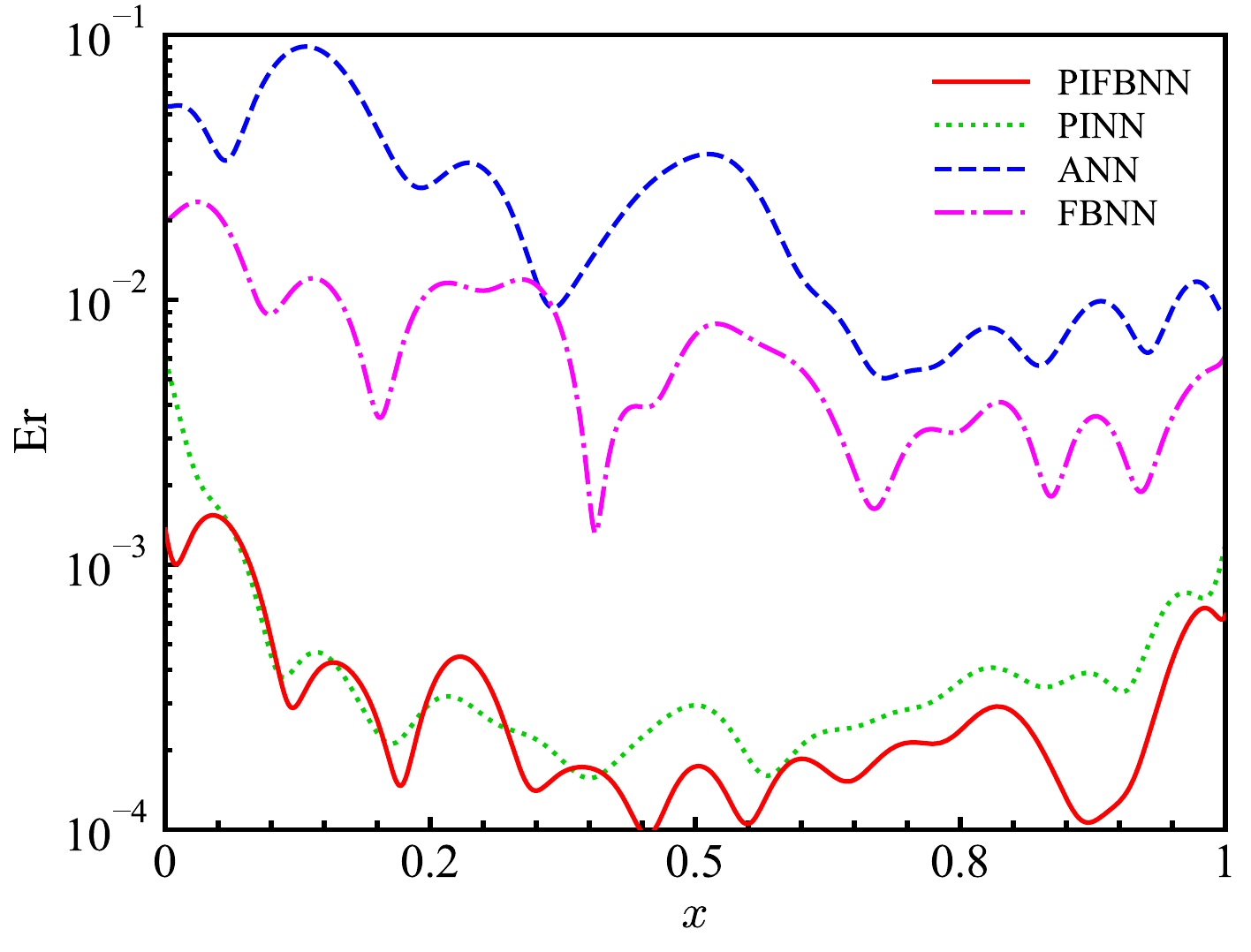}
	}
	
	\caption{Relative errors of different neural networks for Helmholtz equation reconstruction.  }
	\label{hmrecoer}
\end{figure}

 As same as the setup in Section \ref{sec:level1/3}, the neural networks used have 4 layers with 128 neurons per layer, and the ratio of Fourier nodes is 0.6. We randomly sampled 225 points within the domain as labeled data to train neural networks, and 1600 points as test data to evaluate its reconstruction performance. The settings of boundary conditions and the sampling points for calculating the residuals of PINN and PIFBNN are the same as those in Section \ref{sec:level1/3}.
The learning curves of the neural networks used to reconstruct the Helmholtz equation are illustrated in Figure \ref{hmrecoloss}.
Unlike Burgers equation, Helmholtz equation has strong periodicity and nonlinearity, which leads to a significant difference in the training effect of neural networks. FBNN has significantly lower training and testing losses than the ANN, which directly proves that the FBNN architecture has a significantly higher ability to capture nonlinearity and modeling periodicity than the ANN architecture. The PIFBNN, which adds physical information as a training aid, naturally has lower training and testing losses than PINN. The testing loss of PINN with added physical information is much lower than that of FBNN, which highlights the importance of physical information as an auxiliary.
The relative errors along  $x$ of the Helmholtz equation predicted by different neural networks are shown in Figure\ref{hmrecoer}.

\begin{figure}[!htbp]
	\centering
	\subfigure[ANN]{
		\includegraphics[height=4cm]{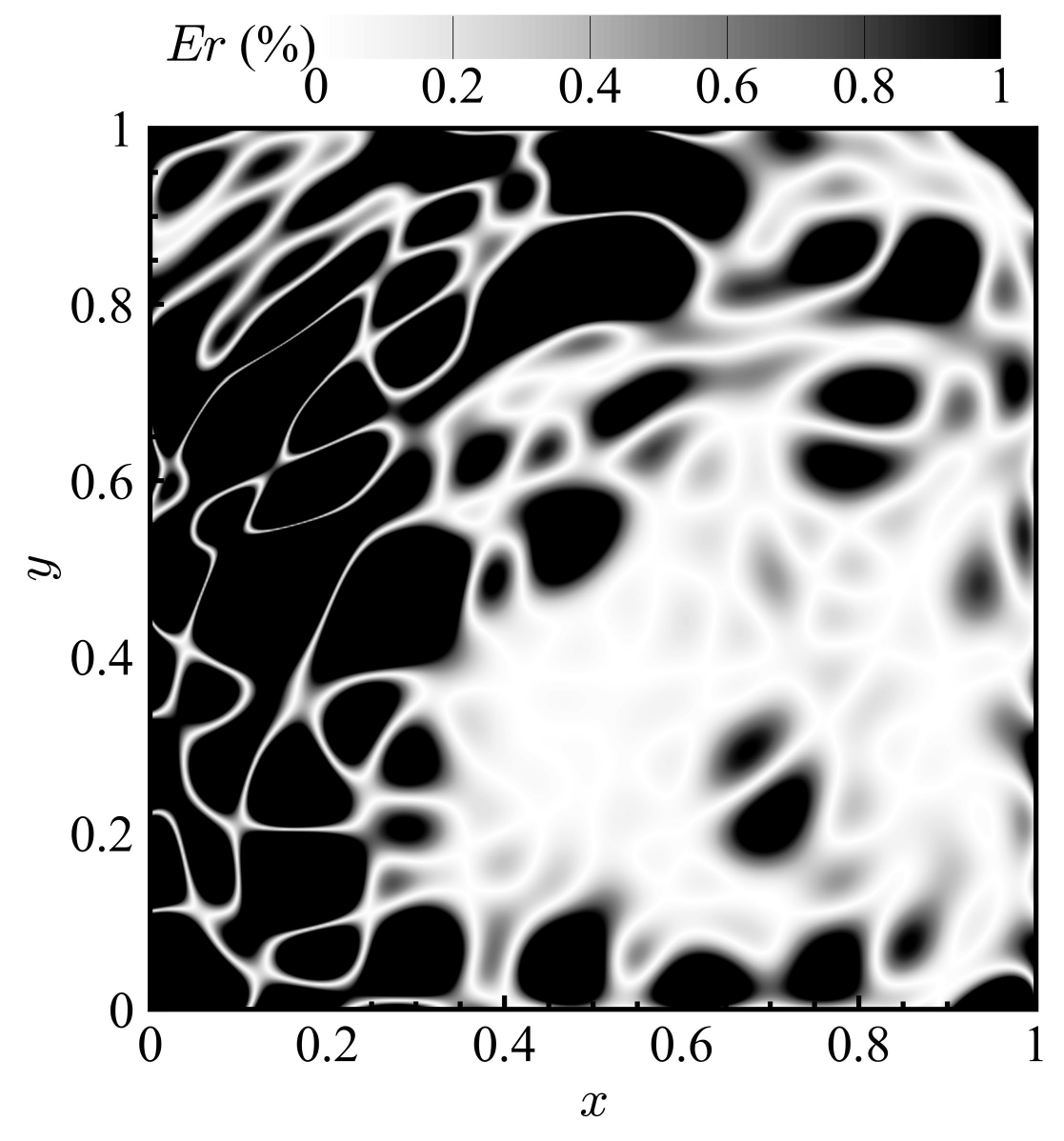}
	}
	\subfigure[FBNN]{
		\includegraphics[height=4cm]{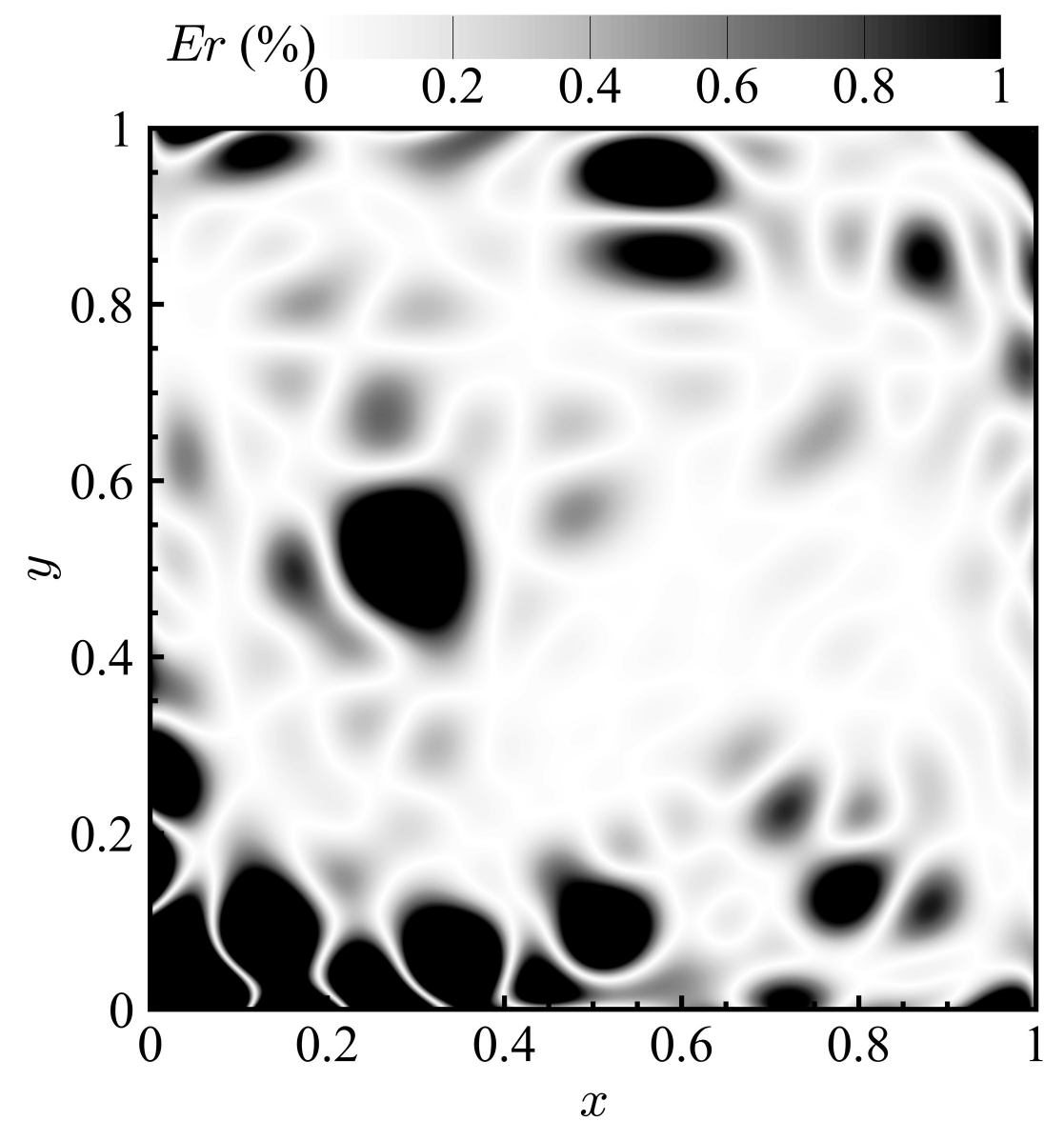}
	}
	\subfigure[PINN]{
		\includegraphics[height=4cm]{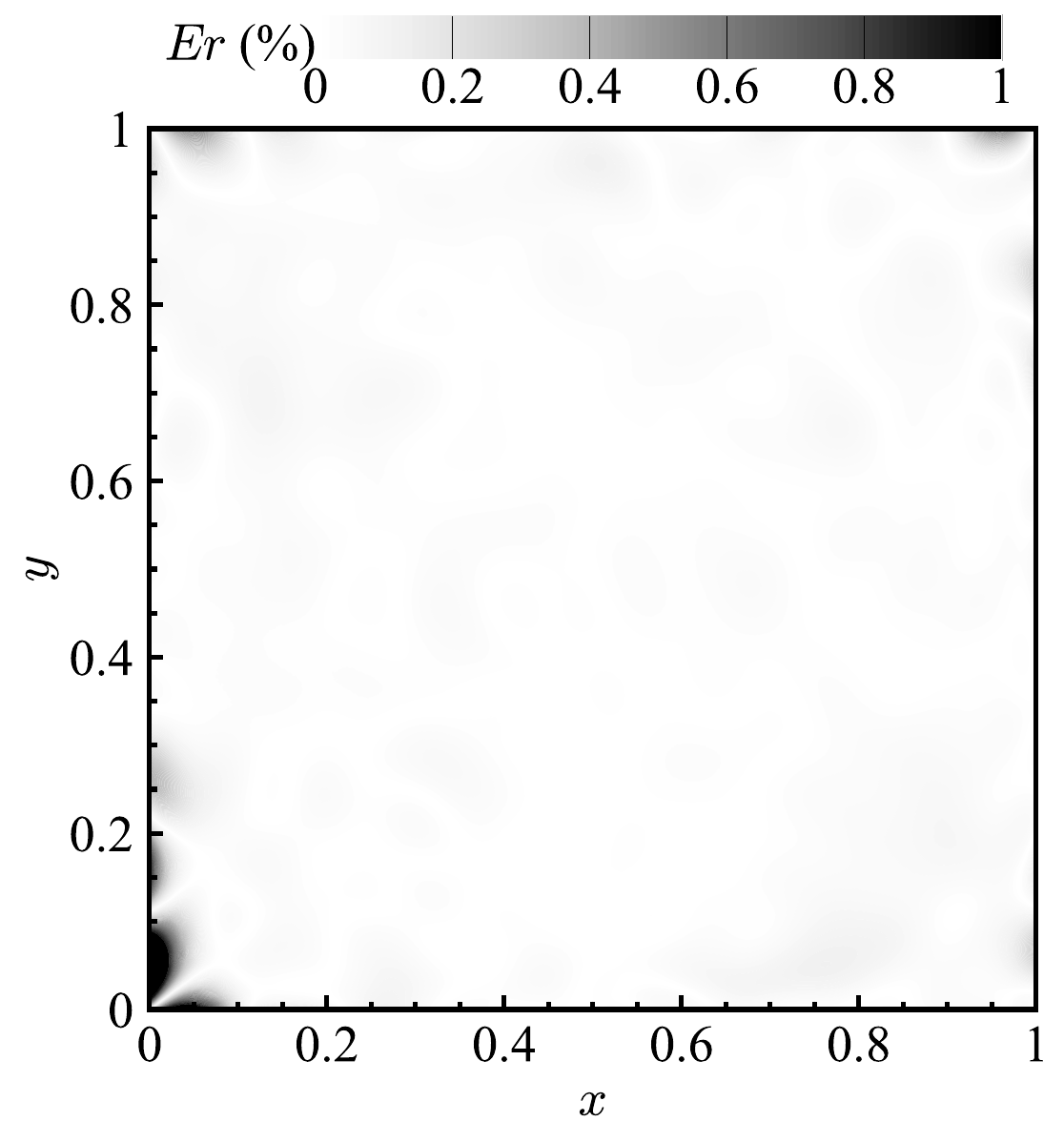}
	}
	\subfigure[PIFBNN]{
		\includegraphics[height=4cm]{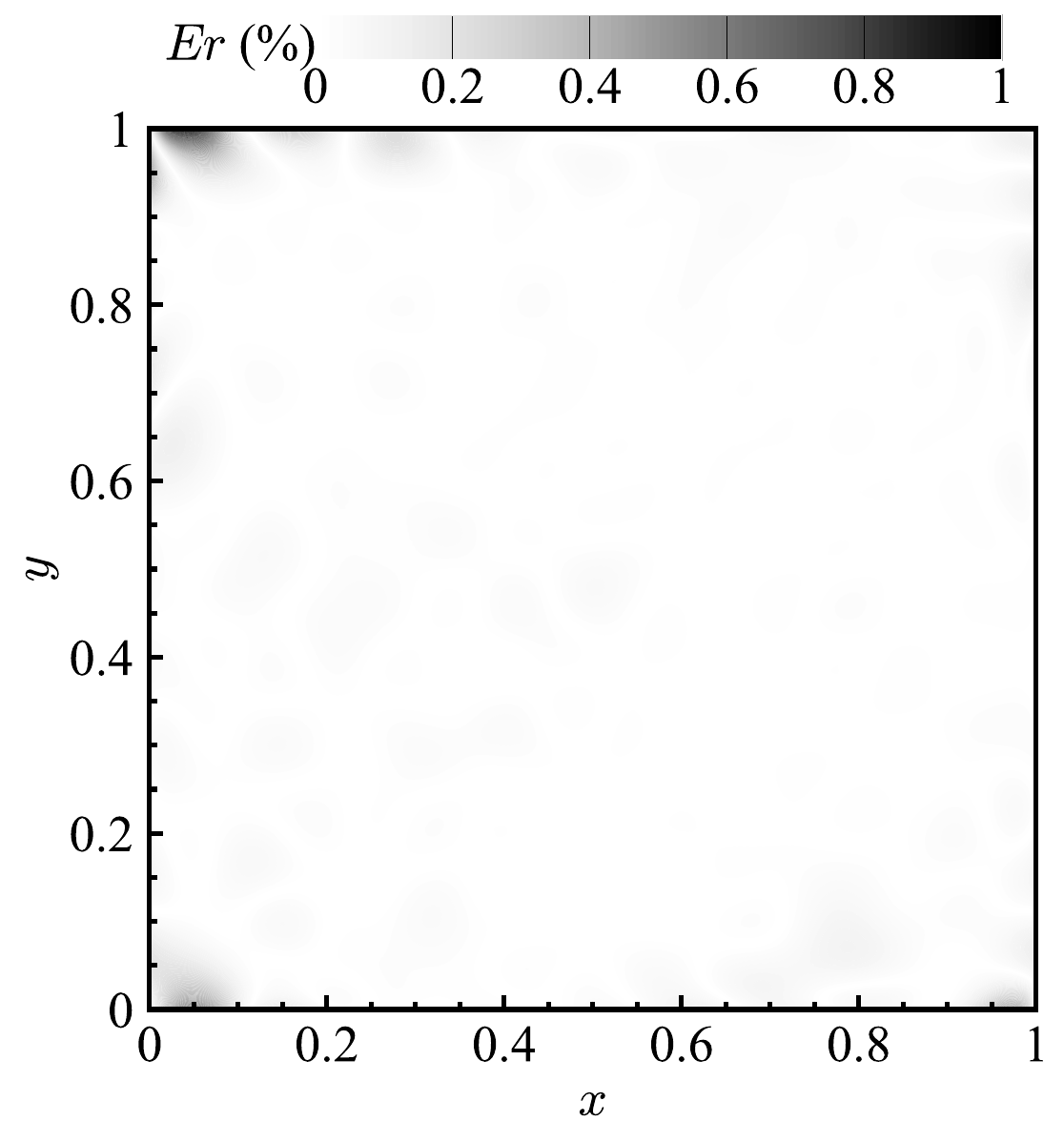}
	}
	\caption{Reconstruction relative error contours of different neural networks for Helmholtz equation.}
	\label{hmrecores}
\end{figure}

Consistent with the above conclusion, the relative error of FBNN is consistently lower than that of the ANN in the entire x-domain, whereas PINN and PIFBNN had sufficient resources for learning because of the dual addition of labeled data and physical information, resulting in generally small errors. PIFBNN exhibits a smaller error than PINN. Although boundary conditions are added, each network exhibits significant errors at the x = 0 and x = 1 boundaries. This may because the addition of labeled data requires more learning resources so neural networks tend to prioritize the learning of labeled data within the domain, while neglecting the learning of boundary conditions. This can be improved by appropriately reducing the labeled data or increasing the weight of boundary condition loss terms.
The contours of relative errors for the Helmholtz equation are demonstrated in Figure \ref{hmrecores}.

By analyzing the distribution and magnitude of the relative errors, the reconstruction error of FBNN is observed to be significantly smaller than that of the ANN and has a sparser distribution. FBNN exhibits obvious errors near the upper and lower boundaries and at a certain period within the domain, while ANN errors are globally distributed. Both PINN and PIFBNN achieve superior overall reconstruction, with a sparse error distribution and minimal error magnitudes. PINN only shows obvious errors in the lower-left corner, whereas the reconstruction results of PIFBNN are near-perfect accuracy.When no special circumstances arise, the FBNN architecture of neural networks can achieve results comparable to that of PINN and be more refined in detail.

\section{\label{sec:level1/6} Conclusion}

In this study, a new PIFBNN was proposed by embedding a rewritten Fourier series and adding physical information to solve PDEs in different fields. We proposed an FBNN framework by adding more learnable weights and biases to the Fourier series and making reasonable parallel operation modifications to make it more in-line with the operational logic of neural networks. The improved Fourier series representation method provides a basis for fitting neural networks compared with the original Fourier series. The resulting network framework yields promising results for computational science problems, such as periodic modeling and strong nonlinear equation prediction, and opens new avenues for fitting more complex nonlinear mathematical problems. Although the FBNN framework relies essentially on the universal approximation theorem, it is a more mathematically logical "black box" compared to the ANN architecture. In the process of PDEs, the presence of Fourier neural nodes can effectively capture the periodicity in PDEs. However, for PDEs with weak periodicity, our network exhibits the same ability as PINN because the fully linked network form is retained on the backbone framework of FBNN. For PDEs with strong periodic solutions, we can enhance their ability to capture periodicity by increasing the proportion of Fourier network nodes, whereas for PDEs with weak periodic solutions, we can appropriately reduce the proportion and allow the fully connected network to learn the equation solution as the main force, while using Fourier nodes for auxiliary learning.

This study considered the boundary and initial conditions to predict the global solution of the equation without the use of labeled data. Compared with the baseline PINN model, the obtained results are better than PINN for all five PDEs considered. PIFBNN offers improved prediction accuracy, enhanced nonlinear processing, periodic modeling capabilities, and effective discontinuous solution prediction. Especially in more challenging cases with stronger nonlinearity and periodicity, the FBNN framework appears to better fit its inherent variation patterns owing to the nonlinear and periodic expression ability of the Fourier series. Therefore, in mathematical terms, FBNN framework is more suitable for periodic modeling and nonlinear fitting than ANN.

As the embedding of Fourier series in neural networks behaves theoretically similarly to activation functions at nodes, FBNN frameworks should have more "benevolent" selection requirements for activation functions. The nonlinear fitting ability of the conventional ANN architecture relies entirely on the nonlinear activation function, and the addition of Fourier nodes is equivalent to embedding trigonometric functions as activations in advance for the neural network. Therefore, two example equations were considered and their performance was compared using two neural networks using different activation functions. The results indicate that compared to the baseline ANN framework, FBNN is not sensitive to the selection of activation functions and performs better than the ANN framework for each type of activation function. When using a PINN based on an ANN architecture, inappropriate selection of activation functions can lead to training failures owing to conflicting optimizations of loss terms. The proposed FBNN contributes to the generalization of neural networks by avoiding the time and resources spent in selecting the most suitable activation function for a particular problem.

To more intuitively highlight the advantages of the FBNN framework and the effect of adding physical information, we combined data-driven methods and compared the sparse data flow-field reconstruction problem using an ANN and FBNN without adding physical information, and PINN and PIFBNN with added physical information. We trained the network using labeled data and boundary or initial conditions, highlighting the fitting advantages of the proposed FBNN framework and the importance of adding physical information. The PIFBNN proposed is a promising alternative for solving PDE prediction or field reconstruction problems.

The introduction of PIFBNN also raises several intriguing questions. How can we determine the optimal proportion of Fourier nodes for different problems or enable the network to automatically learn and adjust this proportion through addition or pruning mechanisms? Because we created more trigonometric basis functions for neural network fitting by improving the Fourier series, can we directly add more basis functions outside of the trigonometric functions to enrich the network's fitting ability? Can we embed a simpler and more effective series into a network to develop more efficient neural networks for different problems? These may be areas of interest for future research. Scientific exploration in this area will undoubtedly benefit from continued collaboration within the research community. We believe that the insights and methods developed in this study will generate lasting impact well beyond its immediate scope.

	\hspace*{\fill} \\
	\textbf{Acknowledgments}
	
	This work was supported by the National Natural Science Foundation of China (NSFC Grant No. 52425111), the Shandong Province Youth Fund Project (Grant No. ZR20240A050), the International Science and Technology Cooperation Project under the Fundamental Research Funds for the Central Universities of Harbin Engineering University (Grant No. 3072024GH2602), the Natural Science Foundation of Hainan Province (Grant No. 425QN376),National Natural Science Foundation of China (Grant No. 52306193), the Basic Product Innovation Research Project (Grant No. KY10100230067), and the Young Scientist Cultivation Fund of Qingdao Innovation and Development Base of Harbin Engineering University.
	
	\bibliographystyle{manuscript}

	\bibliography{myrefpifbnn.bib}
\end{document}